\RequirePackage{fix-cm}%
\documentclass[letterpaper,12pt]{article}
\usepackage[fontsize=11.5pt]{fontsize}%
\usepackage[top=0.90in,bottom=0.90in,left=0.9in,right=0.9in]{geometry}
\usepackage{amssymb,amsmath,amsthm}
\usepackage{bm}
\usepackage{dsfont}
\usepackage[utf8]{inputenc}
\usepackage[T1]{fontenc}
\usepackage{lmodern}
\usepackage{crimson}
\usepackage[kerning]{microtype}
\usepackage[english]{babel}
\usepackage{setspace}
\usepackage{graphicx,booktabs}
\usepackage{threeparttable}
\usepackage{caption}
\usepackage{subcaption,tabularx}
\usepackage{float}
\usepackage[section]{placeins}
\usepackage{enumerate}
\usepackage[dvipsnames]{xcolor}
\usepackage{tikz}
\usetikzlibrary{arrows.meta,positioning}
\usepackage[authoryear]{natbib}
\usepackage{etoc}
\usepackage[colorlinks,urlcolor=NavyBlue,linkcolor=NavyBlue,citecolor=NavyBlue]{hyperref}

\colorlet{revcolor}{black}
\newcommand{\rev}[1]{\textcolor{revcolor}{#1}}

\newcommand{\sym}[1]{\ifmmode^{#1}\else\(^{#1}\)\fi}
\DeclareMathOperator{\diag}{diag}
\DeclareMathOperator{\rowsp}{row}
\DeclareMathOperator*{\argmin}{argmin}

\newtoks\wpAppendixEveryDisplay
\wpAppendixEveryDisplay=\expandafter{\the\everydisplay}
\newskip\wpAppendixTextfloatsep \wpAppendixTextfloatsep=\textfloatsep
\newskip\wpAppendixFloatsep \wpAppendixFloatsep=\floatsep
\newskip\wpAppendixIntextsep \wpAppendixIntextsep=\intextsep
\makeatletter
\let\wpAppendixEndTheorem\@endtheorem
\makeatother
\DeclareMathOperator{\spn}{span}
\newtheorem{athm}{Theorem}[section]
\newtheorem{acor}[athm]{Corollary}
\newtheorem{alem}[athm]{Lemma}
\newtheorem{aprop}[athm]{Proposition}

\newdimen\wsDisplayPadding
\everydisplay\expandafter{\the\everydisplay
  \abovedisplayskip=\wsDisplayPadding
  \belowdisplayskip=\wsDisplayPadding
  \abovedisplayshortskip=\wsDisplayPadding
  \belowdisplayshortskip=\wsDisplayPadding
}
\makeatletter
\let\ws@originalendtheorem\@endtheorem
\newskip\wsStatementGap\wsStatementGap=9pt%
\newskip\wsStatementPostGap\wsStatementPostGap=7.9pt%
\def\thm@space@setup{\thm@preskip=\wsStatementGap\thm@postskip=\wsStatementPostGap}
\def\ws@afterstatementdisplay{%
  \relax
}
\def\@endtheorem{%
  \ifhmode\unskip\par\fi
  \ifvmode
    \ifdim\lastskip=\wsDisplayPadding
      \ws@originalendtheorem
      \aftergroup\ws@afterstatementdisplay
    \else
      \ws@originalendtheorem
    \fi
  \else
    \ws@originalendtheorem
  \fi
}
\makeatother

\newtheorem{theorem}{Theorem}
\newtheorem{proposition}{Proposition}
\newtheorem{lemma}{Lemma}
\newtheorem{assumption}{Assumption}
\newtheorem{definition}{Definition}

\makeatletter
\renewcommand{\@maketitle}{%
  \newpage
  \begin{center}%
    {\fontsize{15.1}{18.1}\selectfont\bfseries \@title \par}%
    \vskip 1.4em%
    {\fontsize{14}{17}\selectfont \begin{tabular}[t]{c}\@author\end{tabular}\par}%
    \vskip 0.5em%
    {\normalsize \@date}%
  \end{center}%
  \par\vskip 0pt}%
\renewcommand\section{\@startsection{section}{1}{\z@}{-2.2ex}%
  {1.0ex}{\normalfont\fontsize{14.1}{17}\selectfont\bfseries}}
\renewcommand\subsection{\@startsection{subsection}{2}{\z@}{-1.9ex}%
  {0.7ex}{\normalfont\fontsize{13.1}{15.7}\selectfont\bfseries}}
\renewenvironment{abstract}{%
  \small\par\centerline{\bfseries\abstractname}\par\vskip 0.35em%
  \list{}{\leftmargin=1.2em\rightmargin=1.2em\listparindent=1.5em\itemindent=0pt%
          \parsep=0pt\topsep=0pt}\item\relax}{\endlist}
\makeatother
\begin{document}

\newcommand{\AppendixURL}{https://www.dropbox.com/scl/fi/4w01zegyg2ue90f2lkpde/appendix_writing_sample_30P.pdf?rlkey=nzn0x7qgq1bve29dra9l02tnz&st=rw9wut6r&dl=0}
\title{Social Networks and the Propagation of Household Shocks to Asset Prices}
\author{Faisal Quaiyyum\thanks{University of Chicago
(faisal.quaiyyum@chicagobooth.edu).}}
\date{November 2026}
\maketitle
\thispagestyle{empty}

\begin{abstract}
\noindent
When can shocks affecting households in a few counties, such as income or wealth losses from storms,
move credit prices? After the shocks take effects, friends and relatives of the affected households in
other counties may send money or save more and this changes their demands for bonds they want to hold.
These household responses can move the aggregate credit premium, depending on intermediary
balance sheet conditions. Tracing the household effects through the whole social network is complex
and I show it is not needed. For given intermediary capital, the household part of the premium depends
only on each county’s response weighted by how strongly its households change their demands. Given
separate estimates of those demand sensitivities, this paper shows that household data recover this summary
exactly when they capture the part of the response that moves bond demand. As responses propagate through
social ties and fade at different rates, a county’s relative importance for credit prices can change over time
and so no fixed connectedness score can generally rank counties at every horizon. This paper shows empirically that social ties to storm-damaged places are associated with lower
spending weeks later outside the damaged areas. Lastly, I use simulations on the actual network to show
that the method recovers this summary.

\end{abstract}
\vspace{0.5pt} %

\onehalfspacing

\section{Introduction}
\label{sec:intro}

When do local household shocks, such as losses of income or wealth, move aggregate credit prices?
The purpose of this paper is to derive conditions under which local shocks move aggregate credit prices. I show that the
financial importance of such shocks depends on two features, where the effects travel over time and
whose investment demand changes. I build an object that captures these conditions: it follows a local shock along social ties to the responses
of households in every county at every horizon and weights each response by how strongly
households there change their desired credit holdings. I call it the
\emph{financially weighted propagation operator} and I call its weights the financial weights (interchangeably demand sensitivities). My main result shows when household measurements are sufficient to
recover this financially relevant response without reconstructing the entire propagation law.
I start from a reduced-form propagation law across counties and connect it to a conditional
credit-market equilibrium.

I develop the mechanism in four steps. First, a local shock (e.g., a storm) lowers income, wealth, and spending
in affected counties. Second, its effects travel along social ties to friends and relatives in other counties
through two broad channels. One is a short lived channel such as news. The other is a long lived channel where
the non-affected households may send money to those who lost homes or jobs, or save more in case of a similar
loss. This channel can persist for months. So, spending falls in counties where shocks did not hit. Third,
households there change their demand for corporate bonds. For example, a family that sends money to their
affected relatives may redeem fund shares and its thinner cushion above usual spending can make risky holdings
less attractive. Moreover, they can also opt to precautionary behavior which can
reinforce or offset this change since a cautious household may move from bonds into safer balances or from
stocks into bonds. Lastly, the
intermediaries and other investors adjust their holdings as household bond demand changes. If the combined
effects of transfers and precautionary behavior reduce household demand on net, these investors would require
higher expected returns to hold the additional bonds, and vice-versa. When intermediary balance sheets are
tight, the return response would become even larger. I link this premium to the part of corporate bond spreads not explained by expected default and
\citet{GilchristZakrajsek2012} called this excess bond premium (EBP). A local shock can hence move national
credit prices months later.

I devise a propagation law to track these effects. I specify this law rather than derive it from
household optimization. The Social Connectedness Index (SCI) of \citet{BaileyCaoKuchlerStroebelWong2018}
supplies the links between counties, and I assume a diffusion law for how responses spread along them and fade.
I then derive equilibrium in a two-date credit market and compare it across household conditions, taking
intermediary capital and payoff moments as given. The operator supplies the demand shift, and the equilibrium determines its price impact.

For credit prices, this approach leads to two questions. The first is what additional financial information is needed to translate household responses into
a change in credit demand since the SCI records social ties, but not the resources or financial claims
exchanged through them.

The second question concerns timing. A regression on a fixed exposure score, such as an SCI-weighted average
of shocks in connected counties, allows for the response to change with the horizon but it also holds
counties’ relative exposure fixed. The question is whether one such score can reproduce locations’ relative
credit-price effects at every horizon. Both questions can be answered by estimating the propagation law
county by county. However, it is not feasible without strong restrictions on it since it links every county
to every other at every horizon. I therefore develop a method that needs only the part of the law that
matters for credit prices.

I model transmission with two channels on one social graph. Each channel assigns a value to every
county, and this value spreads over time to socially connected counties. I call such a channel a
field. Both fields spread along the same social ties and differ in when they start to respond and how quickly they fade.
I write every response as a sum
of regional patterns of the graph, each fading at its own speed. Under the diffusion law, patterns
that vary sharply across strong social ties decay faster than broad ones.

I then aggregate household responses across counties using the financial weights to derive their effect on the credit premium.
Holding intermediary capital and payoff moments fixed, 
the household-demand part of the premium therefore depends on the propagation law only through one financially weighted 
sum of responses at each horizon, given in equation~\eqref{eq:kernel}. Consequently, two propagation laws with the same 
sum imply the same response of this part of the premium, however differently they move individual counties.  

I approach the first question by setting up the exact condition under which household measurements are
sufficient to recover this sum (see Proposition~\ref{prop:recovery}). We need this condition because measurements don't automatically reveal what happens to the bond demand. That being said, the condition is that the measurements must distinguish responses that imply different changes in total bond demand. The data don't need to tell us exactly what happened in each county. It just needs to tell us enough to determine how much more or less households want to hold bonds in total. I take as given how household responses affect measured outcomes (i.e.,
spending) and use independently measured financial weights to connect those responses to bond demand. When
the condition holds, the measurements determine the total demand shift, and the pricing model translates it into a
credit-price response without identifying the propagation law itself.
Consumption measurements, measurements of desired portfolio holdings at given prices, or both together can meet this condition. Additionally,
estimates of household portfolio demand supply the financial weights \rev{and their signs}, which spending alone
cannot reveal, since a household can absorb a loss by cutting spending or by drawing down assets.
\rev{The condition holds regardless of households lowering or raising their bond demand.}

I also show that the response common to all counties needs outside information, since comparisons
across counties within an event reveal only how responses differ. Here the common level matters for prices \rev{whenever the
financial weights do not sum to zero} because
credit demand adds up all counties. Without restrictions on how responses propagate, recovering the
aggregate response \rev{then} requires an independent measurement of the common level or a bound on
it.

I answer the second question with Theorem~\ref{thm:no_scalar} and in general the answer is no.
A location's relative importance for credit prices generically changes with the horizon when the
regional patterns or channels carrying the financially relevant response fade at different speeds.
A location may account for a large share of the credit-price response soon after a shock and a smaller
share later even if its social ties stay the same. One fixed score can't capture this change in relative
importance. The score works across horizons only when the patterns that matter for prices change in the
same proportion over time. The diffusion law, applied to the observed social graph with proxy
financial weights, implies that counties with nearly the same total connectedness can even reverse
their relative importance (Figure~\ref{fig:horizon_importance}).

\looseness=1 I begin the empirical analysis with the spending step of the mechanism, using the SCI and weekly
county spending, with storms as localized disturbances. Spending is lower weeks later in counties
more socially connected to storm-damaged places, although these counties lie outside the storms'
recorded footprints, after accounting for their own weather and for geographic and production links
to the damaged places.

Finally, I use simulations to show that the method can work on the actual network. I use the SCI graph and candidate storm footprints and simulate the household accounts, aggregate
measurements and fund data that recovery requires. When the data follow the specified diffusion
law, the operator predicts the financially weighted responses to held-out storms more accurately
than exposure through direct ties to the hit counties, with near-nominal interval coverage.
Moreover, it keeps this advantage when the scale of the financial weights and the price
impact must be estimated.

\noindent\textbf{Related literature.}\hspace{1em}%
I position this paper alongside demand-system asset pricing and inelastic markets \citep{KoijenYogo2019,KoijenYogo2026,GabaixKoijen2021}, work on the propagation and aggregation of local shocks \citep{Gabaix2011,AcemogluCarvalhoOzdaglarTahbazSalehi2012,CarvalhoNireiSaitoTahbazSalehi2021}, the asset-pricing implications of networks \citep{Herskovic2018} and the identification of aggregate responses from local evidence \citep*{NakamuraSteinsson2014,ChodorowReich2019,Wolf2023,GabaixKoijen2024,chodorowreich2026network}. Since I show when household measurements and independently measured financial sensitivities recover the household-demand component of credit-price responses, this paper also addresses the aggregate implications of social influence noted in \citet{KuchlerStroebel2021}.

Section~\ref{sec:model} develops the propagation law and Section~\ref{sec:pricing} derives the
\rev{conditional EBP response}. Sections~\ref{sec:stage1} and~\ref{sec:stage2} present the spending evidence and its
dynamic counterpart, Section~\ref{sec:finance} financial geography, EBP associations and fund demand,
and Section~\ref{sec:proposal} the recovery program and its simulations. Section~\ref{sec:conclusion}
concludes.

\section{Propagation on a social graph}
\label{sec:model}

\subsection{Economic environment}
\label{subsec:environment}

I start with a two-date endowment economy. Households make their choices at the decision date and receive
payoffs at the payoff date. I use this economy for a conditional comparison at each horizon. The economy has
one consumption good that moves freely across counties. One unit stored at the decision date returns
$R_f>0$ units at the payoff date. There is no other production or physical capital. Households live in counties
$i=1,\ldots,n_C$ and trade a safe asset and risky credit claims in national markets. A borrower sector issues
the credit claims and intermediaries hold part of them.

County $i$ has $n_{if}>0$ households of each type $f\in\mathcal F_i$. A household has current resources
$w_{0,if}$ and future income $Y_{1,if}$. Its consumption commitments $\underline c_{0,if}$ and
$\underline c_{1,if}$ are fixed in advance. It also pays net assistance $\mathsf T_{0,if}$ and
$\mathsf T_{1,if}$ to other households. A negative value means it receives assistance. These payments add up
to zero across households at each date, so assistance redistributes resources. I write net resources as
\begin{equation}\label{eq:env_net}
w^{\mathrm{net}}_{0,if}=w_{0,if}-\underline c_{0,if}-\mathsf T_{0,if},\qquad
Y^{\mathrm{net}}_{1,if}=Y_{1,if}-\underline c_{1,if}-\mathsf T_{1,if}.
\end{equation}
The household chooses consumption above its commitment $c^{\mathrm{sur}}_{0,if}>0$, a safe position
$a^{\mathrm{safe}}_{if}\geq0$ and credit holdings $x_{if}\geq0$ to solve
\begin{equation}\label{eq:env_household}
\begin{split}
\max\quad &u_{if}(c^{\mathrm{sur}}_{0,if})+\varpi_{if}\,E[u_{if}(c^{\mathrm{sur}}_{1,if})],\qquad
u_{if}(c)=\frac{c^{1-\gamma_{if}}}{1-\gamma_{if}},\\
\text{subject to}\quad
&c^{\mathrm{sur}}_{0,if}+a^{\mathrm{safe}}_{if}+Px_{if}=w^{\mathrm{net}}_{0,if},\qquad
c^{\mathrm{sur}}_{1,if}=R_fa^{\mathrm{safe}}_{if}+Vx_{if}+Y^{\mathrm{net}}_{1,if}>0,
\end{split}
\end{equation}
where $\gamma_{if}>0$ is risk aversion and $\varpi_{if}>0$ is the discount factor. I use $u_{if}(c)=\log c$
when $\gamma_{if}=1$. The claim costs $P$ at the decision date and pays $V$ at the payoff date. Households
cannot borrow. Some types do not participate in the credit market, so they solve the same problem with
$x_{if}=0$. Appendix~\ref{app:financialmargin} studies this problem.

A local disturbance lowers resources in the counties it hits. Its effects also reach households elsewhere
through their social ties. I use $z_i$ for the household condition in county $i$ and $L^{\mathrm{dir}}_i$ for
its direct loss. Higher values mean worse conditions. I give all households in a county the same two states
because the data are at the county level. Relative to the path without the disturbance,
\begin{equation}\label{eq:env_condition}
w^{\mathrm{net}}_{0,if}=\bar w^{\mathrm{net}}_{0,if}-m_{0,if}\,z_i-m^{\mathrm{dir}}_{0,if}\,L^{\mathrm{dir}}_i,\qquad
Y^{\mathrm{net}}_{1,if}=\bar Y^{\mathrm{net}}_{1,if}-m_{1,if}\,z_i-m^{\mathrm{dir}}_{1,if}\,L^{\mathrm{dir}}_i,
\end{equation}
where bars denote that path. The four nonnegative coefficients measure dollars of resources per unit of each
state. The condition $z_i$ affects resources after assistance has been paid or received. It does not measure
the transfers themselves. I specify how this condition spreads and fades in Section~\ref{subsec:fields}.
I assume this law rather than derive it from household choices. The insurance and assistance arrangements
behind it remain outside the model.

The borrower sector has a random endowment $Y^B\geq0$ at the payoff date. It has issued a fixed quantity
$\bar B>0$ of claims with face value $F>0$, so each claim pays
\begin{equation}\label{eq:env_claim}
V=\min\{F,\;Y^B/\bar B\}.
\end{equation}
The borrower consumes $P\bar B$ at the decision date and its remaining endowment at the payoff date. Its
random endowment creates the credit risk. I hold its distribution and the quantity issued fixed. So, the local
disturbance changes households' willingness to hold this risk without changing the risk itself. I call the
expected excess payoff $q\equiv E[V]-R_fP$ the premium.

I deem intermediaries as separate investors. They have capital $N>0$ and keep their own profits. An intermediary
uses its capital and safe borrowing to buy $\theta$ claims. Its wealth at the payoff date is
$R_fN+\theta(V-R_fP)$. Its objective is mean wealth less $\gamma_I/(2N)$ times the variance, with
$\gamma_I>0$. A margin of $\chi>0$ per claim limits its position to $0\leq\theta\leq N/\chi$.
I assume $\chi\geq(E[V]-V_{\min})/R_f$, where $V_{\min}$ is the lowest possible payoff. The margin then covers
the largest possible loss on the position and so the intermediary stays solvent and borrowing for it is safe.
Intermediaries hold claims as inventory. They do not originate loans or create deposits.

I take the household conditions, direct losses and intermediary capital as given. The equilibrium price makes
the optimal holdings of households and intermediaries add up to the supply of claims,
\begin{equation}\label{eq:env_clearing}
\sum_{i=1}^{n_C}\sum_{f\in\mathcal F_i}n_{if}\,x_{if}+\theta=\bar B .
\end{equation}
Households' safe positions fund intermediary borrowing and the rest is stored. I require positive storage to
support the safe return $R_f$. Claim payments and assistance cancel when I add up the household, borrower
and intermediary budgets. So, the goods used at each date equal the goods available.

I start from household resources after the disturbance without modelling insurance arranged before it.
The propagation law gives the household conditions at each horizon. I then compare the equilibrium with the
one without the disturbance while keeping intermediary capital and the payoff distribution fixed.
Section~\ref{sec:pricing} derives the premium response. These comparisons don't trace the same households
through time. I don't track how their earlier choices and realized payoffs, or intermediary profits,
change later wealth.

\subsection{Two fields on the social network}
\label{subsec:fields}
\looseness=-1 %
The SCI measures friendship links between counties. I use these links to model how a local shock in one county affects households elsewhere. Social ties also shape investment and the diffusion of news \citep{KuchlerLiPengStroebelZhou2020,HirshleiferPengWang2025}. This paper considers households to either respond briefly to news that friends were hit or change spending and saving for months as they help relatives or prepare for future losses or both. I represent these responses with a fast attention field and a slow demand field. Each field assigns one state to
every county, shared by all households there (Section~\ref{subsec:environment}). I let household types differ in how the state changes their resources and in their preference and their demands add up to a county schedule (see, Section~\ref{sec:pricing}).

The demand field is the household condition of Section~\ref{subsec:environment} with its sign reversed, $\bm d_t=-\bm z_t$. It is not a quantity of bonds demanded. Rather, it is a household state. It changes household resources and through them spending and desired credit holdings. I assume each field's law, so no price enters the dynamics of this section.

I represent the links among $n_C$ counties, indexed by $i=1,\ldots,n_C$, by symmetric weights
$A=(W+W')/2$, where $W\in\mathbb R^{n_C\times n_C}$ contains the retained, nonnegative SCI links,
normalized so that $W\mathbf 1=\mathbf 1$. Here $\mathbf 1\in\mathbb R^{n_C}$ is the vector of ones.
I let $D\equiv\diag(A\mathbf 1)$ record each county's total connectedness, which is positive,
and use the normalized Laplacian $\mathcal L\equiv \mathrm{Id}_{n_C}-D^{-1/2}AD^{-1/2}$ to model diffusion
through these links, with $\mathrm{Id}_{n_C}$ denoting the $n_C\times n_C$ identity matrix.
Since $A$ is symmetric, so is $\mathcal L$. I impose symmetry so that transmission works both ways with the
same weight in normalized coordinates. This excludes one-directional influence and gives the orthogonal
patterns I use below. The household problem doesn't imply this restriction. Market clearing and recovery
also work without it (Appendix~\ref{app:directed} gives the directed case).
\rev{The Laplacian $\mathcal L$ measures the gap between a county's response and the connection-weighted
average among its linked counties, after adjusting for total connectedness. Diffusion gradually closes
these gaps.}

I consider disturbances that are geographically localized at their origin but may have aggregate
consequences through network propagation, as in \citet{CarvalhoNireiSaitoTahbazSalehi2021}. Let
$\bm\varepsilon_{t_0}\in\mathbb R^{n_C}$ denote a fixed vector of local shocks at date $t_0$, with one
entry for each county. For a storm, the entries record the damage it causes in the counties it hits, so
several can be nonzero. I condition on the realized vector without specifying its probability distribution
or assuming independence across counties. I compare two
paths with the same pre-shock history that differ only by this shock at date $t_0$. I report
consumption and field responses relative to the path without the shock. Time is discrete, with running
date $t$, and $h=0,1,\ldots$ measures periods after the shock.

Let $\bm c_t\in\mathbb R^{n_C}$ collect the direct component of county log-consumption responses,
excluding contributions carried by the fields and by changes in the credit premium, with $\Delta\bm c_t\equiv\bm c_t-\bm c_{t-1}$.
Its growth response is $\Delta\bm c_{t_0+h}=\Phi_h\bm\varepsilon_{t_0}$, where
$\Phi_h\in\mathbb R^{n_C\times n_C}$.
I feed the impact response, $\bm\nu_{t_0}\equiv\Phi_0\bm\varepsilon_{t_0}$, into the slow field. This is the direct consumption-growth response at the shock date, holding the credit premium fixed. In the benchmark it is confined to the counties the disturbance hits. I assume that the household condition entering the slow field is proportional to it. This proportionality is a restriction and it doesn't follow from the household problem or from the timing. Observed spending at the shock date can also contain the response to the premium (Section~\ref{subsec:aggregation}). The same
shock generates a signal $\bm s_{t_0}=B\bm\varepsilon_{t_0}$, with $B\in\mathbb R^{n_C\times n_C}$, which
enters the fast field. Both inputs are zero at every other date. I assume the following dynamics, written
for a running date $t$.\footnote{I study propagation from the
impact response. Later direct inputs could make one field look like two, and Appendix~\ref{app:observation}
discusses how to tell them apart.}

\begingroup
\begin{assumption}[Two fields on one graph]
The\label{ass:fields} slow demand field $\bm d_t\in\mathbb R^{n_C}$ and the fast attention field $\bm b_t\in\mathbb R^{n_C}$ evolve as
\begin{align}
\bm d_{t+1}&=A_D\bm d_t+\eta_D\bm\nu_t, & A_D&\equiv\rho_De^{-\kappa_D\mathcal L},\label{eq:slow}\\
\bm b_{t+1}&=A_{\mathrm{att}}\bm b_t+\eta_{\mathrm{att}}\bm s_{t+1}, & A_{\mathrm{att}}&\equiv\rho_{\mathrm{att}}e^{-\kappa_{\mathrm{att}}\mathcal L},\label{eq:fast}
\end{align}
where $\rho_D\in(0,1)$, $\rho_{\mathrm{att}}\in[0,1)$ and $\kappa_D,\kappa_{\mathrm{att}},\eta_D,\eta_{\mathrm{att}}>0$
are scalars and $e^{-\kappa_m\mathcal L}$, $m\in\{D,\mathrm{att}\}$, is a matrix exponential. Each $A_m$ is a
symmetric $n_C\times n_C$ matrix with an eigenvalue
$\rho_me^{-\kappa_m\lambda}$ for each eigenvalue $\lambda\geq0$ of $\mathcal L$. Both fields are updated once
per period. Slow and fast mean that $\rho_De^{-\kappa_D\lambda}>\rho_{\mathrm{att}}e^{-\kappa_{\mathrm{att}}\lambda}$
at the eigenvalues that carry the response, which holds at every eigenvalue when
$\rho_D>\rho_{\mathrm{att}}$ and $\kappa_D\leq\kappa_{\mathrm{att}}$.
\end{assumption}
\endgroup

The attention field responds on impact, with responses $\bm b_{t_0-1}=0$ and
$\bm b_{t_0}=\eta_{\mathrm{att}}B\bm\varepsilon_{t_0}$, while the demand field begins one period later,
with $\bm d_{t_0}=0$ and $\bm d_{t_0+1}=\eta_D\Phi_0\bm\varepsilon_{t_0}$. These are two separate initial
conditions at the shock date, which hold because the two paths coincide before the shock.
Since the laws are linear, the responses to shocks at different dates add up. The
coefficients $\eta_D$ and $\eta_{\mathrm{att}}$ determine how strongly each input enters its field.
\rev{For $m\in\{D,\mathrm{att}\}$, the matrix exponential spreads responses over connected counties
at a rate set by $\kappa_m$, while $\rho_m$ reduces their amplitude each period.}

I choose these forms for three reasons. First, I assume that attention takes the shock's signal in the same
period, while the demand field takes the direct response of the shock date a period later. Second, both
travel through social ties on one graph. Third, an unrestricted transition matrix has $n_C^2$ entries, so I
give each field
one persistence and one spreading parameter, and each spatial pattern of the graph fades at its own rate.

\raggedbottom %
\begin{figure}[h!]
\centering
\includegraphics[width=\textwidth]{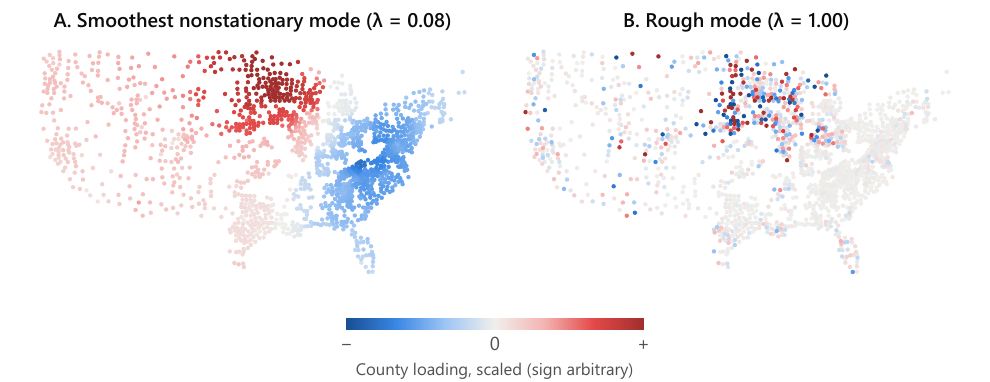}
\caption{Two eigenvectors of the SCI Laplacian on the 1,590 sampled counties, computed from the SCI graph
alone and plotted at county
centroids. Loadings are $D^{-1/2}u_k$, scaled to the 97th percentile of their absolute value. Signs
are arbitrary.}
\label{fig:modes}
\end{figure}

Following the spectral approach to networks of \citet{GaleottiGolubGoyal2020}, I decompose county
responses into spatial patterns called modes, the orthonormal eigenvectors
$U=(u_1,\ldots,u_{n_C})$ of $\mathcal L$, ordered so that their eigenvalues satisfy
$0=\lambda_1\leq\cdots\leq\lambda_{n_C}$.
Each mode gives counties' relative movements, with counties
of the same sign moving together. Its eigenvalue $\lambda_k$, the graph frequency of mode $k$, measures how sharply
the pattern differs across social ties. Smooth modes have low graph frequency and assign similar
degree-adjusted values $D^{-1/2}u_k$ to strongly connected counties. These patterns are computed
from the SCI graph alone, with no calibrated parameter. The graph covers the 1,590 counties in the weekly
spending data of Section~\ref{sec:stage1} that have at least 5,000 residents and lie outside Alaska and
Hawaii. Among each county's 100 largest links by SCI times population, $W$ retains those within this set.
Figure~\ref{fig:modes}
contrasts a smooth mode, which separates the Upper Midwest and
Plains from Appalachia, the Mid-Atlantic and the Carolinas, with a rough mode that alternates in
sign among neighboring counties of the Upper Midwest.

The transition satisfies $A_mu_k=r_{m,k}u_k$, where
$r_{m,k}\equiv\rho_me^{-\kappa_m\lambda_k}$. Without further inputs, each mode therefore keeps
its shape and retains a fraction $r_{m,k}$ of its amplitude each period. Diffusion makes rougher
patterns fade faster when $\rho_m>0$ (Appendix \ref{wp:app:A.4} gives the implied half-lives). The
first mode has $\lambda_1=0$ and is proportional to square-root connectedness,
$u_1\propto D^{1/2}\mathbf 1$. Let's call it the stationary mode. I assume the graph is connected, meaning chains of friends link every two counties. The
stationary mode is then the only mode with $\lambda=0$, and it survives at rate $\rho_m$. The persistence parameter
therefore sets how long the broadest pattern lasts. Calling the demand field slow presumes
$r_{D,k}>r_{\mathrm{att},k}$ for the modes carrying the response, as Assumption~\ref{ass:fields} states.

The regressions in Section~\ref{sec:stage1} measure exposure through direct SCI links. The dynamic
model allows transmission along longer paths and links responses across horizons, so a one-hop
coefficient is not an estimate of $\kappa_D$ (Appendix~\ref{app:directed}). 

I also allow a storm's economic effects to spread through production links. I represent this
channel by a slow field $\bm d_t^P\in\mathbb R^{n_C}$ on a graph built from county industry shares
and the BEA input--output matrix. It follows the slow law
$\bm d_{t+1}^P=A_D^P\bm d_t^P+\eta_D^P\bm\nu_t$, with initial response $\bm d_{t_0}^P=0$ and its
own transition $A_D^P$ and input coefficient $\eta_D^P$ (Appendix~\ref{app:prod}). I include production
exposure as a competing channel because production disruptions can affect household resources
and intermediary capital, and production networks matter for asset prices \citep{Herskovic2018}.

\subsection{The response operator}\label{wp:main:2.3}

To compare channels and counties, I combine the direct consumption effect and the fields into one
response per county, the local valuation $\bm v_t\in\mathbb R^{n_C}$,
\begingroup
\begin{equation}\label{eq:valuation}
\bm v_t\equiv-\gamma\Delta\bm c_t-a_D\bm d_t-a_{\mathrm{att}}\bm b_t-a_D^P\bm d_t^P,\qquad
\gamma,a_D,a_{\mathrm{att}}>0,\ a_D^P\geq0.
\end{equation}
\endgroup
Under power utility with curvature $\gamma$ and no consumption commitments, the direct consumption response contributes
$-\gamma\bm c_t$ to the log-marginal-utility response, and the first term records the change in
this contribution between periods. I choose the signs so that negative direct consumption-growth
responses and adverse field responses raise the index. I interpret a higher index as payoffs
becoming more valuable to households in that county, though not necessarily to the investors who
set prices. The index is a summary of household responses for comparing channels and counties. It is not the pricing equation. Section~\ref{sec:pricing} shows that credit prices load on the slow field, with weights derived from the household problem. I combine county valuations with fixed weights $\pi\in\mathbb R^{n_C}$. Holding these
weights fixed lets me ask whether a source county's relative importance changes across horizons.
The weights can have either sign, so the results below also apply to the financial loadings that
Section~\ref{sec:pricing} derives for credit prices.

\begin{definition}[Response operator and valuation projection]
For\label{def:operator} the one-time shock $\bm\varepsilon_{t_0}$ of Section~\ref{subsec:fields}, the
local response operator is
$\mathcal O_h\equiv\partial\bm v_{t_0+h}/\partial\bm\varepsilon_{t_0}'\in\mathbb R^{n_C\times n_C}$, $h\geq0$.
The valuation projection is the row vector $\Omega_h\equiv\pi'\mathcal O_h\in\mathbb R^{1\times n_C}$.
\end{definition}

Entry $(i,j)$ of $\mathcal O_h$ is the response of county $i$'s valuation $h$ periods after a unit
shock in county $j$, relative to the path without the shock. Because the propagation law is
linear, $\mathcal O_h$ does not depend on the size of the shock. Entry $j$ of $\Omega_h$ adds these
responses across recipient counties with weights $\pi$, so it measures the importance of source
county $j$ at horizon $h$.

\begin{lemma}[Closed form]
Under\label{lem:closed} Assumption~\ref{ass:fields}, $\mathcal O_h=-\gamma\Phi_h+\widetilde{\mathcal O}_h^{SCI}+
\widetilde{\mathcal O}_h^{P}$, where
\begin{equation}\label{eq:closed}
\widetilde{\mathcal O}_h^{SCI}=-a_D\eta_D\mathds 1\{h\geq1\}A_D^{h-1}\Phi_0-a_{\mathrm{att}}\eta_{\mathrm{att}}A_{\mathrm{att}}^hB,
\qquad
\widetilde{\mathcal O}_h^{P}=-a_D^P\eta_D^P\mathds 1\{h\geq1\}(A_D^P)^{h-1}\Phi_0 .
\end{equation}
\end{lemma}

The first term is the direct effect through local consumption. The second carries demand and
attention through social ties and the third carries persistent demand through production links.
In the regressions, I control for measured own weather and include production exposure alongside
social exposure (Section~\ref{sec:stage1}). The rest of this section studies the social part
$\widetilde{\mathcal O}_h^{SCI}$ and its projection
$\widetilde\Omega_h^{SCI}\equiv\pi'\widetilde{\mathcal O}_h^{SCI}$. Section~\ref{sec:pricing}
uses the slow field to derive the household-demand component of the credit-price response
(equation~\eqref{eq:kernel}). All proofs are in Appendix~\ref{app:proofs}.

\subsection{Spectral form and identification}\label{wp:main:2.4}

I now ask whether one fixed exposure score can describe a source county's importance at every
horizon. As modes decay at different rates, the mixture of responses changes even though the
social ties and county weights stay fixed. The same modal form shows when responses across modes and horizons
distinguish a channel's persistence, $\rho_m$, from its diffusion rate, $\kappa_m$.

For a shock shaped like mode $k$, $\bm\varepsilon_{t_0}=u_k$, the response of mode $k$ itself is
$C_k(h)\equiv u_k'\widetilde{\mathcal O}_h^{SCI}u_k$. Lemma~\ref{lem:closed} and
$A_mu_k=r_{m,k}u_k$ give $C_k(h)=C_{D,k}(h)+C_{\mathrm{att},k}(h)$, with
\begin{equation}\label{eq:modal}
C_{D,k}(h)\equiv\mathds 1\{h\geq1\}\,C^{\mathrm{init}}_{D,k}\,r_{D,k}^{\,h-1},\qquad
C_{\mathrm{att},k}(h)\equiv C^{\mathrm{init}}_{\mathrm{att},k}\,r_{\mathrm{att},k}^{\,h},
\end{equation}
where $C_k(h)$ measures how strongly a shock patterned like mode $k$ return as a response in that pattern.  $C^{\mathrm{init}}_{D,k}\equiv-a_D\eta_D\,u_k'\Phi_0u_k$ and $C^{\mathrm{init}}_{\mathrm{att},k}\equiv-a_{\mathrm{att}}\eta_{\mathrm{att}}\,u_k'Bu_k$. The
coefficient $C^{\mathrm{init}}_{m,k}$ sets channel $m$'s initial response, and $r_{m,k}$ sets its decay per
period. The slow channel enters at $h=1$ and the fast channel at $h=0$. 

A shock shaped like one mode can also move other modes, because the direct response $\Phi_0$ and
the signal loading $B$ need not follow the network's patterns. Let $K\in\{1,\ldots,n_C\}$ be the
number of retained modes and $U_K=(u_1,\ldots,u_K)\in\mathbb R^{n_C\times K}$.

\begin{assumption}[Retained-mode diagonality]
On\label{ass:diag} the first $K$ modes, $U_K'\Phi_0U_K$ and $U_K'BU_K$ are diagonal up to small errors, and the
loadings between retained and discarded modes are small (norms in Appendix~\ref{app:diag}).
\end{assumption}

I write $\widetilde{\mathcal O}_h^{SCI}=U\diag(C_k(h))U'+\mathcal E_{h,K}$, where
$\mathcal E_{h,K}$ collects the off-diagonal terms of $\widetilde{\mathcal O}_h^{SCI}$ in the modal
basis, and Assumption~\ref{ass:diag} makes those involving the first $K$ modes small. The
diagonal part of the projection loads on mode $k$ with coefficient $\psi_k(h)\equiv\omega_kC_k(h)$, where
$\omega_k\equiv\pi'u_k$. In the exact diagonal benchmark, $\mathcal E_{h,K}=0$ and
$\widetilde\Omega_h^{SCI}=\sum_k\psi_k(h)u_k'$. In this benchmark, I call channel $m$ active when
$C^{\mathrm{init}}_{m,k}\neq0$ for some mode $k$. With financial weights, a persistent pricing component therefore
needs a mode on which the weights load, $\omega_k\neq0$, and which decays slowly. Within a
channel, smoother modes decay more slowly. A rough mode can still matter when the financial
weights load on it.

\begin{figure}[tp]
\centering
\includegraphics[width=\textwidth]{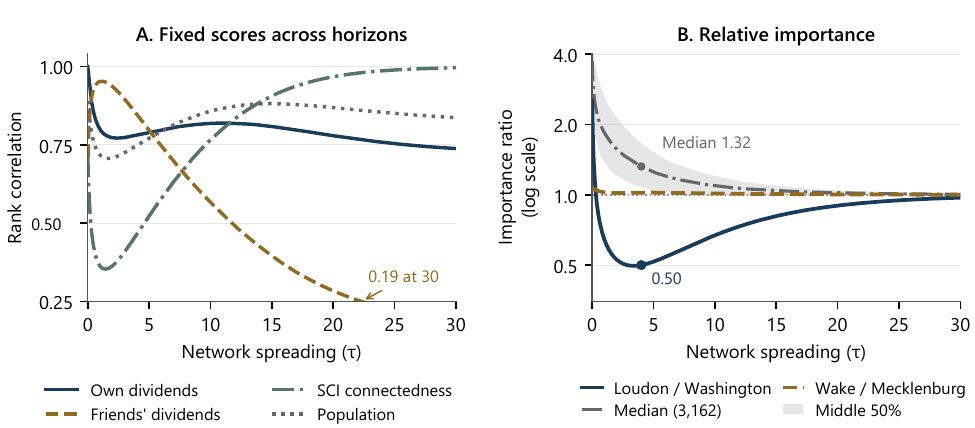}
\caption{Horizon-specific importance on the observed network. Panel A covers 1,590 counties.
The pair distribution in Panel B uses counties with populations of at least 50,000 each.
Its initial upper quartile (6.62) lies above the displayed limit of 4. Results are model-implied
\rev{for equal unit inputs to the slow field $\bm d$ and dividend-proxy weights.} Converting
network spreading $\tau=\kappa_D(h-1)$ into weeks requires the rate $\kappa_D$.
Appendix~\ref{app:split} gives the selection rules and additional comparisons.}
\label{fig:horizon_importance}
\end{figure}
\begingroup\makeatletter
\makeatother

\begin{theorem}[Nonseparability of horizon-specific loadings]
In\label{thm:no_scalar} the exact diagonal benchmark, let $\bar C^{\mathrm{init}}_{m,k}\equiv\omega_kC^{\mathrm{init}}_{m,k}$, so that
$\psi_k(h)=\mathds 1\{h\geq1\}\bar C^{\mathrm{init}}_{D,k}r_{D,k}^{h-1}+\bar C^{\mathrm{init}}_{\mathrm{att},k}r_{\mathrm{att},k}^h$. Suppose two modes
$k\neq\ell$ have nonzero coefficient pairs $(\bar C^{\mathrm{init}}_{D,k},\bar C^{\mathrm{init}}_{\mathrm{att},k})$ and
$(\bar C^{\mathrm{init}}_{D,\ell},\bar C^{\mathrm{init}}_{\mathrm{att},\ell})$ and distinct decay pairs, and both channels are active. Then,
generically, no fixed $\bm s^{\mathrm{fix}}\in\mathbb R^{n_C}$ and scalars $a_h$ satisfy
$\widetilde\Omega_h^{SCI}=a_h(\bm s^{\mathrm{fix}})'$ at every horizon $h\geq0$. Precisely, such a
representation exists if and only if $\psi_k(h)\psi_\ell(j)-\psi_\ell(h)\psi_k(j)=0$ for all
modes $k,\ell$ and horizons $h,j$.
\end{theorem}
\endgroup

A fixed score requires the nonzero weighted modal profiles $\psi_k(h)$ to be proportional across
horizons. One contributing mode satisfies this condition. Different decay rates can break it within
a single channel, while a second channel supplies another route. Appendix~\ref{app:observation} defines the error
of the best fixed-score approximation, and held-out performance measures what additional
dimensions add.

The theorem concerns relative magnitudes, and a reversal of rankings is its most visible form.
Figure~\ref{fig:horizon_importance} shows such reversals on the observed SCI graph. With county
shares of IRS dividend income $w$ as proxy financial weights and \rev{equal unit inputs to $\bm d$
across source counties}, the slow-channel pricing loading of source county $j$ at diffusion time
$\tau=\kappa_D(h-1)$ is $I_j(\tau)=[e^{-\tau\mathcal L}w]_j$, up to a common factor, with $\tau=0$
at $h=1$. Each fixed score in Panel~A tracks only part of the path. Among the 27,169
county pairs whose total SCI connectedness is within 1 percent, 24.5 percent reverse their
initial ordering by $\tau=4$. Panel~B follows a subset of these pairs with an initial loading
ratio of at least two, and their median ratio falls from 3.73 to 1.32 by $\tau=4$. For example,
Loudon, Tennessee, and Washington, New York, reverse their ranking (selection rules in
Appendix~\ref{app:split}).

\begin{theorem}[Bilinear restriction]
Under\label{thm:bilinear} a single active channel with nonzero retained-mode coefficients,
\begin{align}
\log|C_{D,k}(h)|&=\log|C^{\mathrm{init}}_{D,k}|+(h-1)\log\rho_D-\kappa_D(h-1)\lambda_k, && h\geq1,\label{eq:bil_slow}\\
\log|C_{\mathrm{att},k}(h)|&=\log|C^{\mathrm{init}}_{\mathrm{att},k}|+h\log\rho_{\mathrm{att}}-\kappa_{\mathrm{att}} h\lambda_k, && h\geq0,\ 0<\rho_{\mathrm{att}}<1.\label{eq:bil_fast}
\end{align}
Under two active channels, $\log|C_{D,k}(h)+C_{\mathrm{att},k}(h)|$ generically violates this form.
\end{theorem}

With one channel, each mode's log response magnitude follows a straight line in the horizon from
its first response onward, with slope $\log\rho_m-\kappa_m\lambda_k$, so rougher modes decay faster
at a rate set by $\kappa_m$. When both channels contribute to a mode with different rates, they
generally bend its profile (Appendix~\ref{wp:app:A.3}).

\begin{proposition}[What modal decay identifies]
Suppose\label{prop:ident} one channel $m$ is active, $\rho_m>0$, and two modes with distinct
eigenvalues have nonzero responses observed at two distinct horizons each. Call such modes and
their eigenvalues informative. Then horizon variation identifies their decay factors, and
\begin{equation}\label{eq:kappa}
\kappa_m=-\frac{\log r_{m,k}-\log r_{m,\ell}}{\lambda_k-\lambda_\ell},\qquad
\rho_m=r_{m,k}\,e^{\kappa_m\lambda_k}.
\end{equation}
With only one distinct informative eigenvalue, $(\rho_m,\kappa_m)$ cannot be separated. At
$\lambda_k=0$ only $\rho_m$ is identified. Two close eigenvalues can make otherwise identified
parameters weakly determined.
\end{proposition}

Separating two channels is a different problem, which weakens when their rates nearly coincide.
Section~\ref{sec:stage2} applies these results to the storm responses, where the measurement of
spending adds its own restriction.

\subsection{Aggregation to states}\label{wp:main:2.5}

The results so far treat every county as observed, but evidence often comes in coarser units, and the
held-out test of Section~\ref{sec:stage2} uses state averages. County responses of opposite signs
can cancel within a state, even when their financially weighted sum is nonzero. The rough mode in
Panel B of Figure~\ref{fig:modes}, for example, nearly cancels within states. I ask when a few
modes still approximate the state response.

Let $M\in\mathbb R^{n_S\times n_C}$ map county values into the values of $n_S$ states, and let
$R\in\mathbb R^{n_C\times n_S}$ satisfy $MR=\mathrm{Id}_{n_S}$. The matrix $R$ assigns a specified
within-state county profile to each state shock. The state-level network operator is
$\widetilde{\mathcal O}_h^{(S)}\equiv M\widetilde{\mathcal O}_h^{SCI}R$, and its $K$-mode
truncation is $\widetilde{\mathcal O}_{h,K}^{(S)}\equiv MU_K\diag(C_k(h))_{k\leq K}U_K'R$.

\begin{proposition}[Coarse-graining bound]
Under\label{prop:coarse} Assumption~\ref{ass:diag}, with $\|\cdot\|_2$ the spectral norm,
\begin{equation}\label{eq:coarse}
\big\|\widetilde{\mathcal O}_h^{(S)}-\widetilde{\mathcal O}_{h,K}^{(S)}\big\|_2
\leq\|M\|_2\|R\|_2\max_{k>K}|C_k(h)|+\big\|\mathcal E_{h,K}^{(S)}\big\|_2,
\qquad \mathcal E_{h,K}^{(S)}\equiv M\mathcal E_{h,K}R,
\end{equation}
where the maximum is zero when $K=n_C$. In the exact diagonal benchmark, the remainder
$\mathcal E_{h,K}^{(S)}$ vanishes.
\end{proposition}

The first term is the largest response among the discarded modes, scaled by how much aggregation
and allocation can amplify it. The second is the aggregated off-diagonal remainder.
Assumption~\ref{ass:diag} keeps the part of the remainder that involves retained modes small,
while the block among discarded modes stays unrestricted. A few modes therefore suffice when both
terms in the bound are small.

The bound covers shocks whose within-state pattern is the one $R$ assigns. For an actual county
shock, the approximation error also depends on its within-state distribution. The additional
response-error term is $M\widetilde{\mathcal O}_h^{SCI}(\mathrm{Id}_{n_C}-RM)\bm\varepsilon$.
Propagation also reshapes the within-state pattern, so repeating the state-level transition
$MA_DR$ need not reproduce the aggregated county path. After two periods, the gap is
$MA_D^2R-(MA_DR)^2=MA_D(\mathrm{Id}_{n_C}-RM)A_DR$, and Appendix~\ref{app:allocation} gives the full
decomposition. The same logic applies to a graph restricted to sampled counties, since omitted
counties carry transmission paths as well as outcomes.

\section{From propagation to credit prices}
\label{sec:pricing}

This section derives the response of the premium in the economy of Section~\ref{subsec:environment}, taking intermediary capital and
the payoff distribution as given. Households and intermediaries hold a fixed supply of credit claims, and the
premium clears that market. The dynamics come only from the household state,
so the result is a conditional comparison at each horizon and not a joint
solution for saving and future capital. The pricing calculation concerns the net change in desired
holdings of credit claims after insurance and other household adjustments. If this change is nonzero, the
premium adjusts because households and intermediaries have finite willingness to absorb the shift. I do not
derive the insurance arrangements that determine how much of the initial loss households bear.
The section identifies the demand weights and
projection of the slow operator
that determine the credit-premium response. This benchmark isolates the slow household channel.
Recall that $\bm z_t=-\bm d_t$, so that positive values represent adverse household conditions.
This field enters the valuation $\bm v_t$ of Section~\ref{sec:model} as $a_D\bm z_t$. Section~\ref{sec:model}
combined county responses with generic weights $\pi$, and this section derives the weights that the credit
market applies to $\bm z_t$.
With the proportionality restriction $B_D\equiv-\eta_D\Phi_0$ of Section~\ref{subsec:fields},
\begin{equation}\label{eq:state}
\bm z_{t+1}=A_D\bm z_t+B_D\bm\varepsilon_t .
\end{equation}

\vspace{-\dimexpr4pt+2.1ex\relax}%
\subsection{Household demand}\label{wp:main:3.1}

Recall that $P$ is the price of one credit claim, $V$ its payoff and $q\equiv E[V]-R_fP$ the premium.
In the benchmark, future net income is known at the decision date and every household holds a positive safe
position. Problem~\eqref{eq:env_household} then depends on resources only through surplus wealth
$s_{if}\equiv w^{\mathrm{net}}_{0,if}+Y^{\mathrm{net}}_{1,if}/R_f$, and power utility makes choices proportional
to it,
\begin{equation}\label{eq:policies}
c^{\mathrm{sur}}_{0,if}=\varphi^{c}_{if}(q)\,s_{if},\qquad x_{if}=\varphi^{x}_{if}(q)\,s_{if}.
\end{equation}
The functions $\varphi^{c}_{if}$ and $\varphi^{x}_{if}$ depend on preferences, $R_f$ and the payoff distribution
and not on wealth (Appendix~\ref{app:financialmargin}). Holding the direct loss fixed, \eqref{eq:env_condition} gives
$s_{if}=\bar s_{if}-m_{if}z_i$ with $m_{if}\equiv m_{0,if}+m_{1,if}/R_f$. County holdings are
$x_i\equiv\sum_fn_{if}x_{if}$, summed over participating types. Around the reference equilibrium on the path
without the disturbance, with premium $\bar q$, and to first order in the premium change and the household
condition,
\begin{equation}\label{eq:schedule}
x_i=\bar x_i+\zeta_iq-\zeta_i\beta_iz_i,\qquad
\zeta_i\equiv\sum_fn_{if}\,(\varphi^{x}_{if})'(\bar q)\,\bar s_{if},\qquad
\zeta_i\beta_i\equiv\sum_fn_{if}\,\varphi^{x}_{if}(\bar q)\,m_{if}.
\end{equation}
Here $\zeta_i$ is the premium sensitivity, $\beta_i$ is the sensitivity to the local field and $\bar x_i$ is the
intercept of this local schedule. The schedule is exact in the household condition at the reference premium. The
resource-loss benchmark derived here has $\beta_i\geq0$, and with
precaution the sensitivity can have either sign. As in demand-system asset pricing
\citep{KoijenYogo2019,KoijenYogo2026},
holdings respond to the premium and to a demand shifter, which here is the local field.
The same household problem gives the spending response. At a fixed premium, consumption falls by
$\varphi^{c}_{if}(\bar q)\,m_{if}$ per unit of the field, so spending and credit demand are two readings of one
shortfall in resources, with different coefficients. Define
\begin{equation}\label{eq:weights}
\zeta\equiv\textstyle\sum_i\zeta_i,\qquad \bar X\equiv\sum_i\bar x_i,\qquad
\rev{l_i\equiv\zeta_i\beta_i,\qquad \delta\equiv\textstyle\sum_il_i,\qquad S\equiv\bm l'\bm z,}
\end{equation}
so that aggregate household holdings are $X=\bar X+\zeta q-S$. Here $x_i,X,S$ are quantities of
claims, and $S$ is the desired reduction in holdings at a fixed premium.
\rev{A negative $S$ is a desired increase.}
Multiplying by a reference price $\bar P$ converts the shift into dollars. The financial loading
is precisely \rev{$l_i=-\partial x_i/\partial z_i|_q$}.
It weights marginal responses, not simply who is wealthy or owns bonds. These loadings are the financial
weights of the Introduction. They play the role that sales weights play in \citet{Gabaix2011} and
\citet{AcemogluCarvalhoOzdaglarTahbazSalehi2012}, with credit demand in place of output.
\rev{When $\delta\neq0$, the normalized weights are $\pi_i^H\equiv l_i/\delta$, so that $S=\delta\,(\pi^H)'\bm z$.
In the benchmark, $\delta>0$ and $\pi^H$ is a distribution over counties. However, loadings of opposite sign can sum to zero, so that $\delta=0$. Even then, $S$ responds to a geographically uneven field. So, the vector $\bm l$ is the general object and every display written with $\pi^H$ assumes $\delta\neq0$.}

Where do the weights then come from? For a small premium, the demand in \eqref{eq:policies} is close to a mean--variance demand. Consider one participating type and let $s_i>0$ be the surplus it plans to consume at the payoff date. With risk-aversion
parameter $\gamma_i>0$, absolute risk aversion $\gamma_i/s_i$ and claim-payoff variance
$\sigma_i^2$, claim demand is approximately $x_i=s_iq/(\gamma_i\sigma_i^2)$. Suppose the adverse field lowers
surplus from its baseline $\bar s_i$, so $s_i=\bar s_i-m_iz_i$ with $m_i\geq0$.
Linearizing at a premium $\bar q>0$ gives
\begin{equation}\label{eq:surplus}
\zeta_i=\frac{\bar s_i}{\gamma_i\sigma_i^2},\qquad
\beta_i=\frac{\bar q\,m_i}{\bar s_i},\qquad
\zeta_i\beta_i=\frac{\bar q\,m_i}{\gamma_i\sigma_i^2}.
\end{equation}
A smaller surplus cushion raises local risk aversion and lowers desired risky holdings. The consumption commitments play the role of the habit in
\citet{CampbellCochrane1999}. Equation~\eqref{eq:surplus} is accurate only for a small premium, and the exact weights are those in \eqref{eq:schedule}. Within this local
approximation $l_i=\bar q m_i/(\gamma_i\sigma_i^2)$, so households with the same holdings can have
different marginal responses. Nonparticipation is a separate restriction $l_i=0$. Taking
$\bar s_i\to0$ in the interior formula does not produce it, because $\beta_i$ then diverges.
Participation matters \citep{MankiwZeldes1991,VissingJorgensen2002}, but participation data alone
do not identify $m_i$, risk tolerance or payoff exposure. For any proxy
$w$, $(\pi^H)'\bm z=w'\bm z+(\pi^H-w)'\bm z$, so a small spending-weighted average need not mean
a small demand shift.

Selling claims is not the only way to absorb a loss. I measure each named flow as a positive quantity. Then a household's financing identity is
\begin{equation}\label{eq:cashflow}
\text{Sales}=L_Y-L_C-I_{tr}-B_{new}-D_{cash}.
\end{equation}
Here lost income $L_Y$ is met by cutting consumption ($L_C$), receiving insurance or transfers
($I_{tr}$), borrowing ($B_{new}$), drawing down cash ($D_{cash}$) or selling assets. A spending cut
therefore substitutes for liquidity-driven asset sales, holding the other financing sources
fixed. These dollar flows differ from the claim-quantity shift $S$. The identity is general accounting for a household that owns claims. In the economy of Section~\ref{subsec:environment} households buy newly issued claims, so the counterpart of a sale is a smaller purchase than on the path without the disturbance.

\rev{Loss of resources and precaution against background risk are two forces behind $S$.
In the resource-loss benchmark, credit demand rises with available resources, so lost income, current
transfers or future assistance commitments lower desired holdings.} A current transfer lowers spending by
less than the amount sent. Precaution is an extension of the economy of Section~\ref{subsec:environment}. It requires future income to be risky and the household condition to change its distribution, and Appendix~\ref{app:financialmargin} states the conditions under which it lowers or raises credit demand. \rev{Precaution can reinforce this reduction or offset it. I derive both directions for a household with a safe asset, credit and equity in Appendix~\ref{app:financialmargin}. Under its three-state conditions, a
small independent increase in income risk tilts the household away from the claim whose losses are
concentrated in the worst state. So, it lowers holdings of credit that pays in full outside that states whereas credit holdings can rise with other payoff profiles or at a portfolio boundary.}\footnote{\rev{These results are local at the decision date, with prices fixed and certain net income
at the reference point. Appendix~\ref{app:financialmargin} states the remaining conditions and retains the two-asset result with
existing independent income risk, whose proof builds on \citet{Kimball1993}.}}
Lower current commitments can instead reduce spending while raising surplus and credit demand.
Transfers redistribute resources, so net selling pressure requires donors' reductions in desired
holdings to exceed recipients' increases. \rev{Thus the net loading $l_i$ can have either sign, and its sign
cannot be inferred from a spending decline.}

\subsection{Intermediary absorption}\label{wp:main:3.2}

Intermediaries hold the credit claims that households do not hold. I model their inventory
choice for given capital, not loan origination or deposit creation. Their preferences in Section~\ref{subsec:environment} imply a risk cost $\Gamma\equiv\gamma_I\operatorname{Var}(V)>0$.
They choose holdings $\theta$ to maximize $\theta q-\Gamma\theta^2/(2N)$ subject to
$0\leq\theta\leq N/\chi$. Market clearing \eqref{eq:env_clearing} requires $X+\theta=\bar B$. The net supply that
households do not hold
at $q=0$ is $Q\equiv\bar B-\bar X+S>0$, so clearing requires $\zeta q+\theta=Q$.
The domain also requires $E[V]>q$ for a positive claim price.

\begin{proposition}[Conditional equilibrium]
For\label{prop:eq} the linear schedule \eqref{eq:schedule}, the clearing premium is
\begin{equation}\label{eq:G}
q=G(S,N)=
\begin{cases}
\Gamma Q/(N+\zeta\Gamma), & Q\leq(N+\zeta\Gamma)/\chi\quad\text{(slack)},\\
(Q-N/\chi)/\zeta, & Q\geq(N+\zeta\Gamma)/\chi\quad\text{(binding)}.
\end{cases}
\end{equation}
Strictly inside the slack region, $G_S=\Gamma/(N+\zeta\Gamma)>0$, $G_N=-\Gamma Q/(N+\zeta\Gamma)^2<0$
and $G_{SN}=-\Gamma/(N+\zeta\Gamma)^2<0$. Strictly inside the binding region, $G_S=1/\zeta$, $G_N=-1/(\zeta\chi)$
and $G_{SN}=0$. In the household economy of Section~\ref{subsec:environment}, at a reference equilibrium strictly inside either region, the derivatives of the equilibrium premium with respect to the demand shift $S$ and to capital $N$ equal $G_S$ and $G_N$ evaluated there (Appendix~\ref{app:financialmargin}).
\end{proposition}

$G_S$ is the price impact of the household demand shift, measured as the rise in the premium.
In the slack region it is larger when capital is small, because intermediaries with less capital
to bear inventory risk require more compensation for each claim they absorb, as in
\citet{HeKrishnamurthy2013}. Empirical evidence also links intermediary financial conditions to asset prices
\citep{AdrianEtulaMuir2014,HeKellyManela2017,HaddadMuir2021}.
When the margin cap binds, intermediaries absorb no additional
claims, so the premium must rise by $1/\zeta$ per unit of the shift for households collectively
to retain them. The premium is continuous at the boundary, but $G_S$ jumps there by the factor
$(N+\zeta\Gamma)/(\zeta\Gamma)=1+N/(\zeta\Gamma)$, which is larger when intermediaries normally
absorb a large share of the shift. There is no two-sided derivative at the kink.

Clearing also shows why the fixed-price shift and executed selling differ, since
\begingroup\predisplaypenalty=0%
\begin{equation}\label{eq:absorption}
-\frac{dX}{dS}=\frac{d\theta}{dS}=1-\zeta G_S
=\begin{cases}N/(N+\zeta\Gamma),&\text{slack},\\0,&\text{binding}.
\end{cases}
\end{equation}
\endgroup
In the slack region intermediaries absorb part of the shift, while the higher premium induces
households collectively to retain the rest. At the cap, aggregate household holdings do not fall
further, although their composition can change. Fund redemptions and changes in security holdings
therefore differ from the fixed-price shift. A muted price response alone does not establish
intermediary absorption. Similarly, a small observed flow alone does not identify a small
fixed-price shift. With $\bar B$, $\bar X$ and $\zeta$ fixed, clearing gives
$dS=\zeta\,dq+d\theta$, so recovering the shift requires premium changes, aggregate executed
flows and an estimate of $\zeta$.

\subsection{The conditional EBP response}\label{wp:main:3.3}

The EBP is the part of corporate bond spreads not explained by expected default
\citep{GilchristZakrajsek2012,FedEBP2016v18}. I link it to the model's premium with an assumed local
measurement relation, $E^{BP}_t=\alpha_E+b_Eq_t+u^E_t$ with $b_E>0$. Holding
capital, payoff moments, demand parameters and $u^E$ fixed within one smooth regime,
\begingroup\predisplaypenalty=0%
\begin{equation}\label{eq:kernel}
\frac{\partial E^{BP}_{t+h}}{\partial\bm\varepsilon_t'}=b_E\,G_S\,\rev{\bm l'}A_D^{h-1}B_D,
\qquad h\geq1 .
\end{equation}
\endgroup
From right to left, $B_D$ maps the storm into local conditions and
$A_D^{h-1}$ spreads that response over the social graph. \rev{The loadings $\bm l$ then form a signed
weighted sum of the propagated field, which is the fixed-price shift in claim demand. When $\delta\neq0$
this sum equals $\delta\,(\pi^H)'A_D^{h-1}B_D$. With nonnegative loadings, $\pi^H$ forms a
loading-weighted average that the total sensitivity $\delta$ scales into the demand shift.} In turn, $G_S$ converts that shift into a change
in the premium, which $b_E$ maps into the EBP. The scalar multiplier \rev{$b_EG_S$} scales the row
vector \rev{$\bm l'A_D^{h-1}B_D$}, which carries the network geometry.
Up to the scalar $a_D$, this is the slow social part of $\Omega_h$ in Section~\ref{sec:model} with
$\pi=\bm l$, so the criterion of Theorem~\ref{thm:no_scalar} applies.

The total response also contains the direct loss in the counties the disturbance hits, which shifts their credit demand through the same household problem, and capital dynamics,
$b_EG_N\,\partial N_{t+h}/\partial\bm\varepsilon_t'$, which can offset or reinforce the \rev{conditional response}
(Appendix~\ref{app:capital}). If attention shifts desired holdings as well, the \rev{response} gains a term in $A_{\mathrm{att}}^hB$,
and the nonseparability argument of Theorem~\ref{thm:no_scalar} extends to channel-specific
financial loadings under its modal and nondegeneracy conditions (Appendix~\ref{app:target}).
The EBP is an estimated spread residual and not a pure observation of this model's $q$. So, an empirical test must distinguish \rev{this response} from the measurement error $u^E$, expected default, liquidity and other risk-premium channels. Bank-capital evidence can inform absorption capacity or an alternative transmission route. However, it cannot substitute for the missing household loading.

\subsection{Levels and changes}
\label{subsec:levels}

The \rev{conditional EBP response} is a level effect. The monthly regressions instead relate changes in the EBP to lagged
states and to their interaction with intermediary tightness, and the two can have opposite signs.
A positive effect on the EBP level can shrink over time, so its change is negative.
Let $\mathcal T\equiv\bar N-N$ be the capital shortfall relative to a reference level $\bar N$,
and let $\mathcal C_N\equiv\bar N+\zeta\Gamma$, $Q_0\equiv\bar B-\bar X$ and
$q_0\equiv\Gamma Q_0/\mathcal C_N$, with the reference point strictly inside the slack region.

\begin{proposition}[Levels versus changes]
For\label{prop:levels} the linear-demand system of Proposition~\ref{prop:eq}, strictly inside the slack region,
\begin{equation}\label{eq:expansion}
q=q_0+A_SS+A_{\mathcal T}\mathcal T+A_{S\mathcal T}S\mathcal T+A_{\mathcal T\mathcal T}\mathcal T^2
+O(\|(S,\mathcal T)\|^3),
\end{equation}
with $A_S=\Gamma/\mathcal C_N$, $A_{\mathcal T}=\Gamma Q_0/\mathcal C_N^2$, $A_{S\mathcal T}=\Gamma/\mathcal C_N^2$
and $A_{\mathcal T\mathcal T}=\Gamma Q_0/\mathcal C_N^3$.
Let $S_t=\bm l'\bm z_t$, $\bm z_{t+1}=A_D\bm z_t+\bm\eta_{t+1}$ and
$\mathcal T_{t+1}=\rho_K\mathcal T_t+e_{t+1}$, where $\rho_K$ is the persistence of the shortfall
and the innovations are conditionally mean-zero.
For this quadratic approximation
\begin{equation}\label{eq:changes}
\begin{aligned}
E_t\Delta q_{t+1}={}&A_S\bm l'(A_D-\mathrm{Id}_{n_C})\bm z_t
+A_{\mathcal T}(\rho_K-1)\mathcal T_t\\
&+A_{S\mathcal T}\{\mathcal T_t\bm l'(\rho_KA_D-\mathrm{Id}_{n_C})\bm z_t
+\bm l'\operatorname{Cov}_t(\bm\eta_{t+1},e_{t+1})\}\\
&+A_{\mathcal T\mathcal T}\{(\rho_K^2-1)\mathcal T_t^2+
\operatorname{Var}_t(e_{t+1})\}.
\end{aligned}
\end{equation}
\end{proposition}

Proposition~\ref{prop:levels} concerns the linear-demand system. Its second-order coefficients are not second-order approximations to the household economy, because schedule \eqref{eq:schedule} is only a first-order approximation and omits the curvature of household demand.
Within this primitive linear-demand model, scarcer capital amplifies the premium response to a
given demand shift in the slack regime. With fixed capital, nonnegative financial loadings and
$B_D\bm\varepsilon\geq0$, the level response is also nonnegative, although its difference from the
no-diffusion response need not be. For changes, a scalar mean-reversion interpretation additionally
requires $\bm l'A_D=r\bm l'$, so that the aggregate shift follows
$S_{t+1}=rS_t+\bm l'\bm\eta_{t+1}$, even when county responses combine several decay rates.
This restriction is not generic, however, and the vector expression supplies no sign theorem for
the monthly partial regression coefficient. Capital dynamics, regime crossings, payoff news and
observation timing require separate restrictions
(Appendices \ref{app:levels} and \ref{app:capital}).

\subsection{What aggregation can reveal}
\label{subsec:aggregation}

Three results describe what household measurements and credit prices can reveal about
\eqref{eq:kernel}.

\emph{Aggregation can remove sensitivity to diffusion.} Because $A_D$ is symmetric, $A_D=U\diag(r_{D,k})U'$, and
the stationary mode decays at $r_{D,1}=\rho_D$. Hence
\begin{equation}\label{eq:split}
(\pi^H)'A_D^{h-1}B_D=
\underbrace{\big((\pi^H)'u_1\big)\,\rho_D^{\,h-1}\,u_1'B_D}_{\text{filtered incidence}}
+\underbrace{\sum_{k\geq2}\big((\pi^H)'u_k\big)\,r_{D,k}^{\,h-1}\,u_k'B_D}_{\text{nonstationary exposure}} .
\end{equation}
The first term weights the initial field input $B_D\bm\varepsilon$ by square-root degree and
decays at $\rho_D$, independently of the diffusion rate $\kappa_D$.
In contrast, the second term weights nonstationary modes, but is not itself a diffusion effect. It
generally remains nonzero when $\kappa_D=0$. If $\pi^H$ were proportional to $u_1$, the
second term would vanish. More generally, any weight vector with $w'A_D=r_{D,k}w'$ gives
$w'A_D^{h-1}B_D=r_{D,k}^{h-1}w'B_D$ for every $h$, a filtered input with no trace of the other modes.
These are the aggregation weights that restore the single-exposure representation discussed
in Theorem~\ref{thm:no_scalar}. \rev{The same split holds with $\bm l$ in place of $\pi^H$, which covers
loadings of either sign.}

To isolate what diffusion adds, hold $B_D$, \rev{$\bm l$} and $c_E=b_EG_S$ fixed. Equation~\eqref{eq:kernel} can then be expressed as
$K_h=c_E\bm l'A_D^{h-1}B_D$. Let $K_h^0=c_E\rho_D^{h-1}\bm l'B_D$ be \rev{its heterogeneous no-diffusion counterpart}. The diffusion increment is
\begingroup\predisplaypenalty=0%
\begin{equation}\label{eq:diffincrement}
\Delta K_h^{diff}=K_h-K_h^0
=c_E\rho_D^{h-1}\bm l'[e^{-\kappa_D(h-1)\mathcal L}-\mathrm{Id}_{n_C}]B_D .
\end{equation}
\endgroup
It is zero when $\kappa_D=0$ and when $h=1$. Setting $\kappa_D=0$ removes dynamic diffusion, not
any nonlocal mechanism already built into $B_D$. On a two-county graph with eigenvalues $0,2$, let
$\pi^H=(p,1-p)'$, $B_D=\mathrm{Id}_2$ and normalize $c_E\delta=1$. The responses to unit inputs are
\begin{equation}\label{eq:two_region}
K_{1,2}(h)=\frac{\rho_D^{h-1}}2[1\mathbin{\pm}(2p-1)e^{-2\kappa_D(h-1)}].
\end{equation}
At $p=0.8$, the initial 0.8 versus 0.2 contrast is financial heterogeneity before diffusion,
which diffusion subsequently narrows. A test must distinguish that change from the heterogeneous
no-diffusion response, not merely control stationary incidence.

\emph{Within-event contrasts leave the common response unidentified.} Let the county
responses to event $e$ be $\bm y_e=\mathbf 1a_e+\mathcal G_h\bm\varepsilon_e+\bm u_e$, with response
kernel $\mathcal G_h$ and an event-level common component $a_e$. \mbox{A within-event comparison} uses a
contrast $T$ with $T\mathbf 1=0$, and for every row vector $\bm\alpha'$,
\begingroup\predisplaypenalty=0%
\begin{equation}\label{eq:common}
T(\mathcal G_h+\mathbf 1\bm\alpha')=T\mathcal G_h,\qquad
w'(\mathcal G_h+\mathbf 1\bm\alpha')=w'\mathcal G_h+\bm\alpha'\quad\text{if }w'\mathbf 1=1 .
\end{equation}
\endgroup
The common response thus drops out of every within-event contrast but shifts each normalized
aggregate one-for-one. This is the missing-intercept problem of regional and network designs
\citep{NakamuraSteinsson2014,Wolf2023}, which
the network GIV of \citet{chodorowreich2026network} is designed to solve with granular shocks. The common response may include aid or insurance, as well as rebuilding.
Unless the propagation law pins down the missing level, point recovery \rev{of an aggregate that the
common response moves} requires an independent
anchor identifying the aggregate household response. A bound yields an \enlargethispage{5pt}%
interval instead.
If the loading of household measurements on county states, $C_Y$ in the next result, equals $\mathrm{Id}_{n_C}$
and $\ker T=\operatorname{span}\{\mathbf1\}$, \rev{$\bm l$ lies in
$\rowsp(T)$ only if $\bm l'\mathbf1=0$. For unrestricted county responses, within-event spending contrasts
therefore cannot recover the demand shift whenever the loadings do not sum to zero}, however precisely
they are estimated. \rev{Nonnegative nonzero loadings always fall in this case.} Appendix~\ref{app:counterfactual} gives sharp bounds conditional on bounds on the common response.
I present how to develop a relative-asset prediction that cancels it later (see, Section~\ref{sec:proposal}).

\begin{samepage}
\emph{Household measurements recover the pricing functional only under a spanning condition.} Let
household measurements satisfy $\bm y_t=-C_Y\bm z_t+J_Y\bm\varepsilon_t+\bm\upsilon_t$, with signed measurement
loading $C_Y$ and direct input map $J_Y$, and let fixed-price demand pressure be $S_t=\bm l'\bm z_t$\rev{, with loadings $\bm l$ of either sign}.
\end{samepage}

\begin{proposition}[Observability of the pricing functional]
For\label{prop:observability} a given $\bm l$, known $C_Y$ and unrestricted $\bm z$, the functional $\bm l'\bm z$ is identified from noiseless
$C_Y\bm z$ if and only if $\bm l\in\rowsp(C_Y)$.
\end{proposition}

The condition applies to consumption or fixed-price portfolio-demand measurements, separately or stacked together. In the household economy, spending also responds to the premium. This adds one component to spending in every county. When its county loading is common, it is part of the common response in \eqref{eq:common}. Otherwise the measurement loading $C_Y$ must include it for the household condition, and the direct input map $J_Y$ for the direct loss (Appendix~\ref{app:financialmargin}).
For spending, the financing identity \eqref{eq:cashflow} explains why precision alone cannot reveal
the financial response, since spending cuts and asset sales are alternative ways to absorb a loss.
Given independently measured $\bm l$, the measurements must retain the combination that changes
claim demand. Within-event contrasts $TC_Y$ can remove that combination (Appendix~\ref{app:observability}).
Later, I examine the variation retained by specified proxy weights on the sample graph (see,
Section~\ref{sec:finance}) and I extend this recovery condition to the relevant shock-response space in
(see, Section~\ref{sec:proposal}).

\section{Local spending and social exposure}
\label{sec:stage1}

\subsection{Data and design}\label{wp:main:4.1}

The design compares counties outside a storm's recorded footprint that differ in their social
exposure to the affected counties. I use geography and production links \citep{AcemogluCarvalhoOzdaglarTahbazSalehi2012,AcemogluAkcigitKerr2016,CarvalhoNireiSaitoTahbazSalehi2021} as competing exposure measures against the county SCI
\citep{BaileyCaoKuchlerStroebelWong2018}. The storm events are FEMA disaster declarations. Moreover, I link the declarations to NOAA property-damage reports
using FEMA incident dates and designated counties. The common sample contains 48 declarations, 37 onset weeks and 1,574 recipients.
The local weights $W^{loc}$ retain each recipient's 100 largest out-of-county
SCI-times-source-population links, normalized by row. The later spectral analysis uses a separate 1,590-county graph.
Spending coverage is fixed across the local comparisons. Inference also groups
declarations by physical episode and calendar block.

\begingroup
\raggedbottom
Spending is assigned by cardholder residence in the Opportunity Insights tracker
\citep{OIData2026}, and its mapping to consumption depends on the type of purchase
(Appendix~\ref{app:target}). I select Sunday seven-day observations through June 5, 2022 and
published weekly observations thereafter, rather than average overlapping daily
lookbacks. Let $L_{j,t}=\log(1+\mathrm{index}_{j,t})$. I keep only county--event pairs with actual, unclipped levels from week $-13$ through $12$. They aren't interpolated. Weeks $-13$ through $-9$ supply the predated controls. The main post-event outcome is
$Y_{jeh}=100(L_{j,t_e+h}-L_{j,t_e-1})$, where week $t_e$ contains the incident's start.
I use week $-9$ as the base for leads and a common-baseline path. I report the distribution of the exposure indices and the four-week spending change in Table~\ref{tab:summary_stats}.

\begingroup\let\wsOrigTable\table\renewcommand{\table}[1][]{\wsOrigTable[t]}\begin{table}[H]
\setstretch{1}
\setlength{\parskip}{0pt}
\centering
\begin{threeparttable}
\caption{Summary statistics in the county--declaration sample}
\label{tab:summary_stats}
\footnotesize
\setlength{\tabcolsep}{3pt}
\begin{tabular*}{\textwidth}{@{\extracolsep{\fill}}lrrrrrr}
\toprule
Variable & $N$ & Mean & SD & Minimum & Median & Maximum \\
\midrule
Social exposure & 74,779 & 0.059 & 0.523 & 0.000 & 0.000 & 34.782 \\
Gravity exposure & 74,779 & 0.071 & 0.642 & 0.000 & 0.005 & 67.920 \\
Upstream exposure & 74,779 & 1.597 & 0.573 & 0.000 & 1.566 & 8.956 \\
Downstream exposure & 74,779 & 1.615 & 0.548 & 0.000 & 1.583 & 8.184 \\
Four-week spending change & 74,779 & 2.026 & 12.149 & $-167.747$ & 1.506 & 239.770 \\
\bottomrule
\end{tabular*}
\begin{tablenotes}[flushleft]
\footnotesize
\item Notes: Exposures are in index points, and spending is in log percentage points from week $-1$ to week 4. All rows use the same recipient--declaration observations across 48 declarations. Dispersion is unconditional.
\end{tablenotes}
\end{threeparttable}
\end{table}\endgroup

Let's take $\mathcal D_{re}(4)$ as the recorded damage in eligible source county $r$. It is only reasonable that the damage must be dated within the declared incident and assigned no later than the end of week 4. My primary measure uses county-coded reports and their start dates. Let
\begingroup\predisplaypenalty=0%
\begin{equation}\label{eq:localexposure}
s_{re}^{(4)}=\frac{\mathcal D_{re}(4)}{\sum_q\mathcal D_{qe}(4)},
\qquad E_{je}^{SCI}=100\sum_r W^{loc}_{jr}s_{re}^{(4)}.
\end{equation}
\endgroup
The index weights each link by the source county's share of damage. It does not count the share of friends affected. So, normalizing the source vector tells us where the disturbance is but not how large the event is in dollars. I set $s_{re}^{(4)}=0$ when the recorded total is zero. Moreover, coverage flags separate absent records from assessed zero loss. I compare raw-dollar and income-scaled inputs and retain unknown loss coverage in Appendix~\ref{app:localrepair}. It also dates continuing destruction as additional source injections.

For each horizon, holding the week-4 damage shares fixed, I estimate
\begingroup
\begin{equation}\label{eq:lp}
Y_{jeh}=\alpha_{j,h}+\mu_{e,h}
+\beta_h^{SCI}E_{je}^{SCI}+\beta_h^GE_{je}^G
+\beta_h^{up}E_{je}^{up}+\beta_h^{down}E_{je}^{down}
+\bm Z_{je}'\bm\phi_h+u_{jeh}.
\end{equation}
\endgroup
Gravity and payroll-weighted input--output
exposures use the same source input.
$\bm Z$ contains distance bins, a same-hit-state indicator, spending controls dated no later
than week $-9$ and measured recipient weather in weeks $-9$ to $12$ (Appendix~\ref{app:localrepair}).
I use County FEs to absorb persistent recipient differences and event FEs to absorb common calendar shocks and the storm-wide response. I try to do the identification from differences among recipients within an event, conditional on competing exposures. However, to interpret $\beta^{SCI}$ causally, we require social ties not to proxy for direct weather, evacuation, shared news or other recipient-specific shocks. Under that exclusion, the within-event comparison identifies a spatial response contrast. This is the part of propagation that differs across recipients. Recovering its common component is the aggregation exercise I perform in Section~\ref{sec:proposal}. The spatial gradient is a household-response measurement (not a per-dollar source loading or a financial weight).

\par
\endgroup

\subsection{Spending responses and source construction}\label{wp:main:4.2}
\makeatletter\@nobreakfalse\makeatother%

\noindent The four-week gradient is stable across the loss-record constructions in Table~\ref{tab:baseline}.
With incident-aligned county reports in Columns~(3)--(4), a one-percentage-point increase in the
damage-share-weighted link index is associated with 0.353 percent lower spending. Similarly, the 
same one-percentage-point increase is associated with 0.325 percent lower spending when NOAA zone 
reports that map uniquely to eligible source counties are included. Using report end dates instead 
of start dates gives a corresponding spending decline of 0.332 percent. Both retain the negative 
gradient under declaration and episode inference.

All 48 leave-one-declaration-out estimates are negative, ranging from $-0.404$ to
$-0.309$ percent per index point. Restricting onsets to December 2020 onward gives
$-0.319$ percent, while omitting own-weather controls
on the same full sample gives $-0.312$ percent. Thus the four-week result
survives both a change in the observation period and recipient-weather adjustment.
The immediate pre-event comparisons show little difference 
in spending changes by social exposure. I report the corresponding source-sensitivity results, the full event-time profile and the joint lead tests in Appendix~\ref{app:localrepair}.

\begingroup\let\wsOrigTable\table\renewcommand{\table}[1][]{\wsOrigTable[t]}\begin{table}[H]
\setstretch{1}
\setlength{\parskip}{0pt}
\centering
\begin{threeparttable}
\caption{Four-week spending responses across loss-record constructions}
\label{tab:baseline}
\footnotesize\setlength{\tabcolsep}{2pt}
\begin{tabular*}{\textwidth}{@{\extracolsep{\fill}}lcccccccc}
\toprule
 & \multicolumn{2}{c}{Onset reports} & \multicolumn{2}{c}{Incident reports} & \multicolumn{2}{c}{Mapped zones} & \multicolumn{2}{c}{End-dated} \\
\cmidrule(lr){2-3}\cmidrule(lr){4-5}\cmidrule(lr){6-7}\cmidrule(lr){8-9}
 & (1) & (2) & (3) & (4) & (5) & (6) & (7) & (8) \\
\midrule
Social exposure & $-0.336$\sym{**} & $-0.336$\sym{**} & $-0.353$\sym{**} & $-0.353$\sym{**} & $-0.325$\sym{**} & $-0.325$\sym{**} & $-0.332$\sym{**} & $-0.332$\sym{**} \\
 & (0.112) & (0.111) & (0.115) & (0.113) & (0.118) & (0.119) & (0.118) & (0.119) \\
\midrule
SE type & Decl. & Episode & Decl. & Episode & Decl. & Episode & Decl. & Episode \\
Mapped-zone reports & No & No & No & No & Yes & Yes & Yes & Yes \\
Incident-aligned records & No & No & Yes & Yes & Yes & Yes & Yes & Yes \\
Report date & Start & Start & Start & Start & Start & Start & End & End \\
County and declaration FE & Yes & Yes & Yes & Yes & Yes & Yes & Yes & Yes \\
Controls & Yes & Yes & Yes & Yes & Yes & Yes & Yes & Yes \\
Observations & 74,779 & 74,779 & 74,779 & 74,779 & 74,779 & 74,779 & 74,779 & 74,779 \\
\bottomrule\end{tabular*}
\begin{tablenotes}[flushleft]\footnotesize
\item Notes: Coefficients are interpreted as the change in spending from week $-1$ to week 4, in log percentage points, associated with a 1-percentage-point (one-index-point) increase in the damage-share-weighted link index. Controls include own-weather, gravity and production-network measures. Decl. and Episode denote declaration- and physical-episode-clustered wild-bootstrap standard errors using 9,999 Webb draws \citep{Webb2023}. Columns (1)--(2) use onset-week reports, while the others use incident reports dated through week 4, with matched competing exposures. Calendar-block checks preserve the negative signs (Appendix~\ref{app:localrepair}). \mbox{\sym{*} $p<0.10$}, \mbox{\sym{**} $p<0.05$}, \mbox{\sym{***} $p<0.01$}.
\end{tablenotes}\end{threeparttable}\end{table}\endgroup

Table~\ref{tab:calendar_timing} dates repeated damage inputs explicitly. I also exclude counties with their own weather report at the starting week in Columns~(5)--(8). It basically shows how households respond when damage arrives through their connections while their own county has no recorded weather event that week. In this cohort, the four-week gradient is
$-0.457$ percent without controls for subsequent damage and $-0.441$ percent with them.
The four-week gradient and joint weekly
path remain significant under both calendar-block checks. Following counties
through later hits therefore preserves the local pattern while addressing additional shock
arrivals. I show the case of the weekly and accumulated paths and the conditions needed to interpret the future-input controls as isolating a causal pulse in Appendix~\ref{app:localrepair}. Taken together, the spending gradient is stable across loss records and after accounting for
later hits. I deem this relationship between disruptions in socially connected places and local spending as an empirical starting point for the propagation analysis.

\begingroup\let\wsOrigTable\table\renewcommand{\table}[1][]{\wsOrigTable[t]}\begin{table}[H]
\setstretch{1}
\setlength{\parskip}{0pt}
\centering
\begin{threeparttable}
\caption{Four-week spending responses with dated, repeated damage inputs}
\label{tab:calendar_timing}
\footnotesize\setlength{\tabcolsep}{2pt}
\begin{tabular*}{\textwidth}{@{\extracolsep{\fill}}lcccccccc}
\toprule
 & \multicolumn{4}{c}{All counties} & \multicolumn{4}{c}{No own report at origin} \\
\cmidrule(lr){2-5}\cmidrule(lr){6-9}
 & (1) & (2) & (3) & (4) & (5) & (6) & (7) & (8) \\
\midrule
Social exposure & $-0.261$\sym{*} & $-0.261$\sym{***} & $-0.247$\sym{*} & $-0.247$\sym{***} & $-0.457$\sym{**} & $-0.457$\sym{***} & $-0.441$\sym{**} & $-0.441$\sym{***} \\
 & (0.049) & (0.030) & (0.048) & (0.028) & (0.070) & (0.065) & (0.072) & (0.061) \\
\midrule
SE type & 13-week & 26-week & 13-week & 26-week & 13-week & 26-week & 13-week & 26-week \\
Future damage controls & No & No & Yes & Yes & No & No & Yes & Yes \\
Joint weekly-path $p$ & 0.122 & 0.061 & 0.138 & 0.057 & 0.026 & 0.023 & 0.038 & 0.026 \\
Observations & 332,114 & 332,114 & 332,114 & 332,114 & 300,357 & 300,357 & 300,357 & 300,357 \\
\bottomrule\end{tabular*}
\begin{tablenotes}[flushleft]\footnotesize
\item Notes: Coefficients are interpreted as the change in spending from week $-1$ to week 4, in log percentage points, associated with a 1-percentage-point increase in the damage-share-weighted link index. SE types denote calendar-block lengths for the wild bootstrap using 9,999 Webb draws \citep{Webb2023}. All models include county and week effects, own weather, competing inputs, recent spending and damage history. Future damage controls cover weeks 1--12. The origin cohort retains counties hit later. Joint tests cover weeks 0--12. The origin-cohort paths are reported in Appendix~\ref{app:localrepair}. \mbox{\sym{*} $p<0.10$}, \mbox{\sym{**} $p<0.05$}, \mbox{\sym{***} $p<0.01$}.
\end{tablenotes}\end{threeparttable}\end{table}\endgroup
\begingroup\let\wsOrigTable\table\renewcommand{\table}[1][]{\wsOrigTable[t]}\begin{table}[H]
\setstretch{1}
\setlength{\parskip}{0pt}
\centering
\begin{threeparttable}
\caption{Spending and social exposure to federal stimulus payments}
\label{tab:outside}
\footnotesize\setlength{\tabcolsep}{3pt}
\begin{tabularx}{\textwidth}{p{0.4\textwidth} *{4}{>{\centering\arraybackslash}X}}
\toprule
 & \multicolumn{4}{c}{Monthly spending index}\\
\cmidrule(lr){2-5}
 & (1) & (2) & (3) & (4)\\
 & 1 month before & Impact & 1 month after & 2 months after\\
\midrule
Social exposure & $-0.065$ & 0.386\sym{***} & 0.359\sym{***} & 0.268\sym{***}\\
 & (0.114) & (0.060) & (0.069) & (0.086)\\
Production exposure & 0.530 & $-0.444$ & $-0.467$\sym{**} & $-0.033$\\
 & (0.324) & (0.272) & (0.203) & (0.562)\\
\midrule
SE type & Two-way & Two-way & Two-way & Two-way\\
Observations & 26,778 & 28,352 & 26,774 & 25,196\\
\bottomrule\end{tabularx}
\begin{tablenotes}[flushleft]\footnotesize
\item Notes: Coefficients are interpreted as changes in the monthly spending index, in percentage points, associated with a one-standard-deviation increase in exposure. Exposures are standardized within each horizon's sample. Social exposure uses source states at least 300 km away and is residualized against production, distance and home-state payments within each round. All regressions include county and month effects. Two-way standard errors cluster by county and calendar month. Sample moments and round leave-outs are in Appendix~\ref{app:eip}. \sym{*} $p<0.10$, \sym{**} $p<0.05$, \sym{***} $p<0.01$.
\end{tablenotes}\end{threeparttable}\end{table}\endgroup

I would then a complementary test which will ask whether spending rises if the connected places receive cash.
Economic Impact Payments were federal stimulus checks paid in three rounds, and I align them to April 2020,
January 2021 and March 2021. I map payments per resident in distant source states through SCI links. This
separates the exposure from home-state payments and I also account for geography and production links. So,
Table~\ref{tab:outside} asks whether connections to other places receiving transfers predict spending beyond
the county’s own payment exposure.

A one-standard-deviation increase in residual social exposure is associated with a higher spending index of
0.386 percentage points in the payment month and 0.359 points one month later. Both estimates remain positive
and significant when I drop any one payment round (see Appendix~\ref{app:eip}). Since there are only three rounds,
identification requires separating remote exposure from policy and pandemic channels that move at the same
time. Overall, spending falls with social exposure to storm damage and rises with social exposure to payments.
This is consistent with resources lost or received in the source places. Both exercises provide
household-response measurements on the social geometry. Moreover, each input can have its own response
strength and timing.
Section~\ref{sec:proposal} shows how those measurements, combined with financial
weights, yield credit-price predictions. Appendix~\ref{app:localrepair} examines monetary storm
loadings, and Appendix~\ref{wp:app:C.6} reports the historical wage comparison.

\makeatletter\let\ws@secFB\@fb@secFB\let\@fb@secFB\relax\makeatother %
\section{From local exposure to dynamic responses}
\label{sec:stage2}
\makeatletter\let\@fb@secFB\ws@secFB\makeatother

\subsection{Estimating responses across modes and horizons}\label{wp:main:5.1}

The local regressions measure how spending varies with social exposure. The operator also
describes how this response differs across the network's spatial patterns and evolves over time.
With a single active channel, Theorem~\ref{thm:bilinear} makes each latent mode's log response magnitude linear in
the horizon, with a slope that falls with graph frequency.

Applying this to spending requires an observation restriction. Here $B_D=-\eta_D\Phi_0$ maps shocks
into the adverse field $\bm z=-\bm d$, and $C_Y$ maps that field into measured spending losses
(Section~\ref{subsec:aggregation}). The slow-field component of the spending-loss level response
is $H_h^L=C_YA_D^{h-1}B_D$ for $h\geq1$, with $H_0^L=0$. Its first differences are
$H_1^\Delta=C_YB_D$ and $H_h^\Delta=C_Y(A_D-\mathrm{Id}_{n_C})A_D^{h-2}B_D$ for $h\geq2$. Before clipping or other
outcome transformations, the corresponding spending-growth component, which the regressions below
use, is $-H_h^\Delta$. A decaying level therefore need not produce growth of the same sign. Continued inputs, omitted
counties and measurement mixing can also create several apparent decay rates from a single field.
Appendix~\ref{app:observation} states sufficient conditions under which masked, filtered and residualized outcome
coefficients inherit the decay law. These conditions concern the measurement, which
Assumption~\ref{ass:diag} does not cover.

For event $e$, let $\bm b_e^{raw}$ contain in-support county shares of its onset-week recorded damage and
$E_{jke}=u_{jk}(u_k'\bm b_e^{raw})$. I estimate the nonstationary mode coefficients jointly:
\begingroup
\begin{equation}\label{eq:modalreg}
g_{j,t_e+h}=\alpha_{j,h}+\mu_{e,h}+\xi_{t_e+h,h}
+\sum_{k\in\mathcal K}\theta_k^Y(h)E_{jke}+\bm Z_{j,e}'\bm\phi_h+u_{j,e,h},
\end{equation}
\endgroup
where $\mathcal K$ contains 34 nonstationary modes and $g$ is clipped weekly log spending growth. The source construction and controls also differ.
Estimating responses separately across the retained modes and horizons asks more of the storms
than estimating the pooled gradient, and the decay law is not pinned down by these data alone
(Appendix~\ref{app:stage2}). However, credit pricing does not need every modal path. It depends on the financially weighted combination of
responses instead. So, the recovery program targets this combination using restrictions across horizons and
additional household and financial measurements. Modal errors
then matter only through the financial loadings $\bm l'u_k$ (Section~\ref{sec:operator_contribution}).

\subsection{What the transport test measures}\label{wp:main:5.2}

The transport exercise asks whether responses estimated on some storms predict spending
differences across states after other storms. It averages residual weekly spending growth over the
observed non-hit counties within each state, after removing county, event and week effects. The
exercise builds on the modal responses above rather than on the level comparison of
Section~\ref{sec:stage1}. Five purged folds estimate response coefficients and select tuning
settings on training storms, then evaluate them on held-out storms. Held-out predictive
performance is close to the zero-response benchmark for these residual state contrasts
(Appendix~\ref{wp:app:D.2}). Population-weighted state spending averages and credit demand combine county responses
differently. For credit pricing, the relevant weights reflect how strongly households adjust their
desired credit holdings. Section~\ref{sec:finance} examines which spatial responses these financial
weights retain. Section~\ref{sec:proposal} then evaluates recovery of the financially weighted
response in simulation, using an aggregate anchor or restrictions on the common response.

\section{Financial geography and the credit-pricing margin}
\label{sec:finance}

\subsection{Stationary and nonstationary exposure}\label{wp:main:6.1}
\makeatletter\@nobreakfalse\makeatother%

\begin{figure}[t]
\centering
\includegraphics[width=\textwidth]{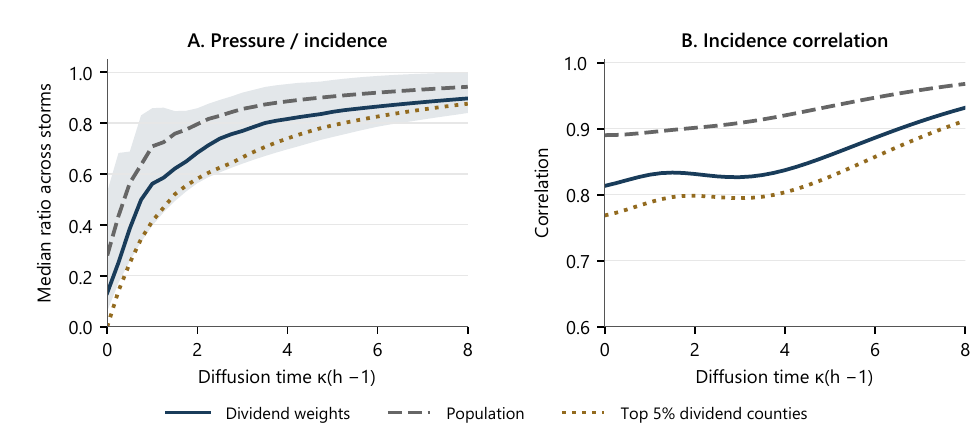}
\caption{Model-implied pressure relative to filtered incidence. Panel~A shows the median
pressure-to-incidence ratio across storms. Shading gives the interquartile range for dividend
weights. Panel~B shows the across-storm correlation between pressure and incidence. Eighteen
candidate storms have no in-support source mass, and 48 retained footprints are partly covered
(Appendix~\ref{app:split}).}
\label{fig:split}
\end{figure}

\noindent Figure~\ref{fig:horizon_importance} shows that diffusion changes the relative importance of
individual sources. How much of that variation survives aggregation over storm footprints? For each
of the 67 storms with damage inside the 1,590-county support, I compute the weighted aggregate
$w'e^{-\tau\mathcal L}\bm b_e=\sum_jb_{ej}I_j(\tau)$, a damage-weighted sum of Figure~\ref{fig:horizon_importance}'s
loadings, using all modes of the retained graph, and compare it with the incidence term
$(w'u_1)(u_1'\bm b_e)$. Here $\bm b_e$ is normalized to sum to one over that storm's in-support
counties, and input loadings are taken proportional to damage. The factor
$\rho_D^{h-1}$ multiplies both terms and cancels, so the comparison needs no estimate of $\rho_D$. Because $\kappa_D$ is not pinned down by these data, Figure~\ref{fig:split}'s horizontal axis measures diffusion time rather than calendar weeks.
Appendix~\ref{app:split} reports source coverage, the dollar-scaled comparison and an implementation check.

Before diffusion ($\tau=0$), dividend-weighted pressure is 0.13 of incidence for the median
storm, reflecting the proxy's low weight on the initial footprint. Under the assumed
diffusion, the median ratio rises to 0.56 at $\tau=1$ and 0.82 at $\tau=4$
(Figure~\ref{fig:split}). For unit in-support footprints, the correlation with incidence
rises from 0.81 to 0.84 over that interval. Scaling each footprint by its in-support damage gives
correlations of 0.97 before diffusion and 0.99 at $\tau=2$. Population weights and weights confined
to the top 5 percent of dividend counties also converge toward incidence (Appendix~\ref{app:split}).
The calculation shows why financial geography matters. The dividend proxy places financial weight away from the initial disturbance and so diffusion changes weighted exposure even when aggregate pressure is highly correlated with incidents. To determine which of these spatial responses survives the projection into credit demand, we just need to focus on measuring the financial weights.

\begingroup
\raggedbottom %

\subsection{Monthly EBP associations}\label{wp:main:6.2}
\makeatletter\@nobreakfalse\makeatother%

\noindent A separate monthly analysis summarizes county spending responses by the amplitude of a low-rank
SCI state and examines its association with EBP changes as intermediary capital tightens. I regress the monthly EBP change, $E^{BP}_{t+1}-E^{BP}_t$, on the lagged state, lagged tightness and their
product. Tightness is the negative of the standardized He--Kelly--Manela capital ratio. The regression also
includes the analogous production state, macro and COVID controls and two lags of the outcome. It covers 49
months of the 2020--2024 sample. With both states standardized, the SCI interaction is $+0.116$ percentage
points of EBP. The production interaction is $-0.010$. The positive interaction indicates that the association
between the SCI amplitude and EBP changes is stronger when intermediary capital is lower. It also appears in
the component of the amplitude not explained by active-storm counts. However, I build the states on the full
sample and select the specification retrospectively. Hence, these are in-sample associations. Their partial
coefficients are also distinct from Proposition~\ref{prop:levels}'s conditional benchmark (see Appendix~\ref{app:finance}).
Moreover, an aggregate association of this kind cannot by itself separate propagation from common storm
timing. In the model, I isolate the increment due to diffusion \eqref{eq:diffincrement} by comparing the
\rev{conditional EBP response} with its no-diffusion counterpart, holding inputs and other parameters fixed.

\subsection{Fund demand and the credit-pricing margin}
\label{sec:fundpricing}
\makeatletter\@nobreakfalse\makeatother%

\noindent Fund data let us examine the market end of the household-demand mechanism.
Table~\ref{tab:creditflow} links quarterly subscriptions to corporate holdings in Panel A
and monthly inflows to EBP changes in Panel B. The pricing instrument uses granular fund-flow
residuals, controlling for two common factors and two lagged EBP changes.

\begin{table}[t]
\setstretch{1}
\setlength{\parskip}{0pt}
\renewcommand{\arraystretch}{0.9}%
\captionsetup{skip=2pt}
\captionsetup[subtable]{position=top,aboveskip=0pt,belowskip=0pt}
\centering
\caption{Fund subscriptions, corporate-bond holdings and credit pricing}
\label{tab:creditflow}
{\footnotesize
\subcaption*{\footnotesize Panel A. Subscriptions and corporate-bond purchases}
\begin{tabularx}{\textwidth}{p{0.4\textwidth} >{\centering\arraybackslash}X >{\centering\arraybackslash}X}
\toprule
 & \multicolumn{2}{c}{Quarterly corporate-bond purchases}\\
\cmidrule(lr){2-3}
 & (1) & (2)\\
 & Including reinvestment & External cash only\\
\midrule
Purchases per dollar subscribed & 0.635\sym{***} & 0.642\sym{***}\\
 & (0.026) & (0.025)\\
\midrule
SE type & Fund cluster & Fund cluster\\
Fund--quarter observations & 6,993 & 6,993\\
\bottomrule
\end{tabularx}\space\space
\subcaption*{\footnotesize Panel B. Fund demand and credit pricing}
\begin{tabularx}{\textwidth}{p{0.4\textwidth} >{\centering\arraybackslash}X >{\centering\arraybackslash}X}
\toprule
 & \multicolumn{2}{c}{Monthly EBP change}\\
\cmidrule(lr){2-3}
 & (1) & (2)\\
 & Including reinvestment & External cash only\\
\midrule
Net inflows & $-4.06$\sym{***} & $-3.96$\sym{**}\\
 & (1.54) & (1.61)\\
\midrule
SE type & HAC (3) & HAC (3)\\
Months & 57 & 57\\
First-stage statistic & 26.6 & 16.4\\
\bottomrule
\end{tabularx}\space\space
\par
\begin{tablenotes}[flushleft]\footnotesize
\item Notes: Panel A uses approximate return-adjusted corporate-bond purchases and subscriptions, scaled by initial assets, with asset weights, quarter effects and fund-clustered standard errors (495 funds). Panel B reports EBP changes in basis points per inflow of 0.01\% of nonfinancial corporate debt securities, using corporate-holdings weights. The pricing instrument follows \citet{GabaixKoijen2024} (Appendix~\ref{app:creditflow}). HAC (3) uses three-month Bartlett weights, conditional on estimated flow factors. \mbox{\sym{*} $p<0.10$}, \mbox{\sym{**} $p<0.05$}, \mbox{\sym{***} $p<0.01$}.
\end{tablenotes}
}
\end{table}

Panel A asks whether money entering funds is reflected in their corporate-bond portfolios.
Each dollar subscribed accompanies \$0.635 of approximate corporate-bond purchases,
or \$0.642 using external cash. At the pricing margin, inflows of 0.01\% of
nonfinancial corporate debt securities coincide with an EBP decline of 4.06 basis points, or 3.96 using
external cash. Together, the two panels document corporate-portfolio adjustment and associated
credit repricing at the fund end of the mechanism. I document the conditions for a structural reading in Appendices~\ref{app:creditflow} and~\ref{app:bridgefund}.

\enlargethispage{2.5pt}%
Households change desired holdings at a fixed price whereas subscriptions measure realized flows after prices adjust. The table records inflows, and the pricing map below is written for selling.
From \eqref{eq:absorption}, selling at reference price $\bar P$ satisfies
$dF^{sell}=\bar P(1-\zeta G_S)dS$. Within a smooth regime with positive absorption, letting
$\vartheta_F=dE^{BP}/dF^{sell}$ gives
\begin{equation}\label{eq:flowbridge}
K_h=\vartheta_F\bar P(1-\zeta G_S)\,
\bm l'A_D^{h-1}B_D .
\end{equation}
The operator and the financial weights locate desired-demand changes. The executed
share and price impact convert them into a premium response. Panel A relates two
observed fund quantities. Household positions and response sensitivities connect
those quantities to the model's desired-demand shift.

Appendix~\ref{app:bridgefund} states the demand-feedback and exclusion restrictions and compares granular
and network moments \citep{chodorowreich2026network}. The 57-month sample leaves the separation of
price impact from demand feedback weakly identified \citep{AndersonRubin1949}.
Changes in desired holdings generate flows.
Cumulating them under static clearing gives \eqref{eq:flowbridge}, while market persistence
adds a temporal filter (Appendix~\ref{app:bridgefund}).
Section~\ref{sec:implementation} demonstrates in simulation how separately estimated household
and market responses combine into a credit-price prediction.

\par
\endgroup
\section{Recovering the financially relevant operator}
\label{sec:proposal}

I then ask which household measurements recover the projection that shifts credit demand. The following program targets that conditional contribution holding capital and payoff moments fixed. A total EBP prediction also needs restrictions on capital and competing channels.

\subsection{From household responses to a pricing projection}\label{wp:main:7.1}

Let the slow field's impulse response be $R_h=A_D^{h-1}B_D$, $h\geq1$, the matrix in
equation~\eqref{eq:kernel}, and the measured household responses
be $H_h=C_*R_h$. Then $\bm l'R_h$ is the financially weighted propagation operator of the Introduction. The known map $C_*$ specifies how household conditions enter each measurement,
including its units and sign. Spending-loss
measurements, i.e., can stack county contrasts $T$ and an independent aggregate with weights $w$:
$C_*=\big((TC_Y)',(w'C_Y)'\big)'$. Consumption and fixed-price portfolio-demand measurements use their own maps and
can be combined (Appendix~\ref{app:target}). In each case, valid restrictions must separate the field's response
from direct damage, aid, migration and other channels. An aggregate measurement must likewise
distinguish the household channel from a competing common response. Thus merely appending national
spending doesn't identify the household channel.

\begingroup %
\begin{proposition}[Recovery of the pricing projection]
Let\label{prop:recovery} $\mathcal R$ contain the field responses to the shocks and horizons being tested, with basis
$R_{\mathcal R}$. Given $\bm l$, $C_*$ and $c_E\equiv b_EG_S>0$, the target $\bm l'\bm z$ is
recoverable from $C_*\bm z$ for every $\bm z\in\mathcal R$ if and only if there is an $a$ satisfying
$\bm l'R_{\mathcal R}=a'C_*R_{\mathcal R}$. The \rev{conditional EBP response} of equation~\eqref{eq:kernel} is then recovered as
\begin{equation}\label{eq:recoverkernel}
K_h\equiv c_E\bm l'R_h=c_Ea'H_h .
\end{equation}
\end{proposition}\endgroup

Under this condition, the pricing projection can be recovered without separately identifying the
propagation and input matrices (proof in
Appendix~\ref{app:target}). \rev{This also holds for mixed-sign loadings. The recovered projection is then a net
response, which is zero when household responses offset.} Recovery weights $a$ can be constructed from the measurement map, independently
measured $\bm l$ and the maintained response subspace. Among valid weights, a precision-aware choice
minimizes projected measurement variance. Response estimates supply $H_h$, and the weights translate
these into changes in desired credit holdings. Market evidence supplies $c_E$.
\rev{Event contrasts need an anchor when the common response affects the target and is not fixed by the
propagation law.}
When household responses are measured as weekly growth, they must first be accumulated consistently.
If county dollar losses satisfy $y_i^*=c_i^Y z_i$ with known
$c_i^Y\ne0$, then $a_i=l_i/c_i^Y$, giving the recovered response the marginal demand weight per
dollar lost. These are losses at a fixed premium. Measured losses also contain the response of spending to the premium, $\bm y^*=[\diag(c_i^Y)-G_S\,\bm\zeta^c\bm l']\bm z$, where $\zeta^c_i$ is the premium sensitivity of county spending. When every county is measured and $G_S\,\bm a_0'\bm\zeta^c\neq1$ for $a_{0,i}=l_i/c_i^Y$, the weights are $\bm a=\bm a_0/(1-G_S\,\bm a_0'\bm\zeta^c)$, the fixed-premium weights times one scalar (Appendix~\ref{app:financialmargin}). This rescaling does not carry over to county contrasts or to an incomplete sample.

For an approximate representation, define $r_l=\bm l-C_*'a$ and assume
$\|R_h\bm\varepsilon\|\leq\bar z_h$. With $a$ and $c_E$ fixed, the forecast
$\hat p_h=c_Ea'\hat H_h\bm\varepsilon$ obeys
\begin{equation}\label{eq:financialbound}
|\hat p_h-K_h\bm\varepsilon|
\leq c_E\|a\|\,\|(\hat H_h-H_h)\bm\varepsilon\|
      +c_E\|r_l\|\bar z_h .
\end{equation}
The first term scales household-response prediction error by the recovery weights. The second captures
a financially relevant direction missing from the measurements, for which only the projection of
$r_l$ onto $\mathcal R$ matters. More precise
measurements can reduce the first. Additionally, portfolio-demand measurements can span the missing
direction. Estimated $a$ and $c_E$ add a further error term (Appendix~\ref{app:target}). Section~\ref{sec:implementation}
estimates household loading scales and the market map from simulated observations.

\subsection{Measuring the household margin}\label{wp:main:7.2}

The financial loading combines changes in investable resources with changes in desired portfolio
shares. At fixed asset price $P_a$,
let $W_i$ be county households' investable wealth and $s_{ia}$ their desired share in credit
asset $a$, so desired dollar holdings are $s_{ia}W_i$. Define
$\chi_i^W=-\partial W_i/\partial z_i$ and $\psi_{ia}^{\mathrm{port}}=-\partial s_{ia}/\partial z_i$.
Differentiating desired dollar holdings gives:
\begingroup
\predisplaypenalty=0%
\begin{equation}\label{eq:margin_v6}
l_{ia}^{\$}\equiv-\frac{\partial(s_{ia}W_i)}{\partial z_i}
=s_{ia}\chi_i^W+W_i\psi_{ia}^{\mathrm{port}},\qquad l_{ia}=l_{ia}^{\$}/P_a .
\end{equation}
\endgroup
The first term captures changes in resources, including accumulated saving. The second is
portfolio rebalancing\rev{ in either direction}, which can occur without changing current consumption. Spending alone
identifies neither margin nor its sign. Account flows and return-adjusted holdings, together
with transfers, borrowing and liquid balances, distinguish them (Appendices~\ref{app:financialmargin} and~\ref{app:visibility}).
Financing assistance requires access to resources, whereas rebalancing can occur within an
account without a withdrawal. Holdings measure
$s_{ia}W_i$, not its derivative, so equal holdings can have different responses. Executed trades
also respond to prices, requiring an identified demand system or valid price controls.

There is a useful intermediate prediction even when the common household response is missing.
After normalization, a response common to all counties moves both assets equally. Their difference
therefore retains the effect of their different investor geography.
For two assets \rev{$a$ and $b$ whose totals $\delta_a^{\$}=\sum_il_{ia}^{\$}$ and $\delta_b^{\$}$ are both
nonzero, let} $\pi_a=\bm l_a^{\$}/\delta_a^{\$}$\rev{, and define $\pi_b$ likewise}. Let the market map $c_a>0$ convert fixed-price dollar demand pressure
into a measured premium, net of capital and payoff news. Here $\Delta p_a$ is the premium-level
response relative to the no-shock path at the horizon being studied. Let $z_\perp$ collect
cross-county differences around the common response $\alpha$. Since both $\pi$'s sum to one,
\begingroup
\begin{equation}\label{eq:relative_v6}
\frac{\Delta p_a}{c_a\delta_a^{\$}}-\frac{\Delta p_b}{c_b\delta_b^{\$}}
=(\pi_a-\pi_b)'z=(\pi_a-\pi_b)'z_\perp,
\quad z=z_\perp+\alpha\mathbf1 .
\end{equation}
\endgroup
Thus the common component $\alpha$ cancels, although the household measurement map must still span
this contrast as in Proposition~\ref{prop:recovery}. \rev{An asset whose loadings sum to zero needs no such
difference, because the common component already cancels from its own demand shift.}
In the equal-degree two-county example, take $\pi_a=(0.8,0.2)'$ and
$\pi_b=(0.2,0.8)'$. A unit shock in county 1 gives a normalized premium gap
$0.6\rho_D^{h-1}e^{-2\kappa_D(h-1)}$, versus $0.6\rho_D^{h-1}$ without diffusion.
Here the prediction is a shrinking gap, not a ranking reversal. Given independently measured financial
weights and market maps, the operator predicts how the normalized premium gap evolves across
horizons. An absolute aggregate premium response additionally requires the common household
component and its market map.

\subsection{An operator estimation program tied to that target}\label{wp:main:7.3}

\textls[-6]{%
The estimation criterion fits household responses while emphasizing the combination that matters
for credit demand.
Let $Y^*_{eh}$ collect the proposed household measurements, $\bm Z_{eh}\bm\phi_h$ the separately
identified competing channels, and $H_h(\vartheta)=C_*A_{\mathrm{tr}}(\vartheta)^{h-1}B_{\mathrm{in}}(\vartheta)$ a parsimonious
candidate with transition $A_{\mathrm{tr}}$ and input map $B_{\mathrm{in}}$.
Define $v_{eh}(\vartheta,\bm\phi)=Y^*_{eh}-H_h(\vartheta)\bm\varepsilon_e-\bm Z_{eh}\bm\phi_h$.
Within each training fold, the criterion~is
}
\begingroup\predisplaypenalty=0%
\begin{equation}\label{eq:targetfit}
(\hat\vartheta,\widehat{\bm\phi})=\argmin_{\vartheta\in\Theta,\bm\phi}
\sum_{(e,h)\in\mathrm{train}}v_{eh}'\big(V_{eh}^{-1}+\omega aa'\big)v_{eh}
+\lambda_{\mathcal P}\mathcal P(\vartheta),\qquad \omega\geq0 .
\end{equation}
\endgroup
Here $V_{eh}$ is the measurement-error covariance, $\mathcal P$ regularizes the operator, and
$\Theta$ imposes stability and a predeclared low-dimensional input-loading family. For diffusion,
$0<\rho_D<1$ and $\kappa_D\geq0$. The $V_{eh}^{-1}$ component rewards accurate prediction of household measurements.
The $\omega aa'$ component gives additional weight to errors in the financial combination recovered by
\eqref{eq:recoverkernel}, with $\omega$ putting that loss in compatible units.
Set $\omega=0$ until $\bm l$, and hence $a$, is independently measured. Proxy choices are
sensitivity exercises, and no financial outcome selects the operator's loadings. Training also
chooses regularization. Held-out storms compare the dynamic operator with one-hop on identical
shock sources, recipient screens and competing exposures. This focus on the financially relevant response connects to 
\citet{AdaoCostinotDonaldson2025}, who evaluate quantitative models against the causal predictions needed for a particular economic question.

\subsection{Implementation on the sample design}
\label{sec:implementation}

\rev{The simulations apply the recovery conditions on the observed SCI graph and all candidate storm
declarations, with financial maps first known and then estimated. I simulate the household
field in log-spending units, so its spending measurement map is known. I use separately
calibrated damage intensity and spending noise rather than estimated event-study or
county-calendar coefficients.} The denoised panel of Table~\ref{tab:recovery} subtracts the tracker's recurring 2019-base calendar pattern
(Appendix~\ref{app:calendar_sensitivity}). The target is the financially weighted response of observed recipient
households to each held-out storm over one to twelve weeks (Appendix~\ref{app:recoverycheck}). \rev{I estimate the operator
under the diffusion law on training storms only and project it with county-resolution loadings.} Its fit
uses the measurement-precision weights of \eqref{eq:targetfit}, without the additional financial-loss
weight or regularization penalty ($\omega=\lambda_{\mathcal P}=0$). A parametric bootstrap that re-estimates it
gives pointwise and twelve-week path bands. Aggregate measurements give each training storm's mean
recipient spending response, including both propagation and the common component, which the model
separates. Under the simulation's measurement and exclusion assumptions, \rev{county spending contrasts
identify spatial response differences, and financial loadings translate spatial and common responses
into credit-demand pressure.}

\begingroup
\raggedbottom %
\begingroup\let\wsOrigTable\table\renewcommand{\table}[1][]{\wsOrigTable[t]}\begin{table}[H]
\setstretch{1}%
\centering
\begin{threeparttable}
\caption{Simulated recovery of held-out financially weighted responses}
\label{tab:recovery}
\footnotesize
\begin{tabular*}{\textwidth}{@{\extracolsep{\fill}}lcc}
\toprule
 & Normalized & Coverage \\
Simulated measurement environment & error & pointwise / path \\
\midrule
Anchor; noise calibrated to the denoised panel; known maps & 0.168 & 0.898 / 0.895 \\
Anchor; loading scales and market map with assumed errors & 0.400 & 0.911 / 0.879 \\
No anchor; noise $\times\frac14$; twice the storms & 0.320 & 0.894 / 0.866 \\
\bottomrule
\end{tabular*}
\begin{tablenotes}[flushleft]\footnotesize
\item Notes: Error relative to the root mean square of the true response of observed recipient households,
horizons 1--12 (0.72 bp with a common response, rows 1--2, and 0.26 bp without, row 3), with 90 percent bands. Without an anchor, the law generates the common response. Row 1 has noise calibrated to
the denoised panel and rows 2--3 design-based noise, so the rows differ in more than one factor. Rows 1 and 3 use known loadings and market map, and noise multipliers scale standard deviations. There are 100 replications on fresh draws, with Monte Carlo s.e.\ at most 0.03.
\end{tablenotes}
\end{threeparttable}
\end{table}\endgroup

The anchored design shows that household measurements support recovery. Its simulated common
response is proportional to damage, and financial maps come from hypothetical auxiliary
measurements, with investment-income geography supplying the assumed\rev{, nonnegative} loading shape. With known maps,
the operator recovers held-out responses with normalized error 0.17, compared with 0.22 for one-hop
and 0.30 for the anchor alone (Table~\ref{tab:recovery}, row 1). \rev{Household measurements therefore
add predictive information beyond the common response.} A \rev{graph polynomial combining
one- to three-link exposures with horizon-varying coefficients has}
error 0.18, but its nominal 90 percent intervals
cover 72 percent of true responses, against 90 percent for the operator (Appendix Table~\ref{tab:recovery_matched}).

\begin{figure}[!t]
\centering
\includegraphics[width=\textwidth]{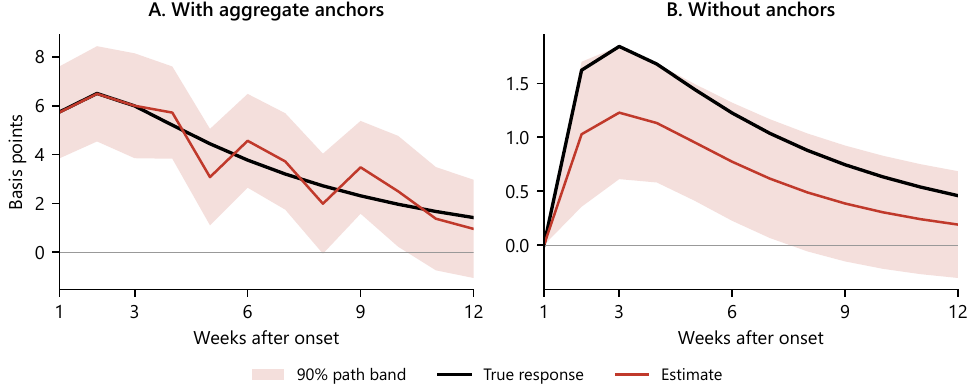}
\caption{Simulated recovery for the held-out storm with the largest true response in one
replication. Panel~A uses training-storm anchors and known financial maps, with noise calibrated
to the denoised panel (Table~\ref{tab:recovery}, row 1). Panel~B uses no anchors, one-quarter of
the design-based noise standard deviation and twice as many storms (row 3). Event details are
in Appendix~\ref{app:recoverycheck}.}
\label{fig:recovery}
\end{figure}

\begingroup\clubpenalty=10000\widowpenalty=10000
Without an anchor, the restrictions must also supply the common response. In the more informative
design of row 3, with less noise and twice the storms, operator error is 0.32, against 0.71 for the
graph polynomial and 0.92 for one-hop. This is the structural route to recovery, in which the law
links responses across counties and horizons and thereby determines their common component. It
relies on that law generating the entire common response (Appendix~\ref{app:recovery_results} examines misspecification).
\par\endgroup

In Figure~\ref{fig:recovery}, the anchored prediction learns the response level from training-storm
aggregates and tracks the true path. The no-anchor prediction instead recovers the path's shape from the propagation law's \enlargethispage{4pt}%
restrictions. Together, these results show that 
the framework can turn household responses into a financial prediction for a storm outside the estimation sample.

\par
\endgroup

\begin{figure}[!t]
\centering
\includegraphics[width=\textwidth]{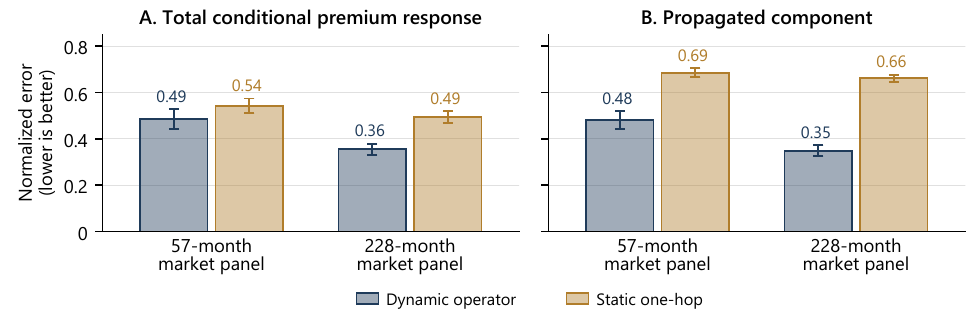}
\caption{Held-out prediction with estimated financial maps. Bars show mean normalized error
across 200 simulated replications. Whiskers give 95 percent Monte Carlo intervals for the mean
error. Each error is normalized by its target's own root mean square over weeks 1--12.
Panel~B excludes the common response added in the simulation.}
\label{fig:bridge}
\end{figure}

Figure~\ref{fig:bridge} asks whether this prediction advantage survives when the household and
market maps must also
be learned. It keeps the first environment of Table~\ref{tab:recovery} and its
target, the held-out desired-selling path valued at the conditional market loading and fixed capital.
Household loading scales are learned from hypothetical account flows, with the within-group shape
still assumed. The market map is learned from an independent simulated fund panel without the
held-out storms, using an instrument built from simulated fund flows rather than the true demand
innovation. The price impact of executed flows is a scenario set at Table~\ref{tab:creditflow}'s
column~(1) estimate, not an identified structural value, so the market stage's precision is
conditional on this calibration.

\looseness=-1 The operator retains a prediction advantage when the maps are estimated. At 57 simulated market
months, its nominal 90 percent pointwise bands cover 88 percent of true total responses, against 49
percent for one-hop. The largest gain concerns the propagated component, where operator error is
about 30 percent lower than one-hop's. Under the simulated diffusion law, the dynamic operator thus
adds predictive information about the propagated component beyond a static exposure measure. With
228 months, its error falls further as the market map is estimated more precisely.\footnote{Pointwise coverage assesses interval calibration. Detecting a nonzero
response is a separate joint test over held-out events and horizons (Appendix~\ref{app:bridgefund}).} Empirically, spending panels supply the household responses $H_h$,\footnote{\interfootnotelinepenalty=10000 The NielsenIQ Consumer
Panel, from 2004, and Retail Scanner data, from 2006 \citep{KiltsNielsen2026v18}, offer a longer
pre-pandemic spending history for $H_h$ (coverage and definitions in Appendix~\ref{app:design}).} household accounts
the financial weights and market data the market map $c_E$.

\subsection{What the operator contributes to the financial prediction}
\label{sec:operator_contribution}

Diffusion affects credit prices by moving household responses across counties whose financial
weights differ.
\rev{Under the diffusion law}, $r_{D,k}=\rho_De^{-\kappa_D\lambda_k}$, so
\begingroup\predisplaypenalty=0%
\begin{equation}\label{eq:kappaprice}
\frac{\partial K_h}{\partial\kappa_D}
=-c_E(h-1)\sum_{k\geq2}\lambda_k(\bm l'u_k)r_{D,k}^{h-1}u_k'B_D .
\end{equation}
\endgroup
Only nonstationary modes enter this derivative. Hence, as we change the diffusion rates, it can affect prices when shocks excite those modes, financial weights load on them and the horizons permit propagation. Its effect does not need to be monotone since diffusion can move pressure toward or away from the marginal investors. What recovery needs is precision in the combinations selected by the financial weights $\bm l'u_k$ and so equal precision in every mode is not needed (see Appendix~\ref{app:target} for details). As Section~\ref{sec:pricing} shows, the magnitude depends on how much households change desired
credit holdings per dollar of spending shortfall and how readily others absorb that shift
(Appendix~\ref{app:visibility}). As in Section~\ref{subsec:aggregation}, a test of the \rev{response's} diffusion-specific EBP restriction
must separate the increment \eqref{eq:diffincrement} from the no-diffusion component and competing
channels (Appendix~\ref{app:diffusionrestriction}).

\section{Conclusion}
\label{sec:conclusion}
\makeatletter\@nobreakfalse\makeatother%

\noindent The main contribution of this paper is a recovery condition for the household-demand component of credit prices.
I take the financial weights as independently measured and the measurement map as known. Then household
measurements recover the pricing projection exactly when they span its financial direction on the relevant shock
responses. So, the pricing projection can be recoverable even when the full propagation law remains
unidentified. Moreover, this paper also shows that the projection's relative weights change with the horizon.
Hence, a single fixed exposure score generally fails to summarize them. In simulations on the observed network data, the operator derived in this paper shows an interesting performance of reducing error on the propagated credit-premium component by about 30\% relative to one-hop under the specified diffusion law. This paper helps me learn the household loading scales and the market map from
simulated accounts and fund flows. The next step is to measure them in the observed data which will contain a household panel that links spending,
balance sheet and portfolio exposures over time. Using that data, we can separate spending responses from
portfolio changes and can test whether the implied demand shift reaches the EBP or not. This test matters because
a small aggregate spending response does not translate into a small shift in bond demand. Finally, the recovery condition is
powerful in the sense that it does not depend on the links being social and so it can be applied with financial
claims too to study the financial architecture directly.  

\singlespacing
\bibliographystyle{aer_doi}
\bibliography{references}

\clearpage
\newgeometry{top=0.9in,bottom=0.9in,left=0.9in,right=0.9in}
\singlespacing
\normalsize
\setlength{\parskip}{0.2\baselineskip plus 1pt}
\flushbottom
\newcommand{\wpOnly}[1]{#1}\newcommand{\appOnly}[1]{}
\renewcommand{\topfraction}{0.95}\renewcommand{\bottomfraction}{0.5}
\renewcommand{\textfraction}{0.05}\renewcommand{\floatpagefraction}{0.9}
\makeatletter
\setlength{\@fptop}{0pt}\setlength{\@fpsep}{12pt plus 1fil}\setlength{\@fpbot}{0pt}
\let\shortOrigTable\table \let\shortOrigEndTable\endtable
\renewenvironment{table}[1][!tbp]{\in@{h}{#1}\ifin@\shortOrigTable[!tbp]\else\shortOrigTable[#1]\fi}{\shortOrigEndTable}
\let\shortOrigFigure\figure \let\shortOrigEndFigure\endfigure
\renewenvironment{figure}[1][!tbp]{\in@{h}{#1}\ifin@\shortOrigFigure[!tbp]\else\shortOrigFigure[#1]\fi}{\shortOrigEndFigure}
\makeatother
\everydisplay=\expandafter{\the\wpAppendixEveryDisplay}
\textfloatsep=\wpAppendixTextfloatsep
\floatsep=\wpAppendixFloatsep
\intextsep=\wpAppendixIntextsep
\captionsetup[figure]{labelsep=period,font=small,skip=10pt}
\makeatletter
\let\@endtheorem\wpAppendixEndTheorem
\def\@spart#1{{\parindent\z@\raggedright\interlinepenalty\@M\normalfont
  \fontsize{15.1}{18.1}\selectfont\bfseries #1\par}\nobreak\vskip 3ex\@afterheading}
\renewcommand\section{\@startsection{section}{1}{\z@}{-3.5ex \@plus -1ex \@minus -.2ex}%
  {2.3ex \@plus .2ex}{\normalfont\fontsize{14.1}{17}\selectfont\bfseries}}
\renewcommand\subsection{\@startsection{subsection}{2}{\z@}{-3.25ex \@plus -1ex \@minus -.2ex}%
  {1.5ex \@plus .2ex}{\normalfont\fontsize{13.1}{15.7}\selectfont\bfseries}}
\makeatother
\appendix
\numberwithin{equation}{section}
\numberwithin{table}{section}
\numberwithin{figure}{section}

\part*{Appendix}
\addcontentsline{toc}{part}{Appendix}
\noindent Equation, theorem, table and figure numbers without a letter refer to the main text.
Appendix~\ref{app:proofs} proves the main-text results. Within it, Appendix~\ref{app:target} derives
the pricing-projection recovery condition, its error bound and a minimum-variance representation.
Appendix~\ref{app:model} collects model details that the main text states briefly.
Appendices~\ref{app:stage1}--\ref{app:finance} report supplementary evidence. Within them,
Appendix~\ref{app:split} details the shock normalization used in the aggregation exercise.
Appendix~\ref{app:design} evaluates a longer spending design
using public data and synthetic experiments. Appendix~\ref{app:localrepair} documents recorded-loss
inputs, source timing, measured-weather controls and event-level inference.
Appendix~\ref{app:operational} derives the financial margin and examines the coverage and
identifying variation available for stronger weather exclusions. Appendix~\ref{app:recoverycheck}
documents the simulated recovery. Within it, Appendix~\ref{app:bridgefund} estimates the financial
maps from hypothetical household accounts and simulated fund observations.

\makeatletter
\let\@fb@secFB\relax
\setlength{\textfloatsep}{20pt plus 12pt minus 4pt}\setlength{\floatsep}{12pt plus 6pt minus 2pt}
\predisplaypenalty=0
\renewcommand{\section}{\@startsection{section}{1}{\z@}{-2.2ex}%
  {1.0ex}{\normalfont\fontsize{14.1}{17}\selectfont\bfseries}}
\renewcommand{\subsection}{\@startsection{subsection}{2}{\z@}{-1.9ex}%
  {0.7ex}{\normalfont\fontsize{13.1}{15.7}\selectfont\bfseries}}
\renewcommand{\subsubsection}{\@startsection{subsubsection}{3}{\z@}{-1.9ex}%
  {0.7ex}{\normalfont\normalsize\bfseries}}
\makeatother
\setstretch{1.12}\advance\baselineskip by 0pt plus 1pt\relax%
\etocsetnexttocdepth{section}
\localtableofcontents
\section{Proofs}
\label{app:proofs}

\subsection{Lemma~\ref{lem:closed}}\label{wp:app:A.1}
In this proof, $t$ denotes the shock date, written $t_0$ in Section~\ref{subsec:fields}. The shocked and comparison paths coincide
before it, so their demand-field difference is zero at $t$ and their attention-field difference
is zero at $t-1$. These are initial responses, not restrictions on the levels at later dates.
Section~\ref{subsec:fields} gives the direct consumption response
$\Delta\bm c_{t+h}=\Phi_h\bm\varepsilon_t$, $h\geq0$. Combining it with the valuation vector
\eqref{eq:valuation} gives the direct term $-\gamma\Phi_h$. The slow field receives the impact innovation
$\bm\nu_t=\Phi_0\bm\varepsilon_t$ at $t+1$ with coefficient $\eta_D$, by \eqref{eq:slow}.
Propagating it $h-1$ further periods through $A_D$ gives $\eta_DA_D^{h-1}\Phi_0\bm\varepsilon_t$ for
$h\geq1$ and nothing at $h=0$. The fast field receives the signal $B\bm\varepsilon_t$ at $t$, by
\eqref{eq:fast}, and propagating it $h$ periods gives $\eta_{\mathrm{att}}A_{\mathrm{att}}^hB\bm\varepsilon_t$. The
production field receives $\bm\nu_t$ at $t+1$ and contributes $\eta_D^P(A_D^P)^{h-1}\Phi_0\bm\varepsilon_t$
for $h\geq1$. Each component enters $\bm v_t$ with coefficient $-\gamma$, $-a_D$, $-a_{\mathrm{att}}$ or $-a_D^P$.
Collecting terms gives \eqref{eq:closed}. \rev{The fields receive inputs only at the shock date,
$\bm\nu_t$ for the slow and production fields and $\bm s_t$ for the fast field. Later direct
responses $\Phi_j\bm\varepsilon_t$, $j\geq1$, do not feed them, so the operators are exact rather
than approximations to the convolution} %
$\eta_D\sum_{j=0}^{h-1}A_D^{h-1-j}\Phi_j$. \qed

\subsection{Theorem~\ref{thm:no_scalar}}\label{wp:app:A.2}
Suppose $\widetilde\Omega_h^{SCI}=a_h(\bm s^{\mathrm{fix}})'$ for all $h$. Under the spectral form, the modal loadings
then satisfy $\psi_k(h)=a_hs_k^{\mathrm{fix}}$ with $s_k^{\mathrm{fix}}\equiv(\bm s^{\mathrm{fix}})'u_k$, so for any two modes $k,\ell$,
\begin{equation}\label{eq:rank}
\psi_k(1)\psi_\ell(2)-\psi_\ell(1)\psi_k(2)=0 .
\end{equation}
From \eqref{eq:modal}, $\psi_k(1)=\bar C^{\mathrm{init}}_{D,k}+\bar C^{\mathrm{init}}_{\mathrm{att},k}r_{\mathrm{att},k}$ and
$\psi_k(2)=\bar C^{\mathrm{init}}_{D,k}r_{D,k}+\bar C^{\mathrm{init}}_{\mathrm{att},k}r_{\mathrm{att},k}^2$, and similarly for $\ell$. Substituting into
\eqref{eq:rank} gives
\begin{equation}
\begin{aligned}
&\bar C^{\mathrm{init}}_{D,k}\bar C^{\mathrm{init}}_{D,\ell}(r_{D,\ell}-r_{D,k})
+\bar C^{\mathrm{init}}_{\mathrm{att},k}\bar C^{\mathrm{init}}_{\mathrm{att},\ell}r_{\mathrm{att},k}r_{\mathrm{att},\ell}(r_{\mathrm{att},\ell}-r_{\mathrm{att},k})\\
&\quad+\bar C^{\mathrm{init}}_{D,k}\bar C^{\mathrm{init}}_{\mathrm{att},\ell}r_{\mathrm{att},\ell}(r_{\mathrm{att},\ell}-r_{D,k})
+\bar C^{\mathrm{init}}_{\mathrm{att},k}\bar C^{\mathrm{init}}_{D,\ell}r_{\mathrm{att},k}(r_{D,\ell}-r_{\mathrm{att},k})=0,
\end{aligned}
\end{equation}
a polynomial in the coefficients and decay factors. It is not identically zero: at
$\bar C^{\mathrm{init}}_{\mathrm{att},k}=\bar C^{\mathrm{init}}_{\mathrm{att},\ell}=0$, $\bar C^{\mathrm{init}}_{D,k}=\bar C^{\mathrm{init}}_{D,\ell}=1$ and $r_{D,\ell}\neq r_{D,k}$ it
equals $r_{D,\ell}-r_{D,k}\neq0$. A polynomial that is not identically zero vanishes only on a
closed set of measure zero, including within the open set where all four coefficients are
nonzero. Away from that set \eqref{eq:rank} fails, so no rank-one representation exists. The same
evaluation shows that distinct decay rates across modes already break rank one within a single
channel. This witness also respects the diffusion-kernel restriction: distinct eigenvalues and
$\kappa_D>0$ imply distinct slow rates, and sufficiently small nonzero fast loadings preserve
the nonzero determinant. More generally the exceptional set is restricted to the admissible
parameter domain, not declared negligible on an arbitrary lower-dimensional restriction.
Conversely, if every two-horizon, two-mode minor vanishes, all nonzero rows are proportional.
Choosing any one as $(\bm s^{\mathrm{fix}})'$ constructs the representation (the all-zero case is trivial).
The augmented two-field state remains a standard linear system. Neither matrix representations
nor ordinal centrality are ruled out. \qed

\begin{acor}[Competing graphs]
Suppose\label{cor:multigraph} the social slow field on $\mathcal L$ and the production slow field on $\mathcal L^P$ are
both active on non-collinear retained modal supports, with nonzero projected loadings and
distinct horizon profiles. Then, generically, no fixed scalar characteristic with a common linear horizon coefficient reproduces pricing relevance
across both graphs and all horizons.
\end{acor}
\begin{proof}
The two fields contribute operator blocks whose decay schedules are governed by different
Laplacians. If the retained modal subspaces are not collinear, a rank-one family would require the
$2\times2$ determinants formed from one social and one production direction to vanish at all
horizon pairs, which again defines a proper algebraic subset of the parameter space. Canonical
correlations and projection distances between the two retained subspaces
(Appendix~\ref{app:prod}) measure how close the data are to the collinear case.
\end{proof}

\subsection{Theorem~\ref{thm:bilinear}}\label{wp:app:A.3}
\begin{figure}[!htbp]
\centering
\includegraphics[width=\textwidth]{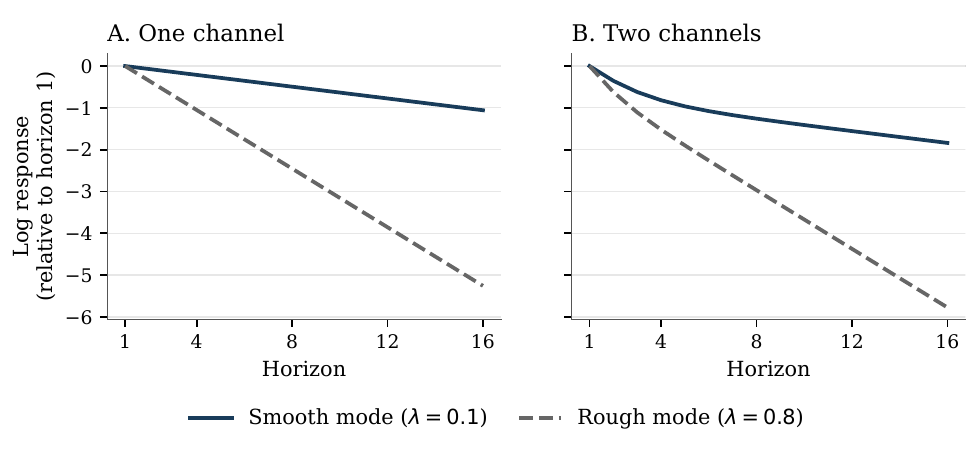}
\caption{The modal decay restriction. Each line is the log response of one mode, normalized at
horizon 1. Panel A uses $r_{D,k}^{h-1}$, whereas Panel B uses
$0.3r_{D,k}^{h-1}+0.7r_{\mathrm{att},k}^{h}$. Illustrative parameters are
$(\rho_D,\kappa_D)=(0.97,0.4)$ and $(\rho_{\mathrm{att}},\kappa_{\mathrm{att}})=(0.55,0.8)$.
Single-channel log paths are straight, whereas a mixture generally bends
(Theorem~\ref{thm:bilinear}).}
\label{fig:modal_illustration}
\end{figure}

For the slow channel, $C_{D,k}(h)=C^{\mathrm{init}}_{D,k}r_{D,k}^{h-1}$ for $h\geq1$. Taking log magnitudes and
substituting $r_{D,k}=\rho_De^{-\kappa_D\lambda_k}$ gives \eqref{eq:bil_slow}. The same steps with
$C_{\mathrm{att},k}(h)=C^{\mathrm{init}}_{\mathrm{att},k}r_{\mathrm{att},k}^h$ give \eqref{eq:bil_fast}. Here $\rho_{\mathrm{att}}=0$ is an impact-only response and
is excluded. Under two channels, $\log|C_{D,k}(h)+C_{\mathrm{att},k}(h)|$ is the log of a sum of two
exponentials in $h$ with mode-specific rates. Its second difference in $h$ is generically nonzero,
while every function of the form $a_k+hc_1+h\lambda_kc_2$ has zero second difference in $h$. Figure~\ref{fig:modal_illustration} illustrates both cases. \qed

\subsection{Proposition~\ref{prop:ident}}\label{wp:app:A.4}
Under one channel, $\log r_{m,k}=\log\rho_m-\kappa_m\lambda_k$. Horizon variation within a mode
identifies $\log r_{m,k}$ as the slope of $\log|C_{m,k}(h)|$ in $h$. Differencing across two
informative modes with $\lambda_k\neq\lambda_\ell$ gives $\kappa_m$, and then $\rho_m$ follows.
\rev{With only one distinct informative eigenvalue, $(\rho_m,\kappa_m)$ cannot be separated, because
every informative mode then has the same $\log r_{m,k}$.} At $\lambda=0$, only
$\rho_m$ is identified. Two distinct but close eigenvalues can instead make identification weak.
Nearly equal channel rates complicate channel separation, a different problem. \qed

\begingroup\color{revcolor}
\paragraph{Half-lives.}\label{app:halflife}
For $\rho_m>0$, mode $k$ of field $m$ survives each period at rate
$r_{m,k}=\rho_me^{-\kappa_m\lambda_k}$, so its half-life solves $r_{m,k}^{t}=1/2$, which gives
\begin{equation}\label{eq:halflife}
t_{1/2}(k)=\frac{\ln 2}{\kappa_m\lambda_k-\ln\rho_m}\ \text{periods}.
\end{equation}
The persistence parameter alone sets the half-life of the stationary mode, $\ln 2/(-\ln\rho_m)$,
and a larger $\kappa_m$ shortens the half-life of rougher modes. When $\rho_{\mathrm{att}}=0$, the attention field has
no persistence beyond impact.\par
\endgroup

\subsection{Proposition~\ref{prop:coarse}}\label{wp:app:A.5}
\rev{In the exact diagonal benchmark}, partition the diagonal as
$\diag(C_k(h))=\Psi_h^{(K)}\oplus\Psi_h^{(tail)}$. Then
$\widetilde{\mathcal O}_h^{(S)}-\widetilde{\mathcal O}_{h,K}^{(S)}=MU_{K^c}\Psi_h^{(tail)}U_{K^c}'R$, and
submultiplicativity with $\|U_{K^c}\|_2\leq1$ gives the first term of \eqref{eq:coarse}, since the
spectral norm of a diagonal matrix is its largest absolute entry, with an empty tail contributing
zero when $K=n_C$. With off-diagonal loadings,
$\widetilde{\mathcal O}_h^{SCI}$ has the additional term $\mathcal E_{h,K}$, whose aggregate is
$\mathcal E_{h,K}^{(S)}=M\mathcal E_{h,K}R$. The triangle inequality adds $\|\mathcal E_{h,K}^{(S)}\|_2$ without a
further $\|M\|_2\|R\|_2$ factor, because the aggregation is already inside it. The direct term
$-\gamma\Phi_h$ is excluded from $\widetilde{\mathcal O}_h^{SCI}$ and hence from this bound by construction.
Under the directed benchmark of Appendix~\ref{app:directed}, the modal tail carries the extra factor
$\operatorname{cond}_2(U)$, the condition number of the eigenbasis. \qed

\begin{alem}[Retained slow and fast components]
Let\label{lem:washout} $\widetilde{\mathcal O}_{m,h}^{(S,K)}\equiv MU_K\diag(C_{m,k}(h))_{k\leq K}U_K'R$ for
$m\in\{D,\mathrm{att}\}$. Then
\begin{equation}
\begin{aligned}
\big\|\widetilde{\mathcal O}_{\mathrm{att},h}^{(S,K)}\big\|_2&\leq\|M\|_2\|R\|_2\max_{k\leq K}\{|C^{\mathrm{init}}_{\mathrm{att},k}|\,r_{\mathrm{att},k}^{h}\},\\
\big\|\widetilde{\mathcal O}_{D,h}^{(S,K)}\big\|_2&\leq\|M\|_2\|R\|_2\max_{k\leq K}\{|C^{\mathrm{init}}_{D,k}|\,r_{D,k}^{h-1}\},\quad h\geq1 .
\end{aligned}
\end{equation}
\end{alem}
\begin{proof}
Submultiplicativity and orthonormality, as above. The two upper bounds do not show that the slow
component dominates. That would also need a lower bound on the aggregated slow term: a nonzero
aggregate slow loading, no cancellation among leading modes, and a slow rate above the fast rates.
Concentration of fast loadings on rough modes is a separate assumption and does not follow from
Assumption~\ref{ass:diag}.
\end{proof}

\subsection{Proposition~\ref{prop:eq}}\label{wp:app:A.6}
In the slack region the intermediary's first-order condition is $q=\Gamma\theta/N$, so
$\theta=Nq/\Gamma$. Clearing, $\zeta q+Nq/\Gamma=Q$, gives $q=\Gamma Q/(N+\zeta\Gamma)$ and inventory
$\theta=NQ/(N+\zeta\Gamma)$, which is below the cap $N/\chi$ exactly when $Q\leq(N+\zeta\Gamma)/\chi$.
When the cap binds, $\theta=N/\chi$ and $q=(Q-N/\chi)/\zeta$. The multiplier on the cap is
$q-\Gamma/\chi$, which is nonnegative exactly when $Q\geq(N+\zeta\Gamma)/\chi$. The two branches agree
at the boundary. The intermediary's objective is strictly concave, so its choice is unique, and
aggregate demand $\zeta q+\theta(q)$ is strictly increasing in $q$, so clearing is unique at fixed
$N$. The derivatives follow by differentiation, using $\partial Q/\partial S=1$. Positive claim
prices require the additional domain restriction $E[V]>q$.

With inherited claims $\theta^-$, capital responds to the price, $N=N^--\ell_{\mathrm{mtm}}(q-q^-)$ with
$\ell_{\mathrm{mtm}}=\theta^-/R_f\geq0$. In the slack region let $\mathcal C_{N,-}=N^-+\ell_{\mathrm{mtm}} q^-+\zeta\Gamma$ and
$\Delta_-=\mathcal C_{N,-}^2-4\ell_{\mathrm{mtm}}\Gamma Q$. The lower root $q_-=(\mathcal C_{N,-}-\sqrt{\Delta_-})/(2\ell_{\mathrm{mtm}})$
is regular for $\ell_{\mathrm{mtm}}>0$ when $\Delta_->0$, capital and price are positive, and the solution
satisfies the slack-regime inequality. For $\ell_{\mathrm{mtm}}=0$, use the fixed-capital solution instead.
The implicit derivative is
$q_S=G_S/(1+\ell_{\mathrm{mtm}} G_N)$ with $1+\ell_{\mathrm{mtm}} G_N=\sqrt{\Delta_-}/(N+\zeta\Gamma)>0$ at that root. In the binding
region $(\zeta\chi-\ell_{\mathrm{mtm}})q=\chi Q-N^--\ell_{\mathrm{mtm}} q^-$, and stability requires $\zeta\chi>\ell_{\mathrm{mtm}}$. \qed

\subsection{Proposition~\ref{prop:levels}}
\label{app:levels}
At a reference point strictly inside the slack region, $q=\Gamma(Q_0+S)/(\mathcal C_N-\mathcal T)$
with $\mathcal C_N=\bar N+\zeta\Gamma$ and $q_0=\Gamma Q_0/\mathcal C_N$. Using the expansion
$(\mathcal C_N-\mathcal T)^{-1}=\mathcal C_N^{-1}(1+\mathcal T/\mathcal C_N+\mathcal T^2/\mathcal C_N^2+\cdots)$ and
collecting terms up to second order gives \eqref{eq:expansion}. The $\mathcal T^2$ term is of the
same order as $S\mathcal T$ and cannot be dropped. For the conditional change, apply $E_t$ to the
quadratic at $t+1$ and subtract its value at $t$, using
$E_t[S_{t+1}\mathcal T_{t+1}]=\rho_K\mathcal T_t\bm l'A_D\bm z_t+
\bm l'\operatorname{Cov}_t(\bm\eta_{t+1},e_{t+1})$ and
$E_t[\mathcal T_{t+1}^2]=\rho_K^2\mathcal T_t^2+\sigma^2_{e,t}$. This gives
\eqref{eq:changes} without imposing scalar closure. The identity is exact for the quadratic approximation, not for the piecewise
equilibrium, and it holds only within one smooth regime: innovations that cross the margin cap
require the piecewise conditional mean. Let
$\sigma_{Se,t}\equiv\bm l'\operatorname{Cov}_t(\bm\eta_{t+1},e_{t+1})$.
If additionally $\bm l'A_D=r\bm l'$, the scalar expression follows:
$A_S(r-1)S_t+A_{\mathcal T}(\rho_K-1)\mathcal T_t+
A_{S\mathcal T}[(r\rho_K-1)S_t\mathcal T_t+\sigma_{Se,t}]+
A_{\mathcal T\mathcal T}[(\rho_K^2-1)\mathcal T_t^2+\sigma^2_{e,t}]$.
This additional invariant-direction restriction generally fails.

\emph{Own persistence.} If, separately justified, the premium follows
$q_{t+1}=\rho_qq_t+\xi_0z_t+\xi_1\mathcal T_tz_t+u_{t+1}$ with a scalar field
$z_{t+1}=rz_t+u_{t+1}^z$ and persistence $\rho_K$ of $\mathcal T$, then one date earlier
\begin{equation}
E_{t-1}\Delta q_{t+1}=\rho_q(\rho_q-1)q_{t-1}+\xi_0(\rho_q+r-1)z_{t-1}
+\xi_1(\rho_q+r\rho_K-1)\mathcal T_{t-1}z_{t-1}+\xi_1\operatorname{Cov}_{t-1}(u_t^z,e_t),
\end{equation}
so, with $\xi_1>0$ and the conditional covariance term held fixed, the coefficient on the lagged
interaction is positive when $\rho_q+r\rho_K>1$, negative when $\rho_q+r\rho_K<1$ and zero at equality.
None of these cases is a sign theorem for a regression that, like the monthly specification,
also conditions on two lagged outcome changes and standardized regressors (Appendix~\ref{app:search}).

\emph{Level sign.} The diffusion kernel, conventionally called the graph heat kernel, is entrywise nonnegative: with
$\mathcal A\equiv D^{-1/2}AD^{-1/2}\geq0$, $e^{-\kappa_D\mathcal L}=e^{-\kappa_D}e^{\kappa_D\mathcal A}=
e^{-\kappa_D}\sum_{n\geq0}(\kappa_D\mathcal A)^n/n!\geq0$. If \rev{$\bm l\geq0$} and $B_D\bm\varepsilon_t\geq0$ (an
adverse storm raises the adverse field in hit counties), every entry of the \rev{conditional EBP response} \eqref{eq:kernel} applied to the storm is nonnegative, so the level response of the EBP is
nonnegative at every horizon, holding capital fixed. \qed

\subsection{Stationary and nonstationary financial exposure}\label{wp:app:A.8}
Because $\mathcal L=U\Lambda U'$, $A_D=\rho_De^{-\kappa_D\mathcal L}=U\diag(r_{D,k})U'$ and
$A_D^{h-1}=U\diag(r_{D,k}^{h-1})U'$, where $\Lambda$ holds the eigenvalues $\lambda_k$ on its diagonal. Hence
$(\pi^H)'A_D^{h-1}B_D=\sum_k((\pi^H)'u_k)\,r_{D,k}^{h-1}\,u_k'B_D$, and separating $k=1$ gives
\eqref{eq:split}. The stationary mode is $u_1\propto D^{1/2}\mathbf 1$, because
$\mathcal LD^{1/2}\mathbf 1=D^{1/2}\mathbf 1-D^{-1/2}A\mathbf 1=D^{1/2}\mathbf 1-D^{-1/2}D\mathbf 1=0$, so
$\lambda_1=0$ and $r_{D,1}=\rho_D$. The first term therefore weights the shock by square-root
degree. The nonstationary term vanishes when $\pi^H\propto u_1$, that is, when
$\pi^H_i\propto D_{ii}^{1/2}$. If $w'A_D=r_{D,k}w'$, then $w'A_D^{h-1}=r_{D,k}^{h-1}w'$ by induction, which
gives the collapse statement in the main text. For symmetric $A_D$, such $w$ lie in a single
eigenspace. As $\tau=\rev{\kappa_D}(h-1)\to\infty$, $e^{-\tau\mathcal L}\to u_1u_1'$, so any weighted aggregate
converges to its incidence term, which is the limit visible in Figure~\ref{fig:split}.
\rev{Every identity in this subsection holds with $\bm l$ in place of $\pi^H$, which covers loadings of either
sign and $\delta=0$.}

Signed responses are invariant to the sign of each eigenvector, because $u_k$ appears twice in each
term. An unsigned norm can also be invariant to eigenvector signs, but discards the economic
direction of a response. It is therefore not the signed pricing projection in
\eqref{eq:kernel}. Main-text equation~\eqref{eq:two_region} follows by using the two orthonormal
eigenvectors $(1,1)'/\sqrt2$ and $(1,-1)'/\sqrt2$, with eigenvalues 0 and 2.

\subsection{The common response and Proposition~\ref{prop:observability}}
\label{app:observability}
The identities \eqref{eq:common} follow from $T\mathbf 1=0$ and $w'\mathbf 1=1$. They are a
missing-direction result rather than a restriction on the common response's sign or magnitude.
An independent aggregate anchor can supply that response. A zero response under one set of weights also need not
mean zero pressure under another. With two counties, $w=(1/2,1/2)'$, $\bm z=(1,-1)'$ and
$\bm l=(1,0)'$ give $w'\bm z=0$ but $\bm l'\bm z=1$.

\emph{Proof of Proposition~\ref{prop:observability}.} If $\bm l=C_Y'a$ then
$\bm l'\bm z=a'C_Y\bm z$. Otherwise the orthogonal decomposition of $\bm l$ with respect to
$\rowsp(C_Y)$ supplies $v\in\ker(C_Y)$ with $\bm l'v\neq0$, and $\bm z$ and $\bm z+v$ have the same
measurement but different targets. \qed

\begin{aprop}[Observability with a history]
With\label{prop:history} known $A_D$, known inputs and observations at dates $0,\ldots,H_{\mathrm{obs}}$, let
$\mathsf{Obs}_{H_{\mathrm{obs}}}\equiv\big(C_Y',(C_YA_D)',\ldots,(C_YA_D^{H_{\mathrm{obs}}})'\big)'$. The initial-state functional
$\bm l'\bm z_0$ is identified if and only if $\bm l\in\rowsp(\mathsf{Obs}_{H_{\mathrm{obs}}})$. Similarly, the final-date
functional $\bm l'\bm z_{H_{\mathrm{obs}}}$ is identified if and only if $(A_D^{H_{\mathrm{obs}}})'\bm l\in\rowsp(\mathsf{Obs}_{H_{\mathrm{obs}}})$, after
known input contributions are removed. If states are restricted to the reachable subspace
$\mathcal R=\spn(B_D,A_DB_D,\ldots)$ with basis $R_{\mathcal R}$, both conditions apply to
$\mathsf{Obs}_{H_{\mathrm{obs}}}R_{\mathcal R}$ and to the target multiplied by $R_{\mathcal R}$.
\end{aprop}
\begin{proof}
Stacking the observations gives $\mathsf{Obs}_{H_{\mathrm{obs}}}\bm z_0$ plus known inputs, and $\bm z_{H_{\mathrm{obs}}}=A_D^{H_{\mathrm{obs}}}\bm z_0$
plus known inputs. Apply the null-space argument above to $\mathsf{Obs}_{H_{\mathrm{obs}}}$, with target $\bm l$ or
$(A_D^{H_{\mathrm{obs}}})'\bm l$.
\end{proof}
This identifies a functional of a known linear system. It does not separately identify $A_D$,
$B_D$ and $C_Y$ from a fitted outcome composite. A residual-maker $T$ replaces $C_Y$ by $TC_Y$ and
can remove observability.

\subsection{Recovery and estimation of the pricing projection}
\label{app:target}

\emph{Proof of Proposition~\ref{prop:recovery}.} Let $\bm z=R_{\mathcal R}x$ and
$C_{\mathcal R}=C_*R_{\mathcal R}$. If $R_{\mathcal R}'\bm l=C_{\mathcal R}'a$, then $\bm l'\bm z=a'C_*\bm z$ for every
$x$. Otherwise there exists an $x_0\in\ker C_{\mathcal R}$ with $\bm l'R_{\mathcal R}x_0\ne0$. States that
differ by $R_{\mathcal R}x_0$ have identical measured responses and different financial targets.
Applying the identity to $\bm z=R_h\bm\varepsilon$ proves \eqref{eq:recoverkernel}. \qed

The response subspace is part of the maintained specification. One must not infer its dimension
or orientation from the same financial outcomes used to test the projection. Observationally
equivalent latent systems can yield the same $H_h$. The proposition concerns their common pricing
functional only when the financial representation is independently justified. It is not a
system-identification theorem for unrestricted $(A_D,B_D,C_Y,\bm l)$.

With full-state spanning, $C_*'a=\bm l$, the representation holds for any propagation path.
Restricting $\mathcal R$ can permit recovery with fewer measurements, but the representation then
inherits those restrictions. Responses outside $\mathcal R$ require a separate check or bound.

\paragraph{Alternative household measurements.}
The map $C_*$ describes an observation of the household state, rather than a particular data
source. If $C_c$ maps that state into consumption responses and $C_p$ into fixed-price portfolio
demand, jointly observing them gives $C_*=\big(C_c',C_p'\big)'$. Proposition~\ref{prop:recovery}
then applies to this stacked map with the same proof. If a measured demand row is $\bm l'$ itself,
it directly reveals the financial target. Otherwise, the spanning condition determines which
combination of observations suffices. Actual holdings require valuation and price-feedback
adjustments before they represent fixed-price demand. Similarly, expenditure and consumption have
different maps when purchases provide services over time. Such known temporal filters can be
written as measurements of an augmented state. The empirical application and recovery simulations
use spending measurements. The conditional holdings problem in Section~\ref{sec:pricing}
provides the local portfolio-demand schedule, taking household circumstances as given.
Appendix~\ref{app:financialmargin} derives local joint consumption--portfolio responses under stated
conditions. Extending those decision-date restrictions across horizons requires additional dynamics.

\paragraph{More than one financially loaded channel.}
The slow-only credit-pricing benchmark sets the financial loading on attention to zero. More
generally, let the fixed-price demand shift be
$S_t=\bm l_D'\bm z_t+\bm l_{\mathrm{att}}'\bm b_t$, allowing signed $\bm l_{\mathrm{att}}$, and put $B_{\mathrm{att}}=\eta_{\mathrm{att}} B$.
For a one-time shock and $h\geq1$, the conditional pricing response is
\[
K_h=c_E\big[\bm l_D'A_D^{h-1}B_D+\bm l_{\mathrm{att}}'A_{\mathrm{att}}^hB_{\mathrm{att}}\big].
\]
At impact only the fast term can respond through these fields. Stacking the two fields and
their measurement maps gives the same recovery condition on their joint response subspace.
In the exact \rev{diagonal} benchmark each mode again has a two-exponential profile, now with coefficients
$c_E(\bm l_m'u_k)(u_k'B_mu_k)$, $m\in\{D,\mathrm{att}\}$. The rank-one-minor criterion of Theorem~\ref{thm:no_scalar} applies
unchanged, and when its nondegeneracy conditions hold for these coefficients, rank one generically fails.
The channel paths here are the maintained laws in Section~\ref{sec:model}. Measuring an
additional outcome changes what can be observed, while assigning it an additional propagation
law is a separate modeling choice. The reported simulations implement the slow-channel benchmark.

\paragraph{Growth responses versus the level target.}
For the same outcome, shock and units, let $H_h^L$ denote its level-loss response and $H_h^\Delta$
its first-difference response. Under no anticipation and a known pre-shock baseline,
\begin{equation}
H_h^\Delta=H_h^L-H_{h-1}^L,\qquad
H_h^L=\sum_{j=0}^h H_j^\Delta,\qquad H_{-1}^L=0 .
\end{equation}
The accumulated estimator's covariance contains all cross-horizon covariances, not just the
sum of horizon-specific variances. The modal-response estimates have the sign of spending
growth, so a loss convention reverses their sign. They also use clipped growth and fixed-effect
residualization. The identity therefore does not turn those transformed estimates into unclipped national
expenditure or restore omitted event means. Recovering the desired $H_h$ requires a consistent
measurement map, outcome definition, temporal aggregation and common-response anchor. A mode's
growth response cannot simply be inserted as its level-demand loading in the \rev{conditional EBP response}.

\paragraph{An aggregate anchor and the nuisance common response.}
For illustration, let $C_Y=\mathrm{Id}_{n_C}$ and let $T$ have rank $n_C-1$ with $T\mathbf1=0$. Then stacking $T$
and a row $w'$ with $w'\mathbf1=1$ gives a full-column-rank measurement map, so the missing mean is
measured. If the observed aggregate also contains unrestricted $a_e$, however, its value is
$w'R_h\bm\varepsilon_e+a_e$, not $w'R_h\bm\varepsilon_e$. The transformation
$R_h\mapsto R_h+\mathbf1\bm\alpha'$ and $a_e\mapsto a_e-\bm\alpha'\bm\varepsilon_e$ preserves the observations.
Thus an aggregate anchor needs an exclusion, a separately measured competing response or a bound
on $a_e$. If only $a_e\in[\underline a_e,\overline a_e]$ is justified, the recovered aggregate
household response is an interval of the same width, not a point estimate.

\paragraph{Error and partial identification.}
With $r_l=\bm l-C_*'a$, subtraction gives
\begin{equation}
\hat p_h-K_h\bm\varepsilon
=c_Ea'(\hat H_h-H_h)\bm\varepsilon-c_Er_l'R_h\bm\varepsilon .
\end{equation}
The triangle and Cauchy--Schwarz inequalities give \eqref{eq:financialbound}. If $P_{\mathcal R}$
is the orthogonal projector onto the response subspace, replace $r_l$ by $P_{\mathcal R}r_l$.
Consequently, independent bounds $\|P_{\mathcal R}r_l\|\leq\bar r$ and
$\|R_h\bm\varepsilon\|\leq\bar z_h$ give a conditional identification interval of half-width
$c_E\bar r\bar z_h$ around $c_Ea'H_h\bm\varepsilon$, before sampling uncertainty. Without such
bounds an unobserved direction cannot be made economically small by better measurement of the
observed directions.
The norm bound allows the least favorable alignment of errors. Independently justified error
bounds make it a quantitative, potentially conservative bound.

When $a$ or $c_E$ is estimated, the plug-in error also contains
$(\hat c_E\hat a-c_Ea)'\hat H_h\bm\varepsilon$. Their uncertainty and any common training-data
dependence must be propagated jointly. Cross-fitting the spending operator alone does not
remove errors-in-variables bias in the financial regression.

\paragraph{Choosing a representation without using the financial outcome.}
The representation $a$ need not be unique. Given $C_{\mathcal R}$, $b=R_{\mathcal R}'\bm l$ and a
positive-definite covariance $V_g$ for an estimated measurement response, a minimum-variance
choice solves
\begin{equation}
\min_a a'V_ga\quad\text{subject to }C_{\mathcal R}'a=b,
\qquad
a^*=V_g^{-1}C_{\mathcal R}(C_{\mathcal R}'V_g^{-1}C_{\mathcal R})^\dagger b .
\end{equation}
The dagger is the Moore--Penrose inverse. Feasibility requires $b\in\rowsp(C_{\mathcal R})$, precisely the
spanning condition. The formula follows from the first-order condition $V_ga=C_{\mathcal R}\mu_{\mathrm{rec}}$ and
the linear constraint. This is a standard constrained quadratic calculation, used here to
target the household-demand projection. It does not estimate an unknown $\bm l$ or authorize
selecting $a$ to fit the EBP. Estimate the covariance and any tuning choices in training data.
Use nonredundant measurement rows, or a justified covariance regularization, in
\eqref{eq:targetfit}, since $V^{-1}$ requires an invertible covariance.

For fixed maps, the covariance criterion selects precision among valid representations, rather
than minimizing coefficient size. Adding valid measurement rows cannot increase the optimal
population variance when the existing measurements and their covariance block remain unchanged,
because the old weights, extended by zeros, remain feasible. Strict improvement requires added rows that load on the response subspace or whose errors correlate with
existing measurement errors.
Large weights may be unavoidable when a financially important state barely moves the measured
outcome. Under approximate spanning, weights must instead balance sampling uncertainty against
the projected residual financial direction in \eqref{eq:financialbound}.

\paragraph{Which modes require precision?}
Conditionally on $a$ and $c_E$, if $\Sigma_h$ is the full covariance of
$(\hat H_h-H_h)\bm\varepsilon$, the variance of the observable projection error is
$c_E^2a'\Sigma_ha$. In the illustrative full-field case $C_*=\mathrm{Id}_{n_C}$ and $a=\bm l$, let that error
be $U\bm e_h^U$ in the orthonormal eigenbasis and put $\bm l^U=U'\bm l$. Then
\begin{equation}
\operatorname{Var}(\hat p_h-p_h)
=c_E^2(\bm l^U)'\operatorname{Var}(\bm e_h^U)\bm l^U
=c_E^2\sum_{k,j}l_k^Ul_j^U\operatorname{Cov}(e_{k,h}^U,e_{j,h}^U).
\end{equation}
A sum of individual mode variances is insufficient unless cross-mode covariances vanish. A
mode with small unweighted measurement variance can matter financially when $|l_k^U|$ is large.

Differentiating $r_{D,k}^{h-1}$ while holding $\rho_D$, $B_D$, $\bm l$ and $c_E$ fixed gives
\eqref{eq:kappaprice}. Since $\lambda_1=0$, the stationary term has zero derivative. The
nonstationary products $(\bm l'u_k)(u_k'B_D\bm\varepsilon)$ can have either sign, so an increase
in diffusion need not increase the financial response even with nonnegative county inputs and
loadings.

\subsection{Observable dynamics and cardinal nonseparability}
\label{app:observation}

\paragraph{Timing and continued forcing.}
\rev{The direct consumption responses $\{\Phi_h\}$ of Section~\ref{subsec:fields} are
assumed summable, $\sum_{h\geq0}\|\Phi_h\|<\infty$.} In the benchmark sequence, the shock at $t$ moves the attention field at $t$ and the slow demand
field at $t+1$, after which propagation is autonomous. A continued-input alternative is
\begin{equation}
\bm d_{t+h}=\eta_D\sum_{j=0}^{h-1}A_D^{h-1-j}\Phi_j\bm\varepsilon_t,\qquad h\geq1.
\end{equation}
This follows by substitution into $d_{t+1}=A_Dd_t+\eta_D\nu_t$ with
$\nu_{t+j}=\Phi_j\varepsilon_t$. If one mode has transition $r$ but is repeatedly forced
at rate $r_{\mathrm{in}}$, its response is proportional to
\[
\sum_{j=0}^{h-1}r^{h-1-j}r_{\mathrm{in}}^j
=\frac{r^h-r_{\mathrm{in}}^h}{r-r_{\mathrm{in}}},\quad r\ne r_{\mathrm{in}},
\]
and equals $hr^{h-1}$ when $r=r_{\mathrm{in}}$. Thus a two-exponential profile need not mean two fields.
Discrimination requires observed inputs, an input-duration exclusion, or a restricted
forcing family tested against the same outcomes. Controls do not make an unrestricted
$\Phi_j$ disappear. Different transmission graphs supply another competing explanation.
For stationary repeated shocks, stability must be combined with suitable stationary inputs
and finite innovation moments, because impulse summability alone is not a stochastic existence proof.

\begin{alem}[Observation restriction]
Suppose $A_D=U\operatorname{diag}(r_{D,k})U'$, $C_Y=U\operatorname{diag}(c_k)U'$ and
$B_D=U\operatorname{diag}(b_k^D)U'$ on all excited and observed directions. Let a known linear
operator $\mathcal F_{e,h}$ apply the actual observation mask, temporal filter and
residualization to the response, and apply exactly the same operator to every modal design
column. Suppose omitted terms are zero or included in the nuisance span, the residualized
design has full column rank, and its columns are orthogonal in conditional expectation to
the observation error. Then joint least squares identifies the coefficients of the
filtered observable model. Without temporal filtering these are
$c_kb_k^Dr_{D,k}^{h-1}$ for the spending-loss level at $h\geq1$.
\end{alem}

\emph{Proof.} Expand $C_YA_D^{h-1}B_D=\sum_k c_kb_k^Dr_{D,k}^{h-1}u_ku_k'$.
Linearity gives the same expansion after applying $\mathcal F_{e,h}$. The rank and
orthogonality conditions identify its coefficients by the normal equations. Missing
counties need not commute with the graph if the same mask is applied to the correct
design. Selecting outcomes or omitting relevant modes can violate orthogonality. \qed

This is a sufficient condition, not a claim that the empirical regressions satisfy it.
With general measurement and input loadings,
\[
(U'H_h^LU)_{kk}=\sum_j (U'C_YU)_{kj}r_{D,j}^{h-1}(U'B_DU)_{jk}.
\]
Even one field can mix several rates in an observed diagonal entry. Exact diagonal input
loading alone simplifies that entry but does not remove all off-diagonal outcome responses
from a masked regression. For county-specific direct impacts
$\Phi_0=\operatorname{diag}(\phi_i)$, commuting globally with a connected Laplacian requires
$(\phi_i-\phi_j)\mathcal L_{ij}=0$ along every edge and hence homogeneous $\phi_i$.
Retained-subspace approximate diagonality is weaker, but needs an explicit residual check.

\paragraph{Levels, growth and the observation filter.}
Take the adverse field to load on the spending-loss level, $H_h^L=C_YA_D^{h-1}B_D$
for $h\geq1$, with $H_0^L=0$. Then
\[
H_1^\Delta=C_YB_D,\qquad
H_h^\Delta=C_Y(A_D-\mathrm{Id}_{n_C})A_D^{h-2}B_D\quad(h\geq2).
\]
A scalar decaying level therefore has an initial jump $C$ followed by
$C(r-1)r^{h-2}$, of the opposite sign when $0<r<1$. Conversely, a growth response
$Cr^{h-1}$ accumulates to $C(1-r^h)/(1-r)$ and leaves a permanent displacement.
Selecting a same-sign growth tail cannot decide between these level models.
If published levels satisfy $\bar y_t=\sum_{j=0}^Jg_j^{\mathrm{filter}}y_{t-j}$, apply this filter before
differencing model predictions. Away from onset a geometric response has multiplier
$\sum_jg_j^{\mathrm{filter}}r^{-j}$, but onset and a change in $g_j^{\mathrm{filter}}$ require the actual filter.
Nonlinear clipping is not a linear observation map. The clipped-growth
surface is therefore retained as an estimator diagnostic, not a \rev{level-demand response}.

\paragraph{How much does rank one miss?}
Let $\mathsf K_{\mathrm{stack}}$ stack the pricing rows $\sqrt{w_h}K_h\Sigma_\varepsilon^{1/2}$ over fixed horizons,
with prespecified nonnegative horizon weights and shock second-moment matrix.
For $\|\mathsf K_{\mathrm{stack}}\|_F>0$, the relative best rank-one error is
\[
\mathcal E_1^2=\frac{\min_{\operatorname{rank}(G)\leq1}\|\mathsf K_{\mathrm{stack}}-G\|_F^2}{\|\mathsf K_{\mathrm{stack}}\|_F^2}
=1-\frac{\sigma_1(\mathsf K_{\mathrm{stack}})^2}{\sum_j\sigma_j(\mathsf K_{\mathrm{stack}})^2}.
\]
This is the standard singular-value approximation identity. Nonseparability says the
error is generically nonzero, not that it is large, precisely estimated, or improves
prediction. It says nothing about ordinal reversals. For example, $(2,1)$ and $(3,1)$ are nonproportional
but retain the same ordering. Estimated response kernels require held-out loss comparisons and joint
uncertainty. With repeated eigenvalues, interpret projected eigenspaces, because individual
eigenvectors within them have no unique orientation.

\subsection{Aggregation allocation and omitted-node memory}
\label{app:allocation}

Let $O_h=\widetilde{\mathcal O}^{SCI}_h$ and let $O_{h,K}$ be its approximation. For an
actual county shock, the exact identity is
\begin{equation}
MO_h\varepsilon-MO_{h,K}RM\varepsilon
=M(O_h-O_{h,K})RM\varepsilon+MO_h(\mathrm{Id}_{n_C}-RM)\varepsilon.
\end{equation}
The first term concerns approximation of the specified lifted state shock. The second
concerns its wrong county allocation. Its norm is at most
$\|MO_h(\mathrm{Id}_{n_C}-RM)\|_2\|\varepsilon\|_2$, but without bounds on allocation this need not be
small. The identity $MR=\mathrm{Id}_{n_S}$ imposes no such bound. Similarly,
\[
MA_D^2R-(MA_DR)^2=MA_D(\mathrm{Id}_{n_C}-RM)A_DR.
\]
Closure requires, for example, $A_DR=R\bar A_D$ with $\bar A_D=MA_DR$. A right inverse alone
does not make state aggregates Markov. The remainder term in Proposition~\ref{prop:coarse}
must also be bounded before that proposition provides a quantitative guarantee.

Partition a national field into observed $\mathcal V_{\mathrm{obs}}$ and omitted
$\mathcal V_{\mathrm{omit}}$ counties. Use transition blocks
$(A_D)_{\mathrm{obs,obs}}$, $(A_D)_{\mathrm{obs,omit}}$,
$(A_D)_{\mathrm{omit,obs}}$, $(A_D)_{\mathrm{omit,omit}}$ and input blocks
$(B_D)_{\mathrm{obs}}$, $(B_D)_{\mathrm{omit}}$. Eliminating the omitted state gives
\begin{align*}
z_{\mathrm{obs},t+1}={}&(A_D)_{\mathrm{obs,obs}}z_{\mathrm{obs},t}
+(B_D)_{\mathrm{obs}}\varepsilon_t
+(A_D)_{\mathrm{obs,omit}}(A_D)_{\mathrm{omit,omit}}^tz_{\mathrm{omit},0}\\
&+\sum_{j=0}^{t-1}(A_D)_{\mathrm{obs,omit}}(A_D)_{\mathrm{omit,omit}}^{j}
  \bigl((A_D)_{\mathrm{omit,obs}}z_{\mathrm{obs},t-1-j}
  +(B_D)_{\mathrm{omit}}\varepsilon_{t-1-j}\bigr).
\end{align*}
This follows by solving the omitted recursion and substituting it into the observed one.
Even a single first-order national field gives the observed subsystem distributed memory.
Inducing and renormalizing a graph on observed outcomes is not equivalent to selecting
observations after propagation. Source completion helps when it restores measured inputs
and paths, but does not certify that a retained graph is the national economic system.

\subsection{No-diffusion comparison, sharp common-direction bounds and relative demand}
\label{app:counterfactual}

Equation~\eqref{eq:diffincrement} follows by subtracting the fixed-input counterfactual
$A_D^0=\rho_D\mathrm{Id}_{n_C}$. In graph coordinates it is
\[
\Delta K_h^{diff}=c_E\rho_D^{h-1}\sum_{k\geq2}
(e^{-\kappa_D(h-1)\lambda_k}-1)(l'u_k)u_k'B_D.
\]
The stationary term is zero, but a nonstationary exposure is not itself this difference.
At $h=1$ the entire difference is zero even with unequal initial financial exposures.
The economic contrast holds inputs, financial loadings and price impact fixed. Thus, if $B_D$
already contains nonlocal transmission, it removes subsequent diffusion only. Changing
topology is a separate intervention and needs a matched degree/input benchmark.

\begin{aprop}[A missing common direction]
Suppose $C_Y=\mathrm{Id}_{n_C}$, $\ker T=\operatorname{span}\{\mathbf1\}$ and the contrast $g=Tz$ is known.
Choose the unique $z_\perp$ with $Tz_\perp=g$ and $\mathbf1'z_\perp=0$. Then every observationally
equivalent field is $z=z_\perp+\alpha\mathbf1$. Given known \rev{$l$ with $\mathbf1'l>0$} and
an independently justified interval $\alpha\in[\underline\alpha,\overline\alpha]$, the sharp
identified interval for demand pressure is
\[
\left[l'z_\perp+(\mathbf1'l)\underline\alpha,
      l'z_\perp+(\mathbf1'l)\overline\alpha\right].
\]
Without a bound on $\alpha$, that functional is unbounded even with error-free contrasts.
\rev{If $\mathbf1'l<0$ the two endpoints change places. If $\mathbf1'l=0$ the functional equals $l'z_\perp$
and is identified without any bound.}
\end{aprop}

\emph{Proof.} The null space characterizes every solution. Substitution gives an affine
\rev{function of $\alpha$ with slope $\mathbf1'l$. It is strictly monotone when the slope is nonzero}, so both endpoints and every point between them are
attained\rev{, and it is constant when the slope is zero}. Sharpness is relative to these restrictions, without additional county bounds. \qed

The interval separates a missing direction from sampling error. It is not an empirical
bound until the common-response bounds and $l$ are independently supported. Observed
national spending with an unrestricted common nuisance cannot supply those bounds.
With additional restrictions such as county-level sign or magnitude limits, intersect the
feasible set before taking extrema. Estimation error in contrasts, loadings and price
impact requires simultaneous confidence sets rather than plugging in point estimates.

For two assets with known claim-demand loadings $\bm l_a,\bm l_b$ and
$\mathbf1'\bm l_b\ne0$, let
\[
r_{ab}^{\mathrm{load}}=\frac{\mathbf1'\bm l_a}{\mathbf1'\bm l_b},\qquad
\Delta\bm l_{ab}=\bm l_a-r_{ab}^{\mathrm{load}}\bm l_b.
\]
Then $\mathbf1'\Delta\bm l_{ab}=0$, so the contrast $(\Delta\bm l_{ab})'z$ is identified
by $Tz$ on this unrestricted state space even when the two aggregate pressures are not.
This is a relative-demand result, not automatically a relative-price result. Let
$c_a^{\mathrm{claim}}=P_ac_a$ and $c_b^{\mathrm{claim}}=P_bc_b$ be the claim-unit market
maps corresponding to the dollar-unit maps in Section~\ref{sec:proposal}. If these maps
are separately identified and competing capital and payoff components are removed,
\[
\frac{\Delta p_a}{c_a^{\mathrm{claim}}}
-r_{ab}^{\mathrm{load}}\frac{\Delta p_b}{c_b^{\mathrm{claim}}}
=(\Delta\bm l_{ab})'z.
\]
A multivariate absorption matrix requires the corresponding known linear transformation.
Dividend income does not supply the required holdings
map, demand sensitivities or absorption. This extension does not rescue national EBP.

\subsection{Assumptions and uncertainty}
\label{app:unitsnotation}

The conditional equilibrium is restricted to positive capital and claim prices, interior
household demand, fixed payoff moments and a specified smooth regime. Its uniqueness at
fixed capital is not global uniqueness with endogenous capital. The surplus-wealth
illustration is first order, since expanding $x_i=(\bar s_i-m_iz_i)q/(\gamma_i\sigma_i^2)$ also
produces $-m_iz_i(q-\bar q)/(\gamma_i\sigma_i^2)$. The second-order coefficients in
Proposition~\ref{prop:levels} belong to the primitive linear-demand schedule, not a
second-order derivation of that illustration. A stationary capital law further requires
$|\rho_K|<1$, suitable innovations and a flow/payout rule consistent with its accounts.

Joint response fitting should minimize a quadratic loss on the outcome coefficients without sign-
or size-based selection, with their full cross-mode and cross-horizon covariance. Near-zero coefficients
are not logged or discarded by sign. Covariance regularization is selected using training
data, and stability constraints are maintained restrictions, not evidence for diffusion.
Whole-calendar blocks preserve common shocks across recipients and overlapping event windows.
However, conditional loss resampling does not cover re-estimation and model selection. A full
procedure must repeat those steps inside every inference or signal-injection replication.
Oracle and feasible power must use the same observation masks, continued-input alternatives
and economic signal scale. An uninformative held-out comparison does not reject propagation.

\section{Model details}
\label{app:model}

\subsection{Retained-mode diagonality}
\label{app:diag}
\begingroup\emergencystretch=3em %
Let $U_K$ collect the first $K$ eigenvectors and $U_{K^c}$ the rest. Assumption~\ref{ass:diag}
states that $U_K'\Phi_0U_K=\diag(\phi_{0,k})_{k\leq K}+E_{\Phi,K}$ and
$U_K'BU_K=\diag(b_k^{\mathrm{att}})_{k\leq K}+E_{B,K}$, with $\|E_{\Phi,K}\|_2$ and $\|E_{B,K}\|_2$ small relative to
the diagonal entries, and that the cross-block terms $U_{K^c}'\Phi_0U_K$, $U_K'\Phi_0U_{K^c}$,
$U_{K^c}'BU_K$ and $U_K'BU_{K^c}$ are small relative to the retained block. The exact diagonal
benchmark sets all off-diagonal terms to zero. The assumption restricts retained and cross-block
terms, not the entire discarded block. Spending-side closure diagnostics are proxies for these
magnitudes rather than identifying every structural entry. The same conditions are assumed for the
production basis.
\par\endgroup

\subsection{The production field}
\label{app:prod}
The competing slow field runs on a symmetric production affinity $A^P$ with positive degrees
$D^P$:
\begin{equation}
\bm d_{t+1}^P=A_D^P\bm d_t^P+\eta_D^P\bm\nu_t,\qquad A_D^P\equiv\rho_D^Pe^{-\kappa_D^P\mathcal L^P},\qquad
\mathcal L^P\equiv \mathrm{Id}_{n_C}-(D^P)^{-1/2}A^P(D^P)^{-1/2},
\end{equation}
with $\rho_D^P\in(0,1)$, $\kappa_D^P>0$, $\eta_D^P\geq0$. The empirical exposures are the directed
operators $W^{up}$ and $W^{down}$ of Section~\ref{sec:stage1}, while the spectral analysis uses their
symmetrized affinity. Excluding production from the household-demand equation is a restriction,
not evidence that production cannot affect capital, payoff risk or credit premia. Social
specificity therefore requires a matched incremental comparison on common shocks, counties,
targets and information sets. Separate identification weakens when the leading modal spaces are
nearly collinear. The canonical correlations between the retained SCI and production modal spaces,
and the projection distance $\|U_KU_K'-U_K^{up}(U_K^{up})'\|_F$ with its downstream analogue,
measure this.

\subsection{The directed benchmark}
\label{app:directed}
The operator formulation does not need a symmetric graph. The directed benchmark replaces
$\mathcal L$ by $\mathrm{Id}_{n_C}-W$. Its eigenbasis is generally bi-orthogonal, and if $W$ is non-normal its
eigenvectors can be ill-conditioned and transient amplification can matter. The diagnostic
$\delta_N(W)\equiv\|WW'-W'W\|_F/\|W\|_F^2$, together with the condition number of the retained basis,
diagnoses that eigenvector representation, not whether the directed transition is usable.
A matrix exponential or Schur calculation avoids inverting an unstable eigenbasis.

Symmetrization imposes an economic restriction on the direction of transmission and changes the
transition matrix. If $A=(W+W')/2$, the reversible walk is
$W^{rev}=D^{-1}A$, and $D^{-1/2}AD^{-1/2}=D^{1/2}W^{rev}D^{-1/2}$. Thus the diffusion kernel's first-order neighbor term is a degree-rescaled version of $W^{rev}$, not the pruned,
row-normalized empirical $W$. A one-hop spending coefficient is not an estimate of $\rev{\kappa_D}$.

\paragraph{Path weights and behavioral strength.}
If $W$ is a neighbor-averaging rule, a two-step response uses
$(W^2)_{ij}=\sum_kW_{ik}W_{kj}$, so link weights multiply along each path and
parallel paths add. For nonnegative row-stochastic $W$, $W^2\mathbf1=\mathbf1$.
Hence a small product on one path does not imply loss of total row weight.
These are weighted walks, including repeated visits, rather than counts of
distinct friends. The SCI becomes such a transition only under a behavioral
interpretation and a specified normalization.

For example, with $z_{t+1}=W_{\mathrm{hop}}z_t$ and
$W_{\mathrm{hop}}=(1-\alpha)\mathrm{Id}_{n_C}+\alpha W$, $0<\alpha<1$, the binomial expansion gives
\begin{equation}
W_{\mathrm{hop}}^h=\sum_{j=0}^h\omega_j^{\mathrm{hop}}(h)W^j,\qquad
\omega_j^{\mathrm{hop}}(h)=\binom{h}{j}\alpha^j(1-\alpha)^{h-j},\qquad
\frac{\omega_2^{\mathrm{hop}}(h)}{\omega_1^{\mathrm{hop}}(h)}
=\frac{(h-1)\alpha}{2(1-\alpha)},\quad h\geq2 .
\end{equation}
Thus relative hop contributions depend on the response speed and horizon,
not an arbitrary discount on a second connection. Multiplying the whole
transition by a persistence factor leaves this ratio unchanged. Halving a
regressor with an unrestricted coefficient merely doubles that coefficient,
while fitted values and its $t$-statistic are unchanged. Distinguishing transmission
laws therefore requires restrictions across horizons or measured behavior,
rather than rescaling an exposure.

\subsection{Capital dynamics}
\label{app:capital}
Intermediary capital obeys the accounting identity
\begin{equation}
N_{t+1}=R_fN_t+\theta_t(V_{t+1}-R_fP_t)+F_{t+1}^K,
\end{equation}
where $F^K$ collects net external flows. At fixed current capital, expected trading earnings
$\theta_tq_t$ rise with pressure in both regions, with $\partial_S(\theta q)=\theta G_S+q\theta_S>0$,
$\theta_S=N/(N+\zeta\Gamma)$ in the slack region and $0$ when the cap binds. Inherited mark-to-market
losses can still reduce capital along an endogenous path. These are different derivatives, and
neither identifies a capital law.

Within one smooth regime, let $k_t$ be the deviation of capital from its reference. Its law and the premium relation are
\begin{equation}
k_{t+1}=\rho_Kk_t+\bm c_K'\bm z_t+D_K\bm\varepsilon_t+u_{t+1},\qquad
\hat q_t=\bm g_q'\bm z_t-b_Kk_t,\qquad \bm g_q=G_S\bm l,\ b_K=-G_N>0 .
\end{equation}
The vector $\bm c_K$ depends on retention, payouts, inherited exposures and capital flows. With
\begin{equation}
A_{\mathrm{aug}}=\begin{pmatrix}A_D&0\\ \bm c_K'&\rho_K\end{pmatrix},\qquad
B_{\mathrm{aug}}=\begin{pmatrix}B_D\\ D_K\end{pmatrix},\qquad
C_{\mathrm{aug}}^E=b_E(\bm g_q',-b_K),
\end{equation}
the EBP impulse response is $IR_h^{EBP}=C_{\mathrm{aug}}^EA_{\mathrm{aug}}^{h-1}B_{\mathrm{aug}}$. With $D_K=0$, block-matrix
powers give
\begin{equation}
IR_h^{EBP}=b_E\Big[\bm g_q'A_D^{h-1}B_D-b_K\sum_{j=0}^{h-2}\rho_K^{h-2-j}\,\bm c_K'A_D^jB_D\Big],
\end{equation}
where the sum is empty at $h=1$. Capital can offset or reinforce the fixed-capital \rev{response}. A scalar
example shows why autocorrelations do not sign the dynamics. Let $q_t=a_zz_t-a_kk_t$ with
$a_z,a_k>0$, $z_{t+1}=rz_t+u_{t+1}^z$ and $k_{t+1}=\rho_Kk_t-c_zz_t+u_{t+1}^K$. Then
\begin{equation}
q_{t+1}=\rho_Kq_t+\big[a_z(r-\rho_K)+a_kc_z\big]z_t+a_zu_{t+1}^z-a_ku_{t+1}^K:
\end{equation}
If $r<\rho_K$, the lagged-field coefficient is negative at $c_z=0$ and becomes positive only if
$a_kc_z>a_z(\rho_K-r)$. Neither coefficient alone signs the total change $q_{t+1}-q_t$.
The autocorrelations of the EBP and of an estimated field do not identify $\rho_K$ or $c_z$.

\subsection{The units bridge}
\label{app:units}
Holdings and $S$ are quantities of the risky claim, while $\bar P S$ is the corresponding fixed-price
dollar demand shift. It is not executed dollar selling, because the price response induces
offsetting household demand as in main-text equation~\eqref{eq:absorption}.
Holding $E[V]$ and $R_f$ fixed in $q=E[V]-R_fP$ gives $\Delta q=-R_f\Delta P$. Modified duration
implies $\Delta\text{yield}\simeq-\Delta P/(PD_{mod})$. Appendix~\ref{app:visibility} combines this
relation with residual absorption to convert a fixed-price dollar demand shift into basis points,
\eqref{eq:units}, and defines its symbols. For EBP measured in basis points,
the local loading in the \rev{conditional EBP response} is $b_E=10^4\varphi_E/(R_f\bar P D_{mod})$ under these
fixed-payoff assumptions. It is not a unit-free equality between EBP and $q$. Market coverage, issuer mix and
market-value conventions must match, and a separate fire-sale multiplier would count feedback
already in $\xi_P$. Economy-wide net trades sum to zero, so total trading volume is not the demand
shift.

In the surplus-wealth illustration \eqref{eq:surplus}, if observed spending loss is $c_i^Y z_i$
with $c_i^Y>0$, the fixed-price credit-demand shift per
dollar of spending is $\lambda^C_i=\bar P\bar qm_i/(\gamma_i\sigma_i^2c_i^Y)$, which requires evidence on
wealth loadings, spending loadings, participation and portfolio risk that is independent of the
EBP. For independently justified nonnegative bounds on the parameters of \eqref{eq:units}, the smallest response uses
the lower bounds on $\varphi_E,\lambda_C,C_{loss}$ and the upper bounds on $A_{mkt},\xi_P,D_{mod}$,
and the largest reverses them. If $\lambda_C$ or $\varphi_E$ can be zero, the lower bound is zero.

\subsection{Power planning}
\label{app:power}
In the Gaussian linear model with $n_{\mathrm{obs}}$ stacked observations,
$y=\mathsf X^{\mathrm{sig}}\theta^{\mathrm{sig}}+\mathsf Z^{\mathrm{ctrl}}\phi^{\mathrm{ctrl}}+\upsilon^{\mathrm{plan}}$
with known $\Sigma=\mathrm{Var}(\upsilon^{\mathrm{plan}})>0$, let
$\mathsf X_w^{\mathrm{sig}}=\Sigma^{-1/2}\mathsf X^{\mathrm{sig}}$ and
$\mathsf Z_w^{\mathrm{ctrl}}=\Sigma^{-1/2}\mathsf Z^{\mathrm{ctrl}}$. Profiling out the nuisance parameters,
$\mathcal I=(\mathsf X_w^{\mathrm{sig}})'(\mathrm{Id}_{n_{\mathrm{obs}}}-P_{\mathsf Z_w^{\mathrm{ctrl}}})\mathsf X_w^{\mathrm{sig}}$, and an unbiased estimator satisfies
$\mathrm{Var}(\widehat{a'\theta^{\mathrm{sig}}})\geq a'\mathcal I^+a$ for
$a\in\mathrm{range}(\mathcal I)$. Noise, weak excitation of the relevant modes and collinearity with
nuisance regressors are separate sources of lost information.

If the estimated state is
$\hat x=x+u$ with independent classical error, unconditional reliability $\mathrm{Rel}$, and controls that
explain a share $R_x^2$ of the true state's variance, then
\begin{equation}
\mathrm{Rel}_{\mathrm{cond}}=\frac{1-R_x^2}{1-R_x^2+(1-\mathrm{Rel})/\mathrm{Rel}},\qquad \mathrm{plim}\,\hat\beta_{naive}=\beta \mathrm{Rel}_{\mathrm{cond}}.
\end{equation}
Regression calibration corrects the attenuation but leaves the zero-slope statistic unchanged.

For a perfectly measured state, effect $B$, outcome noise $\sigma$ and equivalence band
$\pm c_{eq}B$, a normal planning approximation is
\begin{equation}
n\gtrsim\frac{1}{1-R_x^2}\left[\frac{(z_{1-\alpha/2}+z_{0.90})\,\sigma}{c_{eq}B}\right]^2 .
\end{equation}
At $\alpha=0.025$, $c_{eq}=0.5$ and $B=\sigma=10$ basis points this gives about 50 independent
episodes with $R_x^2=0$ and 249 with $R_x^2=0.8$. In a simulation with an imposed 10-basis-point
effect, 10 basis points of noise and 60 independent episodes, joint detection-and-calibration
power is 84.6 percent with a perfect state and orthogonal controls, 0.2 percent when controls
explain 80 percent of the state's variance, and zero when the state's reliability is also 0.03.
These are sensitivities of the stated model, not estimates of this paper's power.

\subsection{Mixing and within-event information}
\label{app:mixing}

The common-response identity in main-text equation~\eqref{eq:common}
has a geometric counterpart. For a propagated field
$z_e(\tau)=e^{-\tau\mathcal L}b_e$ and recipient set $R_e$, an event
intercept removes the fraction
\begin{equation}
f_e^{\mathrm{mean}}(\tau)=\frac{|R_e|\,\bar z_{e,R_e}(\tau)^2}
{\sum_{i\in R_e}z_{ei}(\tau)^2}
\end{equation}
of its squared norm. This is a projection onto a constant within one event,
not the variation removed jointly by all regression controls and fixed effects.
For a connected symmetric normalized Laplacian with positive degrees $d_i^{\mathrm{deg}}$
and a nonzero nonnegative seed, the limiting field is proportional to
$\sqrt{d_i^{\mathrm{deg}}}$. Hence
\begin{equation}
f_e^{\mathrm{mean}}(\infty)=\frac{\left(|R_e|^{-1}\sum_{i\in R_e}\sqrt{d_i^{\mathrm{deg}}}\right)^2}
{|R_e|^{-1}\sum_{i\in R_e}d_i^{\mathrm{deg}}}.
\end{equation}
The limit need not equal one, because the stationary field of this symmetric
normalization is not constant.

Table~\ref{tab:mixing} illustrates this calculation on the 1,590-county
graph, using onset-week recorded-damage shares as seeds and the primary
expanded-footprint recipient screen. Of the paper's 48 declarations, 38 have
positive recorded-damage seed mass on this graph. The other ten, with their
recorded damage outside the retained nodes, are excluded from
this illustration, not from the spending regression. The shares describe
the retained recorded-damage construction, not complete physical shock support.

\begin{table}[htbp]\centering
\begin{threeparttable}
\caption{Event-intercept removal of a diffusion-kernel field}
\label{tab:mixing}
\footnotesize
\begin{tabular*}{\textwidth}{@{\extracolsep{\fill}}lrrr}
\toprule
Diffusion time $\tau$ & Median (\%) & 10th percentile & 90th percentile \\
\midrule
4 & 12.3 & 6.1 & 28.2 \\
8 & 38.4 & 25.4 & 61.8 \\
16 & 79.6 & 69.9 & 86.6 \\
Stationary limit & 91.3 & 91.3 & 91.4 \\
\bottomrule
\end{tabular*}
\begin{tablenotes}[flushleft]\footnotesize
\item Notes: Percentages of squared field norm, across 38 declarations.
All columns are in percent. The operator is $e^{-\tau\mathcal L}$ throughout,
with $A=(W+W')/2$ and $\mathcal L=\mathrm{Id}_{n_C}-D^{-1/2}AD^{-1/2}$.
The limit is analytic. Diffusion time is not a calendar horizon, and no decay
speed or spending response is fitted here.
\end{tablenotes}
\end{threeparttable}
\end{table}

The rising median illustrates how mixing can weaken within-event contrasts
as a field approaches its stationary shape. It complements the aggregation
identity, which shows that weighted aggregates and within-event contrasts retain different
directions of the same field. It is neither a monotonicity result for every
seed nor evidence that the observed spending response follows diffusion.

\section{Data and local evidence}
\label{app:stage1}

\subsection{Sample}\label{wp:app:C.1}
The local spending comparison holds the recipient sample fixed across recorded-loss
constructions. It contains 74,779 county--declaration observations, 1,574 recipient counties
and 48 declarations. Recipients lie outside the expanded recorded storm footprint and have
all required spending levels from thirteen weeks before to twelve weeks after onset.
The index is used without standard-deviation normalization. Its index-point
unit is defined in Appendix~\ref{app:sourceunits}. Table~\ref{tab:baseline} reports the
matched four-week regressions. Declarations sharing a physical episode are grouped in
the episode-bootstrap sensitivity, rather than treated as separate shock realizations.

\begin{table}[!htbp]
\centering
\begin{threeparttable}
\caption{Exposure support and recorded damage across source constructions}
\label{tab:exposure_support_v31}
\footnotesize
\setlength{\tabcolsep}{3pt}
\begin{tabular*}{\textwidth}{@{\extracolsep{\fill}}lrrrrr}
\toprule
Construction & $N$ & Mean & SD & Median & Maximum \\
\midrule
\multicolumn{6}{@{}l}{\textit{Panel A. Social exposure (index points)}}\\
\addlinespace[2pt]
Onset county reports & 74,779 & 0.060 & 0.528 & 0.000 & 34.782 \\
Incident county reports & 74,779 & 0.059 & 0.523 & 0.000 & 34.782 \\
Incident plus mapped zones & 74,779 & 0.059 & 0.524 & 0.000 & 34.782 \\
Mapped zones, end dates & 74,779 & 0.060 & 0.524 & 0.000 & 34.782 \\
\midrule
\multicolumn{6}{@{}l}{\textit{Panel B. Recorded event damage (million 2024 dollars)}}\\
\addlinespace[2pt]
Incident county reports & 48 & 34.69 & 87.53 & 2.80 & 512.82 \\
Incident plus mapped zones & 48 & 40.77 & 87.54 & 4.97 & 512.82 \\
Mapped zones, end dates & 48 & 39.85 & 87.65 & 4.97 & 512.82 \\
\bottomrule
\end{tabular*}
\begin{tablenotes}[flushleft]
\footnotesize
\item Notes: Panel A uses the same recipient--declaration observations in all rows. Zero-exposure shares are 87.35, 86.25, 86.13 and 86.13 percent, respectively. Onset uses original onset-week county reports. The other constructions use eligible reports within the declared incident assigned through week 4. Mapped zones link uniquely to eligible counties, and the last row assigns multiday reports at their end rather than start. Panel B weights each declaration once and gives recorded property-damage subtotals. A zero recorded subtotal for one declaration is not an assessed zero loss, so unknown amounts remain flagged. Currency conversion uses each report month. Original onset loss totals are omitted because their conversion was event-level rather than record-month-level.
\end{tablenotes}
\end{threeparttable}
\end{table}

\subsection{Persistence}\label{wp:app:C.2}
The event-time comparison in Appendix~\ref{app:uncapped_diagnostics} uses the same
source construction and recipient controls as the four-week regression. It holds the
four-week source geography fixed across horizons, while the separate dated-input analysis in
Appendix~\ref{app:datedforcing} distinguishes inputs at onset from later destruction.
The four-week gradient does not by itself determine a decay rate or
the persistence of a single impulse.

\subsection{Identification diagnostics}\label{wp:app:C.3}
Measured own-weather controls retain a broad recipient sample while accounting for observed
rain, temperature, gusts, warnings and reports. Appendix~\ref{app:measuredweather} describes
these variables. The pre-event, leave-one-declaration-out and alternative-baseline
comparisons are reported in Appendix~\ref{app:uncapped_diagnostics}. These diagnostics
assess timing and influential episodes. In contrast, the source-construction agreement in the main table
addresses a different measurement question. Together they support a spatial spending
association, rather than a fully isolated causal social channel.

\subsection{Economic Impact Payments}
\label{app:eip}
The county-month panel uses the 2020--2021 Economic Impact Payment rounds as remote demand inputs.
For this comparison I fix the spatial weighting at
$J_{\mathrm{EIP}}(W)=\sum_{\ell=1}^{6}0.65^{\ell-1}W^\ell$ and normalize each row before aggregating weights to source states.
I then exclude the recipient's state and states whose nearest county centroid is within 300 kilometers of the recipient,
renormalizing over the remaining states before weighting payments per resident.
The same construction applies to production and distance weights.
These fixed weights define the transfer comparison rather than estimate its weekly propagation law.
Within each round, the SCI shift-share is residualized against the production and distance exposures
and direct-home payouts. All regressions have county and month fixed effects and two-way
clustered standard errors. The design needs source-shock and exclusion assumptions, including a
treatment of correlated pandemic and policy exposure, and it says nothing about aggregate pricing.
Table~\ref{tab:eip_summary} reports sample moments, and Table~\ref{tab:remote_rounds}
re-estimates main Table~\ref{tab:outside} without each payment round in turn.

\begin{table}[!htbp]\centering
\begin{threeparttable}
\caption{Summary statistics for the remote-payment comparison}
\label{tab:eip_summary}
\footnotesize\setlength{\tabcolsep}{3pt}
\begin{tabular*}{\textwidth}{@{\extracolsep{\fill}}lrrrrr}
\toprule
Variable & $N$ & Mean & SD & Minimum & Maximum\\
\midrule
Spending index (pp) & 28,352 & $-0.713$ & 14.244 & $-61.553$ & 73.194\\
Social exposure (standardized) & 28,352 & 0.000 & 1.000 & $-11.502$ & 11.446\\
\bottomrule
\end{tabular*}
\begin{tablenotes}[flushleft]\footnotesize
\item Notes: The sample is the impact regression's January 2020--June 2021 county--month sample, including non-payment months. Spending is the monthly spending index in percentage points. Social exposure is standardized over this sample after residualizing remote source-state payments against production, distance and home-state payments within each payment round. Negative standardized values indicate below-average residual exposure.
\end{tablenotes}\end{threeparttable}\end{table}
\begin{table}[!htbp]\centering
\begin{threeparttable}
\caption{Leave-one-round-out stability of the remote EIP design}
\label{tab:remote_rounds}
\footnotesize\setlength{\tabcolsep}{4pt}
\begin{tabular*}{\textwidth}{@{\extracolsep{\fill}}llcccc}
\toprule
Sample & $h$ & SCI residual & Production residual & SCI $-$ production & $N$ \\
\midrule
Full sample & 0 & 0.386\sym{***} & $-0.444$ & 0.830\sym{***} & 28,352 \\
 & & (0.060) & (0.272) & (0.304) & \\
Full sample & 1 & 0.359\sym{***} & $-0.467$\sym{**} & 0.826\sym{***} & 26,774 \\
 & & (0.069) & (0.203) & (0.252) & \\
Drop round 1 & 0 & 0.346\sym{***} & $-0.198$ & 0.544\sym{***} & 26,774 \\
 & & (0.062) & (0.139) & (0.184) & \\
Drop round 1 & 1 & 0.290\sym{***} & $-0.375$\sym{**} & 0.665\sym{***} & 25,196 \\
 & & (0.063) & (0.184) & (0.214) & \\
Drop round 2 & 0 & 0.369\sym{***} & $-0.405$ & 0.774\sym{***} & 26,778 \\
 & & (0.054) & (0.257) & (0.280) & \\
Drop round 2 & 1 & 0.368\sym{***} & $-0.403$\sym{**} & 0.771\sym{***} & 25,200 \\
 & & (0.064) & (0.167) & (0.221) & \\
Drop round 3 & 0 & 0.230\sym{***} & $-0.597$\sym{***} & 0.826\sym{***} & 26,778 \\
 & & (0.051) & (0.132) & (0.164) & \\
Drop round 3 & 1 & 0.226\sym{***} & $-0.521$\sym{***} & 0.746\sym{***} & 25,200 \\
 & & (0.041) & (0.150) & (0.172) & \\
\bottomrule
\end{tabular*}
\begin{tablenotes}[flushleft]\footnotesize
\item Notes: Each row re-estimates the county-month specification after removing one payment round. Coefficients and standard errors are in spending-index percentage points per standard deviation of exposure. The outcome is 100 times OI's monthly mean spending-change index, and exposures are standardized within each estimation sample. Standard errors clustered by county and calendar month appear in parentheses. \mbox{\sym{*} $p<0.10$}, \mbox{\sym{**} $p<0.05$}, \mbox{\sym{***} $p<0.01$}.
\end{tablenotes}
\end{threeparttable}
\end{table}

\subsection{Source responses and aggregation}\label{wp:app:C.5}
The local regression measures differences across observed recipients. A source county's
recorded property loss is an input to that comparison, not an estimate of its consumption
loss. Insurance, transfers, borrowing and rebuilding can change the mapping from destruction
to spending or portfolio demand. Appendix~\ref{app:monetaryseverity} therefore compares
monetary source measures separately. Recovering total household pressure additionally
requires independently measured financial sensitivities and the common-response information in
Appendix~\ref{app:target}. An unsigned sum of county coefficients is not an aggregate multiplier.

\subsection{Pre-2020 labor outcomes}\label{wp:app:C.6}
The pre-2020 exercise reruns the storm design at a quarterly clock on QCEW wages and employment for
658 non-winter FEMA events from 1996 to 2019, using the modern SCI under a topology-stability
assumption. County migration shares correlate 0.784 with log SCI over 2004--2010, which is consistent with that assumption but does
not establish it. Table~\ref{tab:phase3_pre2020_robustness} adds production exposures and a
year-over-year outcome.

\begin{table}[!htbp]\centering
\begin{threeparttable}
\caption{Historical SCI transmission in quarterly county wages, 1996--2019}
\label{tab:phase3_pre2020_stage1lite}
\footnotesize\setlength{\tabcolsep}{3pt}
\begin{tabular*}{\textwidth}{@{\extracolsep{\fill}}lcccc}
\toprule
 & $h{=}1$q & $h{=}2$q & $h{=}4$q & $h{=}8$q \\
\midrule
SCI exposure & 0.0005 & 0.0014\sym{**} & 0.0010 & 0.0014\sym{*} \\
 & (0.0005) & (0.0006) & (0.0007) & (0.0008) \\
Gravity exposure & $-0.0017$\sym{***} & $-0.0014$\sym{**} & $-0.0014$\sym{**} & $-0.0017$\sym{**} \\
 & (0.0005) & (0.0006) & (0.0006) & (0.0008) \\
\midrule
Observations & 1,972,910 & 1,972,214 & 1,968,079 & 1,930,895 \\
\bottomrule
\end{tabular*}
\begin{tablenotes}[flushleft]\footnotesize
\item Notes: Cumulative log county QCEW wage change over $h$ quarters. The sample covers non-winter FEMA storms in 1996--2019 and excludes directly hit counties, with county/event effects and two-way clustered errors. The modern SCI imposes a topology-stability assumption. Migration overlap in 2004--2010 (correlation 0.784) does not validate that assumption for every earlier year. A wage response is not a spending or portfolio-demand response. \mbox{\sym{*} $p<0.10$}, \mbox{\sym{**} $p<0.05$}, \mbox{\sym{***} $p<0.01$}.
\end{tablenotes}
\end{threeparttable}
\end{table}
\begin{table}[!htbp]\centering
\begin{threeparttable}
\caption{Historical wage-response robustness: production controls and seasonality}
\label{tab:phase3_pre2020_robustness}
\footnotesize\setlength{\tabcolsep}{3pt}
\begin{tabular*}{\textwidth}{@{\extracolsep{\fill}}lcccc}
\toprule
 & $h{=}1$q & $h{=}2$q & $h{=}4$q & $h{=}8$q \\
\midrule
\multicolumn{5}{@{}l}{\textit{Panel A. Cumulative wage growth with production controls}} \\
\addlinespace[2pt]
SCI exposure & 0.0005 & 0.0014\sym{**} & 0.0010 & 0.0014\sym{*} \\
 & (0.0004) & (0.0006) & (0.0007) & (0.0008) \\
Upstream production & $-0.0077$ & $-0.0119$\sym{*} & $-0.0105$ & $-0.0113$ \\
 & (0.0056) & (0.0062) & (0.0064) & (0.0077) \\
Downstream production & 0.0076 & 0.0127\sym{**} & 0.0107\sym{*} & 0.0124 \\
 & (0.0055) & (0.0062) & (0.0064) & (0.0076) \\
Observations & 1,972,910 & 1,972,214 & 1,968,079 & 1,930,895 \\
\midrule
\multicolumn{5}{@{}l}{\textit{Panel B. Year-over-year wage growth (seasonality-robust)}} \\
\addlinespace[2pt]
SCI exposure & & 0.0004\sym{*} & 0.0002 & 0.0003 \\
 & & (0.0003) & (0.0003) & (0.0003) \\
Observations & & 1,915,341 & 1,878,421 & 1,841,237 \\
\midrule
\multicolumn{5}{@{}l}{\textit{Panel C. Year-over-year with production controls}} \\
\addlinespace[2pt]
SCI exposure & & 0.0005\sym{*} & 0.0002 & 0.0003 \\
 & & (0.0003) & (0.0003) & (0.0002) \\
Observations & & 1,915,341 & 1,878,421 & 1,841,237 \\
\bottomrule
\end{tabular*}
\begin{tablenotes}[flushleft]\footnotesize
\item Notes: All panels use the pre-2020 storm non-winter event panel (658 events, 1996--2019) with county and event fixed effects and two-way clustered standard errors, hit counties excluded, gravity included throughout. Panel A adds upstream and downstream production exposures built from 2019 CBP county sector shares and the BEA input--output matrix, applied to the historical events under a disclosed share-stability assumption. Panels B and C replace the quarterly-diff cumulative outcome with year-over-year log wage growth at $t{+}h$, which nets out county-quarter seasonality mechanically. \mbox{\sym{*} $p<0.10$}, \mbox{\sym{**} $p<0.05$}, \mbox{\sym{***} $p<0.01$}.
\end{tablenotes}
\end{threeparttable}
\end{table}

\section{Modal surfaces and transport}
\label{app:stage2}

These exercises distinguish the pooled exposure coefficient from recovery of the dynamic
response surface. The modal fits do not pin down a stable decay law, and held-out predictive performance for
residual state contrasts is close to the zero-response benchmark. The target-specific simulation in Appendix~\ref{app:recoverycheck} then evaluates
what additional measurements and restrictions contribute to financial prediction.
The modal and transport exercises use weekly growth and onset-week source shares, rather than
the level changes and through-week-4 source shares in the local-spending comparison.

\subsection{Modal response surfaces and the decay fit}
\label{app:noise}
The modal response surfaces use the 34 lowest nonstationary eigenvectors, excluding the stationary
mode. At each horizon the county-level regression jointly includes all 34 exposure columns
$E_{jke}=u_{jk}(u_k'\bm b_e^{raw})$, county, event and target-week fixed effects, and the controls in
main-text equation~\eqref{eq:modalreg}, so the estimates are jointly fitted exposure
coefficients, not separate regressions of a projected outcome on a projected shock.
Controls include distance bins, a same-hit-state indicator, COVID-19 cases and deaths per capita,
state stringency, economic-support and income-support indices, one-week lags of spending growth
and log spending, and mean spending growth over the previous four weeks.
The outcome is weekly log spending growth, clipped at
the first and ninety-ninth percentiles, rather than the unclipped level changes in Table~\ref{tab:baseline}.
Storm inputs are county shares of onset-week recorded damage restricted to the retained counties.
They are not rescaled to sum to one within that support. Standard errors are clustered by
county and event.

Each surface has 34 modes and 17 weekly horizons, or 578 cells. On the SCI surface, 11.1 percent
of the cells have $p<0.05$, and the ratio of summed squared estimates to summed variances,
$\sum\hat\theta^2/\sum\widehat{\mathrm{se}}^2$, is 1.14. It is about one for a surface
of pure noise. Cells are not independent, so these figures are descriptive rather than a
joint test.

The decay fit uses a stable-sign support. For each retained mode, the support is the longest
contiguous run of horizons with the same sign, subject to a minimum run length and a signal
threshold $c_{\mathrm{sig}}$ fixed in advance. A two-step estimator fits each mode's log-decay slope
on its support by weighted least squares and then regresses the slopes on $\lambda_k$, so that
$\hat\kappa$ is minus the second-step slope and $\hat\rho$ the exponential of its intercept.
Table~\ref{tab:kgrid} repeats the inverse-variance second step for each evaluated choice of $K$ at
$c_{\mathrm{sig}}=0.5$. The estimate of $\kappa$ changes sign across $K$, $\hat\rho$ exceeds one for $K=8$ and
$K=16$, and the largest single-mode weight share is above one half in seven of the ten cases.

At $K=34$, the inverse-variance fit places 99.6 percent of its weight on one mode. With three
cells per mode, a per-mode standard error can be close to zero by chance, which is how one mode
came to carry almost all of the weight. Equal weighting, omitting that mode and pooling the
support cells instead give negative, imprecise diffusion-rate estimates. The result is a
sensitivity of the estimated decay law, rather than an independently replicated structural
parameter, and inference conditions on the estimated, selected surface.

\begin{table}[!htbp]\centering
\begin{threeparttable}
\caption{Inverse-variance decay fit by number of retained SCI modes}
\label{tab:kgrid}
\footnotesize
\begin{tabular*}{\textwidth}{@{\extracolsep{\fill}}lccccc}
\toprule
$K$ & Support modes & $\hat\kappa$ (s.e.) & $p(\hat\kappa)$ & $\hat\rho$ & Largest weight share \\
\midrule
8 & 6 & 15.14 (6.74) & 0.088 & 4.297 & 0.80 \\
16 & 14 & 14.16 (3.62) & 0.002 & 3.844 & 0.50 \\
24 & 20 & 0.24 (1.69) & 0.889 & 0.746 & 0.54 \\
26 & 21 & 2.30 (0.58) & 0.001 & 0.956 & 0.86 \\
28 & 23 & $-1.67$ (2.34) & 0.483 & 0.266 & 0.44 \\
30 & 25 & $-0.37$ (1.91) & 0.848 & 0.479 & 0.30 \\
32 & 27 & 2.36 (1.01) & 0.028 & 0.969 & 0.70 \\
34 & 29 & 2.49 (0.83) & 0.006 & 0.969 & 1.00 \\
36 & 30 & 1.64 (1.33) & 0.229 & 0.821 & 0.78 \\
40 & 32 & $-1.52$ (1.18) & 0.206 & 0.363 & 0.51 \\
\bottomrule
\end{tabular*}
\begin{tablenotes}[flushleft]\footnotesize
\item Notes: Largest weight share is the share of total second-step weight on one mode (0.996 at
$K=34$). Each row uses the same two-step decay estimator at a different mode count.
\end{tablenotes}
\end{threeparttable}
\end{table}

\subsection{Out-of-fold transport}\label{wp:app:D.2}
The transport experiment uses 64 storms, 45 onset dates, 1,590 counties and 34 modes, with five
purged outer folds. Overlapping response windows leave 13 calendar-overlap groups, 12 with complete cumulative
paths. Penalties and aggregation weights are tuned within training folds, while the fixed-decay
benchmark retains the decay shape estimated from the full-sample modal fit. Held-out county, event and week effects use held-out
outcomes while nuisance coefficients use training data, and population-weighted state means are taken over observed
non-hit counties, so the target is not total consumption.

\begin{table}[!htbp]\centering
\begin{threeparttable}
\caption{Out-of-fold aggregate transport with calendar-group resampling intervals}
\label{tab:transport}
\footnotesize\setlength{\tabcolsep}{3pt}
\begin{tabular*}{\textwidth}{@{\extracolsep{\fill}}lcc}
\toprule
Family & Weekly $R^2$ [95\% interval] & Cumulative $R^2$ [95\% interval]\\
\midrule
Fixed decay & $-3.87\!\times\!10^{-5}$ [$-2.19\!\times\!10^{-4},1.53\!\times\!10^{-4}$]
 & $-6.49\!\times\!10^{-5}$ [$-0.00611,0.00318$]\\
Spectral FIR & $-3.91\!\times\!10^{-5}$ [$-1.45\!\times\!10^{-4},2.00\!\times\!10^{-5}$]
 & $-0.000201$ [$-0.00129,0.0000388$]\\
Modal FIR & $-0.000287$ [$-0.000707,0.0000632$]
 & $-0.000273$ [$-0.00882,0.00335$]\\
Aggregation-path FIR & $5.02\!\times\!10^{-7}$ [$-3.03\!\times\!10^{-7},1.66\!\times\!10^{-6}$]
 & $8.04\!\times\!10^{-6}$ [$-0.0000103,0.0000816$]\\
Nested-selected rule & $-0.000126$ [$-0.000264,0.00000783$]
 & $-0.001257$ [$-0.006787,-0.00000917$]\\
\bottomrule
\end{tabular*}
\begin{tablenotes}[flushleft]\footnotesize
\item Notes: $R^2=1-\sum(y-\hat y)^2/\sum y^2$, relative to a zero conditional response. Intervals resample
calendar-overlap groups with fitted predictions fixed, omitting training and selection uncertainty. There are
25,145 weekly and 2,632 cumulative cells, far more than independent shocks. The nested rule selects
zero in two of five folds. The identity $R^2=(2\bm y'\hat{\bm y}-\hat{\bm y}'\hat{\bm y})/\bm y'\bm y$
shows that small predictions give $R^2$ near zero regardless of the true signal.
\end{tablenotes}
\end{threeparttable}
\end{table}

\section{Financial evidence and measurement}
\label{app:finance}

\subsection{Aggregate credit-premium and return associations}
\label{app:ebpassoc}
Table~\ref{tab:ebp_original} reports the aggregate regressions discussed in the main text.
The SCI amplitude is the monthly mean of the absolute slow coordinate from a VAR(2) fitted to
two principal factors of the weekly state-response surface projected onto the SCI basis.
The SCI state and the production-family state (the first principal component
of the upstream and downstream amplitudes) enter jointly with macro and COVID controls. Panel~B
reports a twelve-month intermediary-return regression whose autoregressive columns used
overlapping future-return lags. They are not valid predictive evidence and are shown only for
completeness.

Entering one state at a time under the same monthly design, the tightness interaction is 0.116
for the SCI amplitude, with a recursive-bootstrap $p$-value of 0.042, and 0.104 for a count of the
85 active storm events (0.053). The part of the amplitude not fitted by that count retains an
interaction of 0.106 when both parts enter jointly. A seasonal state, a signed observer that
filters 34 SCI modes at the modal-fit decay rates and a production-network counterpart give
0.054, $-0.049$ and $-0.053$, with bootstrap $p$-values between 0.09 and 0.14. The bootstrap uses
recursive AR(2) null paths with six-month innovation blocks and 4,999 draws, holding the states
fixed. State construction and standardization differ across these comparisons, which are all
descriptive (Appendix~\ref{app:search}).

\begin{table}[!htbp]\centering
\begin{threeparttable}
\caption{Aggregate macro-finance associations}
\label{tab:ebp_original}
\footnotesize\setlength{\tabcolsep}{3pt}
\begin{tabular*}{\textwidth}{@{\extracolsep{\fill}}lccc}
\toprule
 & Static & AR(1) & AR(2) \\
\midrule
\multicolumn{4}{@{}l}{\textit{Panel A. EBP change: interaction with lagged tightness}} \\
\addlinespace[2pt]
SCI state & 0.118\sym{**} & 0.109\sym{**} & 0.116\sym{***} \\
 & (0.050) & (0.044) & (0.039) \\
Production family & $-0.052$\sym{*} & $-0.031$ & $-0.010$ \\
 & (0.030) & (0.034) & (0.041) \\
$T$ & 51 & 50 & 49 \\
$R^2$ & 0.21 & 0.32 & 0.40 \\
\midrule
\multicolumn{4}{@{}l}{\textit{Panel B. Twelve-month intermediary-return sum}} \\
\addlinespace[2pt]
SCI state & 0.050\sym{**} & 0.034\sym{**} & 0.034\sym{**} \\
 & (0.022) & (0.016) & (0.017) \\
Production family & 0.018 & 0.018 & 0.003 \\
 & (0.061) & (0.039) & (0.037) \\
$T$ & 40 & 39 & 38 \\
$R^2$ & 0.75 & 0.88 & 0.88 \\
\bottomrule
\end{tabular*}
\begin{tablenotes}[flushleft]\footnotesize
\item Notes: HAC standard errors in parentheses.
Panel~A outcomes are one-month EBP changes in percentage points.
Panel~B uses the sum of twelve forward monthly He--Kelly--Manela (HKM) value-weighted intermediary investment returns, in decimal units.
Tightness is the negative of the standardized He--Kelly--Manela intermediary capital ratio. Stars
are conventional and do not account for the specification search described in
Appendix~\ref{app:search}. \mbox{\sym{*} $p<0.10$}, \mbox{\sym{**} $p<0.05$}, \mbox{\sym{***} $p<0.01$}.
\end{tablenotes}
\end{threeparttable}
\end{table}

\subsection{Specification selection and inference}
\label{app:search}
The state construction, the tightness interaction and the autoregressive specification were chosen
after several alternatives had been examined, and the pricing model of Section~\ref{sec:pricing}
was written after both the positive association and the later negative observer association were
known. Neither sign is therefore a test of that model. The bootstrap
$p$-values in Appendix~\ref{app:ebpassoc} were also computed after the results were known.

In the AR(2) specification the regressor
$y_{t-1}=E^{BP}_t-E^{BP}_{t-1}$ uses information through month $t$, and regressors were standardized
on the full sample, so the regression is not a real-time forecast. Standardization preserves the
sign of the interaction but not its magnitude or structural interpretation. The storm count explains
62 percent of the monthly amplitude variance. The SCI interaction is 62 percent
spanned by lagged macro and COVID controls (72 percent for the count interaction). Twelve-month
return regressions restricted to information available at the forecast date,
estimated for two alternative states, give negative coefficients, with 40 observations and four non-overlapping
annual windows.

\subsection{Intermediary presence}\label{wp:app:E.3}
The intermediary-presence regression uses county-level 2019 assets from Compustat
headquarters records mapped from ZIP codes to counties. Banks load negatively on both the SCI and
production legs, and fund-trust and real-estate intensity is dominated by administrator addresses
rather than local presence, so the non-bank composite (broker-dealers and insurers) is the headline
object. The conditional model does not uniquely select it. Broker-dealers alone have the same sign
but only 14 counties with positive assets, the largest holding 91 percent
(Table~\ref{tab:mech_concentration}).

Table~\ref{tab:mech} omits lower-order terms, such as exposure-by-local-intensity and
tightness-by-local-intensity, that a test of a triple interaction requires unless the fixed
effects absorb them. Its coefficients are therefore descriptive rather than a test of the fully
hierarchical interaction model.

\begin{table}[!htbp]\centering
\begin{threeparttable}
\caption{Intermediary presence and SCI amplification in the EIP panel}
\label{tab:mech}
\footnotesize\setlength{\tabcolsep}{3pt}
\begin{tabular*}{\textwidth}{@{\extracolsep{\fill}}lccc}
\toprule
\multicolumn{4}{@{}l}{\textit{Panel A. Headline mechanisms}} \\
\addlinespace[2pt]
 & Non-bank composite & Insurance & Broker-dealer \\
\midrule
SCI triple & 0.0008\sym{***} & 0.0008\sym{***} & 0.0007\sym{***} \\
 & (0.0002) & (0.0001) & (0.0002) \\
Production triple & $-0.0006$\sym{***} & $-0.0006$\sym{***} & $-0.0000$ \\
 & (0.0002) & (0.0002) & (0.0001) \\
SCI $-$ production & 0.0014\sym{***} & 0.0014\sym{***} & 0.0007\sym{***} \\
 & (0.0004) & (0.0004) & (0.0003) \\
\midrule
\multicolumn{4}{@{}l}{\textit{Panel B. Alternative intermediary measures}} \\
\addlinespace[2pt]
 & HKM employment & First principal component & Insurance residual \\
\midrule
SCI triple & 0.0003\sym{**} & 0.0003\sym{**} & 0.0008\sym{***} \\
 & (0.0002) & (0.0001) & (0.0001) \\
Production triple & $-0.0008$\sym{***} & $-0.0009$\sym{***} & $-0.0001$ \\
 & (0.0002) & (0.0004) & (0.0001) \\
SCI $-$ production & 0.0012\sym{***} & 0.0013\sym{***} & 0.0009\sym{***} \\
 & (0.0003) & (0.0005) & (0.0002) \\
\bottomrule
\end{tabular*}
\begin{tablenotes}[flushleft]\footnotesize
\item Notes: Dependent variable: one-month-ahead county
log spending. SCI triple and Production triple denote residual social or production exposure
interacted with lagged tightness and the indicated mechanism measure. Each completed interaction
is standardized within its estimation sample.
County and month fixed effects, direct-home payout interactions and remote-source pandemic controls.
Two-way clustered standard errors in parentheses. The county-bootstrap specifications in
Table~\ref{tab:mech_rescue_boot} use different scaling and inference. Their $p$-values are not
interchangeable $p$-values for these coefficients. The non-bank composite rests on 68 counties, the largest holding 64
percent of assets. The insurance residual is insurance intensity orthogonalized with respect to the
other intermediary types. \mbox{\sym{*} $p<0.10$}, \mbox{\sym{**} $p<0.05$}, \mbox{\sym{***} $p<0.01$}.
\end{tablenotes}
\end{threeparttable}
\end{table}

\begin{table}[!htbp]\centering
\begin{threeparttable}
\caption{Restricted-null wild cluster bootstrap for the firm-tier mechanism}
\label{tab:mech_rescue_boot}
\footnotesize\setlength{\tabcolsep}{3pt}
\begin{tabular*}{\textwidth}{@{\extracolsep{\fill}}lcccc}
\toprule
Mechanism & SCI triple & $p_{asymp}$ & $p_{WCR}$ & 95\% wild $t^\ast$ interval \\
\midrule
Broker-dealers (NAICS 5231) & 0.0007\sym{**} & 0.018 & 0.127 & [$-2.97$, 3.09] \\
 & (0.0003) & & & \\
Insurance (NAICS 524) & 0.0007\sym{***} & 0.001 & 0.001 & [$-1.95$, 1.99] \\
 & (0.0002) & & & \\
HKM IB index (BD+bank+ins assets) & $-0.0001$ & 0.740 & 0.748 & [$-1.91$, 1.93] \\
 & (0.0004) & & & \\
PCA-PC1 (shared 5-type variance) & 0.0003 & 0.311 & 0.398 & [$-2.48$, 2.39] \\
 & (0.0003) & & & \\
Non-bank IB (BD+insurance) & 0.0008\sym{***} & 0.001 & $<0.001$ & [$-1.98$, 2.03] \\
 & (0.0002) & & & \\
HKM employment index & 0.0003 & 0.242 & 0.258 & [$-2.04$, 2.03] \\
 & (0.0003) & & & \\
HKM firm-count index & 0.0001 & 0.808 & 0.818 & [$-1.95$, 2.07] \\
 & (0.0003) & & & \\
Insurance $\perp$ other 4 types & 0.0007\sym{***} & 0.001 & 0.001 & [$-1.95$, 2.02] \\
 & (0.0002) & & & \\
\bottomrule
\end{tabular*}
\begin{tablenotes}[flushleft]\footnotesize
\item Notes: County-cluster bootstrap specification, with its own coefficient scaling and standard errors, rather than the two-way covariance fit in the preceding table. Restricted-null bootstrap, $B=1999$ valid draws. The non-bank row has zero exceedances, so $<0.001$ describes Monte Carlo resolution, not an exact zero population $p$-value. The final column is a null $t^\ast$ reference interval, not a confidence interval for the coefficient. These non-hierarchical interactions are descriptive. \mbox{\sym{*} $p<0.10$}, \mbox{\sym{**} $p<0.05$}, \mbox{\sym{***} $p<0.01$}.
\end{tablenotes}
\end{threeparttable}
\end{table}
\begin{table}[!htbp]\centering
\begin{threeparttable}
\caption{County-intensity concentration in the local-home intermediary measures}
\label{tab:mech_concentration}
\footnotesize\setlength{\tabcolsep}{3pt}
\begin{tabular*}{\textwidth}{@{\extracolsep{\fill}}lccccc}
\toprule
Mechanism & Positive counties & Top-1 share & Top-5 share & Top-10 share & Top-25 share \\
\midrule
Broker-dealer & 14 & 0.91 & 0.98 & 1.00 & 1.00 \\
Insurance & 62 & 0.52 & 0.76 & 0.89 & 0.98 \\
Non-bank IB & 68 & 0.64 & 0.80 & 0.90 & 0.98 \\
\bottomrule
\end{tabular*}
\begin{tablenotes}[flushleft]\footnotesize
\item Notes: Shares are computed from the raw 2019 asset aggregates underlying the log-intensity measures.
\end{tablenotes}
\end{threeparttable}
\end{table}
\begin{table}[!htbp]\centering
\begin{threeparttable}
\caption{Drop-top stress test for the local-home broker-dealer mechanism}
\label{tab:mech_drop_top_home}
\footnotesize\setlength{\tabcolsep}{3pt}
\begin{tabular*}{\textwidth}{@{\extracolsep{\fill}}lccc}
\toprule
Drop top-$N$ BD-home counties & SCI $\times\mathrm{Tight}$ & SCI $\times\mathrm{Tight}\times$ log-BD-home & SCI $-$ PROD (triple) \\
\midrule
0 & 0.0007\sym{***} & 0.0007\sym{***} & 0.0007\sym{***} \\
 & (2.91) & (2.97) & (2.61) \\
1 & 0.0007\sym{***} & 0.0007\sym{**} & 0.0010\sym{**} \\
 & (2.88) & (2.25) & (2.32) \\
2 & 0.0007\sym{***} & 0.0007\sym{**} & 0.0007\sym{**} \\
 & (2.84) & (2.18) & (1.96) \\
5 & 0.0007\sym{***} & 0.0007\sym{***} & 0.0009\sym{***} \\
 & (2.80) & (3.39) & (2.62) \\
10 & 0.0007\sym{***} & 0.0006\sym{***} & 0.0007\sym{***} \\
 & (2.72) & (11.07) & (4.41) \\
\bottomrule
\end{tabular*}
\begin{tablenotes}[flushleft]\footnotesize
\item Notes: With only 14 counties having any listed broker-dealer assets in 2019, the broker-dealer result is most vulnerable to dropping the largest broker-dealer headquarters counties. The table re-estimates the headline broker-dealer specification after dropping the top-$N$ broker-dealer-home counties.
Nominal two-way-clustered $t$ statistics in parentheses, not independent
replications. Dropping ten of these 14 counties leaves four.
The resulting $t=11.07$ is not persuasive evidence of robustness with so few exposed clusters.
These non-hierarchical interactions do not identify household demand or EBP transmission. \mbox{\sym{*} $p<0.10$}, \mbox{\sym{**} $p<0.05$}, \mbox{\sym{***} $p<0.01$}.
\end{tablenotes}
\end{threeparttable}
\end{table}

\subsection{Stationary and nonstationary exposure: computation details}
\label{app:split}
The calculation behind Figure~\ref{fig:split} uses the retained SCI graph on the 1,590 support
counties, not a source-complete national graph. The graph is symmetrized as $A=(W+W')/2$ with a
zero diagonal, and the calculation uses its full eigendecomposition. The first eigenvalue is zero to machine precision, the first eigenvector has correlation one with
square-root degree, and the next 34 eigenvalues agree with the modal-regression basis to
$10^{-15}$. The largest eigenvalue is 1.60. Of 85 candidate storms, 18 have no damage inside the
support and drop out. Let $\mathcal D_e$ be event $e$'s total recorded damage and
$\bm b_e^{raw}$ the retained counties' shares of it. The unit footprint is then
$\bm b_e=\bm b_e^{raw}/m_e$ with $m_e=\mathbf1'\bm b_e^{raw}$. Here 48 of the 67 retained
events have $m_e<1$. Damage-weighted comparisons multiply this unit footprint by in-support
damage $m_e\mathcal D_e$, giving $\mathcal D_e\bm b_e^{raw}$. Thus the in-support share enters once. The normalization
changes cross-event weighting but cancels in each event's pressure-to-incidence ratio.
The dividend-based weights put 18.3 percent of their squared norm on the stationary mode,
population weights 32.9 percent, and weights confined to the top 5 percent of dividend counties
10.8 percent. Table~\ref{tab:split} reports selected values.

\begin{table}[!htbp]\centering
\begin{threeparttable}
\caption{Proxy-weighted pressure relative to incidence, by diffusion time}
\label{tab:split}
\footnotesize
\begin{tabular*}{\textwidth}{@{\extracolsep{\fill}}lccccc}
\toprule
 & $\tau=0$ & $\tau=1$ & $\tau=2$ & $\tau=4$ & $\tau=8$ \\
\midrule
\multicolumn{6}{@{}l}{\textit{Median ratio of weighted pressure to incidence}}\\
Dividend shares & 0.13 & 0.56 & 0.68 & 0.82 & 0.90 \\
Population & 0.28 & 0.71 & 0.80 & 0.89 & 0.94 \\
Top 5\% dividend counties & 0.00 & 0.41 & 0.58 & 0.74 & 0.88 \\
\addlinespace
\multicolumn{6}{@{}l}{\textit{Correlation with incidence across storms, unit footprints}}\\
Dividend shares & 0.81 & 0.83 & 0.83 & 0.84 & 0.93 \\
Population & 0.89 & 0.89 & 0.90 & 0.92 & 0.97 \\
Top 5\% dividend counties & 0.77 & 0.79 & 0.80 & 0.80 & 0.91 \\
\addlinespace
\multicolumn{6}{@{}l}{\textit{Correlation with incidence across storms, damage-weighted}}\\
Dividend shares & 0.97 & 0.98 & 0.99 & 1.00 & 1.00 \\
Population & 0.98 & 0.99 & 0.99 & 1.00 & 1.00 \\
Top 5\% dividend counties & 0.97 & 0.98 & 0.99 & 1.00 & 1.00 \\
\bottomrule
\end{tabular*}
\begin{tablenotes}[flushleft]\footnotesize
\item Notes: 67 storms. $\tau=\rev{\kappa_D}(h-1)$. Weighted pressure is $w'e^{-\tau\mathcal L}\bm b_e$, while incidence is
$(w'u_1)(u_1'\bm b_e)$. Square-root-degree weights give a ratio of exactly one at every $\tau$. Computed
from the full eigendecomposition and normalized in-support event footprints.
\end{tablenotes}
\end{threeparttable}
\end{table}

\subsubsection*{Horizon-specific source importance}

The county comparison in Figure~\ref{fig:horizon_importance} evaluates the same symmetric
diffusion operator source by source. Let $\bm e_j$ denote a unit initial slow-field input in
county $j$. With homogeneous input loading, the pricing response is proportional to
\begin{equation}\label{eq:source_importance}
I_j(\tau)=w'e^{-\tau\mathcal L}\bm e_j
         =[e^{-\tau\mathcal L}w]_j,\qquad \tau=\kappa_D(h-1).
\end{equation}
The common price-impact, input-loading and persistence factors cancel in within-horizon
county ratios. The unit is an initial input to the slow field $\bm d$, so
converting a dollar shock into that field requires a separately specified input map.
The calculation uses the same retained graph and dividend-share weights as
Figure~\ref{fig:split}, with all 1,590 modes. Here $\tau=0$ denotes the first slow-field
response, $h=1$. A calendar interpretation additionally requires $\kappa_D$.

\begin{figure}[!htbp]
\centering
\includegraphics[width=\textwidth]{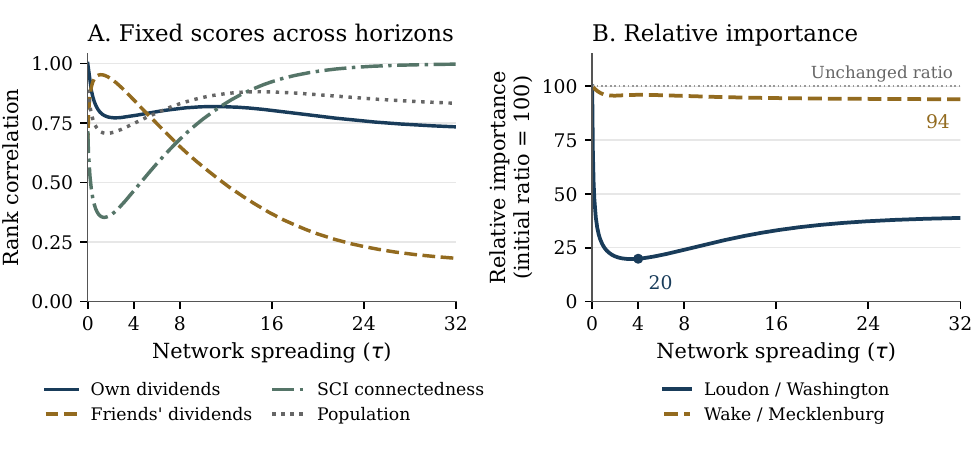}
\caption{Source importance over the full diffusion-time range. Panel A extends main-text
Figure~\ref{fig:horizon_importance}'s comparisons on a zero-based vertical scale and a
numerical diffusion-time axis through 32. Friends' dividends are $D^{-1/2}AD^{-1/2}w$.
Panel B indexes each within-pair ratio to 100 initially. The unindexed Loudon/Washington
ratio has stationary limit 0.999. Model-implied for equal unit inputs to the slow field $\bm d$,
with observed SCI and dividend-proxy weights.}
\label{fig:importance_full}
\end{figure}

Three relations explain why a fixed score tracks different parts of the response:
\begin{equation}\label{eq:importance_stages}
I(0)=w,\qquad
I(\tau)=(1-\tau)w+\tau D^{-1/2}AD^{-1/2}w+O(\tau^2),\qquad
\lim_{\tau\to\infty}I(\tau)=u_1(u_1'w).
\end{equation}
Initially the loading selects a county's own weight. Its first change combines that weight
with its neighbors' weights. Over longer diffusion times the stationary direction,
$u_1\propto\sqrt{\text{degree}}$, governs the limit. Panel A of
Figure~\ref{fig:importance_full} follows these transitions. Own dividend shares have rank
correlation one initially, neighbors' shares reach 0.95 at $\tau=1$, and connectedness reaches
0.997 at $\tau=32$. These benchmark scores contain different information about the
horizon-specific loading.

\subsubsection*{County paths and the pair comparison}

\begin{figure}[!htbp]
\centering
\includegraphics[width=\textwidth]{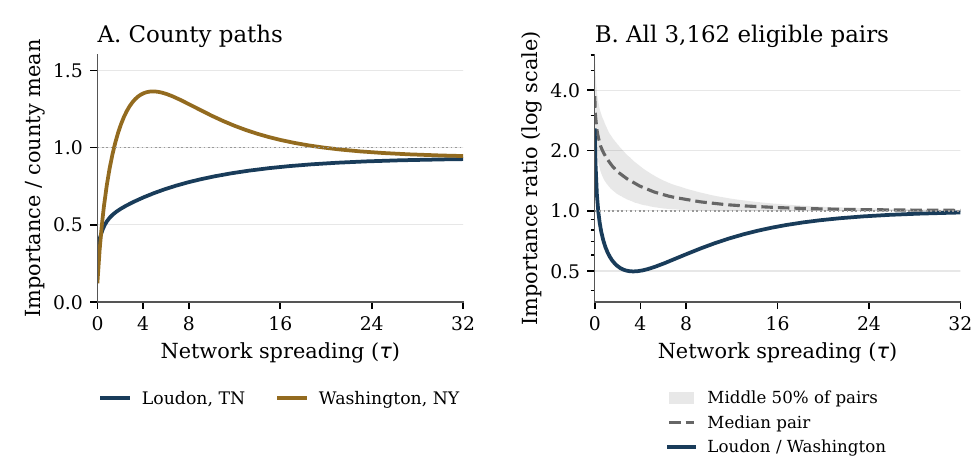}
\caption{County paths and the full example family. Panel A divides each county's importance
by the mean across the retained counties at that diffusion time. Panel B compares the
Loudon/Washington ratio with the median and middle 50 percent of the 3,162 eligible pairs,
and its vertical axis is logarithmic. The horizontal axes measure numerical diffusion time.
All paths use the model diffusion law and dividend-proxy weights.}
\label{fig:importance_pairs}
\end{figure}

Loudon County, Tennessee, and Washington County, New York, have populations of 55,129 and
61,091 in the county frame and total SCI degrees differing by 0.277 percent. Their dividend
weight ratio is 2.52. Their model-implied importance ratio is 0.64 at $\tau=1$ and 0.50 at
$\tau=4$, approaching their square-root-degree ratio, 0.999, in the limit. Both counties gain
importance relative to the county mean over $\tau=0$ to 4, but Washington gains more
(Figure~\ref{fig:importance_pairs}, Panel A). The reversal concerns their relative source
loadings.

The broader comparison includes all 27,169 county pairs satisfying
$|\log d_i-\log d_j|\leq0.01$. Of these, 24.5 percent reverse their initial importance ordering
by $\tau=4$. The displayed example is selected from the 3,162 pairs in that set with both
populations at least 50,000 and an initial importance ratio of at least two, oriented so the
initially larger county is in the numerator. The selection rule chooses the smallest ratio
at $\tau=4$, making this a clear illustration rather than a typical pair. Panel B reports
the entire eligible family's median and interquartile range, and 17.0 percent of that family reverse by
$\tau=4$.

Wake and Mecklenburg counties, North Carolina, provide the comparatively stable case in
Figure~\ref{fig:horizon_importance}. Under the same graph, diffusion law and dividend-proxy
weights, their ratio is 1.065 initially, 1.023 at $\tau=4$ and 1.001 at $\tau=30$. Its largest
relative departure from the initial value is about 6 percent over the plotted range. Their
total SCI degrees differ by 0.0224 percent. I chose this pair after examining the calculated
paths, for comparable connectedness and population and nearly stable relative loadings.

To compare changes across pairs with different initial ratios, Panel B of Figure~\ref{fig:importance_full} displays
\begin{equation}\label{eq:importance_index}
J_{a/b}(\tau)=100\,
\frac{I_a(\tau)/I_b(\tau)}{I_a(0)/I_b(0)}.
\end{equation}
Both indices start at 100. Loudon/Washington reaches 19.87 at $\tau=4$ and 38.59 at
$\tau=30$, while Wake/Mecklenburg is 96.04 and 94.01. An index of 100 means the pair's
relative importance remains at its initial value. Equality of the two counties instead
means the unindexed ratio is one.

\subsubsection*{Alternative weights}

Table~\ref{tab:importance_weights} repeats the calculation with alternative proxy weights.
The changes in relative loadings persist across these choices. The financial geography
determines their magnitude. If $w\propto u_1$, for example, then $I(\tau)=w$ for every $\tau$ and the
stationary direction alone suffices. Thus the comparison illustrates the economic content
of Theorem~\ref{thm:no_scalar}, namely that horizon-specific financial relevance depends jointly on
transmission and the weights selecting household responses.

\begin{table}[!htbp]\centering
\begin{threeparttable}
\caption{Horizon dependence under alternative proxy weights}
\label{tab:importance_weights}
\footnotesize
\begin{tabular*}{\textwidth}{@{\extracolsep{\fill}}lcccc}
\toprule
 & Dividends & Participation & Population & Top 5\% \\
\midrule
County rank correlation, $\tau=0$ vs.\ 4 & 0.78 & 0.80 & 0.79 & 0.37 \\
Best rank-one relative error & 0.39 & 0.34 & 0.33 & 0.43 \\
Near-connectedness pairs reversed (\%) & 24.5 & 23.0 & 22.5 & 32.4 \\
\bottomrule
\end{tabular*}
\begin{tablenotes}[flushleft]\footnotesize
\item Notes: The rank-one error is the best relative Frobenius error after normalizing each
horizon's loading vector to unit Euclidean norm, on $\tau\in\{0,1,2,4,8\}$, and it measures
cardinal approximation error. Participation uses the 2019--22 average number of IRS returns
reporting ordinary dividends. Top 5\% retains 80 counties with positive weight, so its pair
share uses only the 34 near-connectedness pairs with both weights positive (11 reversals).
Its rank correlation among positive-weight counties alone is 0.69. The other pair shares
use the common 27,169-pair comparison. Proxy weights describe scenarios for financial
geography rather than measured marginal bond-demand sensitivities.
\end{tablenotes}
\end{threeparttable}
\end{table}
\subsection{Corporate-bond fund flows and the pricing margin}
\label{app:creditflow}

The exploratory pricing exercise uses SEC N-PORT fund flows and corporate-security
holdings. The fund-level universe permits open-end funds and funds reporting a
fund-of-funds designation, while excluding funds ever classified as ETFs, money-market,
closed-end, interval, insurance-underlying or master-feeder vehicles within the reporting
window. Captive and placeholder series, central-fund names and funds predominantly held
by other reporting funds are excluded. The fund-of-funds treatment and normalized
instrument were considered after the initial construction. This is an exploratory
specification. Fund-months require lagged net assets of at least \$10 million.
Flow rates are winsorized at the first and ninety-ninth percentiles within month.

Let $f_{jt}$ be fund $j$'s net inflow divided by lagged net assets and let
$s_{j,t-1}$ be its lagged corporate-security holdings divided by the market denominator.
The aggregate flow proxy, in percentage units, is
$F_t=100\sum_j s_{j,t-1}f_{jt}$. The denominator is the Federal Reserve's nonfinancial
corporate-business debt-securities series (NCBDBIQ027S), interpolated between quarterly
endpoints. Corporate portfolios can include financial and foreign issuers, so the denominator
and numerator therefore have different scopes. This normalization is a reporting scale,
not an estimated economy-wide demand elasticity.

Principal-component loadings use a balanced July 2020--June 2024 block of 318 funds.
The two factor scores in other months are projections onto those loadings. Regressing
each eligible fund's flow rate on an intercept and these factors over its available
sample gives residuals $u_{jt}$, requiring at least 12 months. For the set $J_t$ with
residuals, the granular instrument is
\begin{equation}
Z_t^{GIV}=100\left\{\sum_{j\in J_t}s_{j,t-1}u_{jt}
 -\left(\sum_{j\in J_t}s_{j,t-1}\right)
  \frac{1}{|J_t|}\sum_{j\in J_t}u_{jt}\right\} .
\end{equation}
This is the size-minus-equal-weighted construction in market units. The aggregate inflow proxy
includes all eligible fund-month flows, including those without enough observations to estimate a
loading.

Factors, fund selection and loadings use the full reporting window.
The exercise relates same-month flows to EBP changes rather than forecasting in real time.
The regression includes an intercept, the two factors and two lagged EBP changes.
HAC calculations use actual calendar-month gaps after observations are removed.
They condition on estimated factors and do not propagate fund-selection or generated-
instrument uncertainty. A strong first stage does not establish exclusion, because residual flows
can still reflect credit news, return chasing or omitted common exposures.

Table~\ref{tab:creditflow_full} reports the fixed measurement and timing comparisons.
Removing reinvested distributions produces the external-cash measure, sales less
redemptions. It leaves the sign and full-sample magnitude close to the original measure.
Using only predated holdings and non-backcast asset estimates has little effect. It does
not make full-sample factor estimates predetermined. The crisis exclusions yield wider
intervals. Omitting one year at a time from the external-cash specification gives
coefficients from $-1.48$ to $-5.76$ basis points, with respective standard errors
1.66 and 1.25. These are endpoints of a sensitivity range, not a confidence interval.

\begin{table}[htbp]\centering
\begin{threeparttable}
\caption{Fund-flow pricing: measurement and sample comparisons}
\label{tab:creditflow_full}
\footnotesize\setlength{\tabcolsep}{5pt}
\begin{tabular*}{\textwidth}{@{\extracolsep{\fill}}lccccc}
\toprule
 & (1) & (2) & (3) & (4) & (5) \\
\midrule
Net inflows & $-4.06$\sym{***} & $-3.96$\sym{**} & $-3.95$\sym{**} & $-3.75$ & $-2.81$ \\
 & (1.54) & (1.61) & (1.63) & (3.54) & (3.73) \\
\midrule
Months & 57 & 57 & 57 & 54 & 48 \\
Flow factors & Yes & Yes & Yes & Yes & Yes \\
Lagged EBP changes & Yes & Yes & Yes & Yes & Yes \\
Reinvestment excluded & No & Yes & Yes & Yes & Yes \\
Predated holdings only & No & No & Yes & No & No \\
March--May 2020 excluded & No & No & No & Yes & Yes \\
July 2020 onward only & No & No & No & No & Yes \\
First-stage statistic & 26.6 & 16.4 & 16.6 & 19.6 & 15.8 \\
\bottomrule
\end{tabular*}
\begin{tablenotes}[flushleft]\footnotesize
\item Notes: Monthly EBP change in basis points per inflow of 0.01\% of the
nonfinancial corporate debt securities denominator. Three calendar-month Bartlett HAC standard
errors in parentheses. The instrument is the holdings-weighted residual flow net of
its equally weighted counterpart, scaled as above. Every column includes two common
flow factors and two lagged EBP changes. Column (3) also excludes backcast asset
estimates. First-stage statistics are squared HAC $t$-statistics, not tests of exclusion. Stars use the
HAC standard errors and normal critical values: \mbox{\sym{*} $p<0.10$}, \mbox{\sym{**} $p<0.05$}, \mbox{\sym{***} $p<0.01$}.
\end{tablenotes}
\end{threeparttable}
\end{table}

\paragraph{From executed flows to the operator.}
\label{app:flowbridge}
In the conditional equilibrium $X=\bar X+\zeta q-S$, $q=G(S,N)$, fix capital,
payoff moments and demand parameters strictly inside one regime. Dollar executed
selling at reference price $\bar P>0$ obeys
\begin{equation}
dF^{sell}=\bar P(-dX)=\bar P(1-\zeta G_S)dS,
\qquad dE^{BP}=b_EG_SdS .
\end{equation}
If $1-\zeta G_S>0$, division yields
\begin{equation}
\vartheta_F\equiv\frac{dE^{BP}}{dF^{sell}}
=\frac{b_EG_S}{\bar P(1-\zeta G_S)},\qquad
\frac{dE^{BP}}{d(\bar P S)}=\frac{b_EG_S}{\bar P}.
\end{equation}
Substitution into $K_h=b_EG_S\bm l'A_D^{h-1}B_D$ gives
\eqref{eq:flowbridge}. In the slack regime
$G_S=\Gamma/(N+\zeta\Gamma)$ and hence
$\vartheta_F=b_E\Gamma/(\bar P N)$. Positive inflows have the opposite
orientation to selling. At the binding cap, where $1-\zeta G_S=0$, the price can change
without a first-order quantity change, and the flow-based inversion is unavailable.
One-sided limits must be used at a regime boundary. Subscriptions multiplied by
corporate portfolio shares are a proxy for executed corporate-claim transactions,
not a measurement of the desired shift $S$. Applying the estimated pricing
coefficient to the household operator would additionally require matching asset
scope, measuring the absorption share and identifying the household loading.
Appendix~\ref{app:bridgefund} implements estimation of the two market terms in simulation and
reports the limits on identifying them in the actual 57-month sample.

\paragraph{Subscriptions and changes in corporate holdings.}
A separate quarterly holdings check supplies an observed quantity margin.
For consecutive three-month reports, approximate purchases equal end-of-quarter
corporate holdings less beginning holdings compounded at the fund's total
return. The corporate sleeve's own return and security-level executions are
not observed. Both purchases and subscriptions are divided by initial net
assets. The regression includes quarter effects, uses initial-net-asset weights
and clusters standard errors by fund.

\begin{table}[!htbp]\centering
\begin{threeparttable}
\caption{Fund subscriptions and approximate corporate-bond purchases}
\label{tab:subscription_holdings}
\footnotesize
\begin{tabular*}{\textwidth}{@{\extracolsep{\fill}}lcc}
\toprule
 & Including reinvestment & External cash only\\
\midrule
Purchases per dollar subscribed & 0.635\sym{***} & 0.642\sym{***}\\
 & (0.026) & (0.025)\\
Mean corporate portfolio share & 0.735 & 0.735\\
Observations & 6,993 & 6,993\\
Funds & 495 & 495\\
\bottomrule\end{tabular*}
\begin{tablenotes}[flushleft]\footnotesize
\item Notes: Consecutive reports exactly three months apart, all three monthly
flow reports present, initial net assets over \$10 million, positive initial
corporate holdings. The sample jointly trims the outer one percent of purchase,
subscription, external-cash and net-asset roll-forward ratios. The portfolio
share is weighted by beginning net assets. Parentheses give fund-clustered
standard errors. Purchases are a return-adjusted holdings proxy, not observed
trades. \mbox{\sym{*} $p<0.10$}, \mbox{\sym{**} $p<0.05$}, \mbox{\sym{***} $p<0.01$}.
\end{tablenotes}
\end{threeparttable}
\end{table}

The subscription slope divided by mean corporate allocation is 0.864, or 0.874
using external cash. This is a proportional-allocation diagnostic, not an
estimate of the equilibrium absorption share $m_F$, and no uncertainty interval for
the ratio is estimated. It connects subscriptions to corporate-portfolio
adjustment without identifying the household-to-fund ownership map or removing
the pricing regression's exclusion concerns.

\paragraph{Conditional power and economic scale.}
An illustrative magnitude calculation assumes modified duration 6.5 and a
unit mapping from the yield change to the EBP. A price multiplier $m_{\mathrm{price}}$ then
implies an absolute slope $100m_{\mathrm{price}}/6.5$ basis points per inflow of 1\% of the
market denominator, which is 15.4 at $m_{\mathrm{price}}=1$ and 76.9 at $m_{\mathrm{price}}=5$, or 0.154 and 0.769
basis points per 0.01\%. These are assumed scenarios, not independently
estimated multipliers or demand elasticities.

The simulation fixes the original-flow design in column (1), including
57 monthly observations, its flows, instruments and factors, and adds
independent Gaussian errors at the standard deviation of EBP changes
residualized on controls alone. At $h=0$ that scale is 22.54 basis points.
Each cell in Table~\ref{tab:flowpower} uses 2,000 draws and the same 2SLS
procedure with three-lag HAC standard errors. The decision rule is
$|t|>1.96$, but its finite-sample null size has not been calibrated. Thus the
table reports conditional rejection frequencies under a nominal 5\% rule,
rather than a validated power guarantee. It does not propagate uncertainty
in factors, fund selection or instrument exclusion. The simulations use
the absolute slope, since changing its sign leaves two-sided power unchanged
under the symmetric error model.

\begin{table}[htbp]\centering
\begin{threeparttable}
\caption{Conditional sensitivity of the original fund-flow design}
\label{tab:flowpower}
\footnotesize
\begin{tabular*}{\textwidth}{@{\extracolsep{\fill}}lccc}
\toprule
Outcome endpoint $h$ (months) & 0 & 1 & 3 \\
\midrule
Assumed multiplier $m_{\mathrm{price}}=1$ & 8.8\% & 9.0\% & 10.7\% \\
Assumed multiplier $m_{\mathrm{price}}=5$ & 13.9\% & 12.9\% & 10.3\% \\
Months & 57 & 57 & 57 \\
\bottomrule
\end{tabular*}
\begin{tablenotes}[flushleft]\footnotesize
\item Notes: Simulated rejection frequencies. The outcome is
$E^{BP}_{t+h}-E^{BP}_{t-1}$, and $h=0$ is the same-month specification in
Table~\ref{tab:creditflow}. Errors, duration and the mapping to EBP are
assumed as described in the text. These simulations use the original-flow
measure, including reinvested distributions, not the external-cash measure.
\end{tablenotes}
\end{threeparttable}
\end{table}

Across the broader set of four fund/instrument constructions and three
endpoints, rejection frequencies at $m_{\mathrm{price}}=5$ range from 10.3\% to 17.4\%.
A separate 4,000-draw calculation for the original-flow, same-month design
at $m_{\mathrm{price}}=5$ gives a mean absolute estimate among significant draws 4.12 times
the imposed slope. This describes selection on significance under the
assumed data-generating process, but it provides no correction factor for
the observed coefficient. As a post-result sample sensitivity, excluding
March--May 2020 from the original-flow specification gives $-4.03$ (3.08)
basis points per 0.01\% inflow over 54 months. The external-cash comparison
and leave-one-year range are reported above. Together these calculations
motivate separating a provisional pricing association from a calibrated
market-wide response.

\section{A longer spending sample: feasibility and power}
\label{app:design}

This planning exercise evaluates a longer spending design and a joint detection-and-calibration
gate. Appendix~\ref{app:recoverycheck} separately evaluates prediction accuracy for a financially weighted target
with hypothetical aggregate and financial measurements. The two exercises answer different
questions and use different data-generating assumptions and success criteria.

As main Section~\ref{sec:implementation} notes, the NielsenIQ Consumer Panel (from 2004) and Retail
Scanner data (from 2006) would add pre-pandemic storms. They measure different objects, so the spending map
must be chosen with care. Store revenue and resident purchases require different maps $C_*$,
because evacuation can relocate sales without an equivalent change in household demand. Coverage, product scope
and household precision must be assessed before choosing a map. Neither panel supplies the financial
loadings $\bm l$, and hence $a$, or the market map $c_E$.

\subsection{Public inputs}\label{wp:app:F.1}
The feasibility of a 2006--2019 event design can be assessed using public inputs,
without observing licensed spending outcomes. Table~\ref{tab:publicdesign} reports the result. The eligibility rules are
research choices (contiguous-U.S.\ county storm declarations, non-winter, complete dates, duration
of at most 14 days and complete support from four weeks before to twelve weeks after onset), not
official definitions of unexpected shocks, and inputs are equal-weighted affected-county incidence
within each declaration rather than damage dollars. NOAA links are candidates, not verified
physical identity, and warning dates and anticipation remain unresolved for the historical events.

\begin{table}[!htbp]\centering
\begin{threeparttable}
\caption{Public-data feasibility of the 2006--2019 event design}
\label{tab:publicdesign}
\footnotesize
\begin{tabularx}{\textwidth}{Xc}
\toprule
Quantity & Result\\
\midrule
Eligible declarations: training / validation & 246 / 127\\
Full $[-4,12]$-week support components: training / validation & 4 / 1\\
Response-only $[4,12]$ components: training / validation & 9 / 8\\
FEMA declarations with NOAA county/date candidate links & 351 of 373\\
Matched NOAA rows with missing or unparsed property damage & 1,568\\
Calendar weeks: full / training / validation & 730 / 406 / 300\\
Residualized network design rank & 102 / 102\\
Column-normalized condition number & 84.1\\
Events with no directly hit county in the 1,590-county support & 43\\
Mean retained within-event designated-county share & 51.9\%\\
\bottomrule
\end{tabularx}
\begin{tablenotes}[flushleft]\footnotesize
\item Notes: Components are transitive overlap groups, not verified independent shocks. Training is
2006--2013 and validation 2014--2019, with a 12-week purge on each side of the split. The 102
columns use 34 nonstationary SCI modes and three temporal terms over lags 4--12, projected
nationally with population-proxy weights.
\end{tablenotes}
\end{threeparttable}
\end{table}

\subsection{Synthetic estimator experiments}\label{wp:app:F.2}
A synthetic design with 18 counties, three states, unequal expenditure weights, serial correlation
0.6, spatially correlated noise and dose-dependent heteroskedasticity tests the nested
cross-fitting estimator. Three outer folds each retune rank, decay, ridge penalty and aggregation
weight within training, and no financial outcome selects an operator. Both a shape oracle, which
knows the response pattern but must estimate its amplitude, and the feasible estimator must beat
incidence, gravity and production benchmarks at two endpoints and satisfy slope and intercept
equivalence.

With calibration evaluated at zero, the strong-signal case passes in 5.25 (oracle) and 9.75
percent (feasible) of 400 draws, and every other scenario, including the incidence and production
nulls, passes in none. An alternative calibration at the training-mean prediction was chosen
after inspecting the first five scenarios and evaluated on 200 new draws. It raises the
strong case to 69.5 percent for both methods, still below an 80 percent target. A linked experiment
that adds a known 10-basis-point financial bridge finds no draw that passes both the spending and
the financial stage. In its first strong case, incidence explains 96.7 percent of the true
aggregate state's training variance, the collapse in \eqref{eq:split} appearing in simulation.
These experiments diagnose the design rather than measure the power of licensed data.

\section{Recorded losses, spending measurement and source timing}
\label{app:localrepair}

\subsection{The source ledger and the exposure unit}
\label{app:sourceunits}

The matched spending comparison uses 48 declarations and the same 74,779
recipient--declaration observations in every column. FEMA supplies incident dates and
designated counties, while NOAA supplies retrospective weather reports and property-damage
amounts. The loss ledger retains 7,014 distinct NOAA report identifiers and the original
684 eligible source county--declaration combinations. It distinguishes positive recorded
amounts, zero-coded reports, blank amounts and absence of a report. Sums of recorded
contributions are not estimates of complete county losses.

Let $\Delta\mathcal D_{re,s}$ be the eligible recorded amount in source county $r$ assigned to response
week $s$, and let $\mathcal D_{re}(h)=\sum_{s\leq h}\Delta\mathcal D_{re,s}$ and
$\mathcal D_e(h)=\sum_r\mathcal D_{re}(h)$. The source shares and the index, in index
points, are
\begin{equation}\label{eq:damage_share_unit}
 s^{(h)}_{re}=\frac{\mathcal D_{re}(h)}{\mathcal D_e(h)},
 \qquad E^{SCI}_{je}(h)=100\sum_rW^{loc}_{jr}s^{(h)}_{re},
\end{equation}
and \eqref{eq:localexposure} is the case $h=4$.
When $\mathcal D_e(h)=0$, set the recorded source vector and $E^{SCI}_{je}(h)$ to zero and retain
the loss-coverage flags. This does not assert zero actual destruction.
As in Section~\ref{sec:stage1}, the local weights $W^{loc}$ retain each recipient's 100
largest out-of-county SCI-times-source-population links, normalized by row. An index
point, one unit of $E^{SCI}$, is one percentage point of the damage-share-weighted link
index. It is not a percentage point of friends affected, since source
weights sum to one across damaged counties, rather than assigning weight one to
each affected county. Multiplying every recorded loss in a declaration by the
same number leaves $E^{SCI}$ unchanged. The index therefore describes the geography of
recorded destruction, while the monetary alternatives below retain its scale.

In \eqref{eq:lp}, the outcome $Y_{jeh}$ is 100 times the log spending change, in log
points, and exposure is in index points. Scaling both by 100 leaves the regression
coefficient unchanged, so it equals the coefficient of the unscaled log change on the
unscaled share index. Thus the primary $-0.352585$ is reported as approximately $-0.353$
percent spending per index point. This change of reporting units leaves the fitted
responses and test statistics unchanged. No exposure cap is applied. The
incident-aligned and onset-week constructions also use different loss-record
windows, so their comparison is not a unit-only rescaling.

The primary input uses county-coded reports beginning within the declared
incident. A coverage comparison adds NOAA zones mapping uniquely to one eligible
source county. Ambiguous multi-county zones remain unallocated. The inherited
zone crosswalk is not a validation of all historical zone boundaries, so direct
county assignment remains primary. The ledger preserves potentially overlapping
reports rather than treating matching amounts or narratives as conclusive duplicates.
The same source definition is applied to the gravity and two production exposures.

\begin{table}[!htbp]\centering
\begin{threeparttable}
\caption{Four-week spending gradients across recorded-loss constructions}
\label{tab:source_comparison_v31}
\footnotesize\setlength{\tabcolsep}{3pt}
\begin{tabular*}{\textwidth}{@{\extracolsep{\fill}}lcccc}
\toprule
 & Spending & \multicolumn{3}{c}{Wild-bootstrap $p$}\\
\cmidrule(lr){3-5}
Loss-record construction & (\%/index pp) & Declaration & Episode & Calendar\\
\midrule
Onset-week county reports & $-0.336$\sym{**} & 0.029 & 0.029 & 0.056\\
Reviewed recorded subtotal & $-0.299$\sym{**} & 0.047 & 0.047 & 0.077\\
Larger recorded amount & $-0.323$\sym{**} & 0.036 & 0.035 & 0.062\\
Onset-week county and mapped zones & $-0.313$\sym{**} & 0.049 & 0.053 & 0.121\\
Incident-aligned county reports & $-0.353$\sym{**} & 0.024 & 0.024 & 0.055\\
Incident-aligned county and mapped zones & $-0.325$\sym{**} & 0.046 & 0.049 & 0.134\\
Mapped reports, end-date assignment & $-0.332$\sym{**} & 0.041 & 0.044 & 0.112\\
Mapped reports, Sumner sensitivity & $-0.326$\sym{**} & 0.046 & 0.049 & 0.134\\
\midrule
County and declaration fixed effects & \multicolumn{4}{c}{Yes}\\
Gravity and production controls & \multicolumn{4}{c}{Yes}\\
Recipient weather and pre-activity controls & \multicolumn{4}{c}{Yes}\\
County--declaration observations & \multicolumn{4}{c}{74,779}\\
\bottomrule
\end{tabular*}
\begin{tablenotes}[flushleft]\footnotesize
\item Notes: The outcome is week-4 log spending relative to week $-1$, multiplied
by 100. Each source construction rebuilds
the social, gravity and production fields on identical recipients. Three
one-way, null-imposed Webb-weight bootstraps use 9,999 draws, with 48 declaration,
39 episode and 13 sixteen-week onset-calendar groups. Stars use declaration
bootstrap $p$-values: \mbox{\sym{*} $p<0.10$}, \mbox{\sym{**} $p<0.05$}, \mbox{\sym{***} $p<0.01$}.
The related comparisons were developed after the original outcomes were examined.
\end{tablenotes}
\end{threeparttable}
\end{table}

The negative four-week gradient is similar across all eight constructions.
The reviewed subtotal retains NOAA property-loss entries and documented county-specific
public amounts, with source links and scope notes recorded in the county--declaration
ledger. It is a partial recorded-property subtotal, not an independent complete
assessment. Here 435 of 684 county--declaration amounts are blank, and two
declarations have no populated revised amount. Taking the larger recorded value
changes only one county relative to the original input. The Sumner comparison
replaces its NOAA county total by a reported \$130 million property assessment,
never adds the two. The source is a news quotation of the county official rather
than a verified primary assessment. These alternatives test sensitivity to
recorded inputs, while the agreement supports the spatial spending gradient.

\subsection{Recipient weather and regression inference}
\label{app:measuredweather}

The regressions retain the recipient sample and control for its measured weather
instead of screening on any report over a long future window. The weather block
contains county precipitation and temperature, heat and cold indicators, station
gust measurements, warning exposure and damaging-weather reports over the aligned
event window. Coverage and missing-value indicators distinguish an unavailable
measurement from measured low weather. The remaining controls are distance bins,
same-hit-state exposure, gravity, upstream and downstream production fields, and
predated spending level and growth. County and declaration fixed effects remove
persistent recipient differences and the response common to a declaration.

Ordinary least squares does not impose a Gaussian regression-error distribution.
Preferred tests impose the tested coefficient's null, multiply restricted
residuals by a common six-point Webb weight within each resampling group, reapply
the nuisance projections, and compare studentized statistics over 9,999 draws.
The weights are $\pm\sqrt{1/2}$, $\pm1$ and $\pm\sqrt{3/2}$, each with probability
$1/6$. This procedure preserves the fitted design's heteroskedastic residual
pattern rather than generating Gaussian observation errors.

Declaration and episode resampling give closely aligned inference for the
primary coefficient. Calendar-block inference is weaker, with $p=0.055$ for the
incident-aligned county input and $0.112$--$0.134$ for the mapped-zone timing
comparisons. The dependence models answer different sampling questions. None
accommodates arbitrary dependence or resolves concentration by construction.
Appendix~\ref{app:uncapped_diagnostics} reports the corresponding pre-event and
influence checks.

\subsection{Monetary severity and the household source response}
\label{app:monetaryseverity}
\label{app:spillover_tracks}

A monetary input asks how much response follows the destruction, rather than
where recorded destruction occurs. In millions of 2024 dollars, let
$E^{\$}_{je}(h)=10^{-6}\sum_rW^{loc}_{jr}\mathcal D^{2024}_{re}(h)$, which retains
total recorded event loss. The income-scaled alternative is
\begin{equation}\label{eq:income_damage_input}
 E^{I}_{je}(h)=100\sum_rW^{loc}_{jr}
          \frac{\mathcal D^{2024}_{re}(h)}{I^{2024}_{r,2019}},
\end{equation}
where $I_{r,2019}$ is observed annual county personal income, converted to the
same prices. All 684 source county--declaration rows match the BEA denominator.
A percentage point of $E^I$ is damage relative to annual source income, not an
observed income loss or a percentage point of friends affected, so the dollar and
income-scaled coefficients cannot be compared by numerical size alone.
Table~\ref{tab:monetary_severity_v31} reports the matched regressions. The negative spatial
gradient is more precisely estimated than a proportional monetary source loading. Locating a
recipient response and calibrating dollars of destruction into a household disturbance
therefore require different information. Conditional tests of proportional damage scaling give
declaration-bootstrap $p$-values of 0.077 on the onset-week recorded input and 0.018 on the
reviewed subtotal, and they do not deliver a specification-independent verdict. The estimates do
not select attention over resource disruption, nor establish that damage has
no role in response magnitude.

\begin{table}[!htbp]\centering
\begin{threeparttable}
\caption{Monetary source inputs on the matched recipient sample}
\label{tab:monetary_severity_v31}
\footnotesize
\begin{tabular*}{\textwidth}{@{\extracolsep{\fill}}lccc}
\toprule
 & County & County and zones & End-date assignment\\
\midrule
Dollar exposure, \% per \$1 million & $-0.054$ & $-0.057$ & $-0.057$\\
Declaration-bootstrap $p$ & 0.268 & 0.257 & 0.257\\
\addlinespace
$\log(1+E^{\$})$, \% per log unit & $-1.474$ & $-1.473$\sym{*} & $-1.478$\sym{*}\\
Declaration-bootstrap $p$ & 0.118 & 0.096 & 0.095\\
\addlinespace
Damage/income exposure, \% per pp & $-0.620$ & 7.133 & 6.821\\
Declaration-bootstrap $p$ & 0.960 & 0.589 & 0.603\\
\midrule
County and declaration fixed effects & Yes & Yes & Yes\\
Gravity and production controls & Yes & Yes & Yes\\
Recipient weather and pre-activity controls & Yes & Yes & Yes\\
County--declaration observations & 74,779 & 74,779 & 74,779\\
\bottomrule
\end{tabular*}
\begin{tablenotes}[flushleft]\footnotesize
\item Notes: Each row is a separate matched regression using incident-aligned
reports assigned through week 4. Every competing exposure uses the same source
input as its social counterpart. Dollars are in
2024 prices. The log-dose model is a different response law, not a change of
coefficient units. Stars and $p$-values use the 9,999-draw declaration wild
bootstrap. No recipient reaches one percentage point of the damage/income input. \mbox{\sym{*} $p<0.10$}, \mbox{\sym{**} $p<0.05$}, \mbox{\sym{***} $p<0.01$}.
\end{tablenotes}
\end{threeparttable}
\end{table}

\subsection{Continuing destruction and response age}
\label{app:datedforcing}

The dated ledger assigns each reported amount to its local beginning date and,
in a separate timing comparison, to its end date. These are weather-report
dates, not observations of within-report loss accrual or real-time publication
of damage estimates. Incident membership is likewise not proof of physical
storm identity. Later reports enter only endpoints on or after their assigned
date, so a whole-incident total is never placed mechanically at onset.

For a dated monetary input $\Delta\mathcal D_{re,s}$ and a response-age bin $B_b$, define
\[
 E^{m,b}_{jeh}
 =\sum_{\substack{s\leq h\\h-s\in B_b}}\sum_rW^m_{jr}\Delta\mathcal D_{re,s},
 \qquad B_b\in\{[0,1],[2,4],[5,12]\},
\]
where $m$ is social, gravity, upstream or downstream. Stacking horizons
$h\in\{1,2,4,8,12\}$ gives
\begin{equation}\label{eq:dated_response_age}
 Y_{jeh}=\alpha_{j,h}+\mu_{e,h}
 +\sum_b\beta_b E^{SCI,b}_{jeh}
 +\sum_{m\ne SCI}\sum_b\phi^{comp}_{m,b}E^{m,b}_{jeh}
 +\bm Z_{je}'\bm\phi_h+u_{jeh}.
\end{equation}
The horizon-specific nuisance terms are residualized within each response block.
Inference groups repeated horizons together. The 373,895 stacked rows are five
measurements of the same 74,779 recipient--declaration pairs, not five times as
many independent shocks. Dollar and income-scaled inputs are estimated separately.

\begin{table}[!htbp]\centering
\begin{threeparttable}
\caption{Dated-dollar spending responses by age of destruction}
\label{tab:dated_response_age_v31}
\footnotesize
\begin{tabular*}{\textwidth}{@{\extracolsep{\fill}}lcc}
\toprule
 & County reports & County and mapped zones\\
\midrule
Response age 0--1 weeks, \% per \$1 million & 0.0060 & 0.0061\\
Declaration-bootstrap $p$ & 0.870 & 0.862\\
Response age 2--4 weeks & 0.0017 & $-0.0003$\\
Declaration-bootstrap $p$ & 0.951 & 0.993\\
Response age 5--12 weeks & $-0.0463$ & $-0.0410$\\
Declaration-bootstrap $p$ & 0.265 & 0.288\\
\midrule
Joint zero-social-response bootstrap $p$ & 0.762 & 0.867\\
Horizon-specific county/declaration effects & Yes & Yes\\
Matched competing fields and controls & Yes & Yes\\
Stacked observations & 373,895 & 373,895\\
\bottomrule
\end{tabular*}
\begin{tablenotes}[flushleft]\footnotesize
\item Notes: Spending levels are relative to week $-1$, with 2024-dollar inputs
dated at report beginning. The three age bins partition the assigned input at
each endpoint. Each coefficient measures a spending-level response at that
response age, not expenditure cumulated across weeks. All tests use 9,999
null-imposed declaration-cluster Webb draws. The joint test re-estimates and
studentizes all three social coefficients together.
\end{tablenotes}
\end{threeparttable}
\end{table}

The dated construction prevents continuing destruction from being mistaken for
slow decay of an initial impulse. The late-age dollar coefficients are negative,
but their bootstrap uncertainty does not establish a persistent severity loading.
For the income-scaled model, joint declaration-bootstrap $p$-values are 0.184
for county reports and 0.122 with mapped zones. These distributed responses
inform source timing, but they are not a new estimate of the graph's decay law.

\subsection{Calendar-pattern measurement and operator comparisons}
\label{app:calendar_sensitivity}

The spending tracker uses a fixed 2019 comparison year. Recurring county-specific
fluctuations can therefore contain a comparison-year component as well as
genuine seasonality. In the schematic observation model
$L^{obs}_{jt}=\ell_{jt}-b_{j,w(t)}+\eta_{jt}$, temporally independent
$\eta$ produces adjacent differenced-noise covariance $-\sigma_j^2$.
A shared noisy baseline also correlates response errors across horizons.
These observations motivate measurement-aware comparisons rather than treating
every short-run change as noise.

The recovery-design noise calibration uses a retrospective calendar adjustment.
Let $HP_{13}(L)$ be the residual from a centered thirteen-week moving average,
computed with at least seven observed weeks. For declaration $e$, donor dates
are $t+52n$, $n\in\{-4,\ldots,4\}\setminus\{0\}$, on or after September 7,
2020, excluding the declaration's weeks $[-19,18]$. With $m$ eligible donors,
\begin{equation}\label{eq:calendar_adjustment}
 L^*_{jet}=L^{obs}_{jt}
 -\frac{m\varrho_{cal}}{m\varrho_{cal}+1-\varrho_{cal}}\,
 \frac1m\sum_{n\in\mathcal N_{jet}}HP_{13}(L^{obs})_{j,t+52n},
\end{equation}
with no subtraction if donors are unavailable. Here $\varrho_{cal}$ is the mean
uncentered cross-cycle correlation at 52, 104 and 156 weeks. Donors can come
from later years. The adjustment is a measurement sensitivity, not real-time
information or an independently observed household response. The local
comparison remains on unadjusted spending levels.

For a fixed linear preprocessing map $F_e^{cal}$, a candidate operator must be
passed through the same map as its observed response:
\begin{equation}\label{eq:filtered_operator}
 F_e^{cal}y_e\quad\text{and}\quad
 F_e^{cal}\mathcal H(\vartheta)\bm\varepsilon_e+F_e^{cal}\bm Z_e\bm\phi,
 \qquad V_e^F=F_e^{cal}V_e(F_e^{cal})'.
\end{equation}
The prediction must cover donor and baseline dates and overlapping inputs.
Filtering can remove observable response directions, so Proposition~\ref{prop:recovery}
applies to the measurement map actually used. More precise measurements and a
valid financial projection play distinct roles in the recovery exercise.

\subsection{Cross-disaster scope and supporting household finance}
\label{app:additional_events}
\label{app:prior_disasters}

A separate exploratory extension examined 67 additional declarations, mostly
floods, hurricanes, winter storms and fires, with a rebuilt exposure and
conditioning design. It did not establish a common negative coefficient across
those disaster types. Source allocations and controls differed from the primary
sample, including equal county allocations when recorded damage was unavailable.
The result confines the demonstrated local gradient to the primary storm setting,
and the extension is not a matched validation of the incident-aligned construction.
Unknown damage allocation alone does not explain the extension's result, since
its recorded-damage subset also lacked a negative association. A cross-disaster
transport test requires comparable source measurement before interpreting
differences as household or operator heterogeneity.

An exploratory monthly mortgage comparison supplies complementary evidence of
household financial strain. Across 433 covered recipient counties and 20,565
county--declaration pairs, the onset-week recorded-loss index predicts
a 0.0208-percentage-point increase per index point in mortgages 30--89 days delinquent,
averaged over months 1--3 relative to month $-1$. Declaration- and episode-bootstrap
$p$-values are 0.038 and 0.040, while sixteen-week calendar-block inference is weaker
(approximately 0.20--0.23). The response is positive under all four compared
loss constructions. The mortgage sample and its weather controls differ from
the spending design, and exposure information is concentrated in a few
declarations. The result is supporting financial-stress evidence, not a measure
of fixed-price portfolio demand or its credit-market price impact.
\subsection{Local-response diagnostics}
\label{app:uncapped_diagnostics}
\begin{table}[!htbp]\centering
\begin{threeparttable}
\caption{Four-week spending gradient: sample and weather comparisons}
\label{tab:uncapped_sensitivity}
\footnotesize\setlength{\tabcolsep}{4pt}
\begin{tabular*}{\textwidth}{@{\extracolsep{\fill}}lccc}
\toprule
 & Declaration & Episode & Calendar \\
\midrule
Measured own weather & $-0.353$\sym{**} & $-0.353$\sym{**} & $-0.353$\sym{*} \\
 & (0.115) & (0.113) & (0.077) \\
Without measured own weather & $-0.312$\sym{**} & $-0.312$\sym{**} & $-0.312$\sym{**} \\
 & (0.112) & (0.113) & (0.067) \\
Onsets from December 2020 & $-0.319$\sym{**} & $-0.319$\sym{**} & $-0.319$ \\
 & (0.111) & (0.107) & (0.086) \\
\midrule
County and declaration effects & Yes & Yes & Yes \\
Matched gravity and production exposures & Yes & Yes & Yes \\
\bottomrule\end{tabular*}
\begin{tablenotes}[flushleft]\footnotesize
\item Notes: Coefficients are interpreted as the change in spending from week $-1$ to week 4, in log percentage points, associated with a 1-percentage-point increase in the damage-share-weighted link index. Columns use declaration, physical-episode and sixteen-week onset-block wild-bootstrap standard errors, respectively (9,999 Webb draws). The first two rows use 74,779 observations and 48 declarations, while the last uses 63,889 observations and 41 declarations. All rows include the same non-weather controls and consistently rebuilt competing exposures. The full sample contains 48 declaration, 39 episode and 13 calendar clusters. \mbox{\sym{*} $p<0.10$}, \mbox{\sym{**} $p<0.05$}, \mbox{\sym{***} $p<0.01$}.
\end{tablenotes}
\end{threeparttable}
\end{table}
The incident-aligned index preserves the ordering of highly exposed recipients. The four-week gradient remains negative in all 48 leave-one-declaration-out samples, with coefficients between $-0.404$ and $-0.309$ per index point and conventional two-way $p<0.05$ in 47 of them, and when the acute pandemic onsets are excluded. Each deletion re-estimates all exposures and controls. Table~\ref{tab:uncapped_sensitivity} separates stability of the point estimate from its precision under alternative dependence structures.

Reported standard errors are the sample standard deviation of 9,999 unrestricted-residual wild-bootstrap coefficient estimates, using the Webb six-point weights \citep{Webb2023}. The original finite-sample variance correction is applied once, as its square root multiplying the residuals. Stars use the corresponding restricted-null, studentized bootstrap tests. The same dependence groups and multiplier family are used for variance estimation and testing. Calendar-panel tests additionally re-estimate Bartlett-HAC studentization in each draw.
For the event-time comparison, the four-week damage geography is held fixed at every horizon. This retrospective construction makes the pre-event coefficients and later spending differences comparable. It is not an information set available before week 4 or an estimated impulse response. The shared week $-9$ baseline in the figure also displays the week immediately preceding onset rather than normalizing it to zero. Its coefficient is $-0.020$ per index point (declaration-bootstrap $p=0.909$). The four-week estimate remains negative under either baseline. Table~\ref{tab:uncapped_event_time} reports the week $-1$ baseline used in the main spending table.
\begin{figure}[!htbp]
\centering
\includegraphics[width=\textwidth]{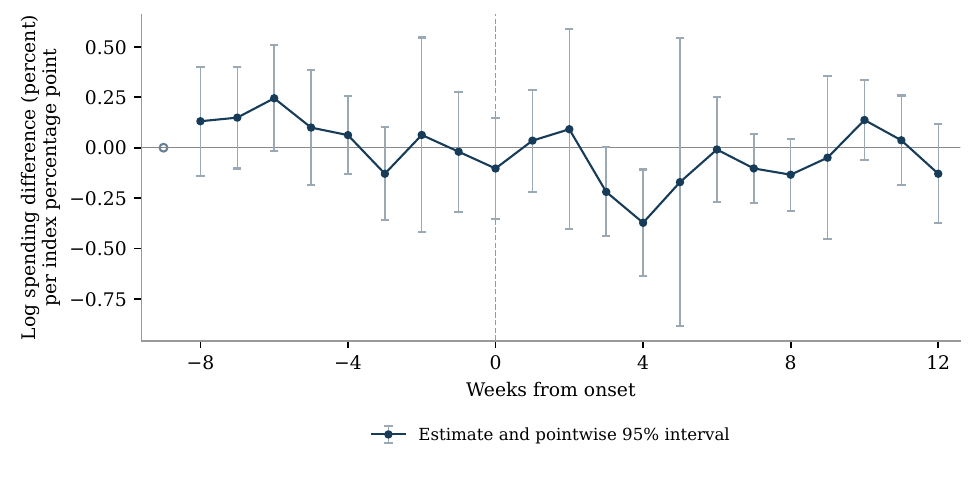}
\caption{Event-time spending gradient for the incident-aligned county exposure.
The same four-week loss geography and 74,779 observations are used at every
date, with spending relative to week $-9$. This is a retrospective event-study
comparison, not a known onset-pulse response. Error bars give 95 percent pointwise
approximate studentized declaration-bootstrap intervals. Activity controls
predate the leads and measured weather covers the path. The main spending table
instead subtracts week $-1$.}
\label{fig:uncapped_local_path}
\end{figure}

\begin{table}[!htb]\centering
\begin{threeparttable}
\caption{Spending differences with a fixed four-week damage geography}
\label{tab:uncapped_event_time}
\footnotesize\setlength{\tabcolsep}{4pt}
\begin{tabular*}{\textwidth}{@{\extracolsep{\fill}}lccc}
\toprule
Week relative to onset & Declaration & Episode & Calendar \\
\midrule
0 & $-0.083$ & $-0.083$ & $-0.083$ \\
 & (0.099) & (0.098) & (0.117) \\
1 & 0.055 & 0.055 & 0.055 \\
 & (0.129) & (0.132) & (0.109) \\
2 & 0.111 & 0.111 & 0.111 \\
 & (0.158) & (0.155) & (0.131) \\
3 & $-0.199$ & $-0.199$ & $-0.199$ \\
 & (0.140) & (0.138) & (0.159) \\
4 & $-0.353$\sym{**} & $-0.353$\sym{**} & $-0.353$\sym{*} \\
 & (0.115) & (0.113) & (0.077) \\
8 & $-0.114$ & $-0.114$ & $-0.114$ \\
 & (0.190) & (0.191) & (0.141) \\
12 & $-0.109$ & $-0.109$ & $-0.109$ \\
 & (0.200) & (0.199) & (0.194) \\
\midrule
Joint pre-event tests ($p$) &  &  &  \\
Weeks $-3$ to $-1$ & 0.702 & 0.713 & 0.774 \\
Weeks $-8$ to $-1$ & 0.856 & 0.855 & 0.951 \\
\bottomrule\end{tabular*}
\begin{tablenotes}[flushleft]\footnotesize
\item Notes: Coefficients are changes in spending, in log percentage points, per point of the damage-share-weighted link index. The post-onset rows subtract week $-1$ and use the same 74,779 observations, fixed effects and controls as Table~\ref{tab:uncapped_sensitivity}. Columns use declaration, physical-episode and sixteen-week onset-block wild-bootstrap standard errors, respectively (9,999 Webb draws). Joint pre-event tests instead use week $-9$ and common bootstrap multipliers across horizons, preserving their covariance. Their reference is the empirical joint Wald distribution, not a normal or chi-squared approximation. All draws are retained. Calendar inference has 13 clusters. The joint tests do not establish parallel trends. The figure uses approximate studentized null-bootstrap-$t$ intervals, not exact confidence-set inversions. \mbox{\sym{*} $p<0.10$}, \mbox{\sym{**} $p<0.05$}, \mbox{\sym{***} $p<0.01$}.
\end{tablenotes}
\end{threeparttable}
\end{table}
The negative gradient is clearest at four weeks. Later point estimates are smaller and their uncertainty is wide, so the event profile supplies evidence about the location and timing of the spending response rather than a measured relaxation rate.
\subsection{Dated repeated inputs and fixed-origin recipients}
\label{app:calendar_timing}

The county-calendar comparison in Table~\ref{tab:calendar_timing} separates
damage arriving at different dates and retains counties after later direct hits.
The full panel has 1,574 counties and 211 common weekly origins, February 17, 2020
through February 26, 2024. Actual spending levels from five weeks before each
origin through twelve weeks after it, and complete seven-day rain and temperature
measurements, give 332,114 county-origin observations. The fixed-origin cohort
excludes a county only if it has an own county-coded NOAA weather report in the
starting week, giving 300,357 observations. Both samples follow later hits.
This is an observational cohort comparison, not randomized intention to treat.

The source universe, the 2,802 counties that supply the local network inputs,
remains separate from spending coverage. The original 48 incident windows contain 2,617
distinct county-coded NOAA reports, with recorded property damage of \$1.592
billion nominal, or \$1.665 billion in 2024 dollars. The entire amount of each
report enters once in its start week. Within-report accrual is unobserved.
Blank amounts remain flagged, and no input is capped or interpolated. Other county-coded
NOAA reports supply background forcing controls, although omitted zone reports and
counties outside the source graph limit coverage.

Let $\Delta\mathcal D_{re,t}$ be new recorded damage for source $r$ in week $t$ and
$\mathcal D_e^{full}=\sum_{r,t\in e}\Delta\mathcal D_{re,t}$ its declaration's fixed full-incident total. Let
\[
 s_{rt}=\sum_e\Delta\mathcal D_{re,t}/\mathcal D_e^{full},\qquad x^{SCI}_{jt}=100\sum_rW^{loc}_{jr}s_{rt}.
\]
For a zero recorded total, define the corresponding $\Delta\mathcal D_{re,t}/\mathcal D_e^{full}$ contribution
as zero and retain its coverage flag. This is not an assessed zero actual loss.
This input dates each report separately and uses a full-incident
denominator, whereas \eqref{eq:localexposure} sums reports through week 4.
Thus the calendar comparison extends the design rather than merely changing
the original coefficient's units.

For every $h=0,\ldots,12$, the outcome is $100(L_{j,t+h}-L_{j,t-1})$.
Current direct and social inputs enter jointly where direct input varies,
alongside current gravity/upstream/downstream exposure and four lags of the direct and
all mapped inputs. The regressions also include the pre-origin spending level and four
growth lags. Additionally, they include background damage with report-presence and
unknown-amount flags, and county and calendar-origin effects. Both specifications
adjust actual own weather before and after origin.
The future-conditioned model additionally includes the next twelve weeks of
direct and all mapped focal/background damage and report flags. Its complete
control set is held fixed at every horizon. As a result, projecting the summed
weekly outcomes gives exactly the sum of endpoint coefficients.

The comparison asks how the conditional path changes when subsequent recorded
arrivals are held fixed. Interpreting it as a causal pulse requires strict
conditional exogeneity and separability of the direct, social and competing
inputs. Future monetary losses need not satisfy these conditions simply because
meteorology is external. Neither the all-county comparison nor the origin cohort
conditions on surviving unhit until the response horizon.

Inference uses common Webb calendar-block multipliers. Standard errors follow
the unrestricted-residual coefficient bootstrap described in
Appendix~\ref{app:uncapped_diagnostics}. Null tests use restricted residuals and
Bartlett time-HAC studentization. Each of 9,999 draws imposes the zero-network-response
null, reprojects the complete nuisance space and re-estimates HAC studentization.
The same multipliers apply across counties and horizons. Thirteen- and
twenty-six-week blocks give 17 and 9 calendar blocks, respectively. The
origin cohort uses its exact unbalanced fixed-effect projection.
These are approximate dependence checks, not exact randomization inference,
and calendar blocks need not coincide with independent physical storms.

\begin{table}[!htb]\centering
\begin{threeparttable}
\caption{Weekly spending response in the fixed-origin cohort}
\label{tab:calendar_weekly}
\footnotesize\setlength{\tabcolsep}{4pt}
\begin{tabular*}{\textwidth}{@{\extracolsep{\fill}}lcccc}
\toprule
 & \multicolumn{2}{c}{History controls} & \multicolumn{2}{c}{Future damage controls} \\
\cmidrule(lr){2-3}\cmidrule(lr){4-5}
Week & 13 weeks & 26 weeks & 13 weeks & 26 weeks \\
\midrule
0 & $-0.207$ & $-0.207$ & $-0.233$ & $-0.233$ \\
 & (0.092) & (0.076) & (0.100) & (0.085) \\
1 & $-0.068$ & $-0.068$ & $-0.063$ & $-0.063$ \\
 & (0.091) & (0.084) & (0.089) & (0.078) \\
2 & 0.160 & 0.160 & 0.151 & 0.151 \\
 & (0.114) & (0.129) & (0.109) & (0.121) \\
3 & $-0.294$ & $-0.294$\sym{*} & $-0.310$ & $-0.310$\sym{*} \\
 & (0.108) & (0.125) & (0.101) & (0.119) \\
4 & $-0.457$\sym{**} & $-0.457$\sym{***} & $-0.441$\sym{**} & $-0.441$\sym{***} \\
 & (0.070) & (0.065) & (0.072) & (0.061) \\
5 & $-0.205$ & $-0.205$\sym{**} & $-0.190$ & $-0.190$\sym{**} \\
 & (0.128) & (0.115) & (0.129) & (0.106) \\
6 & $-0.066$ & $-0.066$ & $-0.054$ & $-0.054$ \\
 & (0.100) & (0.083) & (0.105) & (0.077) \\
7 & $-0.151$ & $-0.151$ & $-0.128$ & $-0.128$ \\
 & (0.117) & (0.094) & (0.117) & (0.099) \\
8 & $-0.214$ & $-0.214$ & $-0.199$ & $-0.199$ \\
 & (0.093) & (0.069) & (0.099) & (0.076) \\
9 & 0.000 & 0.000 & $-0.004$ & $-0.004$ \\
 & (0.089) & (0.063) & (0.097) & (0.076) \\
10 & 0.056 & 0.056 & 0.053 & 0.053 \\
 & (0.060) & (0.056) & (0.069) & (0.066) \\
11 & $-0.033$ & $-0.033$ & $-0.022$ & $-0.022$ \\
 & (0.163) & (0.120) & (0.167) & (0.125) \\
12 & $-0.180$ & $-0.180$ & $-0.191$ & $-0.191$ \\
 & (0.166) & (0.151) & (0.173) & (0.165) \\
\bottomrule\end{tabular*}
\begin{tablenotes}[flushleft]\footnotesize
\item Notes: The 300,357 observations have no own county-coded weather report at origin. Later hits are retained. Coefficients are log percentage points per point of the damage-share-weighted link index. Adjacent columns repeat each specification using 13/26-week calendar-block wild-bootstrap standard errors in parentheses (9,999 Webb draws). Outcomes, controls and source timing follow the text. The whole endpoint path has max-$t$ $p$-values 0.026/0.023 without and 0.038/0.026 with future damage controls, for 13/26-week blocks. \mbox{\sym{*} $p<0.10$}, \mbox{\sym{**} $p<0.05$}, \mbox{\sym{***} $p<0.01$}.
\end{tablenotes}
\end{threeparttable}
\end{table}

The fixed-origin four-week coefficient changes by 3.5 percent when later
recorded arrivals are controlled. The negative timing pattern is retained
without excluding later-hit counties. In the all-county panel, the corresponding
change is 5.5 percent, with more block-length-sensitive inference. This supports
the timing of the local spending gradient while preserving the distinction
between spatial contrasts and a full propagation law.

\begin{table}[!htbp]\centering
\begin{threeparttable}
\caption{Accumulated response in the fixed-origin cohort}
\label{tab:calendar_cumulative}
\footnotesize\setlength{\tabcolsep}{4pt}
\begin{tabular*}{\textwidth}{@{\extracolsep{\fill}}lcccc}
\toprule
 & \multicolumn{2}{c}{History controls} & \multicolumn{2}{c}{Future damage controls} \\
\cmidrule(lr){2-3}\cmidrule(lr){4-5}
Through week & 13 weeks & 26 weeks & 13 weeks & 26 weeks \\
\midrule
4 & $-0.866$\sym{*} & $-0.866$\sym{*} & $-0.896$\sym{*} & $-0.896$\sym{*} \\
 & (0.291) & (0.336) & (0.268) & (0.292) \\
8 & $-1.502$ & $-1.502$\sym{*} & $-1.467$ & $-1.467$ \\
 & (0.516) & (0.473) & (0.536) & (0.439) \\
12 & $-1.658$ & $-1.658$ & $-1.630$ & $-1.630$ \\
 & (0.743) & (0.593) & (0.821) & (0.639) \\
\bottomrule\end{tabular*}
\begin{tablenotes}[flushleft]\footnotesize
\item Notes: Sum of weekly log-level responses, in log-percentage-point-weeks per point of the damage-share-weighted link index, rather than a percentage change in total-window expenditure. Adjacent columns repeat each specification using 13/26-week calendar-block wild-bootstrap standard errors in parentheses (9,999 Webb draws). Averaging the response requires division by the number of weeks. Additionally, expenditure-dollar conversion requires spending bases and weights. \mbox{\sym{*} $p<0.10$}, \mbox{\sym{**} $p<0.05$}, \mbox{\sym{***} $p<0.01$}.
\end{tablenotes}
\end{threeparttable}
\end{table}

A plateau need not reverse accumulated
losses, and a confidence interval including zero does not establish disappearance,
so no exact recovery date is inferred. Similarly, smaller later spending estimates do not determine
the duration of portfolio adjustment, because the separate financial direction in
Section~\ref{subsec:aggregation} may be weakly measured by spending.

\begin{table}[!htbp]\centering
\begin{threeparttable}
\caption{Monetary input alternatives in the calendar panel}
\label{tab:calendar_dose}
\footnotesize\setlength{\tabcolsep}{4pt}
\begin{tabular*}{\textwidth}{@{\extracolsep{\fill}}llcc}
\toprule
Input & Future controls & 13 weeks & 26 weeks \\
\midrule
Damage-share index (points) & No & $-0.261$\sym{*} & $-0.261$\sym{***} \\
 & & (0.049) & (0.030) \\
Damage-share index (points) & Yes & $-0.247$\sym{*} & $-0.247$\sym{***} \\
 & & (0.048) & (0.028) \\
Damage (\$1m) & No & $-0.091$ & $-0.091$ \\
 & & (0.015) & (0.010) \\
Damage (\$1m) & Yes & $-0.084$ & $-0.084$ \\
 & & (0.019) & (0.015) \\
Damage/income (pp) & No & 9.040 & 9.040 \\
 & & (8.650) & (4.156) \\
Damage/income (pp) & Yes & 9.432 & 9.432 \\
 & & (8.242) & (4.104) \\
\bottomrule\end{tabular*}
\begin{tablenotes}[flushleft]\footnotesize
\item Notes: Same 332,114 all-county observations and competing-input definitions. Each row repeats the specification using 13/26-week calendar-block wild-bootstrap standard errors in parentheses (9,999 Webb draws). The damage-share coefficient is per point of the index, while dollar inputs use millions of 2024 dollars. Damage/income is 100 times recorded loss divided by source counties' actual 2019 annual personal income, in common dollars. Normalizations reweight economic inputs rather than simply rescale them. The largest residualized exposure week supplies 84.5 percent of dollar-dose information, 34.7 percent for damage shares and 42.2 percent for income-scaled inputs. Effective exposure weeks are 1.38, 6.83 and 3.43. These are concentration summaries, not independent-storm counts. The original unrestricted-score bootstrap is not used as the main inference in this highly leveraged comparison. \mbox{\sym{*} $p<0.10$}, \mbox{\sym{**} $p<0.05$}, \mbox{\sym{***} $p<0.01$}.
\end{tablenotes}
\end{threeparttable}
\end{table}

The dollar point estimates remain negative with later forcing controls, but the
calendar-block tests and concentrated support do not provide a precise monetary
loading. The income-scaled input has a different sign because it reweights source
locations, not because its reporting units change. Missing recorded losses and
unobserved within-report accrual also prevent these tests from establishing that
true damage size is irrelevant. The location-gradient result and the dollar-dose
question remain distinct.
\section{Financial measurement and a physical-weather design}
\label{app:operational}

\subsection{Household demand: a measurable derivative}
\label{app:financialmargin}

Main equation~\eqref{eq:margin_v6} is a first-order accounting decomposition of a
\emph{fixed-price desired} holding. It does not equate spending losses, cash needs and
executed bond sales. For an asset $a$, let county desired dollar holdings at the
reference price be $V^d_{ia}=s_{ia}W_i$. Differentiation gives
\[
 -dV^d_{ia}=(s_{ia}\chi_i^W+W_i\psi^{\mathrm{port}}_{ia})\,dz_i,
 \qquad \chi_i^W=-\partial W_i/\partial z_i,\quad
 \psi^{\mathrm{port}}_{ia}=-\partial s_{ia}/\partial z_i.
\]
For heterogeneous households this is summed over household-level derivatives. Products
of unweighted county averages generally do not give that sum. Price-adjusted holdings
and account flows measure adjustment after removing valuation changes. Transfers,
insurance, borrowing and cash balances determine whether a spending disturbance depletes
investable resources. A fall in $W$ with stable shares is the cash-resource margin, whereas
a fall in the credit share with stable $W$ is the rebalancing margin. With a complete
asset menu, $\sum_a s_{ia}=1$ implies $\sum_a\psi^{\mathrm{port}}_{ia}=0$, so a lower credit share can mean a
higher cash or Treasury share without net withdrawals from the investment account.

Neither $\chi_i^W$ nor $\psi^{\mathrm{port}}_{ia}$ need be positive. The sign of the adverse field is a
normalization rather than an assumption about a household's financial reaction. If direct payoff
losses also move wealth, they must be separated before attributing $\chi^W$ to the social
channel. If the measured field is rescaled to $\tilde z=bz$, its demand derivative
rescales to $\tilde l^{\$}=l^{\$}/b$, preserving predicted dollar demand. Independently
estimated spending and holdings models must share that normalization.

An executed quantity equation illustrates the measurement problem:
\begin{equation}
\Delta x_{iae}=-l_{ia}z_{ie}+\zeta_{ia}\Delta q_{ae}
                   +\text{payoff and other demand news}_{iae}+u_{iae}.
\end{equation}
Here $\zeta_{ia}$ is the premium sensitivity, not part of the autonomous demand shift.
Household--asset effects remove persistent ownership differences. Asset--event effects
can absorb common prices and issuer news only when their household loadings are common
or modeled. Otherwise, heterogeneous demand slopes leave differential price responses. Instruments
or an independently identified demand system are needed when the remaining price term
is endogenous. Within-asset contrasts still remove a common demand shift. Account-level
budget measurements, an independent aggregate anchor or justified bounds are needed for
absolute national pressure. Return-adjusted trades alone do not recover a fixed-price
demand curve, and delegated managers need not change earnings forecasts for either
household margin to operate.

\medskip
\noindent\textit{A household problem in surplus.}
The two financial margins can be derived without making the spending sign a restriction on
credit demand. Suppress the household index and let $c_0^{\mathrm{sur}}=c_0-\underline c_0$
denote current consumption above a predetermined commitment. Let $\mathsf T_0$ be assistance
paid now and $\mathsf T_1$ a firm future assistance obligation. Net resources are
$w_0^{\mathrm{net}}=w_0-\underline c_0-\mathsf T_0$ and
$Y_1^{\mathrm{net}}=Y_1-\underline c_1-\mathsf T_1$. The household chooses current surplus
consumption, safe investment $a^{\mathrm{safe}}\geq0$ and credit holdings $x\geq0$ to solve
\begin{equation}\label{eq:hh_surplus_problem}
\begin{split}
\max\quad &u(c_0^{\mathrm{sur}})+\varpi E[u(c_1^{\mathrm{sur}})],\qquad
 u(c)=\frac{c^{1-\gamma}}{1-\gamma},\quad \gamma>0,\quad\varpi>0,\\
\text{subject to}\quad
 &c_0^{\mathrm{sur}}+a^{\mathrm{safe}}+Px=w_0^{\mathrm{net}},\qquad
 c_1^{\mathrm{sur}}=R_fa^{\mathrm{safe}}+Vx+Y_1^{\mathrm{net}}>0.
\end{split}
\end{equation}
At $\gamma=1$, $u(c)=\log c$. Prices and the distribution of the claim payoff are fixed.
The payoff has two possible values, $V_L<R_fP<V_H$, each with positive probability.
Background income $Y_1$ is independent of $V$ and can already be risky. Current resources
and both commitments are known at the decision date. Thus future assistance reduces resources
available for own consumption, rather than being assumed to change tastes. Uncertainty about
assistance belongs in net background income instead of being counted again as a firm obligation.

The result below concerns a strictly interior optimum, with positive safe and credit positions
that can both adjust. Surplus consumption stays positive in every state under the local
perturbations, and differentiation can be passed through expectations. An increase in background
risk means adding $\sqrt{\nu}\,\xi$ to $Y_1^{\mathrm{net}}$, where $E\xi=0$, $E\xi^2=1$ and
$\xi$ is bounded and independent of both the existing income risk and $V$. The derivative is
evaluated at $\nu=0$. This gives a local addition to existing risk, not a comparison of arbitrary
finite changes in risk or of risks correlated with the credit payoff.

\begin{aprop}[Local transfer and precautionary responses]\label{prop:hh_signs}
Under these conditions, a marginal increase in either $\mathsf T_0$ or $\mathsf T_1$ lowers
desired credit holdings. The marginal propensity to consume current net resources is strictly
between zero and one. A small independent addition to background income risk lowers current
consumption and credit holdings, while increasing safe investment. These are fixed-price
responses at the decision date.
\end{aprop}

\begin{proof}
Define effective future utility
$\mathcal U^{\mathrm{hh}}(k)=E_Y[u(k+Y_1^{\mathrm{net}})]$ and let
$M_j^{\mathrm{hh}}(k)=E_Y[(k+Y_1^{\mathrm{net}})^{-\gamma-j}]$.
Its absolute risk aversion and absolute prudence are
\[
\mathcal A^{\mathrm{hh}}(k)=\gamma\frac{M_1^{\mathrm{hh}}}{M_0^{\mathrm{hh}}},\qquad
\mathcal P^{\mathrm{hh}}(k)=(\gamma+1)\frac{M_2^{\mathrm{hh}}}{M_1^{\mathrm{hh}}}.
\]
Both are strictly decreasing. For example,
\[
(\mathcal A^{\mathrm{hh}})'=
\gamma\frac{-(\gamma+1)M_2^{\mathrm{hh}}M_0^{\mathrm{hh}}
                 +\gamma(M_1^{\mathrm{hh}})^2}{(M_0^{\mathrm{hh}})^2}<0,
\]
because $(M_1^{\mathrm{hh}})^2\leq M_0^{\mathrm{hh}}M_2^{\mathrm{hh}}$.
The same argument, using $(M_2^{\mathrm{hh}})^2\leq M_1^{\mathrm{hh}}M_3^{\mathrm{hh}}$,
gives $(\mathcal P^{\mathrm{hh}})'<0$. Also $(\mathcal U^{\mathrm{hh}})'''>0$.
This is the decreasing-risk-aversion and decreasing-prudence structure underlying standard
risk aversion \citep{Kimball1993}. The calculation here also allows consumption and saving
to adjust.

Let $e_j^V=V_j-R_fP$ and
$k_j^{\mathrm{fin}}=R_f(w_0^{\mathrm{net}}-c_0^{\mathrm{sur}})+e_j^Vx$.
The two first-order conditions are
\begin{equation}\label{eq:hh_foc}
u'(c_0^{\mathrm{sur}})=\varpi R_f E_V[(\mathcal U^{\mathrm{hh}})'(k^{\mathrm{fin}})],
\qquad E_V[e^V(\mathcal U^{\mathrm{hh}})'(k^{\mathrm{fin}})]=0.
\end{equation}
All moments in the rest of this proof are evaluated at this optimum. To keep them distinct
from the propagation matrices, give them the subscript $\mathrm{hh}$ and let
\begin{align*}
a_{\mathrm{hh}}&=-u''(c_0^{\mathrm{sur}}),&
b_{\mathrm{hh}}&=-E_V[(\mathcal U^{\mathrm{hh}})''],&
k_{\mathrm{hh}}&=E_V[e^V(\mathcal U^{\mathrm{hh}})''],\\
d_{\mathrm{hh}}&=-E_V[(e^V)^2(\mathcal U^{\mathrm{hh}})''],&
u_{\mathrm{hh}}&=E_V[(\mathcal U^{\mathrm{hh}})'''],&
m_{\mathrm{hh}}&=E_V[e^V(\mathcal U^{\mathrm{hh}})'''].
\end{align*}
The determinant of the negative Hessian, after dividing its credit row by $\varpi$, is
\[
\mathfrak D_{\mathrm{hh}}
=(a_{\mathrm{hh}}+\varpi R_f^2b_{\mathrm{hh}})d_{\mathrm{hh}}
 -\varpi R_f^2k_{\mathrm{hh}}^2>0.
\]
Strict concavity gives the inequality. In particular,
$b_{\mathrm{hh}}d_{\mathrm{hh}}-k_{\mathrm{hh}}^2>0$ by the weighted
Cauchy--Schwarz inequality and the two distinct excess payoffs.

For the signs, let $\bar\mu_{\mathrm{hh}}=E_V[(\mathcal U^{\mathrm{hh}})']$ and reweight
each payoff-state probability by its marginal utility divided by $\bar\mu_{\mathrm{hh}}$.
The second condition in \eqref{eq:hh_foc} makes the reweighted mean excess payoff zero.
Let
\[
j_{\mathrm{ret}}=\frac{-e_L^Ve_H^V}{e_H^V-e_L^V}>0,\qquad
\mathcal A_j^{\mathrm{hh}}=\mathcal A^{\mathrm{hh}}(k_j^{\mathrm{fin}}),\quad
\mathcal P_j^{\mathrm{hh}}=\mathcal P^{\mathrm{hh}}(k_j^{\mathrm{fin}}).
\]
Since $x>0$, $k_H^{\mathrm{fin}}>k_L^{\mathrm{fin}}$, and both risk aversion and prudence
are larger in the low-payoff state. Consequently,
\begin{align*}
k_{\mathrm{hh}}&=\bar\mu_{\mathrm{hh}}j_{\mathrm{ret}}
 (\mathcal A_L^{\mathrm{hh}}-\mathcal A_H^{\mathrm{hh}})>0,\\
m_{\mathrm{hh}}&=-\bar\mu_{\mathrm{hh}}j_{\mathrm{ret}}
 (\mathcal A_L^{\mathrm{hh}}\mathcal P_L^{\mathrm{hh}}
  -\mathcal A_H^{\mathrm{hh}}\mathcal P_H^{\mathrm{hh}})<0.
\end{align*}
Differentiating \eqref{eq:hh_foc} with respect to net resources gives
\begin{equation}\label{eq:hh_transfer_derivatives}
\frac{\partial x}{\partial w_0^{\mathrm{net}}}
=\frac{R_fa_{\mathrm{hh}}k_{\mathrm{hh}}}{\mathfrak D_{\mathrm{hh}}}>0,\qquad
\frac{\partial c_0^{\mathrm{sur}}}{\partial w_0^{\mathrm{net}}}
=\frac{\varpi R_f^2(b_{\mathrm{hh}}d_{\mathrm{hh}}-k_{\mathrm{hh}}^2)}
 {\mathfrak D_{\mathrm{hh}}}\in(0,1).
\end{equation}
Current assistance subtracts from $w_0^{\mathrm{net}}$, so its credit-demand derivative
is the negative of the first expression. A future assistance obligation instead gives
\[
\frac{\partial x}{\partial\mathsf T_1}
=-\frac{a_{\mathrm{hh}}k_{\mathrm{hh}}}{\mathfrak D_{\mathrm{hh}}}<0.
\]
Thus current consumption does not absorb the whole current resource loss, and risky investment
is normal at this interior optimum. No fixed portfolio share is assumed.

For the small additional risk,
$\partial_\nu(\mathcal U^{\mathrm{hh}})'=\tfrac12(\mathcal U^{\mathrm{hh}})'''$ at $\nu=0$.
Implicit differentiation then gives
\begin{align}
x_\nu&=\frac{\varpi R_f^2k_{\mathrm{hh}}u_{\mathrm{hh}}
 +(a_{\mathrm{hh}}+\varpi R_f^2b_{\mathrm{hh}})m_{\mathrm{hh}}}
 {2\mathfrak D_{\mathrm{hh}}},\label{eq:hh_credit_risk}\\
(c_0^{\mathrm{sur}})_\nu&=
-\frac{\varpi R_f(d_{\mathrm{hh}}u_{\mathrm{hh}}+k_{\mathrm{hh}}m_{\mathrm{hh}})}
 {2\mathfrak D_{\mathrm{hh}}}.\label{eq:hh_consumption_risk}
\end{align}
The two payoff states imply the identities
\begin{align*}
b_{\mathrm{hh}}(-m_{\mathrm{hh}})-k_{\mathrm{hh}}u_{\mathrm{hh}}
 &=\bar\mu_{\mathrm{hh}}^2j_{\mathrm{ret}}\mathcal A_L^{\mathrm{hh}}
 \mathcal A_H^{\mathrm{hh}}(\mathcal P_L^{\mathrm{hh}}-\mathcal P_H^{\mathrm{hh}})>0,\\
d_{\mathrm{hh}}u_{\mathrm{hh}}+k_{\mathrm{hh}}m_{\mathrm{hh}}
 &=\bar\mu_{\mathrm{hh}}^2j_{\mathrm{ret}}\mathcal A_L^{\mathrm{hh}}
 \mathcal A_H^{\mathrm{hh}}(e_H^V\mathcal P_L^{\mathrm{hh}}-e_L^V\mathcal P_H^{\mathrm{hh}})>0.
\end{align*}
The numerator of \eqref{eq:hh_credit_risk} can therefore be written as
$-a_{\mathrm{hh}}(-m_{\mathrm{hh}})
-\varpi R_f^2[b_{\mathrm{hh}}(-m_{\mathrm{hh}})-k_{\mathrm{hh}}u_{\mathrm{hh}}]<0$.
Both consumption and credit holdings fall. Since current resources have not changed,
$(a^{\mathrm{safe}})_\nu=-(c_0^{\mathrm{sur}})_\nu-Px_\nu>0$.
\end{proof}

\noindent\textit{Interpretation and the boundary.}
Precaution increases total saving in this problem, but the increase is held safely and is
accompanied by a reduction in credit. Saving and portfolio reallocation are chosen together,
not placed in a sequence. No separate or earmarked saving account is needed. Similarly,
the current-transfer result follows from normal risky demand and a consumption response below
one, not from assuming that every transferred dollar was previously invested in bonds.
With CARA utility and independent background risk, the unconstrained risky position has no
wealth effect, so this particular transfer argument vanishes.

Lower current commitments provide another interpretation of a spending decline. Since
$c_0=\underline c_0+c_0^{\mathrm{sur}}$, \eqref{eq:hh_transfer_derivatives} implies, with other inputs fixed,
\[
\frac{\partial c_0}{\partial\underline c_0}
=1-\frac{\partial c_0^{\mathrm{sur}}}{\partial w_0^{\mathrm{net}}}>0,\qquad
\frac{\partial x}{\partial\underline c_0}
=-\frac{\partial x}{\partial w_0^{\mathrm{net}}}<0.
\]
Thus lower current commitments reduce observed consumption while raising surplus resources and
credit holdings at this interior optimum.

Interiority is an economic scope condition. For example, let current resources be 100,
future income initially 50, $P=R_f=\varpi=1$, $\gamma=2$, and $V$ be 0.8 or 1.4 with equal
probabilities. With no borrowing, the optimum has consumption 74.92, credit 25.08 and zero
safe balances. Replacing certain income by independent income of 10 or 90 with equal
probabilities lowers consumption to 63.05 but raises credit to 31.00 and safe balances to
5.95. This finite-change example starts at a binding portfolio boundary and is not covered
by Proposition~\ref{prop:hh_signs}. Observing positive balances in both accounts after the
shock would not reveal that initial constraint.

\medskip
\begingroup\color{revcolor}
\noindent\textit{Equity and the direction of precaution.}
Corporate bonds can be safer than equities without being safe assets. A cautious household could
therefore move into credit as well as out of it. To study this margin, add equity holdings
$x^{E}\geq0$ with price $P^{E}$ and payoff $V^{E}$ to \eqref{eq:hh_surplus_problem}, so that
\[
c_0^{\mathrm{sur}}+a^{\mathrm{safe}}+Px+P^{E}x^{E}=w_0^{\mathrm{net}},\qquad
c_1^{\mathrm{sur}}=R_fa^{\mathrm{safe}}+Vx+V^{E}x^{E}+Y_1^{\mathrm{net}}>0 .
\]
The payoffs take three values across ordered financial states, a worst state ($L$), a downturn
($M$) and a boom ($H$), with probabilities $p_j>0$. Both payoffs are nonnegative and weakly
increasing across these states. The two claims need not be issued by the same firm, and the worst
state need not be an issuer's default. Let $\Delta_1V=V_M-V_L$ and $\Delta_2V=V_H-V_M$, define
$\Delta_1V^{E}$ and $\Delta_2V^{E}$ in the same way, and let
\[
\mathfrak s_{\mathrm{hh}}=\Delta_1V\,\Delta_2V^{E}-\Delta_2V\,\Delta_1V^{E}\neq0,
\]
so that the three assets span the three states. When $\mathfrak s_{\mathrm{hh}}>0$, credit loses a
larger part of its payoff range than equity between the downturn and the worst state, so its losses
are concentrated in the worst state. Credit with a capped payoff, which pays the same amount outside
the worst state ($\Delta_2V=0<\Delta_1V$), is the leading case. When $\mathfrak s_{\mathrm{hh}}<0$,
credit loses a smaller part of its payoff range there than equity does. Net future income
$Y_1^{\mathrm{net}}$ is certain at the reference point. The optimum is strictly interior, with
positive safe, credit and equity positions that can all adjust. Let $c_j^{\mathrm{sur}}$ denote
future surplus in state $j$ at this optimum, and let the added income risk be $\sqrt\nu\,\xi$ as
before.

\begin{aprop}[Precaution with credit and equity]\label{prop:hh_three_assets}
Under these conditions, a small independent addition to income risk lowers current consumption and
raises future surplus in the worst state. Let
$\mathfrak g_1=1/(c_L^{\mathrm{sur}}c_M^{\mathrm{sur}})+\mathfrak t_{\mathrm{hh}}$ and
$\mathfrak g_2=1/(c_M^{\mathrm{sur}}c_H^{\mathrm{sur}})+\mathfrak t_{\mathrm{hh}}$, with
$\mathfrak t_{\mathrm{hh}}>0$ given in the proof, so that $\mathfrak g_1>\mathfrak g_2>0$. The two
risky holdings respond as
\begin{equation}\label{eq:hh_three_asset_responses}
\begin{aligned}
x_\nu&=-\frac{\gamma+1}{2}\Big[x\,\mathfrak g_2
 +\frac{\Delta_2V^{E}\,(c_M^{\mathrm{sur}}-c_L^{\mathrm{sur}})(\mathfrak g_1-\mathfrak g_2)}
       {\mathfrak s_{\mathrm{hh}}}\Big],\\
x^{E}_\nu&=-\frac{\gamma+1}{2}\Big[x^{E}\mathfrak g_2
 -\frac{\Delta_2V\,(c_M^{\mathrm{sur}}-c_L^{\mathrm{sur}})(\mathfrak g_1-\mathfrak g_2)}
       {\mathfrak s_{\mathrm{hh}}}\Big].
\end{aligned}
\end{equation}
The first terms lower both holdings at a common proportional rate. The second terms tilt the
portfolio away from the claim whose losses are concentrated in the worst state, relative to the
other claim.
(i) If $\mathfrak s_{\mathrm{hh}}>0$, credit holdings fall, and by a larger proportion than equity
holdings. With a capped credit payoff, equity holdings also fall and safe investment rises.
(ii) If $\mathfrak s_{\mathrm{hh}}<0$, equity holdings fall, and credit holdings fall by a smaller
proportion or rise.
A marginal increase in $\mathsf T_0$ or $\mathsf T_1$ lowers credit and equity holdings in the same
proportion. These are fixed-price responses at the decision date.
\end{aprop}

\begin{proof}
The three assets span the payoff states and $Y_1^{\mathrm{net}}$ is certain. The household therefore
chooses $c_0^{\mathrm{sur}}$ and the three values $c_j^{\mathrm{sur}}$ subject to
$c_0^{\mathrm{sur}}+\sum_j\varsigma_jc_j^{\mathrm{sur}}=w_0^{\mathrm{net}}+Y_1^{\mathrm{net}}/R_f$,
where the state prices $\varsigma_j>0$ are fixed by $(R_f,P,P^{E})$ and equal
$\varpi p_j(c_0^{\mathrm{sur}}/c_j^{\mathrm{sur}})^{\gamma}$ at the optimum. The added income risk is
not spanned. With it, the condition for state $j$ is
$\varpi p_jE_\xi[u'(c_j^{\mathrm{sur}}+\sqrt\nu\,\xi)]=\varsigma_ju'(c_0^{\mathrm{sur}})$, and
$\partial_\nu E_\xi[u'(c+\sqrt\nu\,\xi)]=u'''(c)/2$ at $\nu=0$. Differentiating these conditions and
the budget gives
\[
(c_0^{\mathrm{sur}})_\nu=-\frac{\gamma+1}{2}\,\mathfrak t_{\mathrm{hh}}\,c_0^{\mathrm{sur}},\qquad
(c_j^{\mathrm{sur}})_\nu=\frac{\gamma+1}{2}\Big(\frac1{c_j^{\mathrm{sur}}}
 -\mathfrak t_{\mathrm{hh}}\,c_j^{\mathrm{sur}}\Big),\qquad
\mathfrak t_{\mathrm{hh}}=\frac{\sum_j\varsigma_j/c_j^{\mathrm{sur}}}
 {c_0^{\mathrm{sur}}+\sum_j\varsigma_jc_j^{\mathrm{sur}}}.
\]
Since $\mathfrak t_{\mathrm{hh}}<1/(c_L^{\mathrm{sur}})^2$, surplus in the worst state rises. For
any two states,
$(c_j^{\mathrm{sur}}-c_i^{\mathrm{sur}})_\nu=-\tfrac{\gamma+1}{2}
(c_j^{\mathrm{sur}}-c_i^{\mathrm{sur}})[1/(c_i^{\mathrm{sur}}c_j^{\mathrm{sur}})+\mathfrak t_{\mathrm{hh}}]$.
Precaution thus narrows every gap in surplus between a better and a worse state, and it narrows the
gap to the worst state by the larger proportion. The positions determine the two gaps through
\[
c_M^{\mathrm{sur}}-c_L^{\mathrm{sur}}=\Delta_1V\,x+\Delta_1V^{E}x^{E},\qquad
c_H^{\mathrm{sur}}-c_M^{\mathrm{sur}}=\Delta_2V\,x+\Delta_2V^{E}x^{E},
\]
both of which are positive. Differentiating these identities and solving for the two responses gives
\eqref{eq:hh_three_asset_responses}. In case (i) the bracket for credit exceeds $x\,\mathfrak g_2$,
and the bracket for equity is at most $x^{E}\mathfrak g_2$, with equality for a capped credit
payoff. In case (ii) the two roles are reversed. Safe investment satisfies
$R_fa^{\mathrm{safe}}_\nu=(c_L^{\mathrm{sur}})_\nu-V_Lx_\nu-V^{E}_Lx^{E}_\nu$, which is positive when
both holdings fall. Finally, at fixed prices CRRA utility makes $c_0^{\mathrm{sur}}$ and every
$c_j^{\mathrm{sur}}$ proportional to surplus wealth $w_0^{\mathrm{net}}+Y_1^{\mathrm{net}}/R_f$. Both
risky positions are linear in the two gaps, so
$\partial x/\partial w_0^{\mathrm{net}}=x/(w_0^{\mathrm{net}}+Y_1^{\mathrm{net}}/R_f)$, and the same
expression holds for $x^{E}$. Current assistance lowers $w_0^{\mathrm{net}}$, and a future
obligation lowers $Y_1^{\mathrm{net}}$.
\end{proof}

\noindent\textit{Existing income risk.}
When $Y_1^{\mathrm{net}}$ is already risky, the same steps apply to effective utility
$\mathcal U^{\mathrm{hh}}$ and to financial wealth
$k_j^{\mathrm{fin}}=R_fa^{\mathrm{safe}}+V_jx+V^{E}_jx^{E}$. They give
$(k_j^{\mathrm{fin}})_\nu=\tfrac12\mathcal P_j^{\mathrm{hh}}
-\Lambda_{\mathrm{hh}}/\mathcal A_j^{\mathrm{hh}}$, with
$\Lambda_{\mathrm{hh}}=-\gamma(c_0^{\mathrm{sur}})_\nu/c_0^{\mathrm{sur}}>0$. Risk aversion and
prudence are both decreasing, so this response is larger in worse states and both gaps still narrow,
\[
\Delta_1V\,x_\nu+\Delta_1V^{E}x^{E}_\nu<0,\qquad
\Delta_2V\,x_\nu+\Delta_2V^{E}x^{E}_\nu<0 .
\]
The household therefore lowers its exposure to the worst state and cannot raise both holdings. With
a capped credit payoff, equity holdings fall. The division of the reduction between credit and
equity is pinned down by \eqref{eq:hh_three_asset_responses} only when net future income is certain.
In particular, with three assets credit demand need not rise with resources once income is already
risky, so the transfer statement of the proposition is a benchmark under certain income.

\medskip
\noindent\textit{Three examples.}
Let $P=P^{E}=R_f=\varpi=1$, $\gamma=2$ and current resources be 100, and let the worst state, the
downturn and the boom have probabilities 0.2, 0.3 and 0.5. First, let credit have a
capped payoff, paying 0.65 in the worst state and 1.15 otherwise, and let equity pay
0.30, 0.90 and 1.60. With certain future income of 10, the optimum has
consumption 55.22, safe balances 26.38, credit 9.09 and equity 9.31. Replacing
certain income by independent income of 2 or 18 with equal probabilities lowers
consumption to 54.35, credit to 8.55 and equity to 8.93, and raises safe balances to
28.17. Credit falls by 5.9 percent and equity by 4.1 percent. Second, let credit
lose no additional value in the worst state, paying 0.95, 0.95 and 1.10, while
equity pays 0.30, 0.90 and 1.80. The optimum has consumption 55.48, safe
balances 22.90, credit 5.52 and equity 16.10. The same change in income lowers
consumption to 54.59 and equity to 15.25 but raises credit to 6.46 and safe balances to
23.69. This household moves from stocks into bonds although its safe balance can adjust.

Third, return to the capped credit payoff and let future income be 60. With no borrowing, the
optimum has consumption 80.22, credit 3.32, equity 16.47 and zero safe balances.
Replacing certain income by 50 or 70 lowers consumption to 79.31 and equity to
15.26 but raises credit to 5.43, with safe balances still at zero. Credit is then the safest
claim that can adjust, and precaution moves the household into it even though its losses are
concentrated in the worst state. This last case starts at a binding boundary and is not covered by
Proposition~\ref{prop:hh_three_assets}. In all three examples a loss of one unit of current
resources lowers credit holdings.

The household problem holds all asset prices fixed. It therefore delivers a conditional shift in
credit demand, which the pricing map of Section~\ref{sec:pricing} takes as given. It is not a
joint equilibrium of the credit and equity markets.
\endgroup

Equation~\eqref{eq:surplus} remains a local surplus-wealth illustration, not the exact
global demand function of \eqref{eq:hh_surplus_problem}. Surplus consumption in utility and
wealth above the present value of commitments are distinct quantities. The optimization
supplies conditions for the sign of the local fixed-price derivative. \rev{Under their respective
conditions, the two propositions give assistance needs that rise with $z_i$ a nonnegative term in
$l_i$ at the reference state. If independent background risk also rises with $z_i$, its term is
nonnegative in the two-asset problem and in case (i) of Proposition~\ref{prop:hh_three_assets}, which
assumes certain net income. It can be negative in case (ii) or at a portfolio boundary. The net
loading $l_i$ is the sum of these terms and can therefore have either sign. The propositions supply}
neither an empirical value for $l_i$ nor a duration for which
that derivative stays fixed. \rev{Correlated income risk, more payoff states than assets} and later
balance-sheet accumulation require separate restrictions.

\medskip
\begingroup\color{revcolor}
\noindent\textit{Signed loadings in the pricing results.}
Proposition~\ref{prop:recovery} and its proof in Appendix~\ref{app:target} place no restriction on
the sign of $\bm l$. When households raise desired credit holdings as conditions worsen, the
corresponding entries of $\bm l$ are negative. The target remains $\bm l'\bm z$, and the same recovery
condition applies to these signed loadings. The recovery weights change with the loadings. They are
computed from independently measured signed loadings and are not chosen to fit the EBP.
Signed loadings require three qualifications. First, the normalization $\pi^H=\bm l/\delta$ requires
$\delta\neq0$, and the \rev{conditional EBP response} can always be written with $\bm l$ directly. Loadings that sum to zero
still respond to a geographically uneven field. Second, the level-sign result of
Appendix~\ref{app:levels} requires $\bm l\geq0$. Without it the \rev{response} can change sign across
horizons, because diffusion moves the response between counties whose loadings differ in sign. In
the two-county example of \eqref{eq:two_region}, loadings $(l_1,l_2)$ and a unit input in county 1
give
\[
K_1(h)=\frac{c_E\rho_D^{h-1}}{2}\big[(l_1+l_2)+(l_1-l_2)e^{-2\kappa_D(h-1)}\big].
\]
For $\kappa_D>0$ and $0<\rho_D<1$, this response switches sign once over the full horizon path if
$l_1(l_1+l_2)<0$, and does not switch otherwise. The switch need not fall inside the horizons
studied, and without diffusion there is none. Third, with $C_Y=\mathrm{Id}_{n_C}$ and $\ker T=\spn\{\mathbf1\}$,
$\rowsp(T)$ is the set of vectors orthogonal to $\mathbf1$. For unrestricted county responses,
within-event contrasts therefore recover the target exactly when $\bm l'\mathbf1=0$, that is, when a
response common to all counties leaves credit demand unchanged. On a restricted response space,
Proposition~\ref{prop:recovery} gives the corresponding condition.
Theorem~\ref{thm:no_scalar} uses the weights only through the modal coefficients $\omega_k$, whose
signs are unrestricted, and applies unchanged. When household responses offset at a horizon,
$\bm l'\bm z=0$ and the household-demand component of the premium does not move at that horizon,
holding the other channels fixed. This is a zero demand shift. It differs from the attenuated
response that arises when intermediaries absorb part of a nonzero shift.
\endgroup

\begin{table}[htbp]
\centering
\caption{Liquid balances among holders of non-retirement credit proxies}
\label{tab:scf_household_scope}
\begin{singlespace}
\footnotesize
\setlength{\tabcolsep}{4pt}
\begin{tabular}{p{0.37\textwidth}rrrr}
\toprule
& \multicolumn{3}{c}{Share of asset dollars (\%)} & Share of \\
\cmidrule(lr){2-4}
Liquid-balance criterion & Direct bonds & Bond funds & Combined & holders (\%) \\
\midrule
Positive transaction balances & 100.0 & 100.0 & 100.0 & 100.0 \\
At least 1 month of normal income & 87.0 & 81.8 & 83.1 & 81.9 \\
At least 3 months of normal income & 75.9 & 75.6 & 75.7 & 65.2 \\
At least 6 months of normal income & 33.1 & 54.8 & 49.5 & 44.0 \\
\bottomrule
\end{tabular}
\par\vspace{3pt}
\begin{minipage}{\textwidth}
\textit{Notes:} The 2022 Survey of Consumer Finances, 4,595 public-use families and five implicates (184--196 holders of these categories per implicate). Survey-weighted ratios are computed in each implicate and averaged. Direct bonds are corporate or other bonds (X7639). Bond funds are other bond mutual funds and ETFs (X3828). Both exclude separately recorded retirement accounts, but neither category isolates corporate credit. The combined column sums the two categories. Transaction balances sum checking, savings, money-market accounts and funds, brokerage call accounts and prepaid accounts. Months divide these balances by one-twelfth of annual normal income, in common dollar units. Asset columns use all dollars in the named category as the denominator. The last column uses all families holding either category. Nonpositive normal income does not satisfy a months threshold. These are descriptive point estimates, not estimates of portfolio interiority or adjustment capacity. Source: \href{https://www.federalreserve.gov/econres/scf_2022.htm}{Federal Reserve Board, 2022 SCF}.
\end{minipage}
\end{singlespace}
\end{table}

\medskip
\noindent\textit{Netting donors and recipients.}
Transfers redistribute resources. Let
$g_i^{\mathrm{cred}}=P\,\partial x_i/\partial w_{i0}^{\mathrm{net}}$ be the household's
fixed-price marginal credit demand per dollar of current net resources. A transfer of
$t_{dr}\geq0$ from donor $d$ to recipient $r$ lowers donor desired dollar holdings by
$g_d^{\mathrm{cred}}t_{dr}$ and raises recipient desired dollar holdings by
$g_r^{\mathrm{cred}}t_{dr}$. With all other inputs fixed, its net selling pressure is
\begin{equation}\label{eq:hh_transfer_netting}
S_{\mathrm{tr}}^{\$}=\sum_{d,r}
 (g_d^{\mathrm{cred}}-g_r^{\mathrm{cred}})t_{dr}.
\end{equation}
Thus transfers generate positive aggregate selling pressure only when the transfer-weighted
donor response exceeds the recipient response. Recipient spending or low credit participation
can make this condition plausible, but do not establish it. Recipients remain in the financial
aggregation even when directly hit counties are excluded from a spending comparison. The \rev{conditional EBP
response} applies to the net fixed-price demand shift, not gross donor sales. This is a static
accounting restriction and does not require a new propagation law.

\medskip
\noindent\textit{Observable scope.}
Table~\ref{tab:scf_household_scope} describes liquid balances among households holding two
non-retirement credit proxies in the 2022 SCF. The categories are held by about 1.35 percent
of families and do not cover all household credit
exposure. They are used because the public survey separates them from retirement accounts,
not because they isolate the relevant corporate claims perfectly.

These observations make liquidity and account access measurable dimensions of the sign
condition. They do not classify an optimum as interior. Transaction needs, committed balances
and trading restrictions can bind even when both balances are positive. Moreover, withdrawal
access matters for financing a transfer, whereas precautionary rebalancing can occur within a
retirement account without a withdrawal. Measuring those restrictions alongside the derivatives
in main equation~\eqref{eq:margin_v6} is what separates the two routes empirically.

\medskip
\noindent\textit{The benchmark economy of Section~\ref{subsec:environment}.}
Problem~\eqref{eq:env_household} is problem~\eqref{eq:hh_surplus_problem} for a household of type $f$ in
county $i$. The main text uses a benchmark in which net future income $Y^{\mathrm{net}}_{1,if}$ is known at the
decision date. This part derives its results. Surplus wealth is
$s_{if}\equiv w^{\mathrm{net}}_{0,if}+Y^{\mathrm{net}}_{1,if}/R_f$, and $\bar s_{if}$ is its value on the path
without the disturbance. Let $\theta_q$ and $\theta_N$ denote the derivatives of the intermediary position
$\theta=\min\{\max\{Nq/\Gamma,0\},N/\chi\}$. Strictly inside the slack region $\theta_q=N/\Gamma$ and
$\theta_N=q/\Gamma$, and strictly inside the binding region $\theta_q=0$ and $\theta_N=1/\chi$. A reference
equilibrium is an equilibrium on the path without the disturbance, with premium $\bar q>0$ and claim price
$\bar P$, in which every household holds a strictly positive safe position, every participating household holds
a strictly positive quantity of claims, storage is strictly positive, the intermediary position lies strictly
inside one of the two regions, and $\zeta+\theta_q>0$.

\begin{aprop}[The benchmark economy]\label{prop:hh_benchmark}
Let net future income be known at the decision date. At a reference equilibrium:
\begin{enumerate}[(i)]
\item For premiums and surplus wealth near their reference values, a participating household chooses
$c^{\mathrm{sur}}_{0,if}=\varphi^c_{if}(q)\,s_{if}$ and $x_{if}=\varphi^x_{if}(q)\,s_{if}$. The functions
$\varphi^c_{if}$ and $\varphi^x_{if}$ depend on $\gamma_{if}$, $\varpi_{if}$, $R_f$ and the distribution of $V$,
and not on wealth. A non-participating household chooses
$c^{\mathrm{sur}}_{0,if}=\varphi^{c,n}_{if}s_{if}$ with
$\varphi^{c,n}_{if}=[1+(\varpi_{if}R_f)^{1/\gamma_{if}}/R_f]^{-1}$, which does not depend on the premium.
\item For a participating household,
$(\varphi^c_{if})'(q)/\varphi^c_{if}(q)=\frac{\gamma_{if}-1}{\gamma_{if}}\,\varphi^x_{if}(q)/R_f$.
\item The equilibrium premium is differentiable in the household conditions and in capital, with
$\partial q/\partial z_i=G_S\,l_i$ and $\partial q/\partial N=G_N$. Here $l_i=\zeta_i\beta_i$ and
$\zeta=\sum_i\zeta_i$ are those of \eqref{eq:schedule} and \eqref{eq:weights}, and $G_S$ and $G_N$ are the
expressions of Proposition~\ref{prop:eq} evaluated at the reference equilibrium.
\item Let $c_i\equiv\sum_fn_{if}c^{\mathrm{sur}}_{0,if}$ be county spending above commitments and $\bm c$ the
vector of these. Its Jacobian with respect to the household conditions is
$-\diag(c_i^Y)+G_S\,\bm\zeta^c\bm l'$. Here $c_i^Y\equiv\sum_fn_{if}\varphi^c_{if}(\bar q)m_{if}$, with
$\varphi^{c,n}_{if}$ in place of $\varphi^c_{if}$ for non-participating types, and
$\zeta^c_i\equiv\sum_fn_{if}(\varphi^c_{if})'(\bar q)\bar s_{if}$, summed over participating types.
\item Let $M\equiv\diag(c_i^Y)-G_S\,\bm\zeta^c\bm l'$, so that county spending losses are $M\bm z$ to first order,
and let $a_{0,i}\equiv l_i/c_i^Y$ with every $c_i^Y>0$. If $G_S\,\bm a_0'\bm\zeta^c\neq1$, then
$\bm a\equiv\bm a_0/(1-G_S\,\bm a_0'\bm\zeta^c)$ satisfies $\bm a'M=\bm l'$.
\end{enumerate}
\end{aprop}

\begin{proof}
(i) Replace the safe position by $\tilde a\equiv a^{\mathrm{safe}}_{if}+Y^{\mathrm{net}}_{1,if}/R_f$. The
constraints of \eqref{eq:env_household} become $c^{\mathrm{sur}}_{0,if}+\tilde a+Px_{if}=s_{if}$ and
$c^{\mathrm{sur}}_{1,if}=R_f\tilde a+Vx_{if}$, and the no-borrowing constraint becomes
$\tilde a\geq Y^{\mathrm{net}}_{1,if}/R_f$. Drop that constraint. The remaining constraints are homogeneous of
degree one in the choices and $s_{if}$ together. The objective is homogeneous of degree $1-\gamma_{if}$ in the
choices, or shifts by a constant under log utility, and it is strictly concave. The solution for surplus wealth
$s_{if}$ is therefore $s_{if}$ times the solution for unit surplus wealth, which defines $\varphi^c_{if}$ and
$\varphi^x_{if}$. At the reference equilibrium this solution satisfies the dropped constraint strictly and has
$x_{if}>0$, so by continuity it solves the original problem nearby. For a non-participating household
$x_{if}=0$, and the condition $u_{if}'(c^{\mathrm{sur}}_{0,if})=\varpi_{if}R_fu_{if}'(c^{\mathrm{sur}}_{1,if})$
with $c^{\mathrm{sur}}_{1,if}=R_f(s_{if}-c^{\mathrm{sur}}_{0,if})$ gives the stated share.

(ii) Suppress the indices. Under log utility $c^{\mathrm{sur}}_0=s/(1+\varpi)$ at every price, so both sides are
zero. For $\gamma\neq1$, homogeneity gives the value function $v(s,P)=s^{1-\gamma}v(1,P)$. The envelope condition
for wealth is $v_s=u'(c^{\mathrm{sur}}_0)=(\varphi^cs)^{-\gamma}$, so
$v(1,P)=(\varphi^c)^{-\gamma}/(1-\gamma)$. The envelope condition for the price and the first-order condition for
consumption give $v_P=-x\,u'(c^{\mathrm{sur}}_0)=-\varphi^xs\,(\varphi^cs)^{-\gamma}$. Differentiating
$s^{1-\gamma}(\varphi^c)^{-\gamma}/(1-\gamma)$ in $P$ and equating the two expressions gives
$\varphi^c_P/\varphi^c=\frac{1-\gamma}{\gamma}\varphi^x$. Since $q=E[V]-R_fP$, a derivative in $q$ is $-1/R_f$
times a derivative in $P$, which gives the identity.

(iii) With the policies of (i), market clearing \eqref{eq:env_clearing} reads
$\sum_i\sum_fn_{if}\varphi^x_{if}(q)(\bar s_{if}-m_{if}z_i)+\theta(q,N)=\bar B$, with the inner sums over
participating types. The left side is continuously differentiable near the reference equilibrium, and its
derivative in the premium is $\zeta+\theta_q>0$. The implicit function theorem gives
$\partial q/\partial z_i=l_i/(\zeta+\theta_q)$ and $\partial q/\partial N=-\theta_N/(\zeta+\theta_q)$. In the slack
region $1/(\zeta+\theta_q)=\Gamma/(N+\zeta\Gamma)$ and $\theta_N/(\zeta+\theta_q)=\bar q/(N+\zeta\Gamma)$. In the
binding region the two are $1/\zeta$ and $1/(\zeta\chi)$. These are $G_S$ and $-G_N$ of
Proposition~\ref{prop:eq}, because $q=\Gamma Q/(N+\zeta\Gamma)$ in its slack region.

(iv) County spending above commitments is
$c_i=\sum_fn_{if}\varphi^c_{if}(q)(\bar s_{if}-m_{if}z_i)$, with the constant $\varphi^{c,n}_{if}$ for
non-participating types. Differentiate it in $z_j$ and use (iii) for the premium.

(v) By construction $\bm a_0'\diag(c_i^Y)=\bm l'$. Hence
$\bm a_0'M=\bm l'-G_S(\bm a_0'\bm\zeta^c)\bm l'=(1-G_S\,\bm a_0'\bm\zeta^c)\bm l'$, and dividing by the scalar
$1-G_S\,\bm a_0'\bm\zeta^c$ gives $\bm a'M=\bm l'$.
\end{proof}

\noindent\textit{Scope of the benchmark.}
The proportional policies hold for any distribution of the payoff. They need not hold when net future income is
risky or when a portfolio constraint binds, which are the cases studied in
Propositions~\ref{prop:hh_signs} and~\ref{prop:hh_three_assets} and in the boundary examples above. The safe
position implied by the policies is
$a^{\mathrm{safe}}_{if}=\bar s_{if}[1-\varphi^c_{if}(\bar q)-\bar P\varphi^x_{if}(\bar q)]
-\bar Y^{\mathrm{net}}_{1,if}/R_f$, so the benchmark requires net future income to be small enough relative to
surplus wealth. With certain income, $\varphi^x_{if}(q)>0$ exactly when $q>0$. The loadings of this benchmark are
nonnegative. Signed loadings arise in the extensions with precaution above.

\medskip
\noindent\textit{The local schedule.}
County holdings are $x_i=\sum_fn_{if}\varphi^x_{if}(q)(\bar s_{if}-m_{if}z_i)$. Expanding around the reference
premium and a zero condition,
\[
x_i=\bar x_i+\zeta_iq-l_iz_i-(q-\bar q)\,z_i\sum_fn_{if}(\varphi^x_{if})'(\bar q)\,m_{if}+O\big((q-\bar q)^2\big),
\qquad \bar x_i\equiv\sum_fn_{if}\varphi^x_{if}(\bar q)\bar s_{if}-\zeta_i\bar q .
\]
Schedule \eqref{eq:schedule} drops the last two terms. It is a first-order approximation in the premium change
and the household condition jointly, and it is exact in the household condition at the reference premium. The
derivatives in Proposition~\ref{prop:hh_benchmark} do not depend on this approximation. The closed-form level of
Proposition~\ref{prop:eq} and the second-order coefficients of Proposition~\ref{prop:levels} do, because the
dropped terms are of second order.

\medskip
\noindent\textit{Measured spending and the direct component.}
Measured county spending is $\mathcal C_i\equiv c_i+\sum_fn_{if}\underline c_{0,if}$, with reference value
$\bar{\mathcal C}_i$. A direct loss $L^{\mathrm{dir}}_i$ lowers surplus wealth by
$m^{\mathrm{dir}}_{if}L^{\mathrm{dir}}_i$, where
$m^{\mathrm{dir}}_{if}\equiv m^{\mathrm{dir}}_{0,if}+m^{\mathrm{dir}}_{1,if}/R_f$. Its spending and credit loadings
$c^{Y,\mathrm{dir}}_i$ and $l^{\mathrm{dir}}_i$ are defined as $c_i^Y$ and $l_i$, with $m^{\mathrm{dir}}_{if}$ in
place of $m_{if}$. To first order and at fixed capital,
\[
d\log\mathcal C_i=-\frac{c_i^Y}{\bar{\mathcal C}_i}\,dz_i
-\frac{c^{Y,\mathrm{dir}}_i}{\bar{\mathcal C}_i}\,dL^{\mathrm{dir}}_i
+\frac{\zeta^c_i}{\bar{\mathcal C}_i}\,dq,\qquad
dq=G_S\,\big(\bm l'd\bm z+(\bm l^{\mathrm{dir}})'d\bm L^{\mathrm{dir}}\big).
\]
The model gives levels at the decision date, and division by $\bar{\mathcal C}_i$ converts them into the log
responses of Section~\ref{sec:model}. The direct component $\bm c_t$ of that section is the second term, taken at
a fixed premium. The response is zero before the shock date, so growth at that date equals the log response, and
$\Phi_0\bm\varepsilon_{t_0}=-\diag(c^{Y,\mathrm{dir}}_i/\bar{\mathcal C}_i)\,\bm L^{\mathrm{dir}}_{t_0}$. Holding
the premium fixed does not confine this response to the counties the disturbance hits. In the benchmark the
direct loss of a county depends only on its own entry of $\bm\varepsilon_{t_0}$, so that
$\Phi_0=\diag(\phi_i)$, the county-specific case of Appendix~\ref{app:observation}. The input
$B_D\equiv-\eta_D\Phi_0$ of the slow field is an assumed proportionality between the household condition that
enters the field and this response. It is not derived from the household problem or from the timing. Observed
spending at the shock date also contains the third term, which reaches counties outside the footprint of the
disturbance, and the measurement equation of Section~\ref{subsec:aggregation} carries it in the direct input
map $J_Y$. The loading $\zeta^c_i/\bar{\mathcal C}_i$ of that term
equals $\frac{\gamma-1}{\gamma}\varphi^x(\bar q)/R_f$ in every county when preferences are common, every household
participates and there are no commitments. It differs across counties when commitments are different shares of
spending or when participation differs. Within-event contrasts remove the term only when the loading is common.

\medskip
\noindent\textit{Feasibility and solvency.}
Add the first constraint of \eqref{eq:env_household} over all households, with weights $n_{if}$, to the
intermediary's budget $N=P\theta+a^I$, where $a^I$ is its safe position. Assistance adds up to zero and the
credit market clears, so
\[
\sum_i\sum_fn_{if}\,(c^{\mathrm{sur}}_{0,if}+\underline c_{0,if})+P\bar B+K=\sum_i\sum_fn_{if}\,w_{0,if}+N,\qquad
K\equiv\sum_i\sum_fn_{if}\,a^{\mathrm{safe}}_{if}+a^I,
\]
where $P\bar B$ is the borrower's consumption and $K$ is storage. Adding the second constraint to the
intermediary's wealth $R_fa^I+V\theta$ and the borrower's residual $Y^B-\bar BV$ gives, for every payoff,
\[
\sum_i\sum_fn_{if}\,(c^{\mathrm{sur}}_{1,if}+\underline c_{1,if})+(Y^B-\bar BV)+(R_fa^I+V\theta)
=\sum_i\sum_fn_{if}\,Y_{1,if}+Y^B+R_fK .
\]
Goods are therefore feasible at both dates when $K\geq0$, and a strictly positive $K$ supports the safe return
$R_f$. For every position $0\leq\theta\leq N/\chi$ the intermediary's wealth satisfies
$R_fN+\theta(V-R_fP)\geq R_fN-\frac N\chi\max\{R_fP-V_{\min},0\}$. Since $P\leq E[V]/R_f$ at a nonnegative
premium, the margin condition $\chi\geq(E[V]-V_{\min})/R_f$ of Section~\ref{subsec:environment} makes this
wealth nonnegative, so loans to intermediaries are safe.

\subsection{Relative identification and the role of the operator}
\label{app:relative}

Let $S_a^\$=(\bm l_a^{\$})'\bm z$ be fixed-price dollar demand reduction. At a specified reference
state, suppose $\Delta p_a=c_a S_a^\$+n_a$, where $n_a$ collects other premium movements.
Equation~\eqref{eq:relative_v6} applies after $n_a,n_b$ are independently removed and
$c_a,c_b$ are known positive scalars\rev{, with both totals $\delta_a^{\$}$ and $\delta_b^{\$}$ nonzero}. If cross-asset absorption is a matrix $C^{\mathrm{mkt}}$, the
corresponding identified linear transformation of $\Delta p-n$ must be used. Separate
division by two diagonal entries is generally invalid. Identical investor loadings give
zero relative signal, however much county spending moves.

Cancellation is not sufficient when the spending map is incomplete. With observed
$TC_Yz$ and $\Delta\bm\pi_{ab}=\pi_a-\pi_b$, the contrast is recoverable on $\mathcal R$ only if
$(\Delta\bm\pi_{ab})'R_{\mathcal R}=a'TC_YR_{\mathcal R}$ for some independently justified $a$.
When $C_Y=\mathrm{Id}_{n_C}$ and $\ker T=\operatorname{span}\{\mathbf1\}$ this follows from $(\Delta\bm\pi_{ab})'\mathbf1=0$.
Otherwise Proposition~\ref{prop:recovery} imposes an additional measurement condition.

For the two-county example, let
$\mathcal L=\left(\begin{smallmatrix}1&-1\\-1&1\end{smallmatrix}\right)$,
$B_D=\mathrm{Id}_2$, $\pi_a=(0.8,0.2)'$, $\pi_b=(0.2,0.8)'$, and the shock be $e_1$.
At $h\geq1$, let $n=h-1$ denote elapsed periods, distinct from diffusion time $\tau=\kappa_D n$.
\begin{align*}
z_h&=\frac{\rho_D^n}{2}
 (1+e^{-2\kappa_D n},\,1-e^{-2\kappa_D n})',\\
(\pi_a-\pi_b)'z_h&=0.6\rho_D^n e^{-2\kappa_D n},\\
\text{relative diffusion increment}
 &=0.6\rho_D^n(e^{-2\kappa_D n}-1).
\end{align*}
An arbitrary additive common field cancels in the difference. The increment is zero
at the first slow-field response, $h=1$, and at $\kappa_D=0$, and negative thereafter for this origin and loading
configuration. Reversing the shock origin reverses the relative sign. Those restrictions
are declared by the model before observing asset returns. They do not predict a universal
positive diffusion effect or reverse the counties' ranking. Unknown $\rho_D$, changing
loadings or differential capital news could mimic the path, so the spending and financial
measurements must distinguish them.

This example shows what success would mean qualitatively. Holding initial dose fixed,
remote adverse responses should occur where the spending operator predicts, with stronger demand
adjustment where independently measured $l^{\$}$ is larger. The relative premium should then
follow the resulting horizon profile after the heterogeneous
no-diffusion response and competing news are controlled. In the scalar slack-capital
benchmark, the same positive fixed-price demand reduction raises the level premium more
when capital is low, since $G_{SN}<0$. At the cap this derivative changes, and endogenous
capital can alter the total path. There is still no sign theorem for the old lagged-state
EBP-change regression.

The absolute EBP route requires the measured dollar pressure to satisfy the units bridge
\eqref{eq:units} below. An observed zero net withdrawal is compatible with
portfolio rotation, whereas zero \emph{fixed-price credit-demand adjustment} rules out that
margin of the proposed mechanism. A null relative-price response is informative only
when demand, absorption and the design's power are known. Kilts can improve spending measurement, not automatically
the investor map or absorption elasticity.

\subsection{Conditional financial visibility}
\label{app:visibility}

The two household margins need not coincide. Ignoring valuation, financial accumulation
satisfies the event-change budget identity
\[
\Delta F=\Delta Y+\Delta T+\Delta B-\Delta C-\Delta I^{nonfinancial},
\]
where transfers and net borrowing are included, all terms cover the same interval, and
internal financial-account transfers cancel. If the other components are unchanged,
lower consumption increases financial accumulation. Thus a remote spending decline does
not establish a cash need. Portfolio rotation remains a distinct testable margin.

For specified counterfactual expenditures $C^0_{i,t+h}$, the dollar shortfall is
\[
C_{loss}(H)=-\sum_{h=0}^H\sum_i C^0_{i,t+h}\big(e^{g_{i,h}}-1\big),
\]
where $g_{i,h}$ is a log-level response. Accumulated growth alone is not this dollar sum. Let $\lambda_C$ be the fixed-price
credit-demand reduction per dollar of that shortfall, $A_{mkt}$ the matched market value, $\xi_P$ the
residual absorption elasticity, $D_{mod}$ modified duration and $\varphi_E$ the loading of the EBP on
yields. Residual holders absorb the fixed-price demand shift $\lambda_CC_{loss}$ with
$-\Delta P/P\simeq\lambda_CC_{loss}/(A_{mkt}\xi_P)$. The duration relation of Appendix~\ref{app:units},
$\Delta\text{yield}\simeq-\Delta P/(PD_{mod})$, converts this price change into a yield change, and
$\varphi_E$ maps the yield change into the EBP:
\begin{equation}\label{eq:units}
\Delta E^{BP}\ [\text{bp}]=10^4\,\frac{\varphi_E\,\lambda_C\,C_{loss}}{A_{mkt}\,\xi_P\,D_{mod}} .
\end{equation}

Table~\ref{tab:visibility} makes the economic hurdle explicit. Every input is illustrative,
including a \$1 trillion matched credit market, duration five and unit yield-to-EBP loading. Positive
pressure denotes a fixed-price reduction in desired credit holdings, not executed turnover.
The resource and share components are first-order changes, whereas their product is second order.

\begin{table}[!htbp]\centering
\begin{threeparttable}
\caption{Conditional household demand and EBP magnitudes}
\label{tab:visibility}
\footnotesize
\begin{tabularx}{\textwidth}{Xccc}
\toprule
Illustrative household adjustment & Demand & \multicolumn{2}{c}{EBP (bp)}\\
\cmidrule(lr){3-4}
 & (\$bn) & $\xi_P=1$ & $\xi_P=0.2$\\
\midrule
\$1bn resource loss, 10\% credit share, no rotation & 0.1 & 0.2 & 1.0\\
0.5 pp credit-share fall on \$100bn wealth; resources fixed & 0.5 & 1.0 & 5.0\\
2.5 pp credit-share fall on the same wealth; resources fixed & 2.5 & 5.0 & 25.0\\
First row plus 0.1 pp credit-share increase on \$100bn & 0.0 & 0.0 & 0.0\\
\bottomrule
\end{tabularx}
\begin{tablenotes}[flushleft]\footnotesize
\item Notes: Equation~\eqref{eq:units} with assumed inputs, not estimated responses
or achieved power. Wealth must belong to the exposed households, and absorption cannot be
chosen to fit EBP. The last row cancels at first order. Large movements require checking
the local-linear price approximation. ``pp'' denotes percentage points.
\end{tablenotes}
\end{threeparttable}
\end{table}

For illustration, a five-basis-point increase in a \$1 trillion market requires
\$2.5 billion of demand reduction at unit absorption, \$2.50 per dollar of a \$1 billion shortfall, or
\$0.5 billion at $\xi_P=0.2$. At the 57-month mean of the debt-securities denominator of
Appendix~\ref{app:creditflow}, about \$7,970 billion, the same increase would require about \$20 per dollar. A lower EBP loading raises those requirements inversely.
The demand object is a change in desired holdings at the endpoint, not the sum of the same
persistent adjustment each week. Relating it to cumulative spending requires matched
timing and population. A statistically detectable household response may still be too
small to explain national EBP. Additionally, strong absorption can mute prices.

For a single unbiased normal estimate of aggregate demand pressure, the approximate
80-percent-power, two-sided 5-percent-test requirement is
\[
\operatorname{SE}(\hat S_a)\leq
\frac{S_a^*}{z_{0.975}+z_{0.80}}.
\]
The respective illustrative SE ceilings are \$0.892 billion and \$0.178 billion. These
do not establish power for the joint validation criteria. Financial sampling, episode dependence, operator
estimation, price endogeneity and uncertain scaling require full-design simulation.
The appropriate first financial observations are therefore resource changes and
valuation-adjusted credit-share rotation, followed by the independently weighted operator
contrast in Appendix~\ref{app:relative}. No such response has been established in the available panel.

\subsection{Physical-weather coverage and recipient measurement}
\label{app:weatherplan}

The measured-weather adjustment in Appendix~\ref{app:measuredweather} accounts for observed
county precipitation and temperature, station gusts, warnings and reports. A stronger
physical exclusion would require high-resolution rainfall and wind coverage, relevant
flood indicators, and reconciliation of FEMA declarations with NOAA physical episodes.
Warning geometry measures threatened areas, rather than realized damage, and station
coverage alone cannot certify the absence of local destructive weather.

Sources include NOAA MRMS precipitation, IEM archives of NWS warning polygons, NOAA
ISD/ASOS gust observations, HRRR wind fields and USGS stage or discharge observations.
Product versions and archived coverage matter when comparing early 2020 with later
events. A proposed low-hazard recipient design would retain explicit coverage flags
and align each hazard window to the spending endpoint. Absence of a report would not
be converted into measured zero exposure. Sources and recipients remain separate, so
a revised footprint must be passed through the social, gravity, production and direct
fields before selecting observed recipients.\footnote{Sources are documented at
\href{https://www.nssl.noaa.gov/projects/mrms/}{NOAA MRMS},
the \href{https://mesonet.agron.iastate.edu/archive/mrms.php}{IEM MRMS archive},
the \href{https://mesonet.agron.iastate.edu/request/gis/watchwarn.phtml}{IEM warning archive},
\href{https://www.ncei.noaa.gov/products/land-based-station/integrated-surface-database}{NOAA ISD},
\href{https://emc.ncep.noaa.gov/emc/pages/numerical_forecast_systems/hrrr.php}{NOAA HRRR}
and the \href{https://api.waterdata.usgs.gov/}{USGS Water Data APIs}.}

\subsection{Warning geography and the cost of long exclusion windows}
\label{app:weathercomparison}

An outcome-blind warning-coverage exercise reconciles 23,544 selected warning keys
to archived geometries, with no missing key or null geometry. Keys include issuing
office, year, warning number, phenomenon, zone and issue/expiration time. The
4,638 distinct geometries are intersected with 2020 Census county boundaries in
an equal-area projection. The IEM exporter uses the warning's archived geography
identifier rather than today's zone code. This resolves the geography of the selected
warnings, not complete physical-hazard coverage.

Table~\ref{tab:weather_coverage_v31} reports sample sizes only. These are intersections
of recorded-weather, county-warning and archived zone-warning screens on the earlier
75,237-row candidate sample, not additional spending regressions or the current
74,779-row measured-weather sample. These sample counts describe warning coverage,
not identifying variation in the spending regression.

\begin{table}[!htbp]\centering
\begin{threeparttable}
\caption{Recipient coverage under alternative warning-exclusion windows}
\label{tab:weather_coverage_v31}
\footnotesize
\begin{tabular*}{\textwidth}{@{\extracolsep{\fill}}lc}
\toprule
Screening weeks & Retained county--declaration rows\\
\midrule
$-9$ through 12 & 4,372\\
$-9$ through 4 & 9,536\\
$-1$ through 4 & 18,581\\
$-1$ through 12 & 7,440\\
\bottomrule
\end{tabular*}
\begin{tablenotes}[flushleft]\footnotesize
\item Notes: All windows exclude expanded storm support, county-coded acute
NOAA reports, selected county warnings, positive archived zone-warning area contact
and unknown warning timing. These outcome-blind sample counts do not estimate a
spending response, its identifying variation or achieved power.
\end{tablenotes}
\end{threeparttable}
\end{table}

The comparison explains the use of measured own-weather controls in the main
application, because a long future-weather screen sharply changes the recipient population
and can exclude counties for conditions occurring after the outcome endpoint.
Conservative warning times also use a plus/minus-one-day UTC envelope rather than
county-local conversion. Remaining hazard gaps belong in the interpretation of the
local association. Financial recovery still requires the distinct demand and price
measurements derived in Appendices~\ref{app:financialmargin}--\ref{app:visibility}.

\section{Simulated recovery on the sample design}
\label{app:recoverycheck}

This appendix documents the implementation summarized in Section~\ref{sec:implementation} and
Table~\ref{tab:recovery}. It is a simulation on the observed network and storm design with
hypothetical financial measurements, so it does not recover the actual mechanism. Code, frozen
protocols, saved replications and the full run record are kept with the replication files.

\subsection{Design}
\label{app:recovery_design}

The simulation uses the 2,793-county source-complete SCI graph, in which 1,574 counties are observed
recipients, and all 85 candidate storm declarations (79 episodes), not only the 48 declarations of the
empirical sample, on one weekly calendar from January 2020 to June 2024. Sixty-nine storms are held
out across three calendar folds. Training weeks exclude every held-out storm's 52-week response
support, and footprint counties are masked for 52 weeks after onset. The household field follows the
slow operator of Section~\ref{sec:model} with $\rho_D=0.85$ and $\kappa_D=1$ per week, simulated in
coordinates in which county log spending is the state, an added measurement assumption. The input is
local damage relative to one week of baseline spending.

The target is the financially weighted response of observed recipient households for each held-out
storm at horizons one to twelve. Loadings are proportional to county dividend and interest income,
measured in five groups, and the market map is $10^9c_E^{\$}=0.48$ basis points per billion dollars of desired
selling, where $c_E^{\$}=c_E/\bar P$ is the map per dollar at reference price $\bar P$. In the \emph{anchor route}, a common household response proportional to damage doubles the
average recipient response. A hypothetical event-level aggregate anchor measures each storm's
recipient-mean response path, propagated field plus common response, with a reported standard
deviation equal to half these paths' root mean square over all storms and horizons. In the \emph{structure route}, the
data-generating process has no common household response outside the propagation law, and no anchor
is used. That restriction cannot be tested from spending contrasts.

\subsection{Estimator, comparisons, inference and the planning diagnostic}
\label{app:recovery_inference}

The structural estimator is the diffusion-kernel operator under criterion \eqref{eq:targetfit}
in its simplest case, with $\omega=0$ and $\lambda_{\mathcal P}=0$. Its
persistence is continuous on $[0.50,0.985]$, and its diffusion rate is profiled on a 13-point grid that
excludes the scenario value and is then refined. Its amplitude is estimated jointly with a
geographic nuisance and, when an anchor exists, the common response. Held-out anchors are not used. Instead, held-out common
responses are predicted from damage with the per-dollar profile estimated on training anchors, a form
correct in this scenario. Spending contrasts are weighted
by inverse county noise variance estimated in training, and anchors by their reported precision. The
comparisons use the same criterion with graph-polynomial features (ridge penalty chosen by inner
blocked cross-validation), static one-hop features and a shape oracle that knows the true field
shape. An earlier two-step regression and an anchor-only prediction are also reported.

The financial target is projected with county-resolution loadings, each equal to the group scale, known or
estimated from hypothetical auxiliary measurements with assumed errors, times the county income shape within the group. Uncertainty comes from a parametric bootstrap
under a declared measurement-error model, county AR(1) noise plus a state AR(1) factor, fitted to
training residuals by the method of moments with an indirect-inference correction for the
fixed-effect projection. Anchors use their reported standard deviation and the financial and market
maps their reported errors. Each of 199 bootstrap samples re-estimates the operator with the training
weights held fixed. The bands are pointwise Wald intervals and sup-$t$ bands over each storm's
twelve-week path, both centred at the estimate.

For a held-out storm $e$, the estimation error decomposes as
$\hat{\bm k}_e-\bm k_e=\bm b_e+\bm s_e+G_e\bm m$, the sum of specification bias, sampling error with covariance
$\Sigma_e$, and the first-order effect of the error $\bm m$ in the financial and market maps, with
sensitivity $G_e$ and covariance $V_m$. If $\bm s_e$ and $\bm m$ have mean zero and the three
components are uncorrelated, then to first order
\begin{equation}\label{eq:accuracy}
\mathbb E\|\hat{\bm k}_e-\bm k_e\|^2=\|\bm b_e\|^2+\operatorname{tr}\Sigma_e
+\operatorname{tr}\big(G_eV_mG_e'\big).
\end{equation}
Under a normal approximation to the estimation error, a calibrated pointwise interval at horizon $h$
covers with probability
$\Phi(z-\beta_{eh})-\Phi(-z-\beta_{eh})$, where $z$ is the normal critical value,
$\beta_{eh}=b_{eh}/\sigma_{eh}$, and $\sigma_{eh}$ is the total standard error at that horizon,
including map error. A bound on the bias gives a bound on the mean squared error, not its value. A
path-level bound $\|\bm b_e\|\leq\varrho\|\bm k_e\|$ implies only $|\beta_{eh}|\leq\varrho\|\bm k_e\|/\sigma_{eh}$, which can be
uninformative, because a horizon can
have a small standard error even when the path's total uncertainty is large. Coverage predictions
therefore use $\beta_{eh}$ horizon by horizon. Heterogeneous loadings enter through their covariance and the
target's sensitivity to them, $G_eV_mG_e'$.

Equation~\eqref{eq:accuracy} supplies a planning diagnostic. For each design, the sampling term is the
spread of the estimate across 40 noise draws, and the bias term is the error of a noiseless fit. Both
use the scenario's assumed signal magnitudes and specification errors. Pooling over held-out storms
and horizons, with $\mathrm{SNR}$ the target's root mean square over the root-mean-square standard error and
$\varrho$ the relative bias, the predicted normalized error is
$(\mathrm{SNR}^{-2}+\varrho^2+\sigma_a^2+\sigma_c^2)^{1/2}$, where the map terms $\sigma_a=0.25/\sqrt5$ and
$\sigma_c=0.38$ enter only when the maps carry assumed errors. They are this study's planning approximation for five
equally precise group scales and a log-normal market map, not the general term $\operatorname{tr}(G_eV_mG_e')$
for arbitrary loadings. The diagnostic is scenario-based. It plans
measurement, but it does not certify that a given data set will recover the response. Across the
nineteen non-null designs of Table~\ref{tab:recovery_full}, Panel A, predicted and achieved log
errors correlate at 0.998, and the prediction falls on the correct side of the 0.5 threshold in
eighteen.

\subsection{Measurement-noise models}
\label{app:recovery_noise}

Panel A uses a design-based noise model, estimated from raw residuals of the county panel after
county and week effects, that combines county AR(1) noise with coefficient 0.26 and a state factor taken as the
state-week mean residual. Panel B uses noise calibrated more carefully to the panel after September
7, 2020, with storm-affected cells excluded. Estimated by within-state cross-county covariances, the
regional factor carries about 6 percent of the noise variance, whereas the raw state-mean factor of Panel A
carries 27 percent, inflated by averaging over a few counties in small states. The residual
autocorrelation decays slowly, 0.15 at four weeks and 0.09 at thirteen in the raw panel and 0.22 and
0.12 after denoising. A two-component model fits it, with white noise plus a persistent county drift
whose autoregressive coefficient is 0.99. The drift carries about 45 percent of the variance in the denoised
panel. Denoising subtracts the 2019-base calendar pattern (Appendix~\ref{app:calendar_sensitivity}) and
lowers the white component, but it leaves the drift. County piecewise-linear trends with knots every
26 weeks, removed identically from the data and from every model column, take out the drift's
low-frequency part. Without an anchor at the 85 declarations, they lower the predicted error from 2.91
to 1.79. Their spacing was chosen by the planning diagnostic, before the Panel B evaluation draws.

\begin{table}[htbp]\centering
\begin{threeparttable}
\caption{Simulated recovery on the sample design: all environments (structural estimator)}
\label{tab:recovery_full}
\footnotesize\setlength{\tabcolsep}{3pt}
\begin{tabular*}{\textwidth}{@{\extracolsep{\fill}}lccccc}
\toprule
 & Predicted & Normalized & \multicolumn{2}{c}{Coverage, pointwise / path} & \\
\cmidrule(lr){4-5}
Environment & error & error & final inference & predeclared & Label \\
\midrule
\multicolumn{6}{@{}l}{\textit{Panel A. Design-based noise model}} \\
\addlinespace[2pt]
\multicolumn{6}{@{}l}{Anchor route (event-level aggregate anchor)} \\
\quad Anchor SD $1\times$, current noise & 0.185 & 0.170 & 0.90 / 0.87 & 0.80 / 0.81 & recovered \\
\quad Anchor SD $2\times$ & 0.365 & 0.335 & 0.90 / 0.87 & 0.78 / 0.83 & recovered \\
\quad Anchor SD $3\times$ & 0.546 & 0.502 & 0.89 / 0.87 & 0.77 / 0.80 & not recovered \\
\quad Anchor SD $10\times$ & 1.802 & 1.705 & 0.90 / 0.89 & 0.83 / 0.73 & not recovered \\
\quad Aggregate-innovation calibration ($87.5\times$) & 15.70 & 14.72 & 0.90 / 0.87 & 0.79 / 0.47 & not recovered \\
\quad Anchor SD $1\times$, maps with assumed errors & 0.437 & 0.400 & 0.91 / 0.88 & 0.88 / 0.85 & recovered \\
\quad Anchor SD $3\times$, noise $\times0.5$, $+85$ storms & 0.215 & 0.206 & 0.88 / 0.86 & 0.82 / 0.66 & recovered \\
\quad Anchor SD $10\times$, noise $\times0.25$, $+255$ storms & 0.419 & 0.385 & 0.92 / 0.91 & 0.83 / 0.67 & recovered \\
\multicolumn{6}{@{}l}{Structure route (no common response outside the law)} \\
\quad Current spending noise & 3.384 & 3.048 & 0.96 / 0.98 & 0.80 / 0.70 & not recovered \\
\quad Noise $\times0.5$ & 1.346 & 1.461 & 0.95 / 0.98 & 0.74 / 0.61 & not recovered \\
\quad Noise $\times0.25$ & 0.739 & 0.715 & 0.94 / 0.95 & 0.71 / 0.50 & not recovered \\
\quad Noise $\times0.5$, $+85$ storms & 0.761 & 0.574 & 0.94 / 0.97 & 0.72 / 0.52 & not recovered \\
\quad Noise $\times0.5$, $+255$ storms & 0.548 & 0.478 & 0.91 / 0.92 & 0.73 / 0.41 & recovered \\
\quad Noise $\times0.25$, $+85$ storms & 0.403 & 0.320 & 0.89 / 0.87 & 0.70 / 0.28 & recovered \\
\quad Noise $\times0.25$, $+255$ storms & 0.280 & 0.257 & 0.89 / 0.86 & 0.72 / 0.23 & recovered \\
\quad \begin{tabular}[t]{@{}l@{}}Noise $\times0.25$, $+85$ storms,\\ maps with assumed errors\end{tabular} & 0.565 & 0.505 & 0.91 / 0.89 & 0.82 / 0.59 & not recovered \\
\quad Noise outside the bootstrap family & 0.386 & 0.346 & 0.86 / 0.85 & -- & recovered \\
\multicolumn{6}{@{}l}{Misspecified law and geography (two timescales, squared geographic kernel)} \\
\quad Anchor route & 0.637 & 0.626 & 0.34 / 0.00 & 0.42 / 0.05 & not recovered \\
\quad Structure route & 3.726 & 3.464 & 0.52 / 0.00 & 0.38 / 0.00 & not recovered \\
\midrule
\multicolumn{6}{@{}l}{\textit{Panel B. Noise calibrated to the denoised county panel}} \\
\addlinespace[2pt]
\quad Anchor SD $1\times$; two-way fixed effects & 0.187 & 0.168 & 0.90 / 0.90 & -- & recovered \\
\quad Anchor SD $1\times$; county trends & 0.184 & 0.169 & 0.89 / 0.87 & -- & recovered \\
\quad No anchor; county trends; actual storms & 1.787 & 1.633 & 0.95 / 0.96 & -- & not recovered \\
\quad No anchor; county trends; $+85$ storms & 0.839 & 0.686 & 0.89 / 0.87 & -- & not recovered \\
\bottomrule
\end{tabular*}
\begin{tablenotes}[flushleft]\footnotesize
\item Notes: The estimator is the parsimonious diffusion-kernel operator under criterion \eqref{eq:targetfit}, fitted on training storms and evaluated on 69 held-out storms over horizons 1--12. Normalized error is the root-mean-squared error over the root mean square of the true response. Coverage is for nominal 90 percent bands. \emph{Final inference}: county-resolution loadings and a parametric bootstrap that re-estimates the operator (Appendix~\ref{app:recovery_inference}). \emph{Predeclared}: the first inference, a Gauss--Newton sandwich with group-mean loadings (Appendix~\ref{app:recovery_history}). Labels apply the study's diagnostic criteria to the final inference, under which a design counts as recovered if normalized error is at most 0.50, pointwise coverage at least 0.85, path coverage at least 0.80 and failures at most 2 percent of replications, for nominal 90 percent bands. Predicted errors come from the planning diagnostic of Appendix~\ref{app:recovery_inference}, evaluated at the scenario's signal magnitudes. Under the nulls, with no diffusion ($\kappa_D=0$) the composition band excludes zero for 0.7 percent of storms, and with zero financial loading it excludes zero for 0.0 percent. Panel A uses 100 replications in the leading environments and 40 in the others, and Panel B uses 100 (anchor) and 40 (no anchor).
\end{tablenotes}
\end{threeparttable}
\end{table}

\subsection{Results}
\label{app:recovery_results}

\begin{figure}[htbp]
\centering
\includegraphics[width=\linewidth]{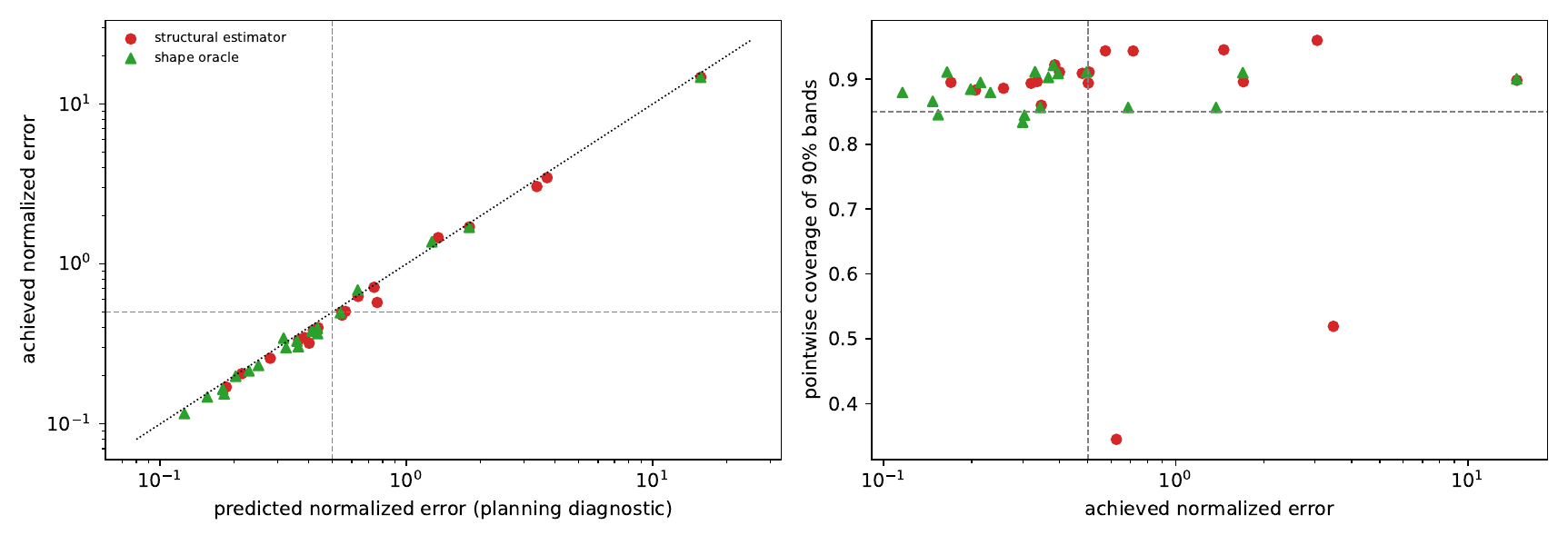}
\caption{Predicted versus achieved normalized error (left) and achieved pointwise coverage (right)
across the designs of Table~\ref{tab:recovery_full}, Panel A, for the structural estimator and the
shape oracle. Dashed lines mark the recovery thresholds.}
\label{fig:recovery_map}
\end{figure}

Table~\ref{tab:recovery_full} reports every design, and Figure~\ref{fig:recovery_map}
compares predicted and achieved accuracy across them.

With an anchor at the assumed precision, the operator recovers the response at current
spending noise (normalized error 0.170), under the panel-calibrated noise (0.168), and with loading
scales and market map that carry assumed errors (0.400). It narrowly misses when the anchor's error is three times that
precision (0.502, Monte Carlo s.e.\ 0.010) and fails under an aggregate-innovation calibration (14.7),
in which the anchor's standard deviation equals the innovation standard deviation of an AR(1) benchmark
for the constructed spending aggregate, 87.5 times the assumed precision (128 times over all weeks). That
innovation includes genuine aggregate movements, so it is not a floor on the error of every event-attribution
procedure. Precise spending and more storms substitute for a weaker anchor (0.206 and 0.385). Under the same
draws and final inference (Table~\ref{tab:recovery_matched}), an anchor-only prediction has normalized error
0.30 with known maps and 0.51 with maps carrying assumed errors, and its pointwise bands cover 24 and 48 percent of
true values, against 24 and 49 percent under the earlier predeclared inference. The operator's
advantage is therefore not an artifact of the inference method. The operator also estimates the cross-county
composition, about 4 percent of the target. With panel-calibrated noise, this composition component has
normalized error 0.38, against 1 for predicting zero, with coverage 0.88 pointwise and 0.85 by path.

Without an anchor, recovery requires a correctly specified common response together
with a quarter of the design-based noise and twice the storms (0.320), or half the noise and four
times the storms (0.478). With panel-calibrated noise, denoising and county trends, the error is 1.63
at the 85 declarations and 0.69 with twice the storms, although the bands remain honest (pointwise coverage
0.95 and 0.89).

The comparison estimators use the same final inference (Table~\ref{tab:recovery_matched}). Without an
anchor, at a quarter of the noise and twice the storms, graph-polynomial features reach 0.71 and the
one-hop response 0.92, against 0.32 for the structural operator, and their bands cover 14 and 8 percent of
true values. With an anchor the criterion-based estimators are closer (0.17 to 0.22 in the first row's
design), but only the structural operator's bands are calibrated. Under the predeclared inference the
two-step regression had error 1.00 with an anchor and 0.47 without, the latter with bands covering 52 percent.

In the joint two-timescale and geographic-misspecification
scenarios, where the true geographic nuisance uses the squared kernel at three times the size while
every estimator fits the first-order kernel, the tested feasible estimators (structural,
graph-polynomial, one-hop, two-step and anchor-only) fail, with errors of 0.6 or more with an anchor
and 1.1 or more without, as the specification error predicts. These scenarios do not isolate the
number of temporal channels.
Without diffusion, the composition band excludes zero for 0.7 percent of storms, and with zero financial
loading, it never excludes zero.

When the noise is more persistent than the bootstrap family
allows (AR(2) dynamics with Student-$t$ innovations), pointwise coverage is 0.860, close to the loss
implied by a 22--27 percent understatement of the long-run variance. The shape oracle's bands
undercover slightly without an anchor (0.85 in the leading design and 0.82 under panel-calibrated noise
with twice the storms), while the structural estimator's remain calibrated.

\begin{table}[htbp]\centering
\begin{threeparttable}
\caption{Matched comparison: every estimator under the final inference on the same draws}
\label{tab:recovery_matched}
\footnotesize\setlength{\tabcolsep}{3pt}
\begin{tabular*}{\textwidth}{@{\extracolsep{\fill}}lcccc}
\toprule
 & \multicolumn{3}{c}{With anchors} & No anchor \\
\cmidrule(lr){2-4}\cmidrule(lr){5-5}
 & (1) & (2) & (3) & (4) \\
Estimator & Panel noise & Design noise & Maps with assumed errors & $\frac14$ noise, $2\times$ storms \\
\midrule
Anchor only & 0.30 / 0.24 & 0.30 / 0.24 & 0.51 / 0.48 & -- \\
Static one-hop & 0.22 / 0.45 & 0.22 / 0.47 & 0.44 / 0.73 & 0.92 / 0.08 \\
Graph polynomial & 0.18 / 0.72 & 0.19 / 0.72 & 0.41 / 0.87 & 0.71 / 0.14 \\
Structural operator & 0.17 / 0.90 & 0.17 / 0.90 & 0.40 / 0.91 & 0.32 / 0.89 \\
Shape oracle (infeasible) & 0.16 / 0.90 & 0.16 / 0.91 & 0.40 / 0.91 & 0.15 / 0.85 \\
\bottomrule
\end{tabular*}
\begin{tablenotes}[flushleft]\footnotesize
\item Notes: Normalized error / pointwise coverage of nominal 90 percent bands for the recipients' financially weighted response, horizons 1--12, 69 held-out storms. Columns (1), (3) and (4) are the designs of main Table~\ref{tab:recovery}, rows 1--3, and column (2) is row 1's design with design-based noise. Each column uses the 100 production replications of Table~\ref{tab:recovery_full}, the same data for every estimator. Every estimator uses county-resolution loadings and the parametric bootstrap of Appendix~\ref{app:recovery_inference} (B = 199) with its own fitted noise model, while the graph-polynomial and one-hop fits hold their training-fold ridge penalty fixed in the bootstrap. The structural operator has the lower error in 100 percent of paired replications against anchor only and 100 percent against one-hop in column (1), and in 97 percent against the graph polynomial in column (4). Monte Carlo standard errors are at most 0.035 for errors and 0.03 for coverage.
\end{tablenotes}
\end{threeparttable}
\end{table}

\subsection{Development history}
\label{app:recovery_history}

The procedure was developed in stages. Each stage was evaluated on fresh simulation draws under a
protocol frozen beforehand. Development used separate draws, and every earlier result is retained.
In the first stage, under the predeclared inference, the two-step regression gives normalized error
1.00 in the baseline anchored design. Without an anchor, at a quarter of the spending noise and twice
the storms, its error is 0.47 and its nominal 90 percent pointwise bands cover 52 percent of true responses.

In the second stage, the target-oriented and structural estimators under \eqref{eq:targetfit}, with a
predeclared Gauss--Newton sandwich inference and group-mean loadings, achieved the predicted accuracy, but
pointwise coverage was 0.70 to 0.83 in most designs. A diagnosis on separate development draws then found
four sources of the undercoverage. First, group-mean loadings leave a deterministic resolution error of
2.6 to 5.3 percent of the target. Second, a few high-damage storms dominate the common-response
information. Third, serial dependence crosses the two-week clusters. Finally, percentile intervals are
miscentred through the nonlinearity in persistence and diffusion.

The final inference, county-resolution loadings with the parametric bootstrap, was then frozen and
evaluated on fresh draws (Panel A), and later under the panel-calibrated noise (Panel B). It was chosen
after the predeclared inference failed, so only its evaluation is fresh.

\subsection{A diffusion-specific EBP restriction}
\label{app:diffusionrestriction}

Apply one dated linear contrast $T_e^{\mathrm{time}}$, the mean over three full months after onset minus the mean
over three full months before, to both the observed EBP and the predicted path.
Construct the basis-point path $p_t=\sum_{e'}K_{t-t_{e'}}\bm\varepsilon_{e'}$, including overlapping
shocks, and set $p_e=T_e^{\mathrm{time}}\{p_t\}$. Under no anticipation, $K_h=0$ for $h<1$.
Equation~\eqref{eq:diffincrement} gives the decomposition $p_e=p_e^0+p_e^{diff}$. The proposed validation is
\begin{equation}\label{eq:validation}
Y_e=\alpha+\beta p_e^{diff}+\gamma_0p_e^0+Z_e'\bm\gamma+u_e,\qquad
\beta=1\quad\text{under the complete, correctly measured bridge}.
\end{equation}
The complete model also implies $\gamma_0=1$. The nuisance span includes an intercept, $p^0$,
and predetermined production, geography, seasonality, macro and insurer channels. If $M_{[1,p^0,Z]}$ projects off that span, identifying
diffusion requires $(p^{diff})'M_{[1,p^0,Z]}p^{diff}>0$, so stationary incidence alone is
insufficient. The lead contrast tests anticipation and timing. Parameter uncertainty in
$\hat H_h$, $a$ and $c_E$ must enter inference, since an unbiased unit coefficient is not guaranteed by
plugging noisy estimates into a regression. Capital or other omitted responses can invalidate the
restriction even when spending transport succeeds, and the monthly change regression does not test
it. Appendix~\ref{app:power} illustrates how noise, weak excitation of the relevant modes and collinearity
with nuisance controls determine the power of this proposed test.
\subsection{Recovering the credit-premium response with an estimated market map}\label{app:bridgefund}

This section extends the first design of Table~\ref{tab:recovery}, which treats both financial maps as known, by
estimating them from simulated data. In contrast, the second design draws them with assumed errors. Household loading scales are estimated from
hypothetical county account flows, and the market map from an independently simulated panel of fund observations whose
instrument is built from the simulated fund flows, not supplied as the true demand innovation.

\paragraph{The fund demand--price system.}
Let $f_{jt}$ be fund $j$'s net subscriptions over lagged net assets, $w_{jt}^{\mathrm{hold}}$ its lagged corporate-security holdings as a share of the
market denominator of Appendix~\ref{app:creditflow}, $F_t=100\sum_jw_{jt}^{\mathrm{hold}}f_{jt}$ the holdings-weighted subscription proxy for aggregate purchases, in percent of that
denominator, and $\Delta E_t^{BP}$ the monthly EBP change in basis points. With common flow factors $\bm f_t^{\mathrm{com}}$ and controls $\bm Z_t^{\mathrm{mkt}}$ (an
intercept and two lagged EBP changes),
\[
\begin{aligned}
f_{jt}&=\alpha_j^F+\zeta_j^F\Delta E_t^{BP}+(\bm L_j^F)'\bm f_t^{\mathrm{com}}+u_{jt},\\
\Delta E_t^{BP}&=\phi F_t+(\bm Z_t^{\mathrm{mkt}})'b+g'\bm f_t^{\mathrm{com}}+u_t^{\mathrm{mkt}} .
\end{aligned}
\]
With $\zeta_{\mathrm{agg}}^F=100\sum_jw_j^{\mathrm{hold}}\zeta_j^F$, aggregate purchases are $F=\zeta_{\mathrm{agg}}^F\Delta E^{BP}+U^{\mathrm{pct}}$, where $U^{\mathrm{pct}}$ denotes desired purchases at fixed
prices, so
\[
\Delta E^{BP}=\phi m_FU^{\mathrm{pct}}+m_F((\bm Z^{\mathrm{mkt}})'b+g'\bm f^{\mathrm{com}}+u^{\mathrm{mkt}}),\qquad
m_F=(1-\zeta_{\mathrm{agg}}^F\phi)^{-1}.
\]
For desired selling $S^{\mathrm{pct}}=-U^{\mathrm{pct}}$ the premium loading is $c_E^{\mathrm{sell}}=-\phi m_F$.

The market-percentage units give $S^{\mathrm{pct}}=100\bar PS/A_{\mathrm{mkt}}$, where $S$ is the model's claim-quantity shift and
$A_{\mathrm{mkt}}$ is dollar market value. Thus $\phi=-(A_{\mathrm{mkt}}/100)\vartheta_F$ for the dollar-flow slope in
Appendix~\ref{app:flowbridge}, and $m_F=1-\zeta G_S$ is the executed share. The maps satisfy
\[
c_E^{\$}=\frac{100}{A_{\mathrm{mkt}}}c_E^{\mathrm{sell}}
=\vartheta_F(1-\zeta G_S)=\frac{b_EG_S}{\bar P},\qquad
c_E=\bar P c_E^{\$}=\frac{100\bar P}{A_{\mathrm{mkt}}}c_E^{\mathrm{sell}}.
\]
The final expression is the factor in \eqref{eq:flowbridge}, now with the market-percentage conversion explicit.
Matching $\zeta_{\mathrm{agg}}^F=100\bar P\zeta/(A_{\mathrm{mkt}}b_E)$ assumes that the observed funds are the only price-elastic
non-intermediary demand. A granular instrument $Z^{\mathrm{GIV}}$ has first-stage slope
\[
m_F\,\mathrm{Cov}(Z^{\mathrm{GIV}},U^{\mathrm{pct}})/\mathrm{Var}(Z^{\mathrm{GIV}}),
\]
which equals $m_F$ only under the unknown normalization of $Z^{\mathrm{GIV}}$ in units of desired demand. An estimated first-stage
coefficient is not by itself an estimate of the executed share.

\paragraph{Identification of the two market terms.}
The construction is the granular instrumental-variables approach of Gabaix and Koijen (2024), applied to the two
equations of the system. Let $w_j^{\mathrm{prec}}\propto1/\sigma_j^2$ be precision
weights normalized to sum to one, with $\sigma_j^2$ the idiosyncratic variance of fund $j$, and let $\widetilde w_{jt}^{\mathrm{prec}}$ be the corresponding precision-weighted
shares, normalized so that $\sum_j\widetilde w_{jt}^{\mathrm{prec}}=\sum_jw_{jt}^{\mathrm{hold}}$. The instrument is $Z_t^{\mathrm{GIV}}=100\sum_j(w_{jt}^{\mathrm{hold}}-\widetilde w_{jt}^{\mathrm{prec}})\hat u_{jt}$,
where $\hat u_{jt}$ are fund residuals after removing factors estimated from the cross-sectionally demeaned panel. With a
common $\zeta^F$, the demeaned panel contains neither the price term nor any common rate. The demand equation uses the
precision-weighted average of fund-demeaned flows, $f_{\mathrm{prec},t}$, whose idiosyncratic part is $u_{\mathrm{prec},t}$. Two moments follow.
The instrumental-variables regression of $\Delta E^{BP}$ on $F$ with instrument $Z^{\mathrm{GIV}}$ identifies $\phi$ when $\mathrm{Cov}(Z^{\mathrm{GIV}},u^{\mathrm{mkt}})=0$
and $\mathrm{Cov}(Z^{\mathrm{GIV}},F)\neq0$. Similarly, the regression of $f_{\mathrm{prec}}$ on $\Delta E^{BP}$ with the same instrument identifies $\zeta^F$ when
$\mathrm{Cov}(Z^{\mathrm{GIV}},u_{\mathrm{prec}})=0$ and $\mathrm{Cov}(Z^{\mathrm{GIV}},\Delta E^{BP})\neq0$. Both require the common factors to be removed validly.

The second exclusion condition holds in population under the following assumptions. Let the weights be fixed or
predetermined relative to current idiosyncratic shocks, conditional on the factor and information structure. Let the
conditional covariance of the idiosyncratic shocks across funds be diagonal, with $\mathrm{Var}(u_j)=\sigma_j^2$. Also let
the normalized population precision weights be $w_j^{\mathrm{prec}}=\varpi/\sigma_j^2$, with $\sum_j\widetilde w_j^{\mathrm{prec}}=\sum_jw_j^{\mathrm{hold}}$. Then
\[
\begin{aligned}
\mathrm{Cov}(Z^{\mathrm{GIV}},u_{\mathrm{prec}}\mid\text{information})
&=100\sum_j(w_j^{\mathrm{hold}}-\widetilde w_j^{\mathrm{prec}})\,w_j^{\mathrm{prec}}\sigma_j^2\\
&=100\,\varpi\Big(\sum_jw_j^{\mathrm{hold}}-\sum_j\widetilde w_j^{\mathrm{prec}}\Big)=0 .
\end{aligned}
\]
This is a population statement, not an exact finite-sample property of estimated residual weights. The cancellation need
not hold with residual common or category shocks, misspecified weights, heterogeneous feedback that the factors do not
remove, or factors and weights estimated in a finite sample. The simulations and exclusion diagnostics below examine
these departures. Equal weights generally violate the condition, and in simulated 228-month panels their median estimate of
$\zeta^F$ is 2.0 times the truth.

\paragraph{Household stocks, fund flows and the premium path.}
Fix the intermediary regime and payoff news, and express the household responses
at monthly measurement dates. Let $S_m^{\$}=(a^{\$})'H_m\varepsilon$ be the
reduction in desired credit holdings at fixed reference prices, in dollars.
$a^{\$}$ contains the separately measured financial sensitivities and any
observed-sector pass-through. Let $X_m^{\$}$ be the deviation in actual holdings
net of valuation gains and $E_m^{BP,\mathrm{dev}}$ the premium-level deviation in basis points.
With a fixed price-responsive demand slope $\zeta^{\$}=\bar P\zeta/b_E$,
\[
\begin{aligned}
X_m^{\$}&=\zeta^{\$}E_m^{BP,\mathrm{dev}}-S_m^{\$},\\
F_m^{buy}=\Delta X_m^{\$}&=\zeta^{\$}\Delta E_m^{BP,\mathrm{dev}}-\Delta S_m^{\$}.
\end{aligned}
\]
Thus monthly purchases respond to changes in desired holdings, not their level.
Under static inverse absorption $E_m^{BP,\mathrm{dev}}=\phi_{\$}X_m^{\$}$, with $\phi_{\$}<0$ and
$1-\zeta^{\$}\phi_{\$}>0$,
\[
\begin{aligned}
m_F&=(1-\zeta^{\$}\phi_{\$})^{-1},&\qquad
F_m^{buy}&=-m_F\Delta S_m^{\$},\\
\Delta E_m^{BP,\mathrm{dev}}&=c_E^{\$}\Delta S_m^{\$},&
c_E^{\$}&=-\phi_{\$}m_F>0.
\end{aligned}
\]
Starting from zero deviations, cumulation gives the exact level identity
$E_m^{BP,\mathrm{dev}}=c_E^{\$}S_m^{\$}$. Let $R_m$ denote the field response evaluated at the monthly
measurement date, so $H_m=C_*R_m$. In the main text's claim units,
$S_m^{\$}=\bar P\bm l'R_m\varepsilon$, yielding \eqref{eq:flowbridge}.
If purchases are recorded as $F_m^{pct}=100F_m^{buy}/A_{\mathrm{mkt}}$, for dollar market size
$A_{\mathrm{mkt}}$, the coefficient per one percent of the market satisfies
$\phi_{\$}=100\phi/A_{\mathrm{mkt}}$ and $c_E^{\$}=-100\phi m_F/A_{\mathrm{mkt}}$.
Identifying weighted subscriptions with purchases is a separate observation
restriction, examined through the holdings diagnostic in Appendix~\ref{app:creditflow}.

When lagged price terms represent structural market adjustment, let the
cleared response satisfy
\[
\Delta E_m^{BP,\mathrm{dev}}=c_E^{\$}\Delta S_m^{\$}
             +d_1\Delta E_{m-1}^{BP,\mathrm{dev}}+d_2\Delta E_{m-2}^{BP,\mathrm{dev}}.
\]
If the original inverse-absorption equation has lag coefficients $b_1,b_2$ on
price changes, simultaneous clearing gives $d_j=m_Fb_j$, not simply $b_j$.
Integrating with zero initial deviations yields
\[
\begin{aligned}
E_m^{BP,\mathrm{dev}}&=c_E^{\$}S_m^{\$}+d_1E_{m-1}^{BP,\mathrm{dev}}+d_2E_{m-2}^{BP,\mathrm{dev}},\\
K_m^{market}&=c_E^{\$}(a^{\$})'\sum_{j=0}^{m}\psi_j^{\mathrm{mkt}}H_{m-j},
\end{aligned}
\]
where $\psi_0^{\mathrm{mkt}}=1$, $\psi_j^{\mathrm{mkt}}=d_1\psi_{j-1}^{\mathrm{mkt}}+d_2\psi_{j-2}^{\mathrm{mkt}}$ and negative-index
coefficients are zero. The stable permanent-shift multiplier
$(1-d_1-d_2)^{-1}$ does not replace this finite-horizon convolution for a
transient household response. The scored conditional impact-loading target
below is distinct from the dynamically filtered premium-level path.

Weekly measurements are aligned to monthly flows by differencing the holdings stock at the
month's boundaries and summing matched premium changes, since summing weekly holdings levels
would create a stock--flow mismatch. The household measurement and financial-sensitivity
conditions remain necessary under either static or dynamic clearing.

\paragraph{Simulated fund observations.}
The market stage uses simulated fund observations with the sample's design, namely the 510 funds, their monthly shares and
reporting pattern (79 percent of fund-months observed), flow volatilities (median 3.4 percent per month) and fund mean
flows, factor loadings and persistence, the price equation's lag and factor coefficients, and the calm-month EBP-change
standard deviation (19.8~bp). The outcomes and structural parameters are simulated. The demand feedback is homogeneous,
and $(\zeta_{\mathrm{agg}}^F,\phi)$ are set to the scenario values $c_E^{\mathrm{sell}}=277.5$~bp per desired selling equal to 1 percent of the
market denominator and $m_F=0.683$, the reduced form with its sign reversed and the first stage of column~(1) of
Table~\ref{tab:creditflow}. They are scenario values, not estimates of either quantity. The executed-purchase
coefficient is $\phi\approx-406$~bp per 1 percent, matching the magnitude and sign of that column's estimate. Its
magnitude corresponds to a price multiplier of about 26 on the conversion of Appendix~\ref{app:creditflow}, against 1 and
5 in that section's power calculation. The reported precision is conditional on this price-impact calibration.
Performance at smaller impacts requires a separate sensitivity exercise. Idiosyncratic shocks are Gaussian and independent across funds given two factors,
household demand passes through the observed funds in proportion to their size, and the simulation identifies the
holdings-weighted subscription proxy with executed corporate-security purchases, assuming proportional portfolio
allocation. At the 57-month mean market value of \$7,970 billion, the scenario's loading is 3.48~bp per billion dollars of
desired selling, against 0.48 in Appendix~\ref{app:recovery_design}. Household loading scales are reduced by the same
factor, 0.48/3.48, so that the target keeps the first design's root mean square, 0.72~bp in total and 0.26~bp for the
propagated part.

\begin{table}[htbp]\centering
\begin{threeparttable}
\caption{The premium loading estimated from simulated fund observations: market-only simulations}
\label{tab:market_sim}
\footnotesize
\begin{tabular*}{\textwidth}{@{\extracolsep{\fill}}lcccccc}
\toprule
Scenario & True $c_E^{\mathrm{sell}}$ & \multicolumn{2}{c}{From fund data} & Oracle & \multicolumn{2}{c}{90\% coverage} \\
\cmidrule(lr){3-4}\cmidrule(lr){6-7}
& & median (bias) & RMSE & RMSE & bootstrap & AR set \\
\midrule
Valid, 57 months & 277.5 & 272.0 ($-3.3$) & 144.4 & 127.3 & 0.93 & 0.78 \\
Heavy-tailed fund shocks & 277.5 & 274.1 (2.8) & 155.1 & 132.9 & 0.90 & 0.81 \\
Large valid shocks in stress months & 277.5 & 260.3 ($-15.0$) & 118.1 & 105.9 & 0.89 & 0.81 \\
Category flows coincide with news & 277.5 & 515.8 (246.1) & 295.0 & 179.0 & 0.64 & 0.81 \\
Same-size independent news & 277.5 & 270.3 (64.8) & 997.6$^a$ & 146.9 & 0.88$^a$ & 0.80 \\
Factors correlated with news & 277.5 & 283.5 (13.5) & 134.3 & 136.5 & 0.88 & 0.83 \\
Amplifying feedback ($m_F=1.19$) & 253.3 & 234.9 ($-17.9$) & 149.5 & 136.1 & 0.90 & 0.81 \\
Valid, 114 months & 277.5 & 253.9 ($-14.8$) & 93.9 & 94.1 & 0.92 & 0.85 \\
Valid, 228 months & 277.5 & 273.9 (0.6) & 63.8 & 62.2 & 0.93 & 0.88 \\
\bottomrule
\end{tabular*}
\begin{tablenotes}[flushleft]\footnotesize
\item Notes: 200 independent simulated panels per scenario, in basis points per 1\% of the market of desired selling. The
bias is the mean error. ``Oracle'' observes the true granular desired-demand innovation and uses the same
estimated-factor controls. ``Bootstrap'': share of panels whose 90\% percentile interval for $c_E^{\mathrm{sell}}$, from a
bootstrap that regenerates the fund panel and repeats the whole construction, contains the true value. The bootstrap is not
robust to weak identification. ``AR set'': share of panels whose joint Anderson--Rubin (1949) set for $(\zeta^F,\phi)$, with the
$\chi^2_2$ critical value, contains the true values. Its projection onto $c_E^{\mathrm{sell}}$ is bounded in only 0.48 of valid
57-month panels.
\item $^a$ One panel has a near-singular $1-\hat\zeta_{\mathrm{agg}}^F\hat\phi$ and $\hat c_E^{\mathrm{sell}}=14{,}171$. Without it the RMSE is
173.8. Its bootstrap failed numerically (177 of 199 draws), so its interval is not used. Bootstrap coverage is therefore over the 199
panels with usable inference, and over all 200 attempted panels it lies between 0.875 and 0.880.
\end{tablenotes}
\end{threeparttable}
\end{table}

\paragraph{Market-only results.}
Table~\ref{tab:market_sim} reports the market stage alone. In the valid 57-month scenario the estimate from fund data has
no detectable bias, and its root-mean-square error is 144.4 against 127.3 for the oracle. Pooled over every independent
57-month sample (800 panels), the fund-data estimate's root-mean-square error is 9 percent above the oracle's. On the same
pooled panels, the oracle's bias is $-21.0$~bp per 1\% (Monte Carlo s.e.\ 4.6), against $1.0$ (5.0) for the fund-data
estimate, so the oracle's advantage is in variance. With the true factors as controls its bias is $1.5$ (9.0) on 200 of
these panels, so the bias comes from the estimated factor controls. When short-duration category flows coincide with
credit news in the stress months, the exclusion fails and the estimate is biased upward. The joint Anderson--Rubin set
undercovers at this sample length. At 57 months it contains the true values in 0.78 of valid panels at nominal 90
percent, and the shortfall persists when the instrument, factors and weights are replaced by their true values (0.79, with the
lag controls and HAC(3) still estimated) and with no HAC lags (0.81). A critical value calibrated by the same bootstrap is
a partial, design-specific correction, giving 0.85 at 57 months and 0.90 and 0.91 at 114 and 228.

\paragraph{The bridge.}
The spending stage is the first design of Table~\ref{tab:recovery}, namely the structural operator on held-out storms, with the
same anchor serving the common response in every method. Household loading scales for the five groups are estimated from
hypothetical county account flows, with noise of 0.1 percent of holdings for the median county, by regression on the recovered state change in
purged training months, and are refitted on every spending-bootstrap draw. The within-group loading shape remains the
supplied income shape, so estimating the five scales does not learn an unrestricted county-to-fund ownership map. The market
stage estimates $c_E^{\mathrm{sell}}$ from an independently simulated 57-month calibration panel with the sample's fund geometry.
It builds the precision-weighted granular instrument, estimates the factors, $\zeta^F$ and $\phi$, and forms
$c_E^{\mathrm{sell}}=-\phi m_F$. Every bootstrap draw regenerates that panel and repeats each step, so the uncertainty of the
construction is carried into the response. The oracle benchmark instead observes the true granular desired-demand
innovation, with the same estimated-factor controls.

The scored target is
\[
K^R_{e,h}=c_E^{\mathrm{sell}}\,S^{R,\mathrm{pct}}_{e,h},\qquad h=1,\ldots,12,
\]
where $S^{R,\mathrm{pct}}_{e,h}$ is the desired-selling stock, in percent of the market denominator, of observed recipients $h$ weeks after the onset of storm $e$, for observed
recipient counties outside the storm footprint. Footprint households and unobserved counties are excluded.
It uses the contemporaneous pricing map at fixed capital, the conditional within-period premium response. The calibration
panel's outcome is a monthly EBP change with lag dynamics, and the storms do not enter that independent panel. In this
scenario the price equation's lags imply a long-run multiplier of about 0.81 on a permanent desired-selling stock, so the
scored path is not the complete cumulative EBP response that the lag equation would generate. The results are not
rescaled by 0.81, which would change the estimand without supplying the horizon-dependent response.

\begin{table}[htbp]
\centering
\caption{The credit-premium response recovered with estimated financial maps (simulation)}\label{tab:bridge_fund}
\begin{threeparttable}
\footnotesize
\begin{tabular*}{\textwidth}{@{\extracolsep{\fill}}lcccc}
\toprule
& \multicolumn{4}{c}{Market map} \\
\cmidrule(lr){2-5}
Spending stage & True & Oracle instrument & From fund data & From fund data \\
& & (57 months) & (57 months) & (228 months) \\
\midrule
Structural operator & 0.299 & 0.486 & 0.487 & 0.355 \\
 & (0.85) & (0.79) & (0.88) & (0.84) \\
Static one-hop & 0.465 & 0.551 & 0.542 & 0.495 \\
 & (0.28) & (0.40) & (0.49) & (0.33) \\
Anchor only & -- & -- & 0.970 & 0.802 \\
\midrule
Propagated part: operator / one-hop & 0.289 / 0.649 & 0.486 / 0.696 & 0.481 / 0.685 & 0.349 / 0.662 \\
\bottomrule
\end{tabular*}
\begin{tablenotes}[flushleft]\footnotesize
\item Notes: Mean normalized error of the conditional within-period premium response of observed recipients, weeks 1--12
after onset, held-out storms, 200 simulated replications. Coverage of nominal 90\% pointwise bands is in parentheses. Path
coverage with the 57-month fund-data market map is 0.84. With the true loadings and market map the operator's error is
0.171, the counterpart on new simulation draws of 0.168 in Table~\ref{tab:recovery}, row 1. Household loading scales are estimated from hypothetical account flows in every column. ``Oracle instrument''
observes the true granular demand innovation, with the same estimated-factor controls. Its bands use normal-approximation
draws of the market map, so their coverage is not comparable with the other columns. ``From fund data'' uses only
simulated fund flows, shares and EBP changes on an independent calibration panel with the sample's fund geometry, and its
bootstrap regenerates that panel. The 228-month panel tiles that geometry. ``Anchor only'' is the anchor-only prediction of
Table~\ref{tab:recovery_matched}, the anchor's common response with no propagated part. Here its loading scales are fitted
to that common-response state, are more dispersed than the operator's, and give a heavy-tailed error (median 0.73). Monte Carlo standard errors of the errors are at most 0.022, except 0.053 and
0.039 for anchor only. The paired one-hop
minus operator difference with the 57-month fund-data map is 0.056 (0.022) for the total and 0.204 (0.018) for the
propagated part. The simulated propagation law is the operator's diffusion kernel.
\end{tablenotes}
\end{threeparttable}
\end{table}

\paragraph{Bridge results.}
With propagation following the operator's diffusion kernel, its advantage over one-hop is concentrated in the propagated part of the response
(Table~\ref{tab:bridge_fund}). Removing information one stage at a time, the normalized error is 0.171 with the true
loadings and market map, 0.299 with loading scales estimated from accounts, 0.486 or 0.487 with the market map from the
oracle instrument or from fund data, and 0.975 when the demand feedback is ignored. The largest part of the remaining error
therefore comes from the market map's sampling error. Indeed, replacing the constructed instrument by the oracle changes the bridge's error
by $-0.001$ (Monte Carlo s.e.\ 0.018), no detectable difference, although not a demonstration of equivalence. Calibration
panels that tile the sample's fund geometry over 114 and 228 months lower the error to 0.388 and 0.355. These are
scenario comparisons, not a forecast of what additional real months would identify. Account precision matters as well.
With account noise of 0.05 or 0.25 percent of holdings instead of 0.1 percent, the error is 0.451 or 0.754.

On its point estimate the fund-data error, 0.487, meets the 0.50 criterion of Table~\ref{tab:recovery_full}. Its Monte
Carlo 95 percent interval, 0.44 to 0.53, includes that threshold. Pointwise coverage of nominal 90 percent bands is 0.881, with a Monte Carlo 95 percent interval from 0.850 to 0.913, and
path coverage is 0.844. Coverage is near nominal at 57 months partly through compensation. It is 0.890 with the true
loadings and market map, 0.847 with estimated loadings and the true market map, and 0.842 with the 228-month market panel.
The account stage's bands are too narrow, and a longer market panel improves prediction without, by itself, calibrating
the bands.

With zero financial loading, the 5 percent joint test, a 95 percent simultaneous band over every held-out event and
horizon, rejects in 0 of 200 replications (exact 95 percent interval from 0.000 to 0.018). Without diffusion, where the
true propagated part is zero, its estimate has root-mean-square error 0.038~bp and its joint test rejects in 0 of 100
replications. These nulls
show only that the test does not over-reject. Where the response is not zero, the same test rejects in 8 of 200
replications with a 57-month market map (0 for the propagated part), in 122 of 200 with 228 months, and in all 200 with the
true market map. At 57 months the point estimate is informative, but the joint bands rarely exclude a zero response.

In a near-noiseless fixture, with no spending noise, accounts with standard deviation $10^{-6}$ and a 1,140-month
calibration panel with 2~bp price news (fund shocks remain), the error is 0.019, against 0.001 with the true maps and
0.011 with the true market map alone. Large valid granular shocks in the stress months do not detectably change accuracy
(error 0.481). When category flows coincide with credit news the procedure is biased (error 0.842). Its pointwise bands
still cover 0.854 of true values, and the joint Anderson--Rubin set of the market stage contains the true values at
least as often as in the valid world, so neither reveals the violation.

Numerical failures are reported separately from point estimation, and the baseline bridge panels have none. In the
no-diffusion family, one large-shock sensitivity panel has 183 of 199 failed market-bootstrap draws and an
inadmissible point estimate, $\hat m_F=-16.6$. Its inference is treated as numerically failed rather than imputed,
and this sensitivity enters no table. In the market-only scenarios, the panel of note $a$ in
Table~\ref{tab:market_sim} is the only one with failed draws, and three point estimates (one with factors correlated with
news, two with category news) have $\hat m_F\leq0$. All point estimates are retained in the error statistics.

The historical window cannot serve as its own calibration sample. Purging each fold's held-out storm windows (52-week
support) leaves 24, 24 and 38 of the 57 months when the outcome and both lag months must be clean (26, 26 and 38 if only
the outcome month must be), and the first fold loses both stress months. The design also builds in favourable features.
The simulated propagation law is the operator's diffusion kernel, and the simulated fund panel shares the estimator's assumptions
(two factors, homogeneous $\zeta^F$, independent idiosyncratic shocks). Resampling whole real residual cross-sections
instead lowers $c_E^{\mathrm{sell}}$ by about 16 percent, and by about 14 percent with the oracle instrument, so independence
across funds matters for its level.

\paragraph{The actual 57-month sample.}
Applied to the actual fund panel (precision weights, two factors and fund fixed effects in the demand average, including
reinvested distributions), the construction gives $\hat\zeta^F=-2.26\times10^{-4}$ per bp (Wald $t=-4.9$),
$\hat\phi=-299$, $\hat m_F=1.99$ and $\hat c_E^{\mathrm{sell}}=594$. Taken at face value, flows chase returns and the system
amplifies desired selling. Inference robust to weak identification does not support this reading. The Anderson--Rubin
set for the demand feedback is the whole real line, and the projection of the joint set onto $c_E^{\mathrm{sell}}$ is unbounded.
With the $\chi^2$ critical value these sets are, if anything, too small at this sample length, as the undercoverage in
the simulated 57-month panels shows, and a size-corrected critical value can only enlarge them. March~2020 carries 69
percent of the demand moment and 65 percent of the price moment. Refitting without March~2020 gives
$\hat c_E^{\mathrm{sell}}=275$, and without both stress months 95. Instruments from short-duration funds alone give 1{,}530, the
direction of the upward bias that the category-news simulation produces, although the simulation does not reproduce the
category split itself. The external-cash definition gives the same conclusions. The network estimating equation of
Chodorow-Reich, Gabaix and Viviano (2026) is no more informative here. Its iteration reaches no root, minimum distance
finds several far-apart roots, one at essentially zero price impact, and its joint set accepts every admissible point.
The sample supports the pricing association of Table~\ref{tab:creditflow}, not the demand--price decomposition.

\end{document}